\documentclass[twocolumn]{aastex63}
\usepackage{amssymb}
\usepackage{graphicx,graphics}			
\usepackage{epsfig}						
\usepackage{natbib}
\usepackage{times}
\usepackage{wrapfig}
\usepackage{comment}

\usepackage{pslatex}
\usepackage{graphicx}

\begin{document}

\title{Cosmological Simulations of Galaxy Groups and Clusters-III: Constraining Quasar Feedback Models with the Atacama Large Millimeter Array }

\author{Avinanda Chakraborty}
\affiliation{School of Astrophysics, Presidency University, Kolkata, 700073, India}
\affiliation{Department of Physics, Presidency University, Kolkata, 700073, India}
\author{Suchetana Chatterjee}
\affiliation{School of Astrophysics, Presidency University, Kolkata, 700073, India}
\affiliation{Department of Physics, Presidency University, Kolkata, 700073, India}
\author{Mark Lacy}
\affiliation{National Radio Astronomy Observatory, 520 Edgemont Road, Charlottesville, VA 22903, USA}
\author{Soumya Roy}
\affiliation{Inter University Centre for Astronomy and Astrophysics, Pune, 411007, India}
\affiliation{Center for Astrophysics, Cambridge, MA, USA}
\author{Samrat Roy}
\affiliation{Department of Physics, Presidency University, Kolkata, 700073, India}
\author{Rudrani Kar Chowdhury}
\affiliation{Department of Physics, The University of Hong Kong, Pok Fu Lam Road, Hong Kong}


\begin{abstract}
The thermal Sunyaev-Zeldovich (SZ) effect serves as a direct potential probe of the energetic outflows from quasars that are responsible for heating the intergalactic medium. In this work, we use the GIZMO meshless finite mass hydrodynamic cosmological simulation SIMBA (\citealt{dave19}), which includes different prescriptions for quasar feedback, to compute the SZ effect arising from different feedback modes. From these theoretical simulations, we perform mock observations of the Atacama Large Millimeter Array (ALMA) in four bands (320 GHz, 135 GHZ, 100 GHz and 42 GHz) to characterize the feasibility of direct detection of the quasar SZ signal. Our results show that for all the systems we get an enhancement of the SZ signal, when there is radiative feedback, while the signal gets suppressed when the jet mode of feedback is introduced in the simulations. Our mock ALMA maps reveal that, with the current prescription of jet feedback, the signal goes below the detection threshold of ALMA. We also find that the signal is higher for high redshift systems, making it possible for ALMA and cross SZ-X-ray studies to disentangle the varying modes of quasar feedback and their relative importance in the cosmological context.
\end{abstract}

\section{Introduction}\label{section_1}
Through a series of observations in the last two decades, it has been established that supermassive black holes (SMBH) residing at the centers of galaxies play a significant role in cosmic evolution of structures in the Universe \citep[e.g.,][]{D&R88, Kormendy93, k&h00, F&M00, Gebhardt00, Graham01, h&r04, Cattaneo09, K&H13, Salviander15, Fiore17, MP18, Schutte19, deNicola19, Marsden20, magorrianetal98, richstoneetal98, ferrareseetal05, s&r98, gebhardtetal00, m&f01, tremaineetal02, haringtal04, dimatteoetal05, alleretal07, gittietal12} and hence modeling the effect of SMBH on galaxy evolution has emerged as a frontier in studies involving structure formation \citep[e.g.,][]{dimatteoetal05, dimatteoetal08, sijackietal07, sijackietal15, vogelsbergeretal14, khandaietal15, liuetal16, dave19, RKC19}. 

Effects of SMBH feedback (or active galactic nuclei; AGN feedback) have been directly observed in galaxies and clusters using multi-wavelength datasets \citep{SRC07, K&H13, Fiore17, SRC17, Harrison17, Schutte19, Roy21a, Roy21c}. The effect of SMBH feedback on several observables has been explored in the literature, including the $L_{x}-T$ relation in galaxy clusters and groups \citep[e.g.][]{puchweinetal10, Andersson09, Maughan12, Molham20}, absence of cooling flow in galaxy clusters \citep[e.g.,][]{David01, Peterson03}, Sunyaev-Zeldovich (SZ; \citealt{s&z72}) profiles \citep[e.g.,][] {chatterjeeetal08}, SZ power spectrum \citep[e.g.,][]{c&k07, scannapiecoetal08}, and star-formation properties of galaxies \citep[e.g.,][] {hopkinsetal06, Vitale13, Costa15, Harrison17}.

\begin{figure*}
    \begin{center}
        \begin{tabular}{cccc}
        \hline
         {No}&{No-Jet}&{No X-ray}&{All}\\[0.1pt]
         \hline
        \includegraphics[width=4.5cm]{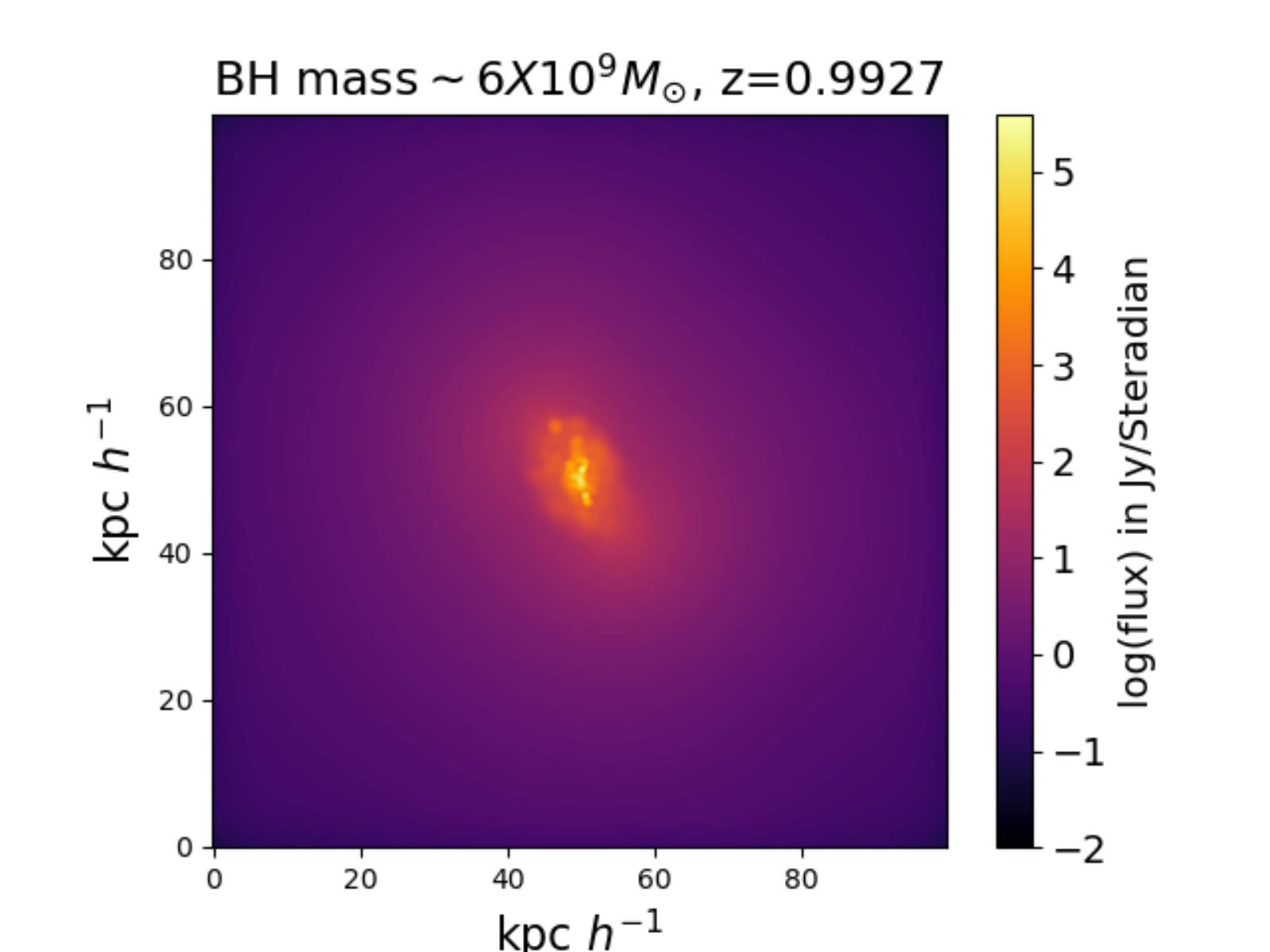}&\includegraphics[width=4.5cm]{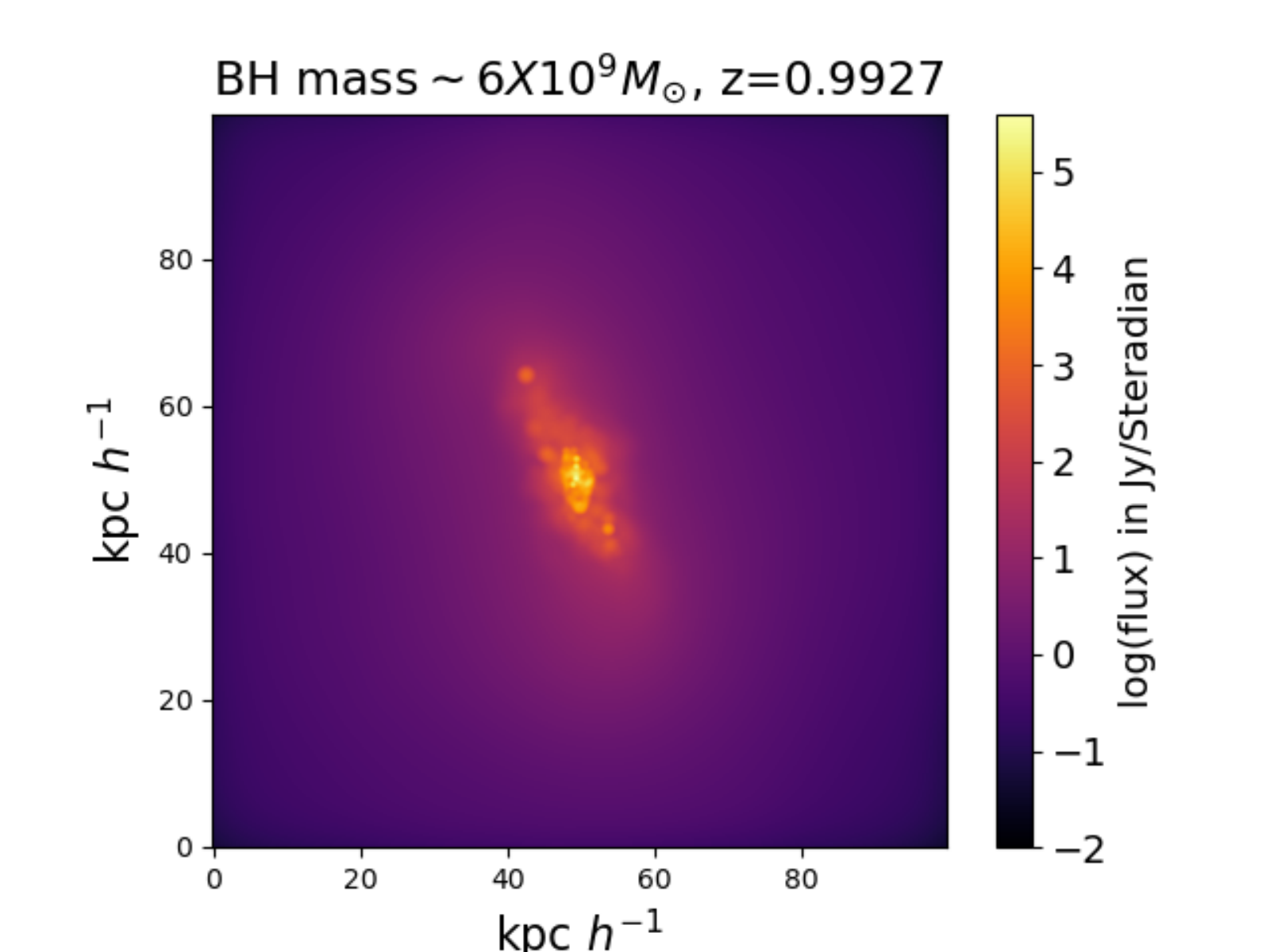}&\includegraphics[width=4.5cm]{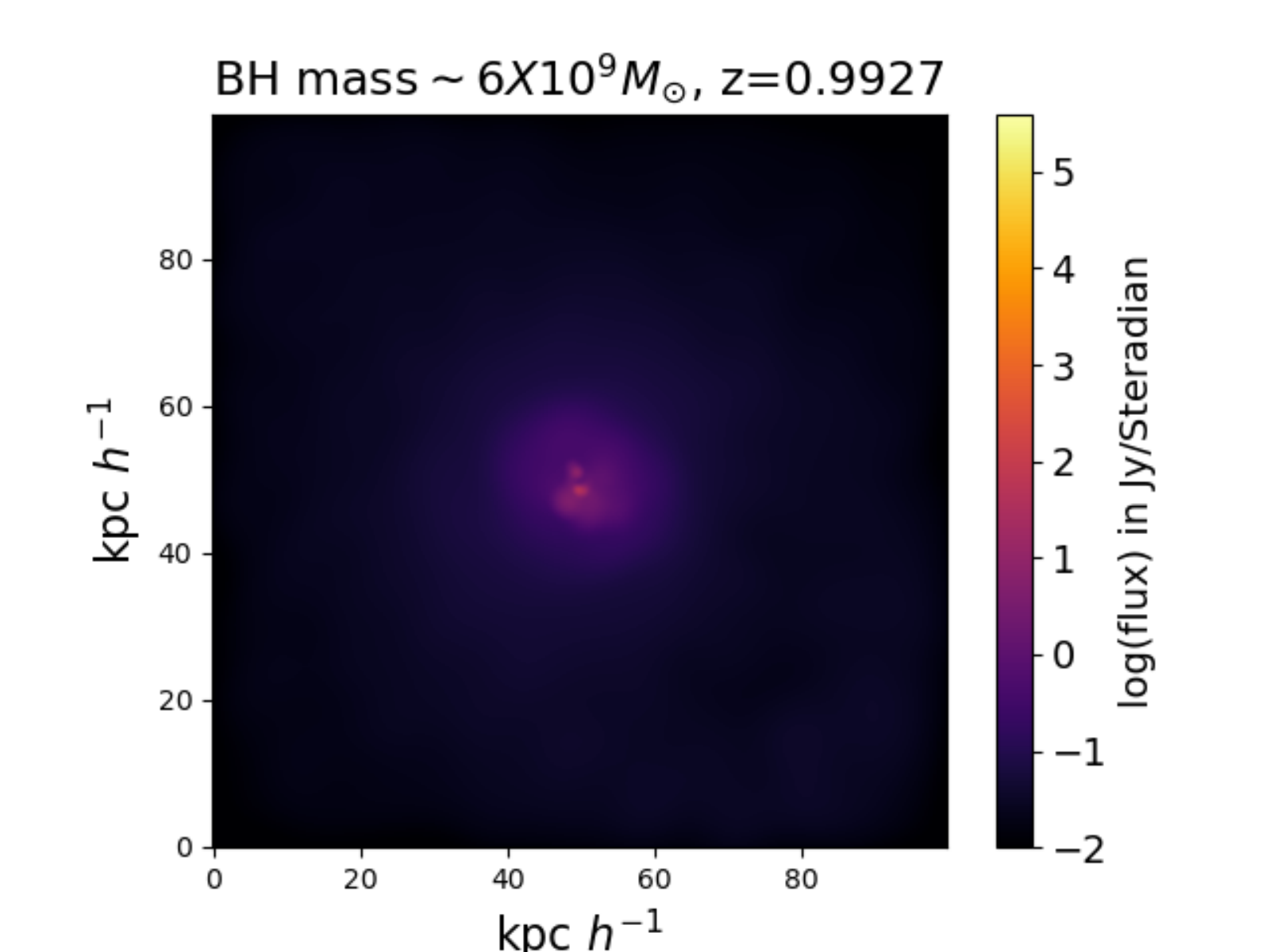}&\includegraphics[width=4.5cm]{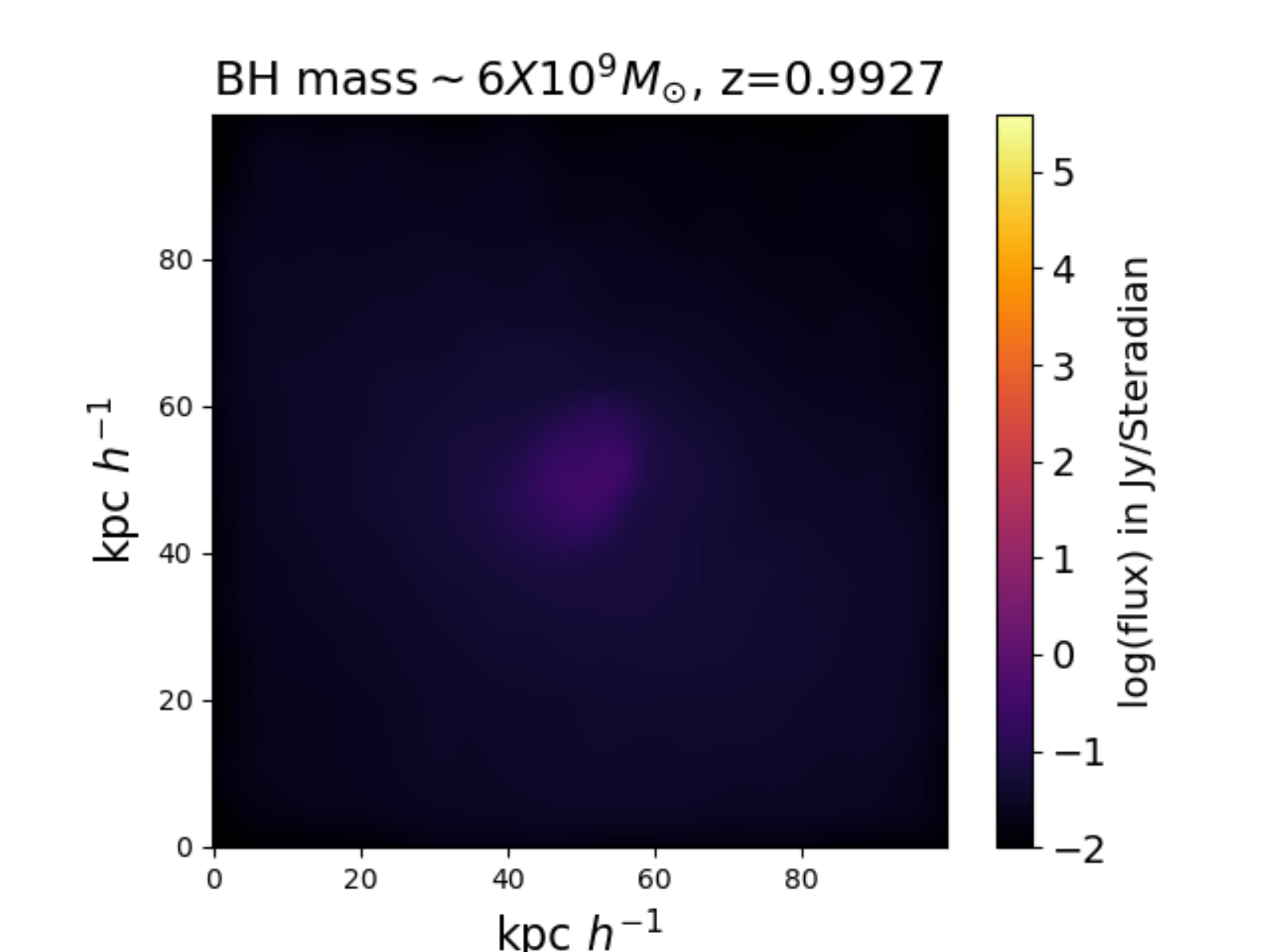}\\
       \includegraphics[width=4.5cm]{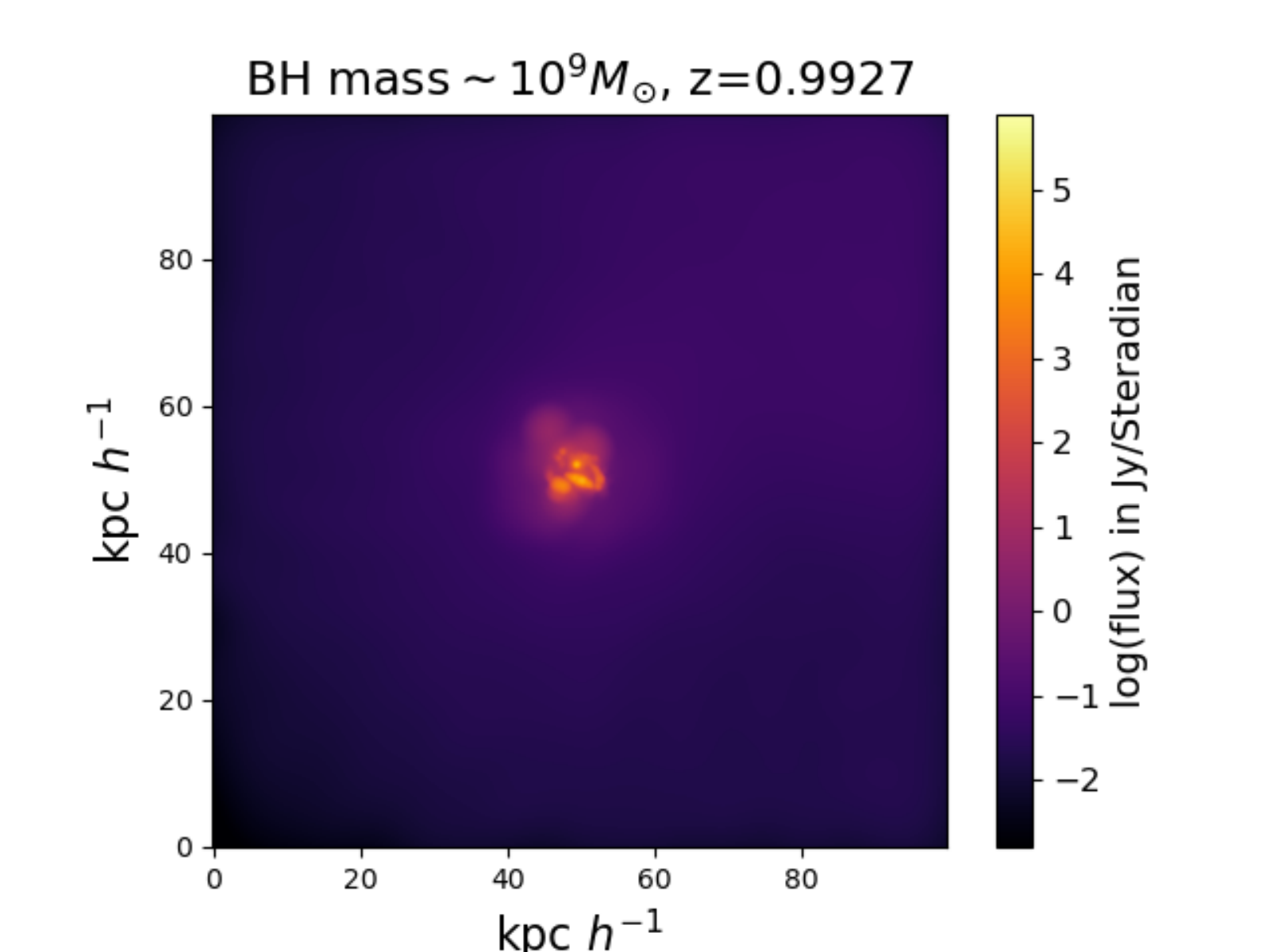}&\includegraphics[width=4.5cm]{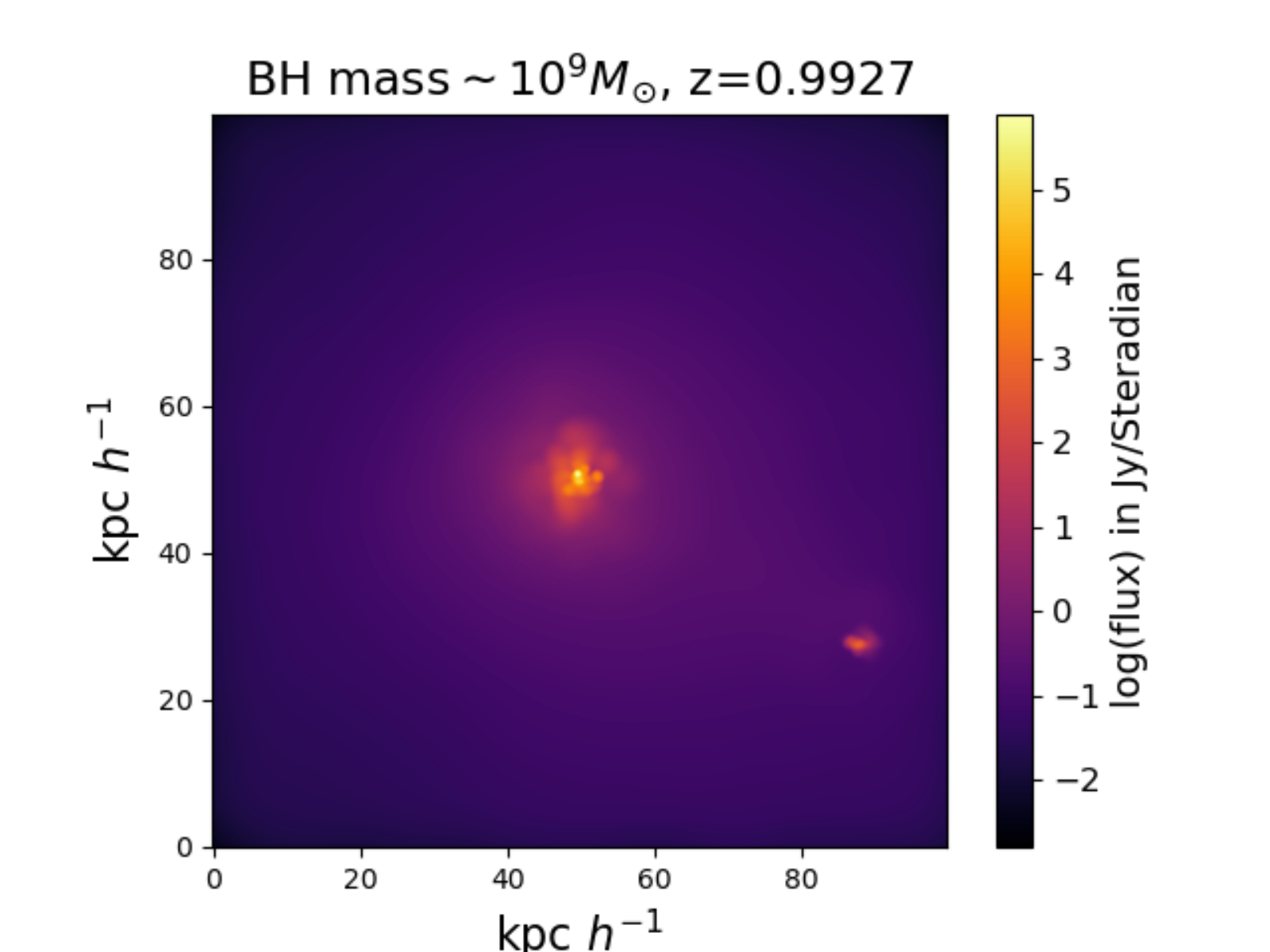}&\includegraphics[width=4.5cm]{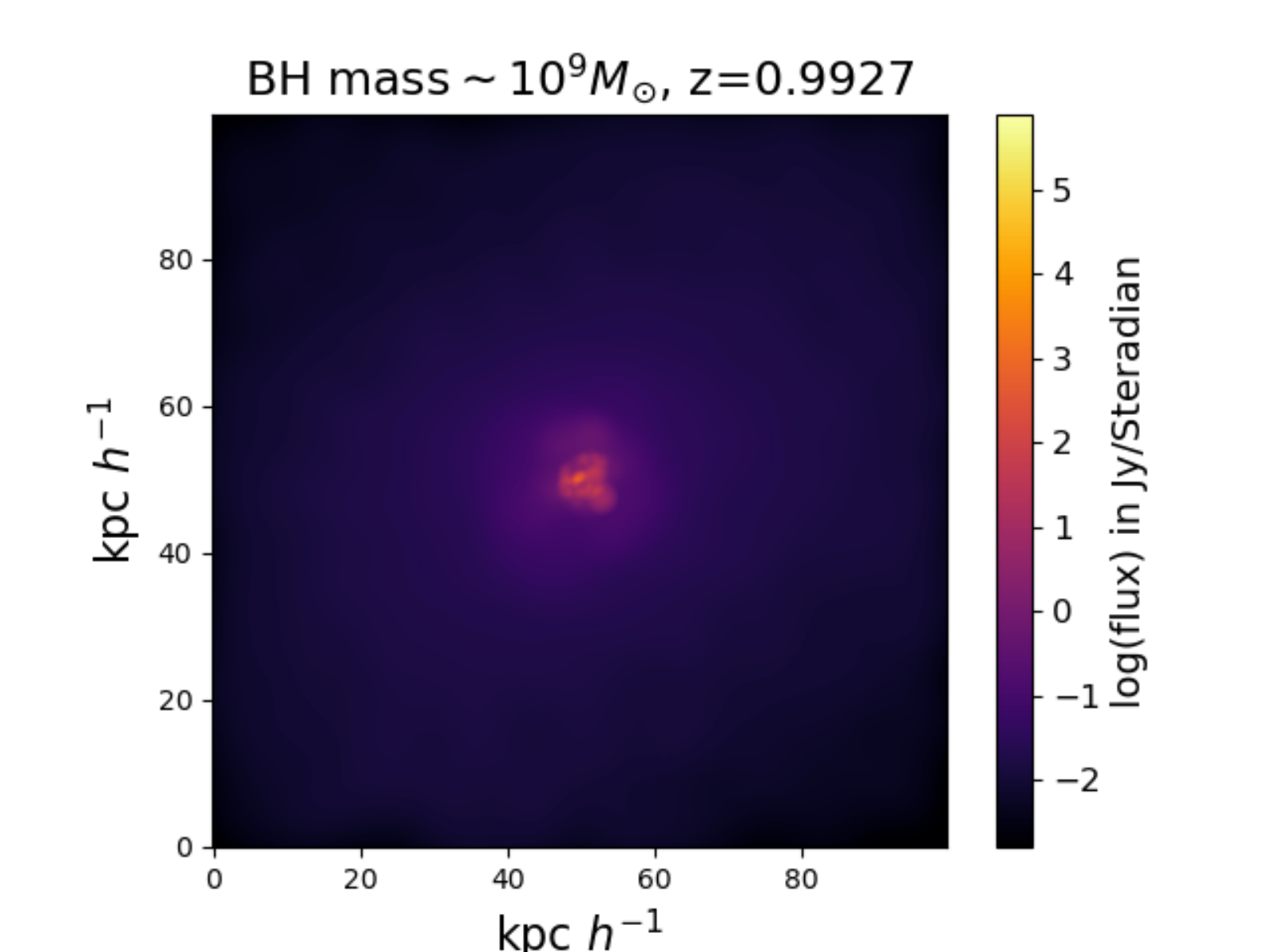}&\includegraphics[width=4.5cm]{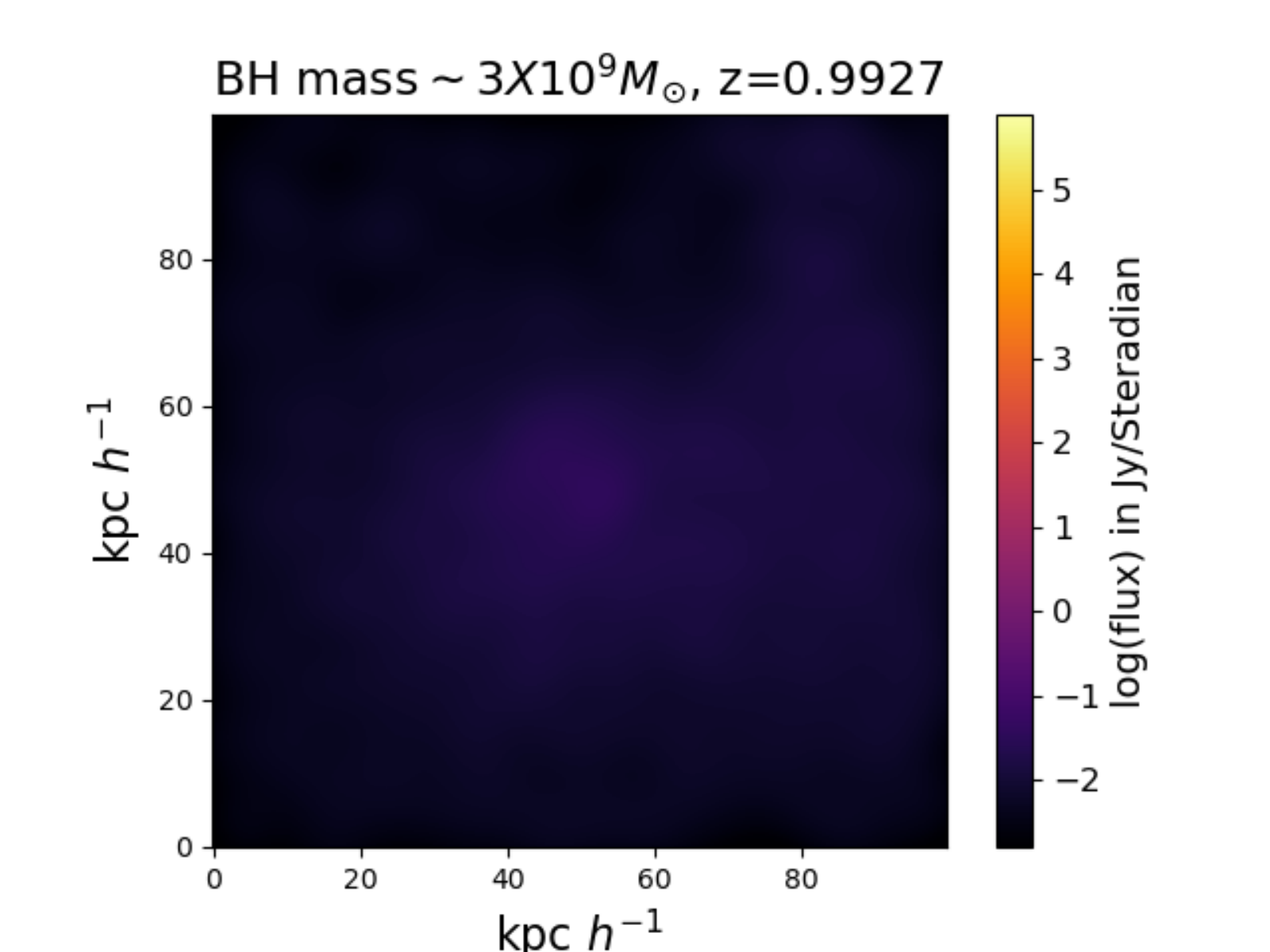}\\
      \end{tabular}
        \caption{The theoretical simulated tSZ maps for different feedback modes (left most: no feedback, second left: no jet feedback, second right: no X-ray feedback, rightmost: all feedback)  around similar BHs at z$\sim$1 from \cite{dave19}. {\bf Top Panel} Simulated tSZ map at 320 GHz of the most massive black hole for no feedback, no jet feedback, no X-ray feedback and all feedback modes respectively. {\bf Bottom Panel} Same as the top panel but now for most active black hole. From the figure we see that addition of radiative feedback and jet feedback have opposing effects (enhancement versus decrement) in the SZ signal. Table 1 and 2 summarizes the feedback model nomenclature and the black hole properties respectively.}
        \label{fig:1}
    \end{center}
\end{figure*}

In the literature, AGN feedback has been broadly divided into two main modes. The scenario where a very strong quasar outburst launches hot winds in short timescales, is generally classified as the ``quasar" or the ``radiative" mode while comparatively lower power outflows arising from jets or relativistic plasma which operate on much longer timescales, are classified as ``radio" or ``kinetic" mode. But the specific or distinct roles that these modes play in the evolution of the host galaxy, are still unclear. As proposed before, an effective way to detect hot outflows can be the SZ effect \citep{n&s99, aghanimetal99, yamadaetal99, lapietal03, plataniaetal02, roychowdhuryetal05, c&k07, scannapiecoetal08, chatterjeeetal08, zannietal05, sijackietal07}. The thermal Sunyaev-Zeldovich (tSZ) effect is the spectral distortion of the cosmic microwave background (CMB) radiation arising from the inverse Compton scattering of the CMB photons by the high-energy electrons present along the line of sight \citep{s&z72}. It serves as a probe for accumulations of hot gas in the Universe (see \citealt{d&c17} and references therein)

Previous studies predicted that the integrated SZ effect gets enhanced due to the presence of hot gas in the vicinity of a quasar owing to its feedback mechanism \citep{n&s99, c&k07, scannapiecoetal08, chatterjeeetal08}. The effect was predicted to be directly detectable using mm wave interferometric experiments like the Atacama Large Millimeter Array (ALMA) and statistical techniques using CMB temperature maps \citep{c&k07}. \citet{chatterjeeetal10} reported a $1.5\sigma$ lower limit of the signal using the quasar catalog of SDSS and WMAP CMB maps. In the same study, it was predicted that, high-resolution CMB experiments will provide better constraints on this effect .

Following the first study, several other teams have tried to detect this signal using the Planck Surveyor Satellite \citep{ruanetal15, verdieretal16}, the Atacama Cosmology Telescope (ACT; \citealt{crichtonetal16}) and the South Pole Telescope \citep[SPT;][]{spaceketal16}. Moreover, in addition to these cross-correlation studies we now have the very first detection of SZ effect from quasar feedback using the ALMA compact configuration \citep{lacyetal19}. There are a number of upcoming proposals to employ this new technique in understanding the connection between SMBH and the gas distribution in their host galaxies \citep[e.g.,][]{mroczkowskietal19}. In this work, we employ high resolution cosmological simulations from \citet{dave19} to test for the feasibility of direct detection of AGN feedback using SZ observations. 


Using the same set of simulations, in a companion paper \citet{RKC22} report that the jet/kinetic mode of feedback plays the most significant role in suppressing the X-ray signal in groups and clusters, but the relative contribution from the radiative and the kinetic modes can not be fully determined through X-ray observations only. In this work, we construct the SZ distortion maps and provide a robust machinery to evaluate the relative contributions of radiative and kinetic modes through current and future observations. Our technique also provides a feasible tool to study signatures of AGN feedback in high redshift systems. The paper is organized as follows. In \S 2 we briefly discuss the simulation and the methodology for constructing the tSZ maps and present our results in \S 3. In \S 4 we discuss the implications of our results in light of current and future observations. 

\begin{figure*}
    \begin{center}
        \begin{tabular}{cccc}
        \hline
         {No}&{No-Jet}&{No X-ray}&{All}\\[0.1pt]    
         \hline
      \includegraphics[width=4.5cm]{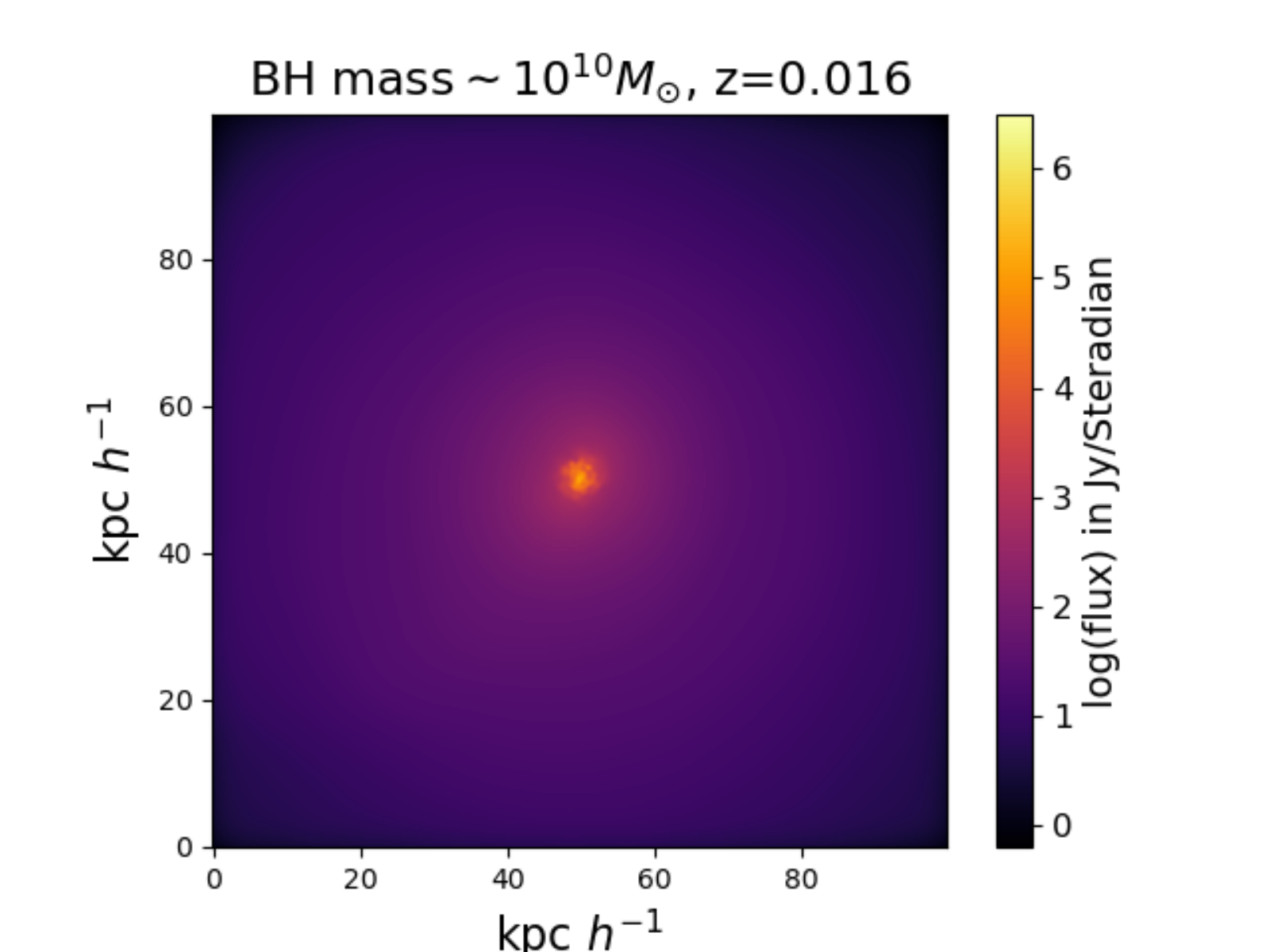}&\includegraphics[width=4.5cm]{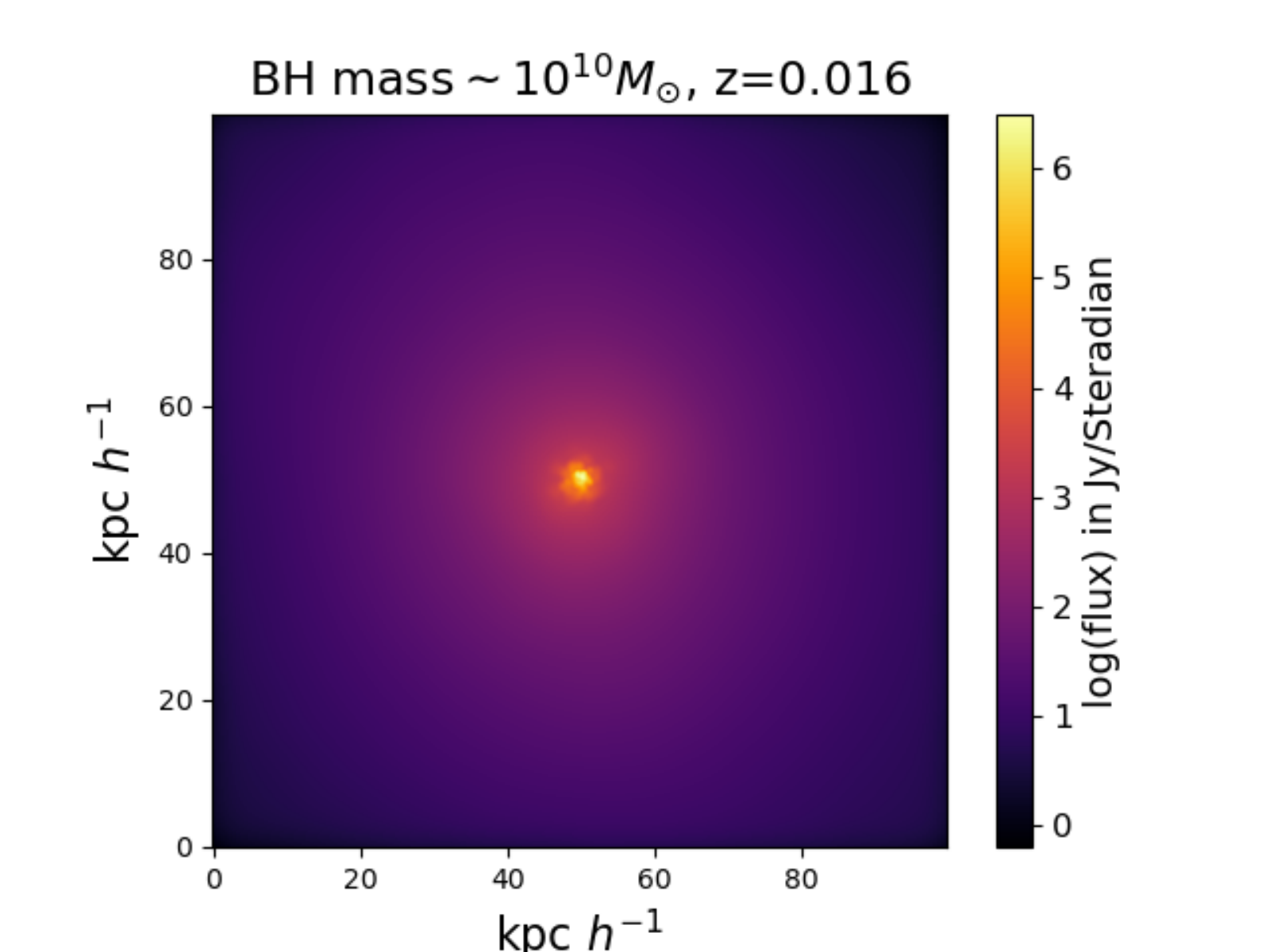}&\includegraphics[width=4.5cm]{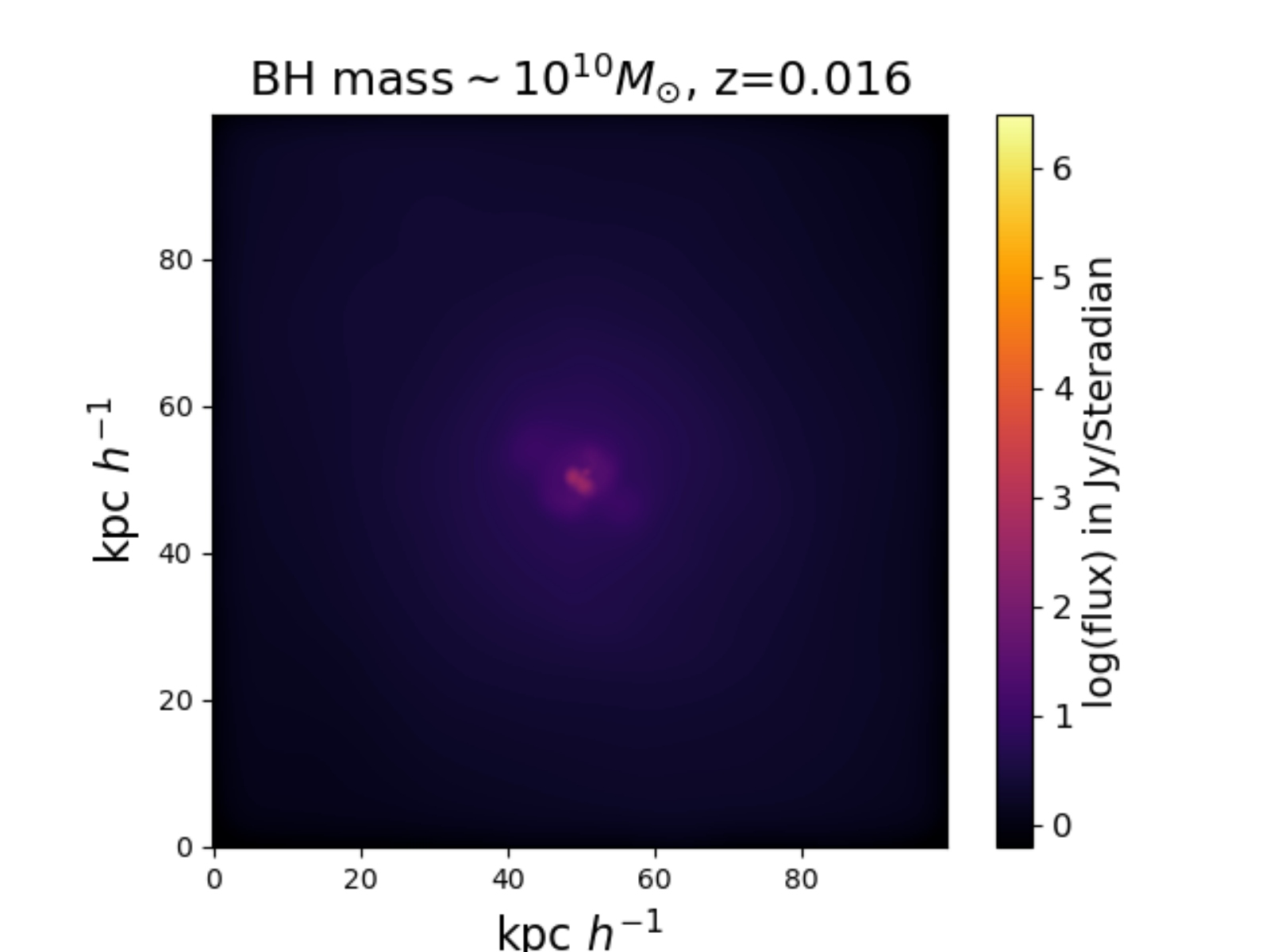}&\includegraphics[width=4.5cm]{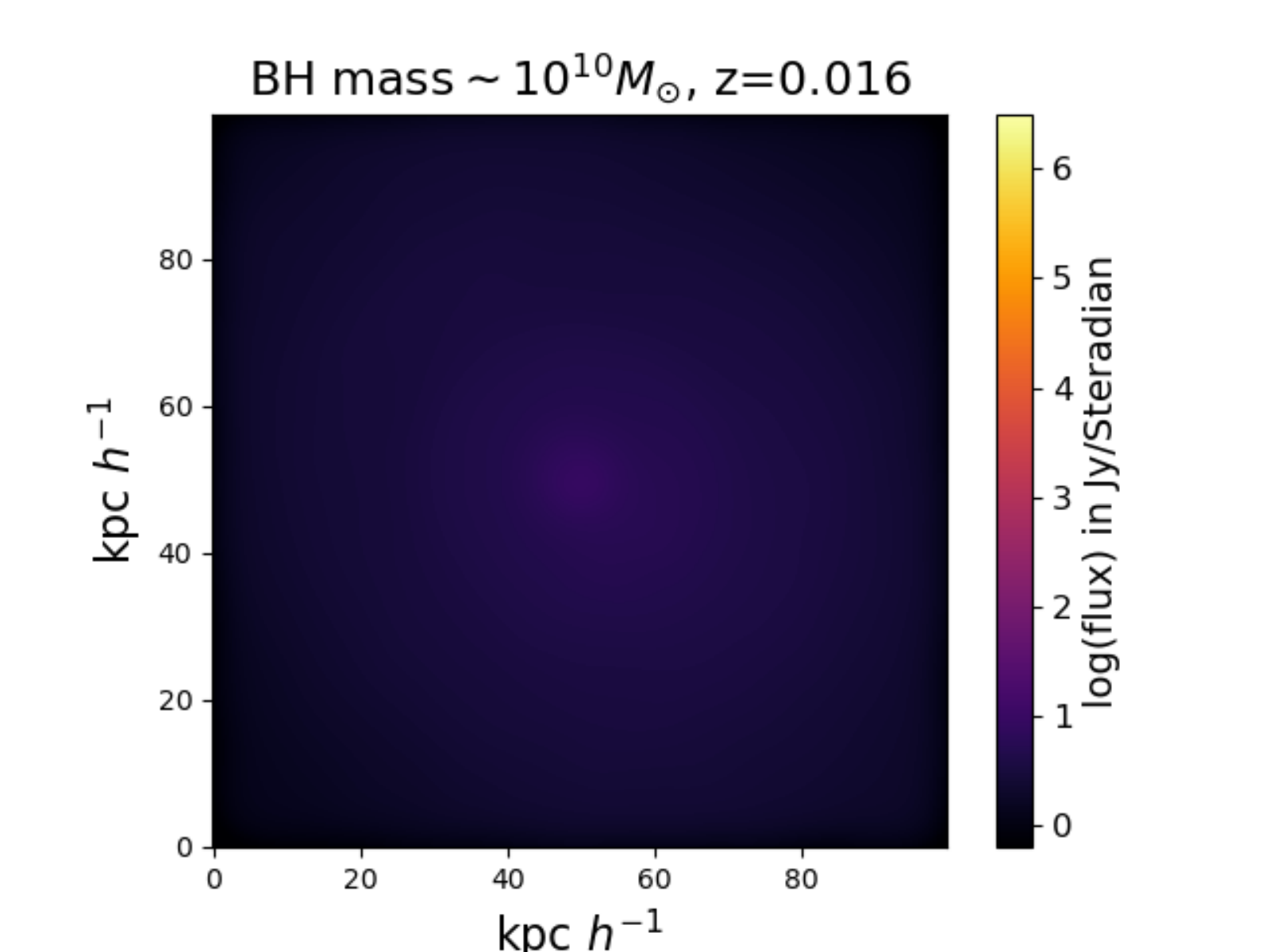}\\     
      \includegraphics[width=4.5cm]{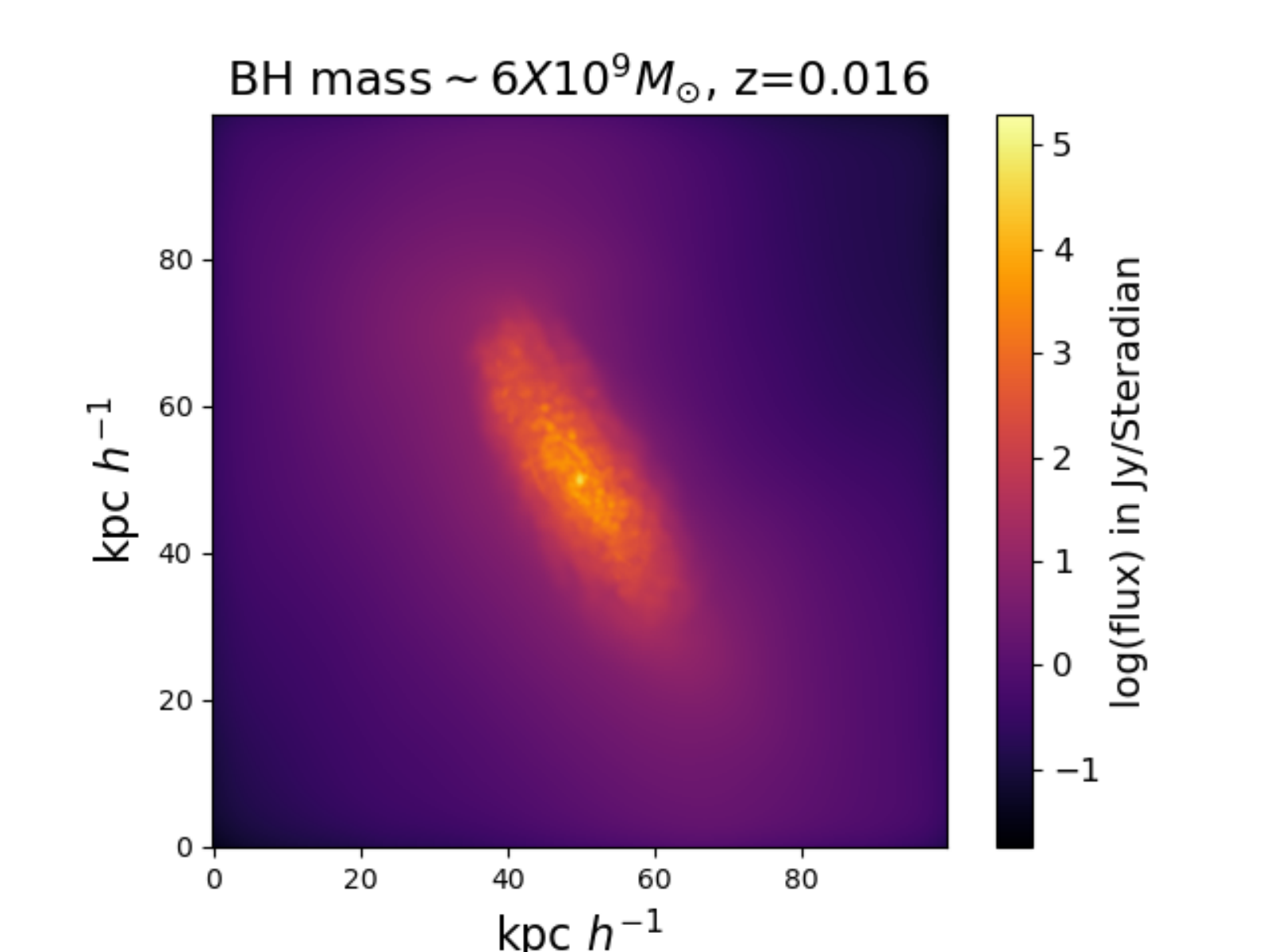}&\includegraphics[width=4.5cm]{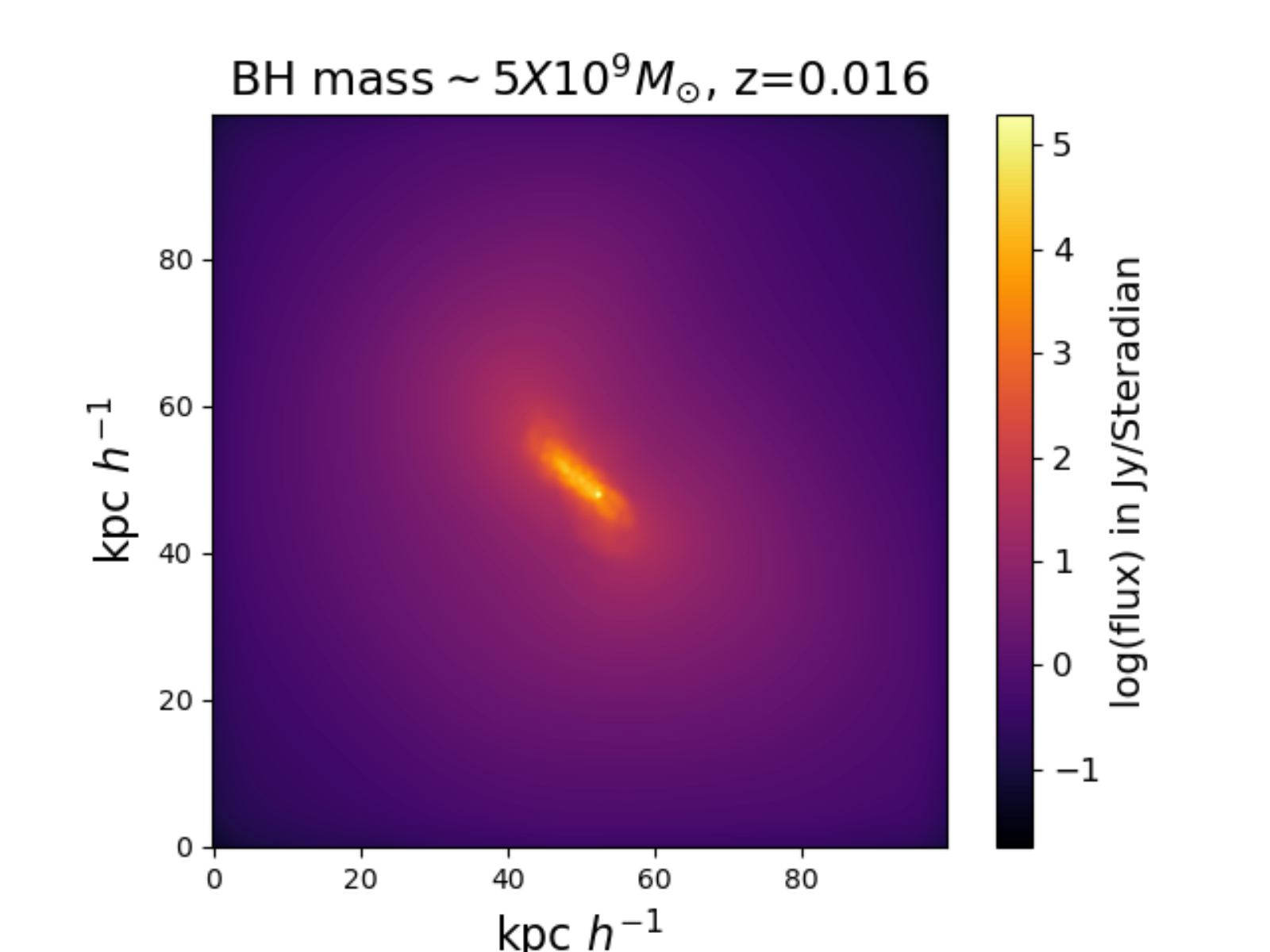}&\includegraphics[width=4.5cm]{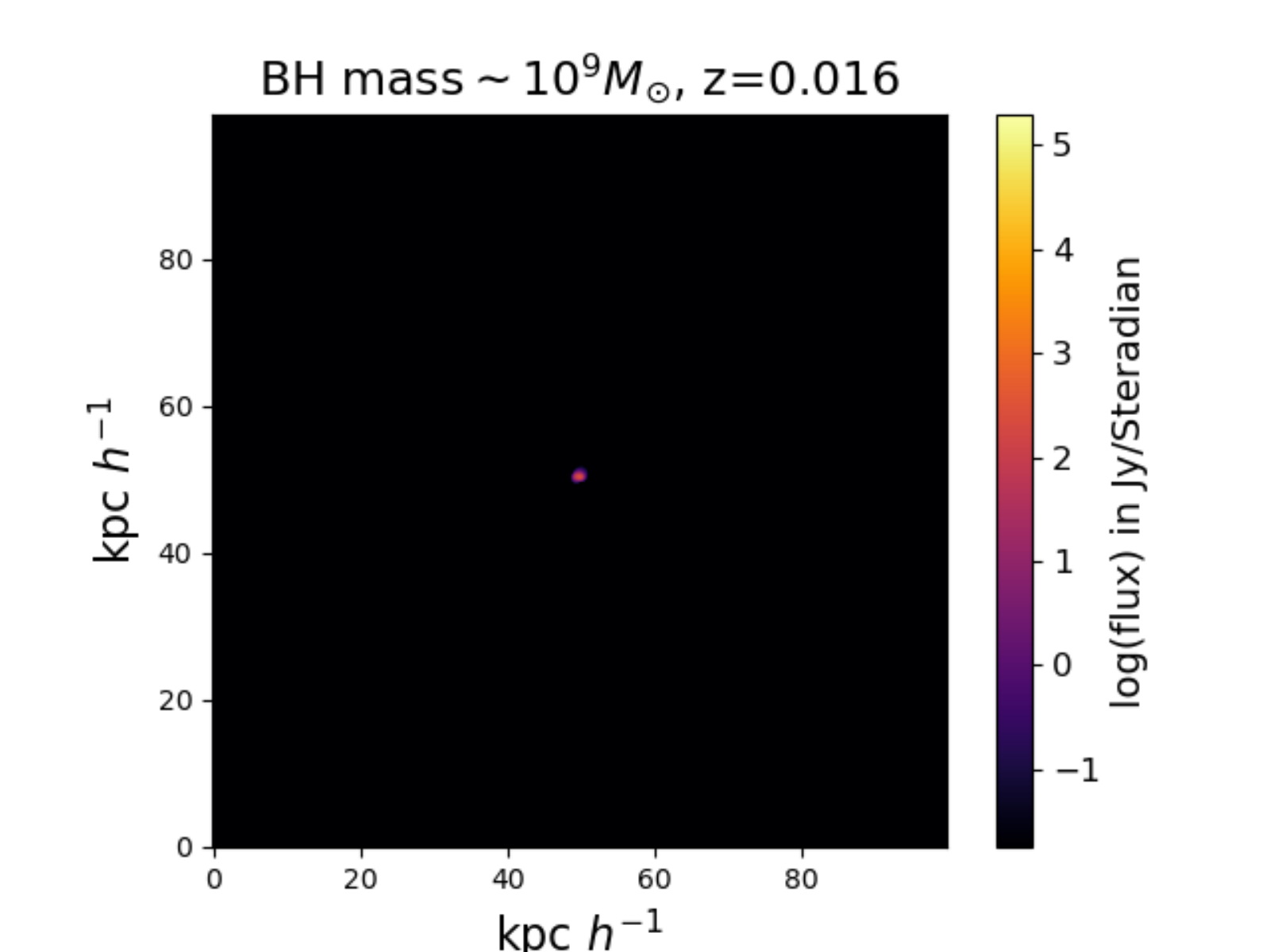}&\includegraphics[width=4.5cm]{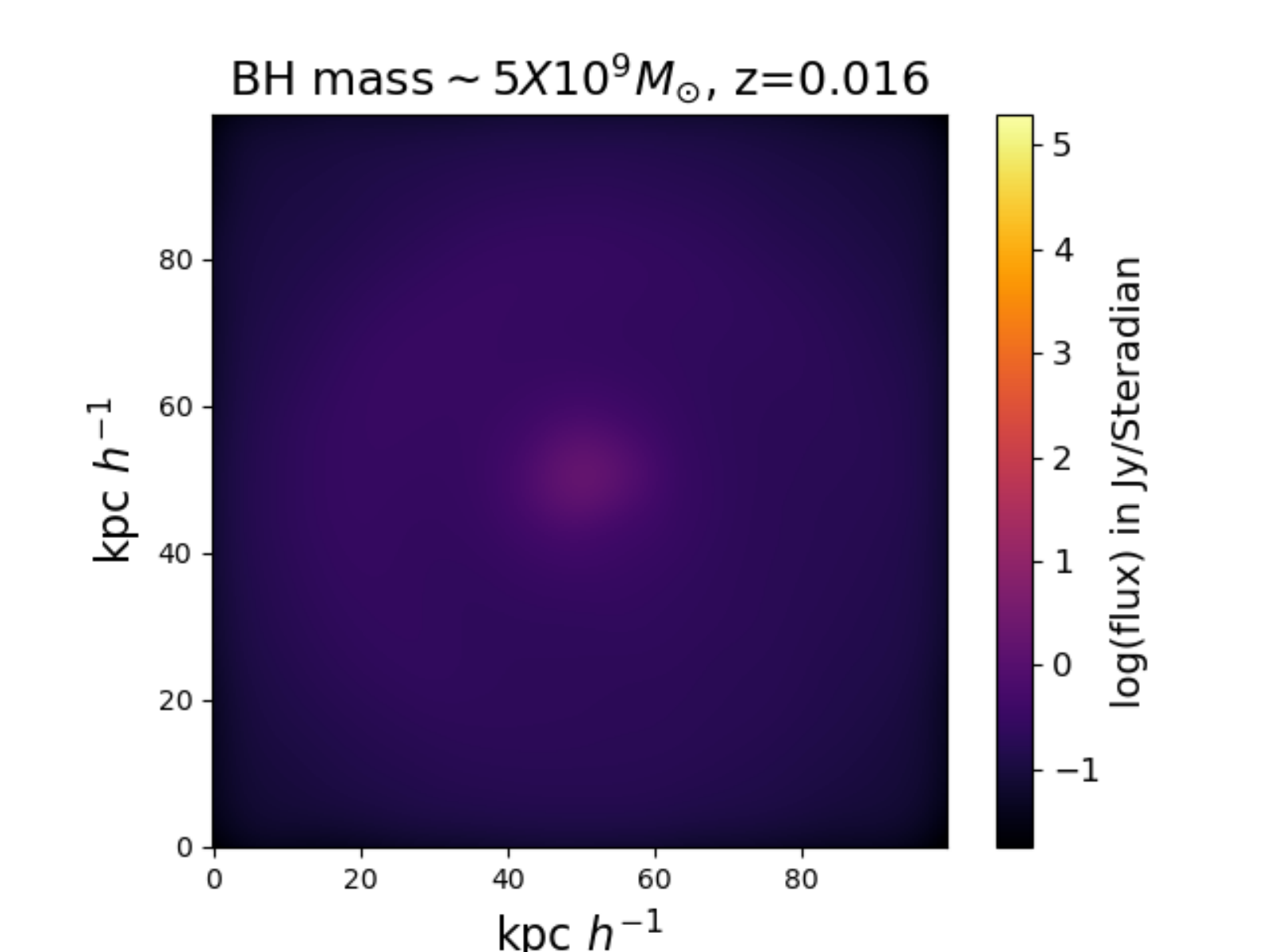}\\
      \end{tabular}
        \caption{Same as in Fig.\ 1 but at z=0.016. {\bf Top Panel} Simulated tSZ map at 320 GHz of the most massive black hole for no, no-jet no X-ray and all feedback modes respectively. {\bf Bottom Panel} Same as the top panel but now for most active black hole. Table 1 and 2 summarizes the feedback model nomenclature and the black hole properties respectively. The results are same as we find in Fig.\ 1 for the high redshift blackholes. Inclusion of the radiative mode of feedback enhances the SZ signal while adding the jet mode decreases it. This is in stark contrast to the X-ray signal where both modes suppresses the X-ray flux (see \citealt{RKC22})}
        \label{fig:2}
    \end{center}
\end{figure*}

\section{Simulation}\label{section_2}

For this work, we have used the SIMBA simulation \citep{dave19}. SIMBA is one of the most updated cosmological simulations which is the next generation of the MUFASA cosmological galaxy formation simulations that runs with GIZMO’s meshless finite mass hydrodynamics. Cosmological parameters of the simulation are adopted from \citep{planck16}. The simulation box is 50$h^{-1}$ Mpc

\begin{table}
\caption{Different Feedback Modes} 
\centering 
\begin{tabular}{c c } 
\hline \hline
Feedback mode & Configuration\\ [0.5ex] 
\hline 
All feedback & Radiative+jet+X-ray\\
No-jet feedback & Radiative\\
No X-ray feedback & Radiative+jet\\ [0.5ex] 
\hline 
\end{tabular}
\end{table}

\begin{table}
\caption{Black hole Parameters for Fig.\ 1 and Fig.\ 2} 
\centering 
\begin{tabular}{c |c c c c c} 
\hline \hline
& Feedback & z & BH Mass & Accretion & Halo Mass\\ 
&  &  & ($M_{\odot}$) & ($M_{\odot}$/yr) & ($log_{10}(h^{-1}M_{\odot})$)\\ 
[0.5ex] 
\hline 
Massive & No & 0.99 & $\sim 6\times10^9$ & 0.07 & 13.6\\
& & 0.016 & $\sim 10^{10}$ & 0.032 & 13.9\\
& No-jet & 0.99 & $\sim 6\times10^9$ & 0.01 & \\
& & 0.016 & $\sim 10^{10}$ & 0.029 & \\
& No X-ray & 0.99 & $\sim 6\times10^9$ & 0.0003 & \\ 
& & 0.016 & $\sim 10^{10}$ & 0.00036 & \\
& All & 0.99 & $\sim 6\times10^9$ & 0.002 & \\
& & 0.016 & $\sim 10^{10}$ & 0.000024 & \\[0.5ex] 
\hline 
Active & No & 0.99 & $\sim 10^9$ & 0.13 & 13.4\\
 & & 0.016 & $\sim 6\times10^{9}$ & 0.083 & 13.6\\
& No-jet & 0.99 & $\sim 10^9$ & 0.11 & \\
& & 0.016 & $\sim 5\times10^{9}$ & 0.12 & \\
& No X-ray & 0.99 & $\sim 10^9$ & 0.07 & \\ 
 & & 0.016 & $\sim 10^9$ & 0.004 & \\
 & All & 0.99 & $\sim 3\times10^9$ & 0.05 & \\
 & & 0.016 & $\sim 5\times10^{9}$ & 0.0065 & \\[0.5ex] 
\hline\hline
\end{tabular}
\end{table} 

\subsection{Modeling AGN Feedback Modes}\label{section_2.1}
The black holes (BHs) are seeded and considered to be collisionless sink particles, which can grow by accreting the surrounding gas or by merging with other BHs. The accretion rate of the BHs is modelled via two processes, one is torque-limited accretion model \citep{angle17} and the other is the Bondi accretion model \citep{Bondi52, hoyleetal39}. Hence the total accretion rate for a given black hole is given as,
$$\dot{M_{BH}} = (1-\eta)\times(\dot{M_{Torque}}+\dot{M_{Bondi}})$$ where $\eta$ is the radiative efficiency of accretion (see Appendix B for more discussions on the accretion models). The accretion rate in the torque-limited model can be mildly super Eddington, but the value is never allowed to exceed 3 times the Eddington rate so that it remains consistent with other works \citep{mart18, jiang14} but for the Bondi accretion model, black holes are not allowed to exceed the Eddington limit. See \citep[D19 hereafter]{dave19} and \cite{RKC22} for more details. 

A kinetic subgrid model has been incorporated for black hole feedback along with X-ray energy feedback (see Appendix B for more details). The motivation for the kinetic feedback model comes from the observed dichotomy which includes a “radiative mode” at high Eddington ratios and a “jet mode” at low Eddington ratios in black hole growth modes that is reflected in their outflow characteristics \citep[e.g.,][]{heckman14}. The kinetic mass outflow rate is taken directly from the Feedback in Realistic Environments \citep[FIRE;][]{hop14, hop18, mur15} high-resolution simulations which provides a synergy between cosmological-scale simulations of galaxy populations and the interstellar medium (ISM) resolving simulations of individual galaxies. Apart from kinetic feedback, energy input into surrounding gas for the photoionization heating from the X-ray photons coming from the accretion disk of the AGN is also included \citep{choi12}.

AGN feedback can be modelled and tested by various observations, but the origin of the seed SMBH remains unknown. Also, we are limited by the resolution of the simulation for probing the relevant length scales. Hence, to seed black holes in galaxies dynamically during the simulation a on-the-fly Friends-of-Friends (FoF) algorithm is used \citep[e.g.,][]{dimatteoetal08, angle17}. The star particle closest to the center of mass of a galaxy is converted into a black hole particle if the galaxy does not already contain a black hole and reaches a stellar mass $M_{*} > \gamma_{BH} \times M_{seed}$. Here $M_{seed} = 10^4 M_{\odot}/h$ and $\gamma_{BH} = 3 \times 10^5$, and the threshold stellar mass is $M_{*} \geq 10^{9.5} M_{\odot}$ for the fiducial simulations. The nomenclature for different feedback models is summarized in Table 1. 

\begin{figure}
    \begin{center}
        \begin{tabular}{c}
      \includegraphics[width=4cm]{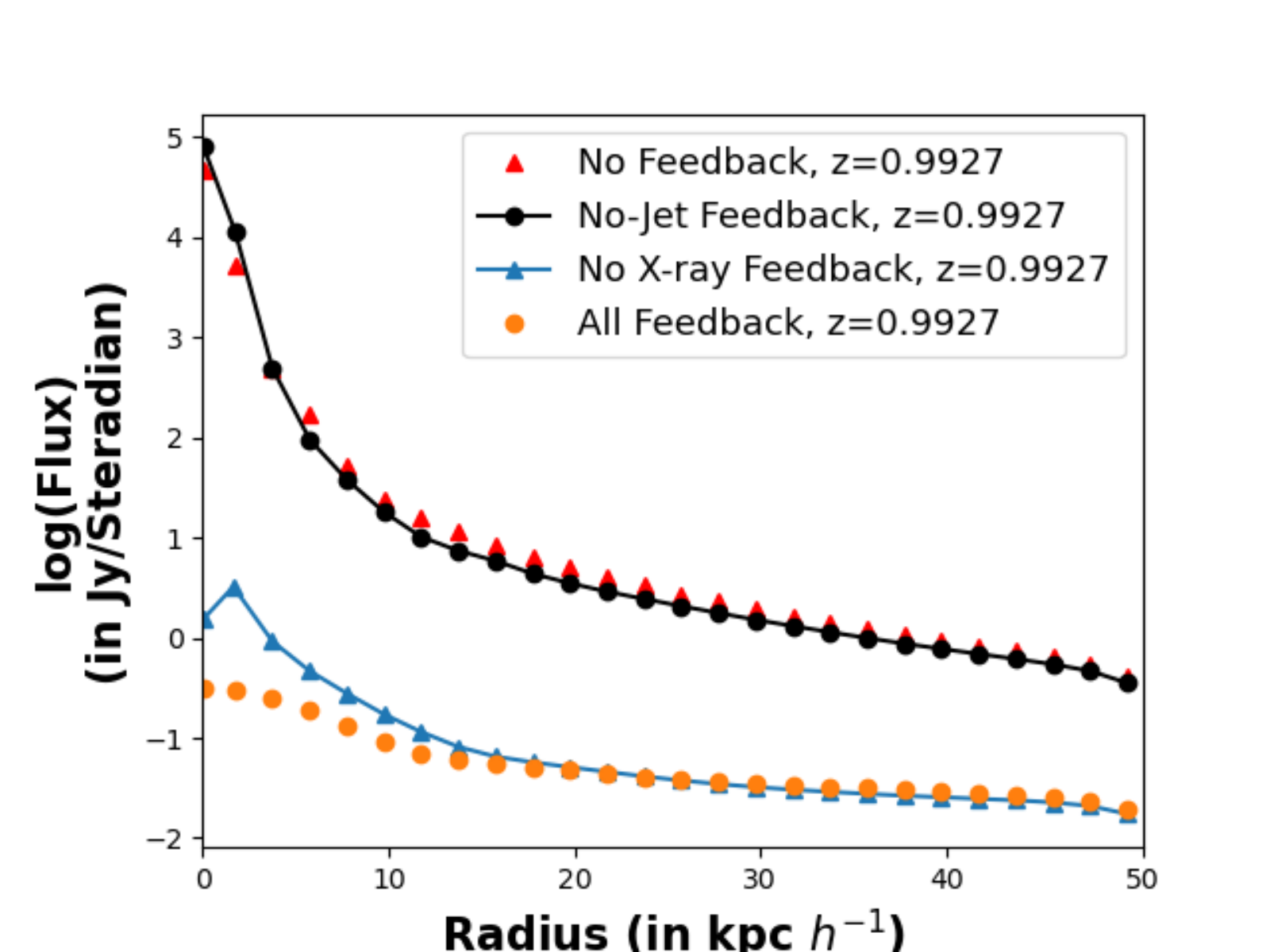}
       \includegraphics[width=4cm]{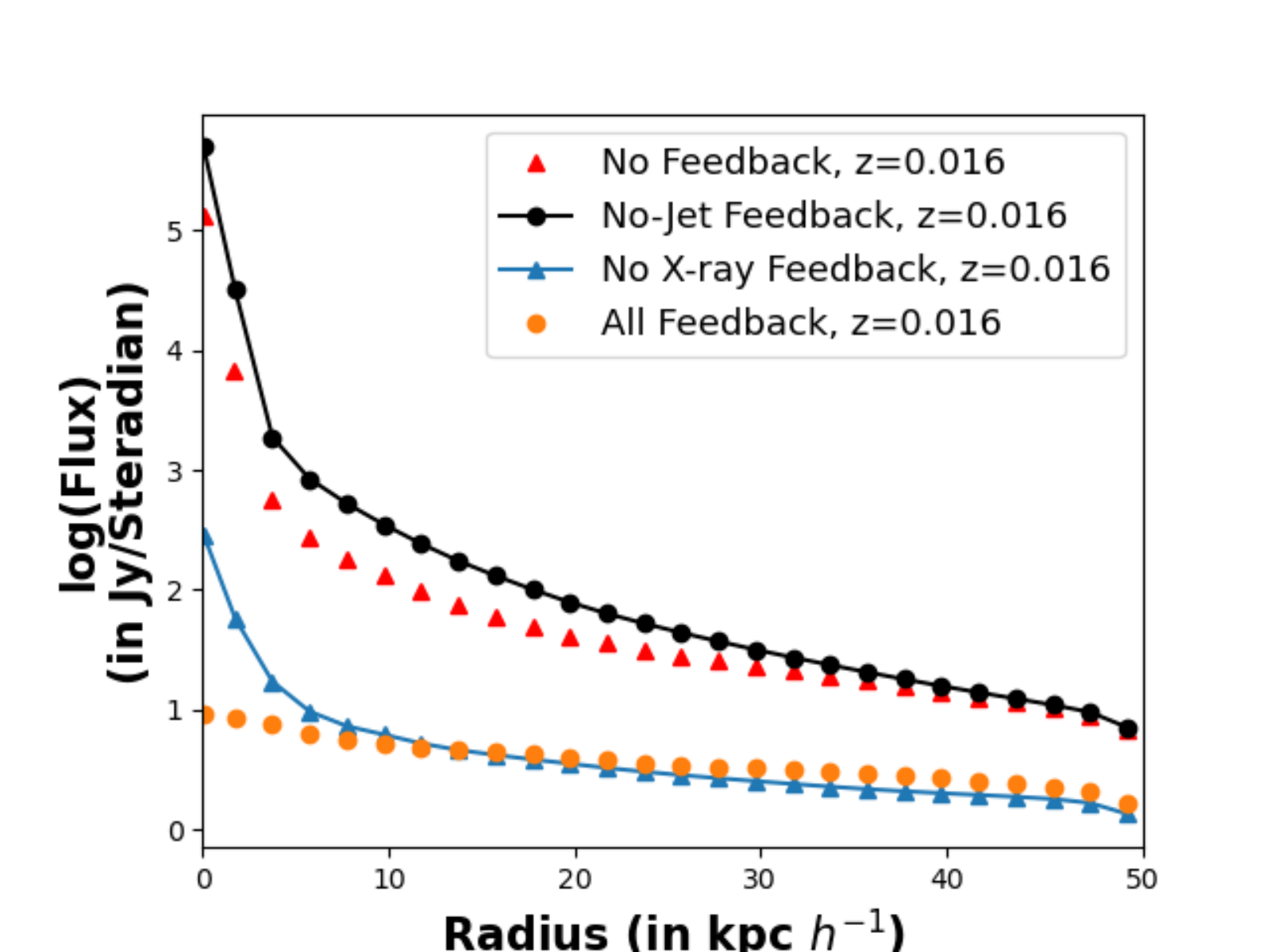}\\
       \includegraphics[width=4cm]{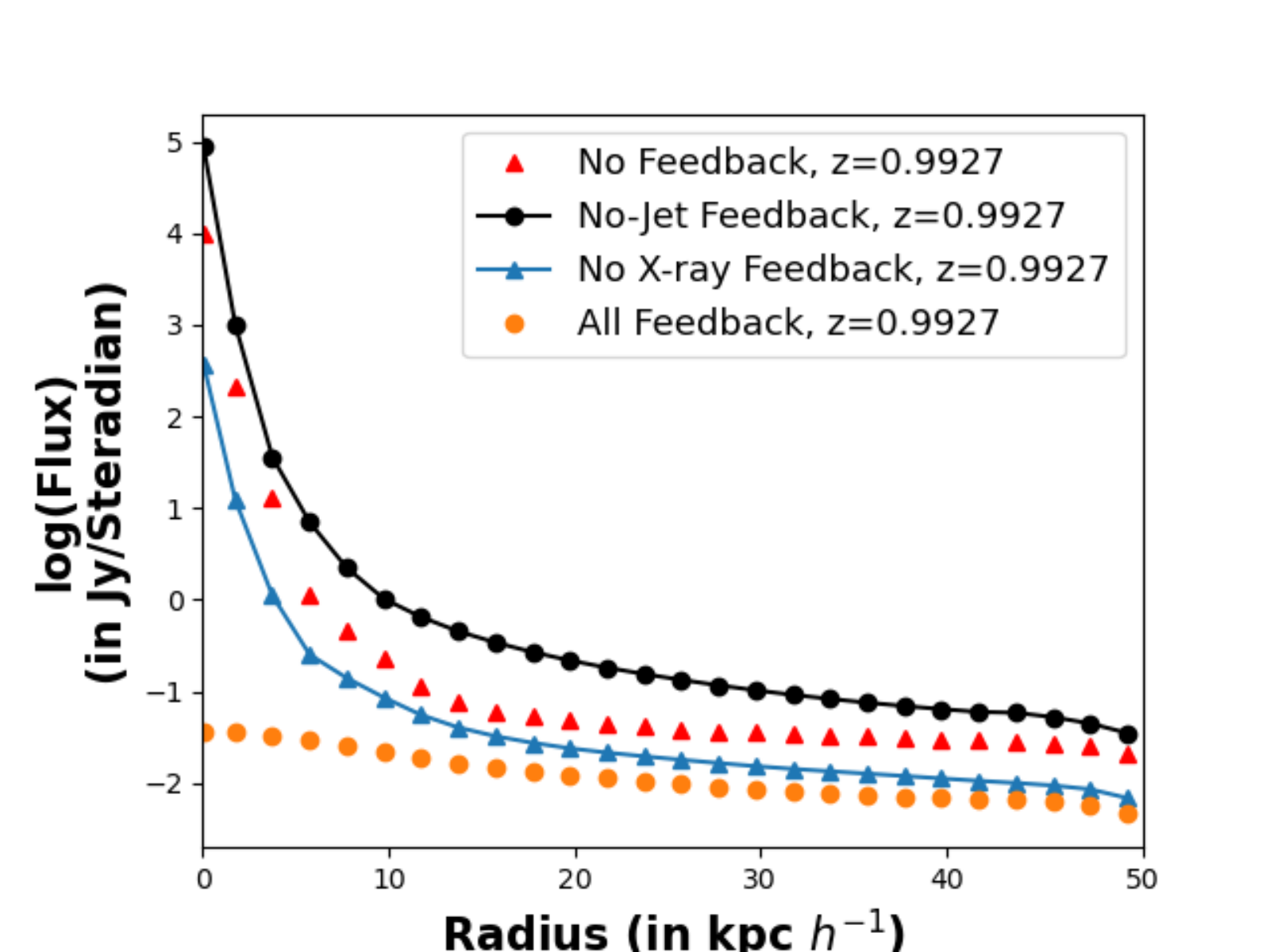}
       \includegraphics[width=4cm]{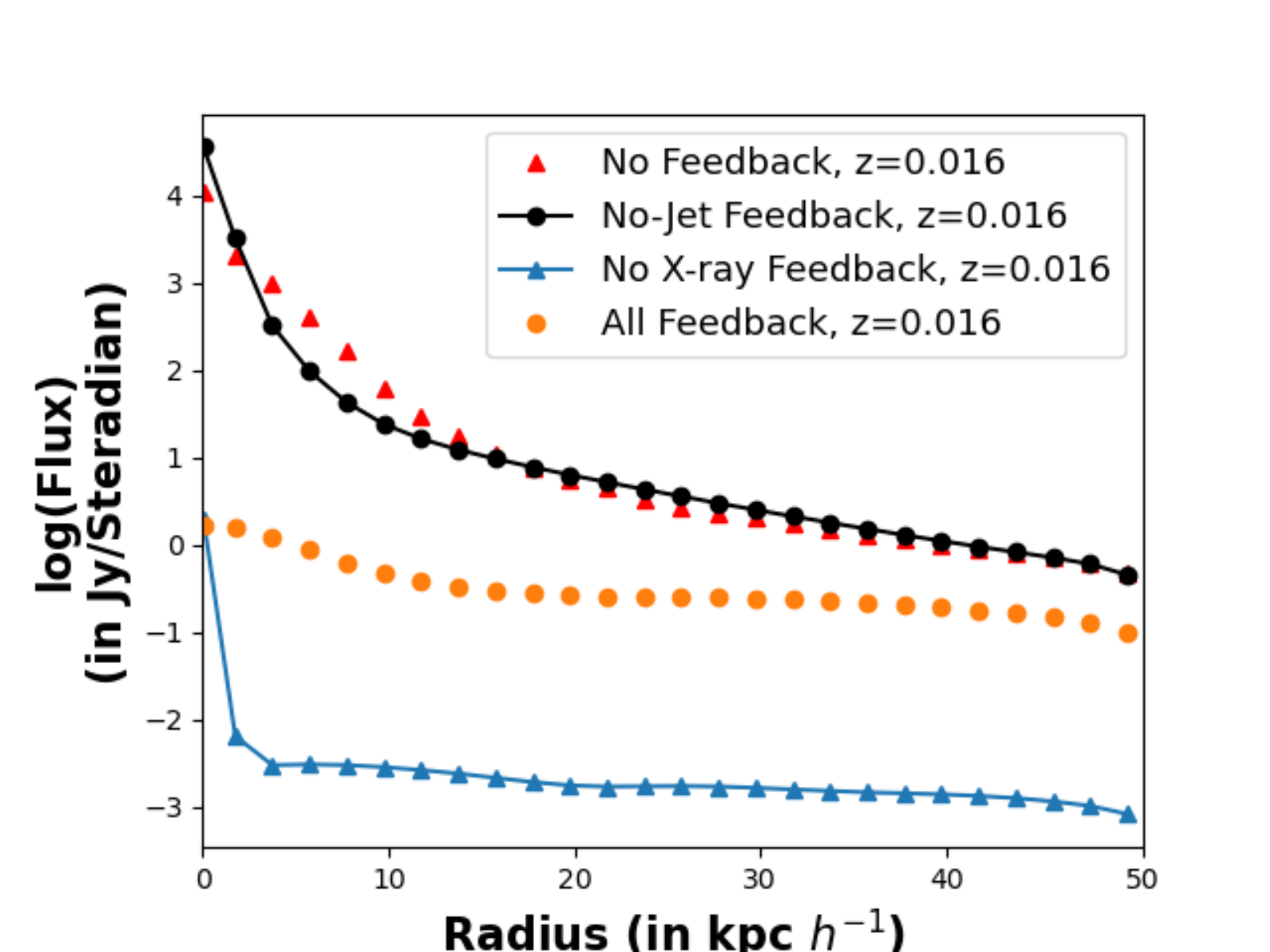}\\
      \end{tabular}
        \caption{The theoretical SZ radial profiles of the most massive and most active BHs corresponding to Figs.\ 1 and 2. {\bf Top Left } Radial profile for the most massive BH for no, no-jet, no X-ray, and all feedback modes respectively at z$\sim$1 (see Table 1 for nomenclature and Table 2 for black hole properties). {\bf Top Right} Same as the upper left panel but now at z=0.016. {\bf Bottom Left} Same for the most active BH at z$\sim$1. {\bf Bottom Right} Same as the lower left panel but now at z=0.016. We note that for all four cases we hardly see any significant difference between no feedback and no-jet feedback modes but no X-ray feedback and all feedback modes are separable. A significant suppression of the signal occurs when the jet mode of feedback is introduced in the model.}
        \label{fig:3}
    \end{center}
\end{figure}

\begin{table}[h]
\caption{CASA Parameters for `simalma'} 
\centering 
\begin{tabular}{c c } 
\hline \hline
Parameters & Values\\ [0.5ex] 
\hline 
incell & 0.5 arcsec\\
incenter & 320 GHz/135 GHz/100 GHz/42 GHz\\
inwidth & 7.5 GHz\\
integration & 30s\\
mapsize & 10 arcsecs\\
antennalist & alma.cycle 8.1.cfg\\
totaltime & 3h\\
pwv & 0.5\\
imsize & 300\\
cell & 0.1 arcsec/0.23 arcsec/0.32 arcsec/0.76 arcsec\\
niter & 1000\\[0.5ex] 
\hline 
\end{tabular}
\end{table} 

\subsection{Construction of Sunyaev-Zeldovich Maps}\label{section_2.2}
One of the main goals of our work is to simulate the SZ effect arising from different feedback modes and study their redshift evolution. 

\subsubsection{Sample Selection} 
To perform our analysis we have made use of the SIMBA halo catalog. We have restricted our analysis to central black holes within the halos. The identification of central and satellite black holes are done based on the SIMBA selection criterion, where SMBHs residing in central galaxies are considered to be central black holes while SMBHs residing in satellite galaxies are considered to be satellite black holes. We have selected our central black holes to have mass $\ge 10^{7}$h$^{-1}$M$_{\odot}$, residing in halos of mass $\ge 10^{12}$h$^{-1}$M$_{\odot}$ at all redshifts \citep{RKC22}. The cut-off mass are motivated from the mass resolutions in the simulation. The distribution of halo mass are shown in \citep{RKC22}. To ensure that the same black hole is identified for different feedback mode runs, we used the method of tracking the host halo in the SIMBA catalog. The host halo mass of the black holes remain same for different feedback modes. In this work we selected two representative black holes from the simulation, the most massive (highest mass) and the most active (highest accretion rate) ones at z$= 0.016$ and z$=1.0$. The properties of the black holes are summarized in Table 2. 

\begin{figure*}
    \begin{center}
        \begin{tabular}{cccc}
        \hline
         {No}&{No-Jet}&{No X-ray}&{All}\\[0.1pt]
         \hline
      \includegraphics[width=4.5cm]{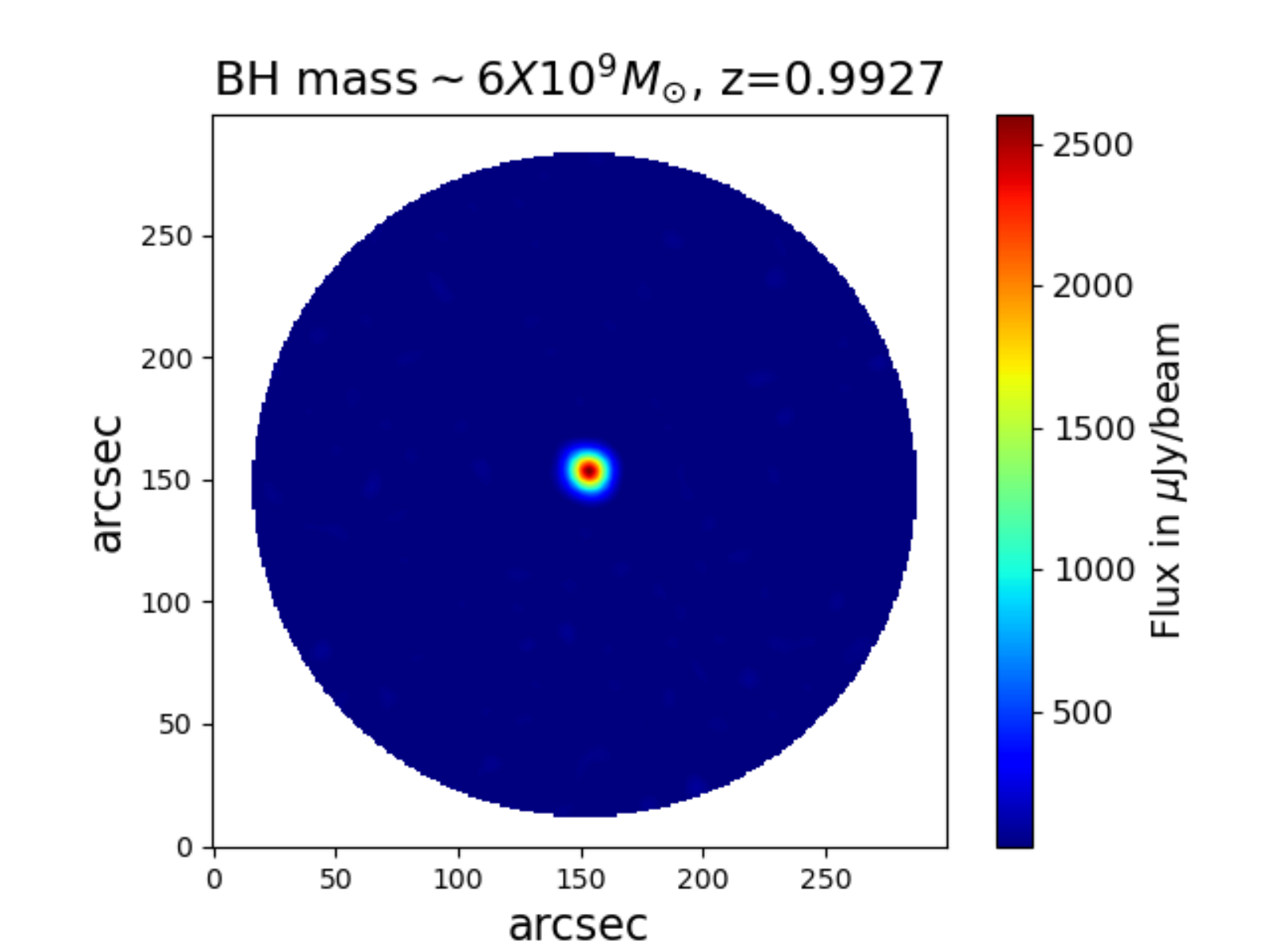}&\includegraphics[width=4.5cm]{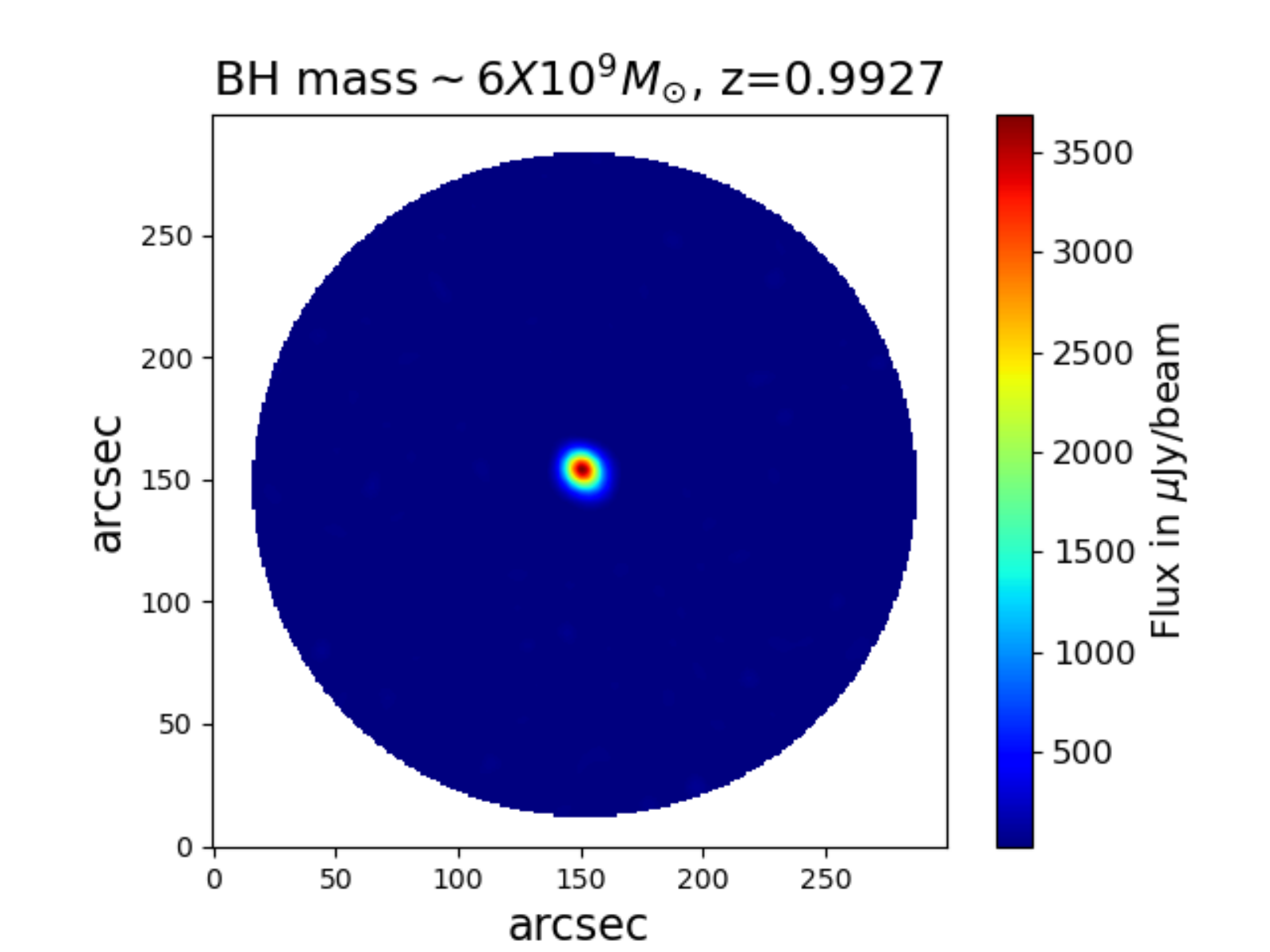}&\includegraphics[width=4.5cm]{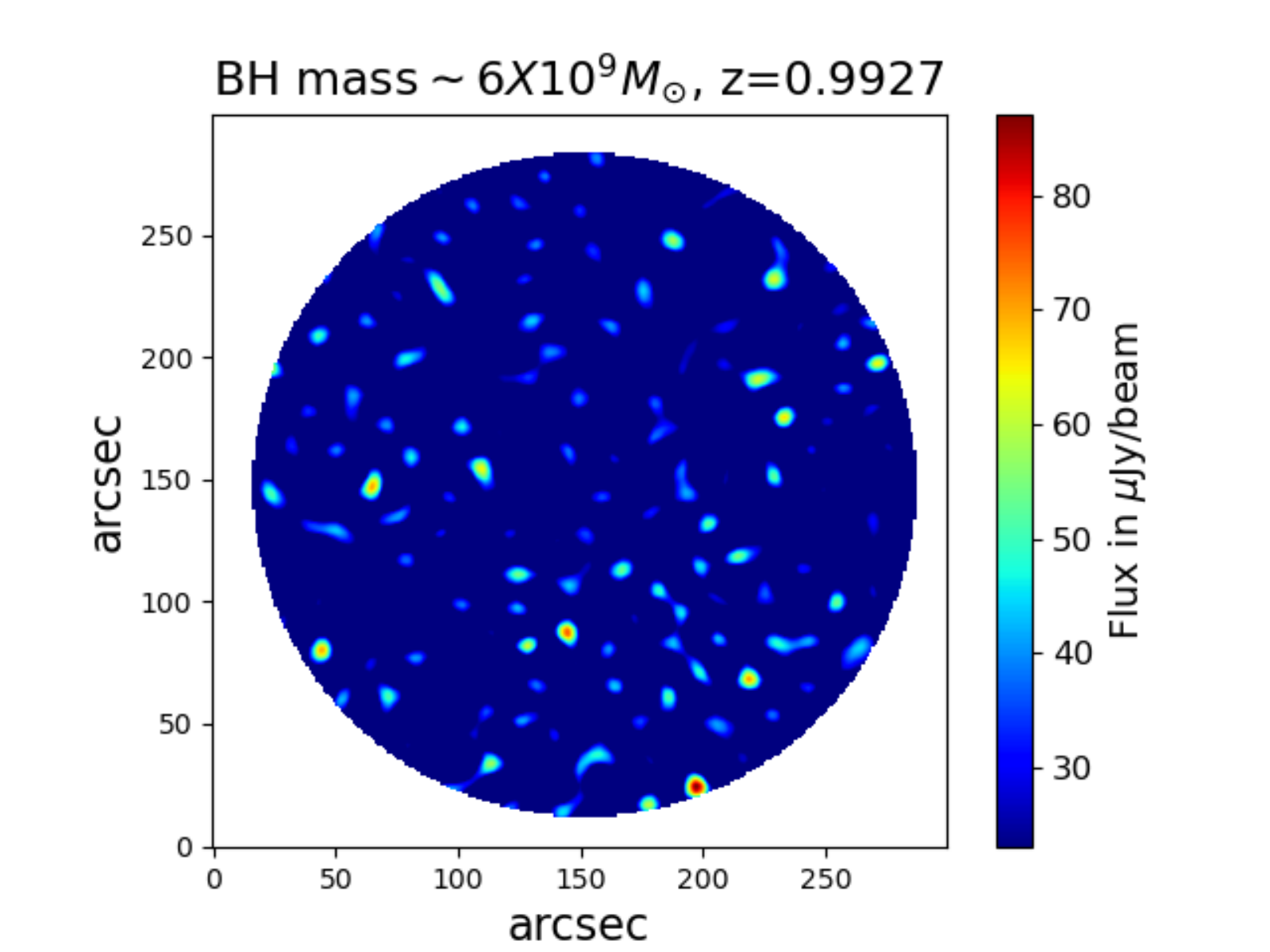}&\includegraphics[width=4.5cm]{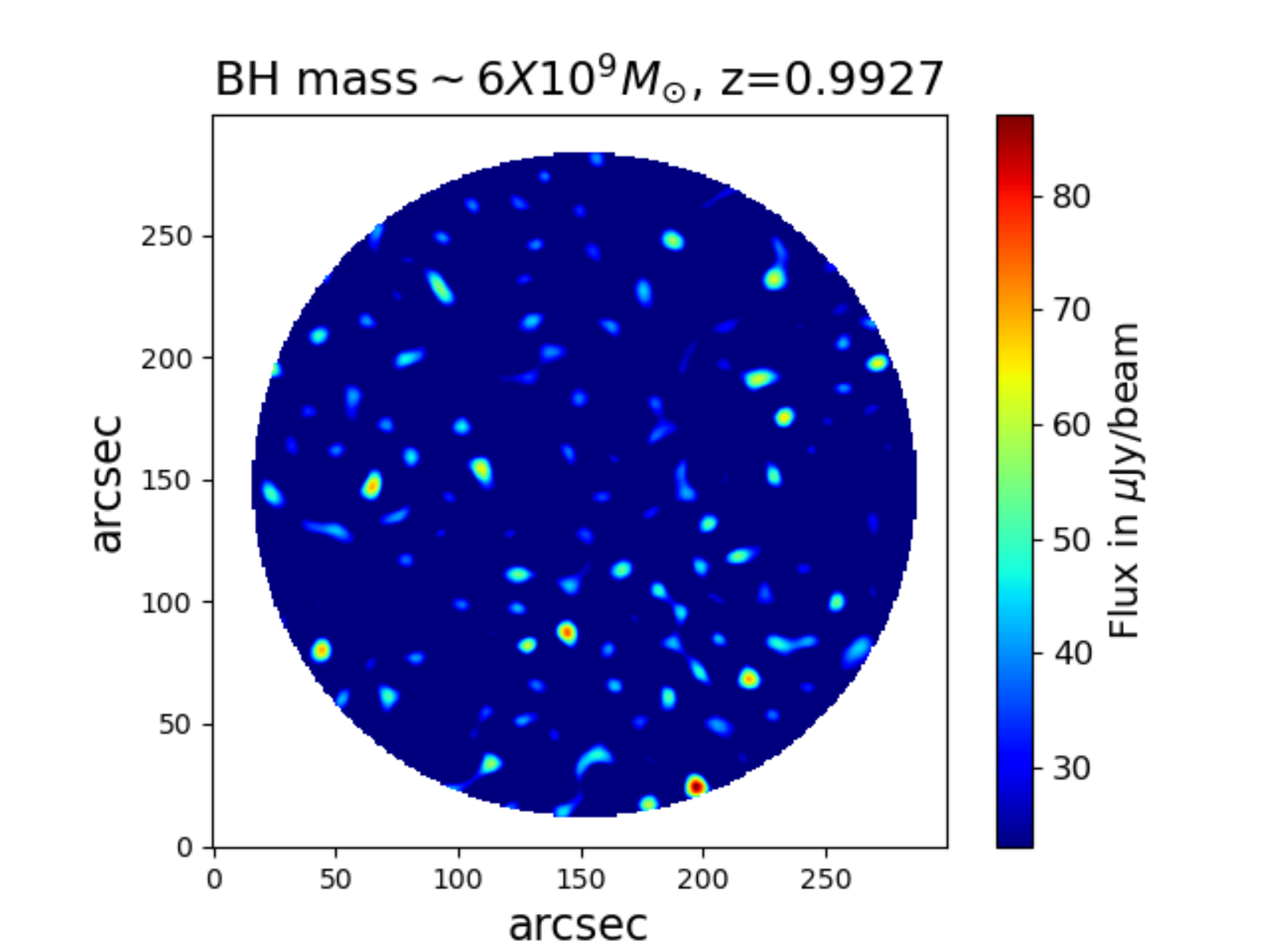}\\
      \includegraphics[width=4.5cm]{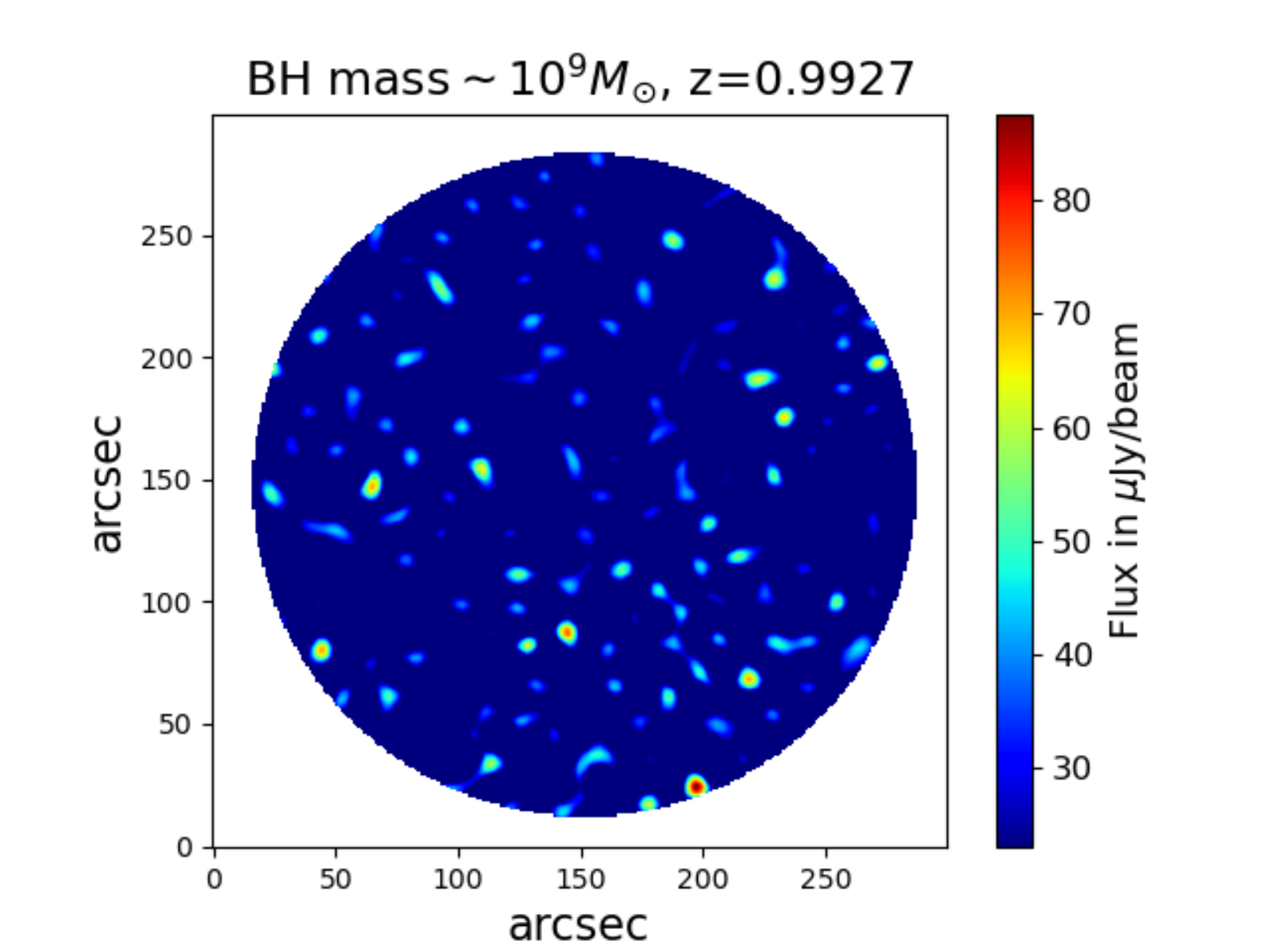}&\includegraphics[width=4.5cm]{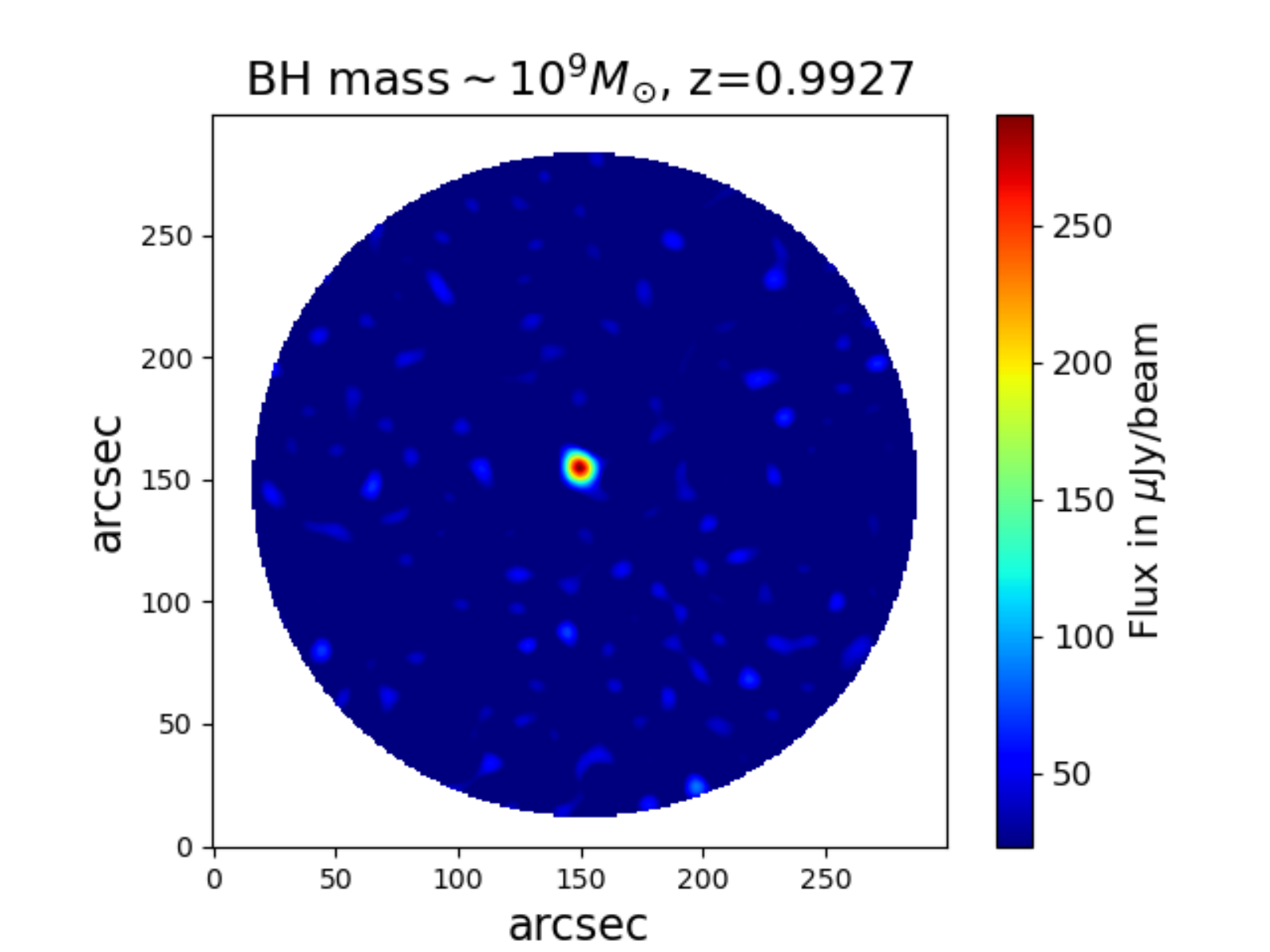}&\includegraphics[width=4.5cm]{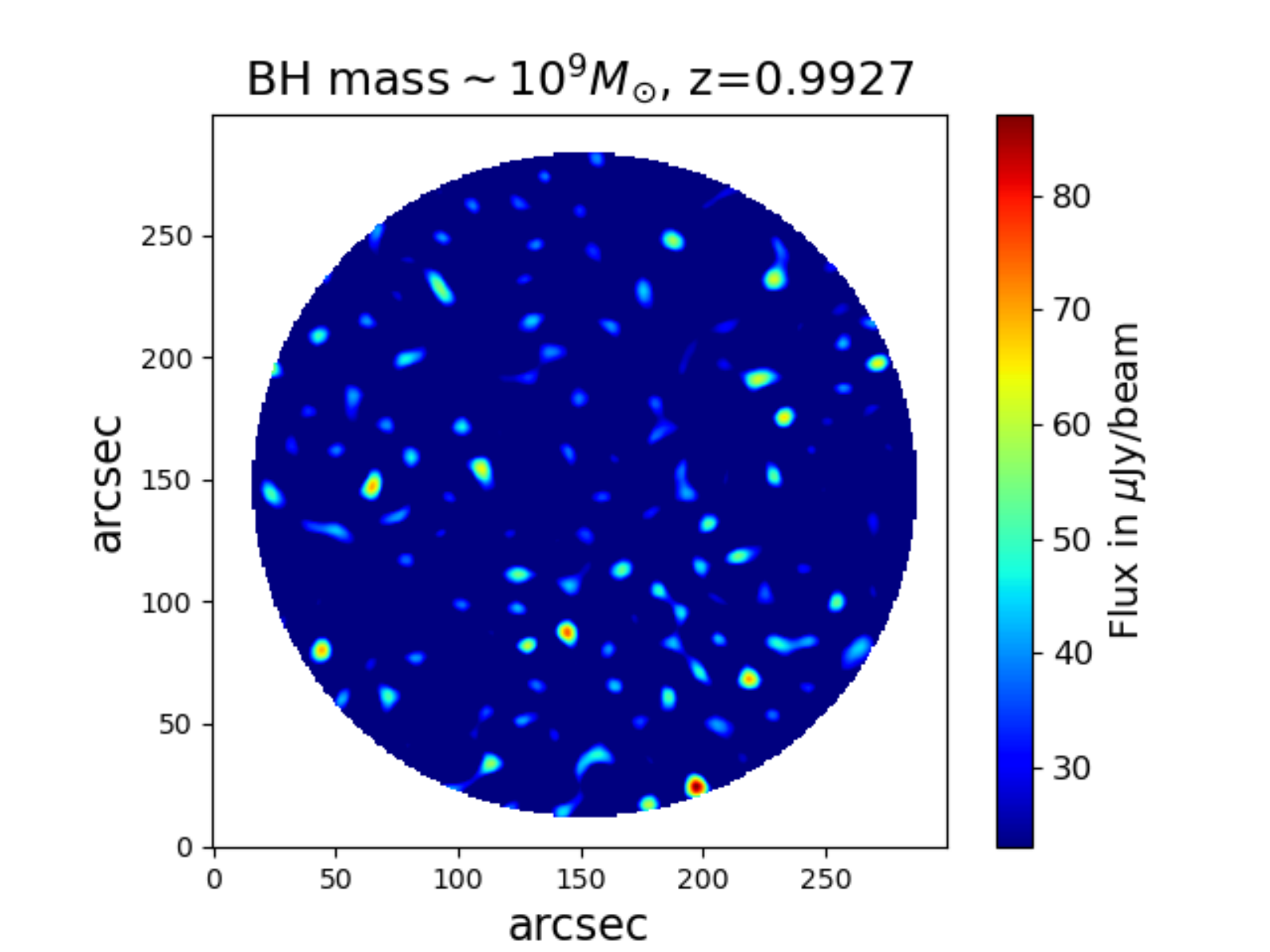}&\includegraphics[width=4.5cm]{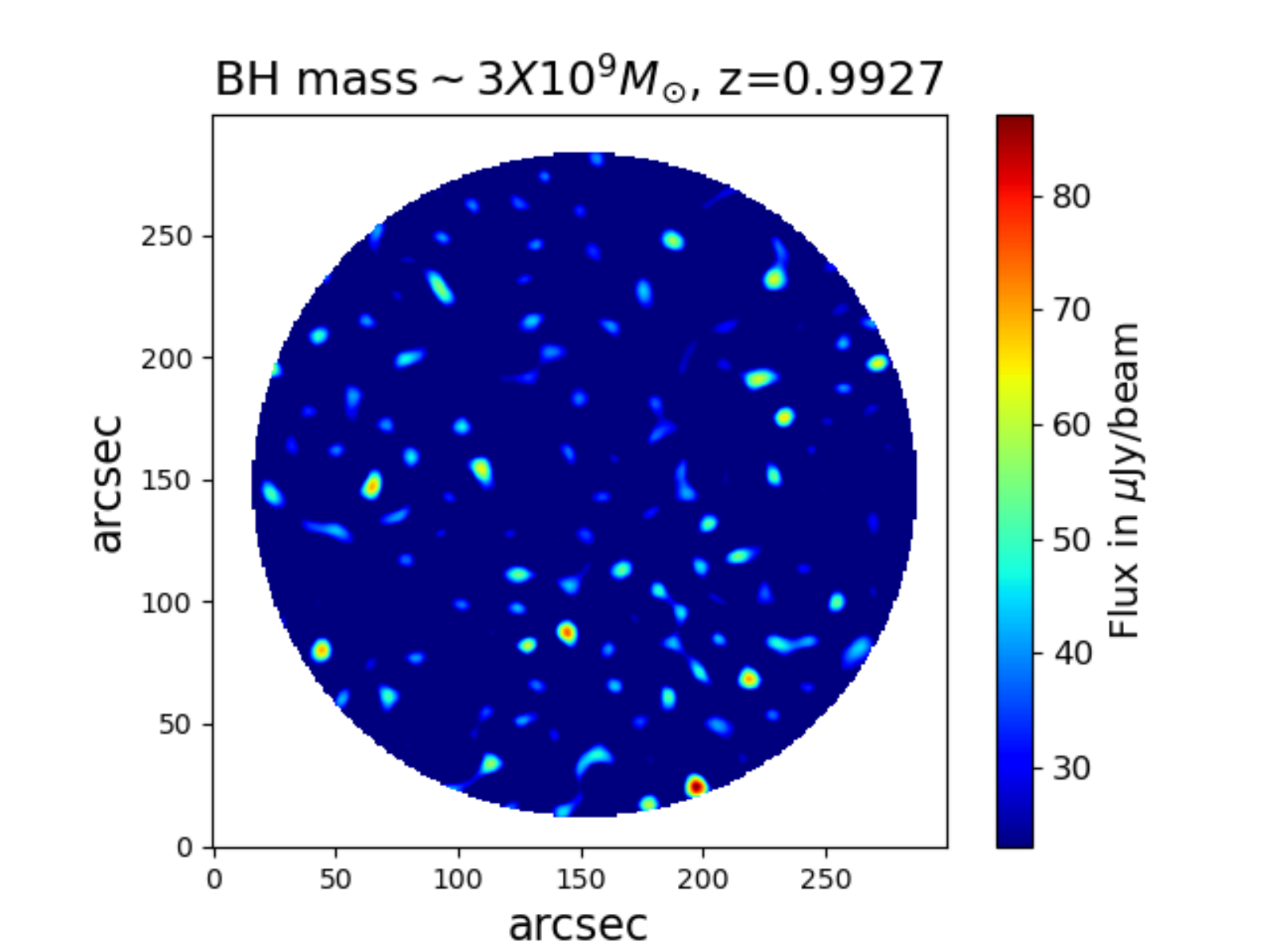}\\
       \includegraphics[width=4.5cm]{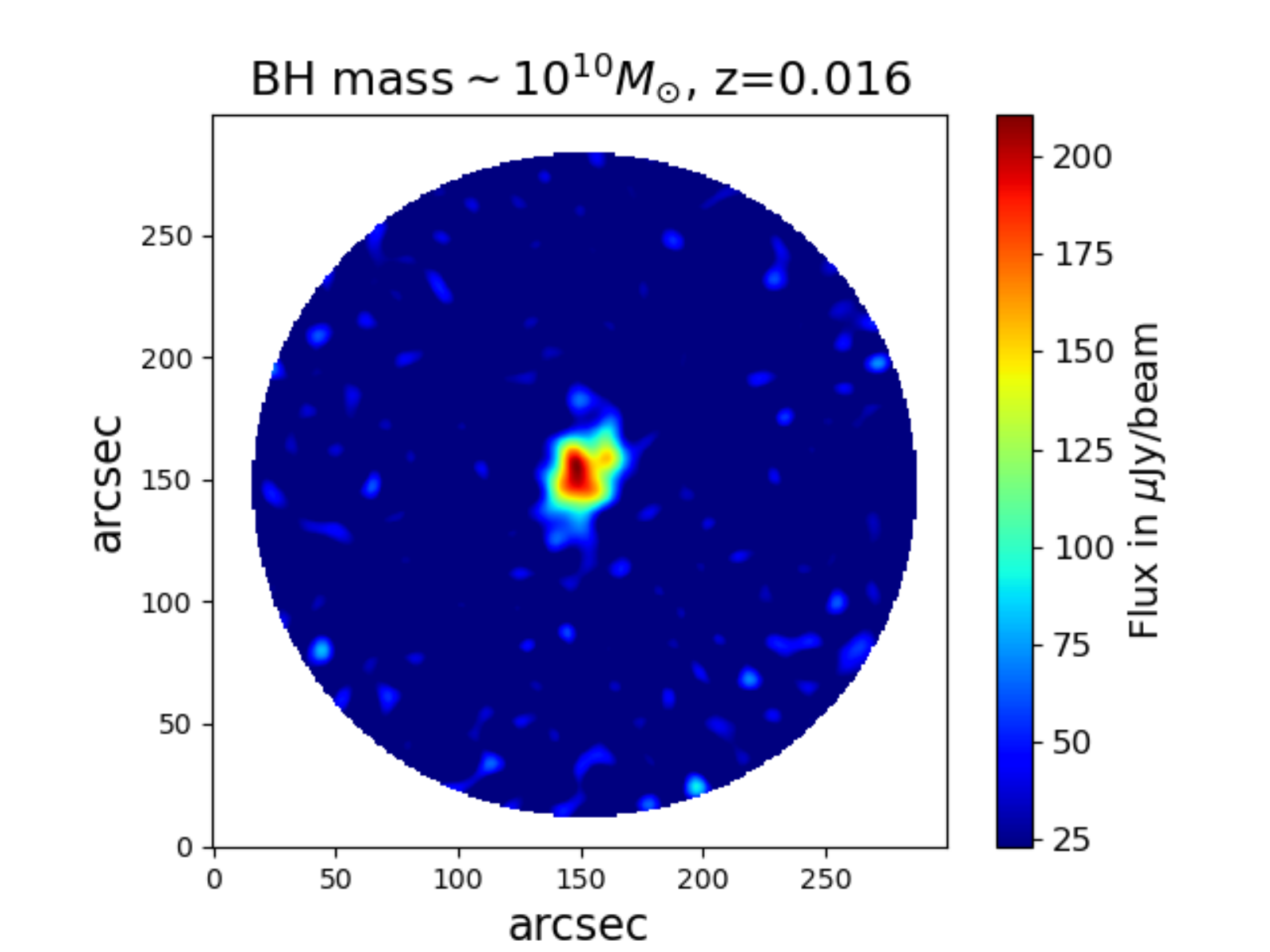}&\includegraphics[width=4.5cm]{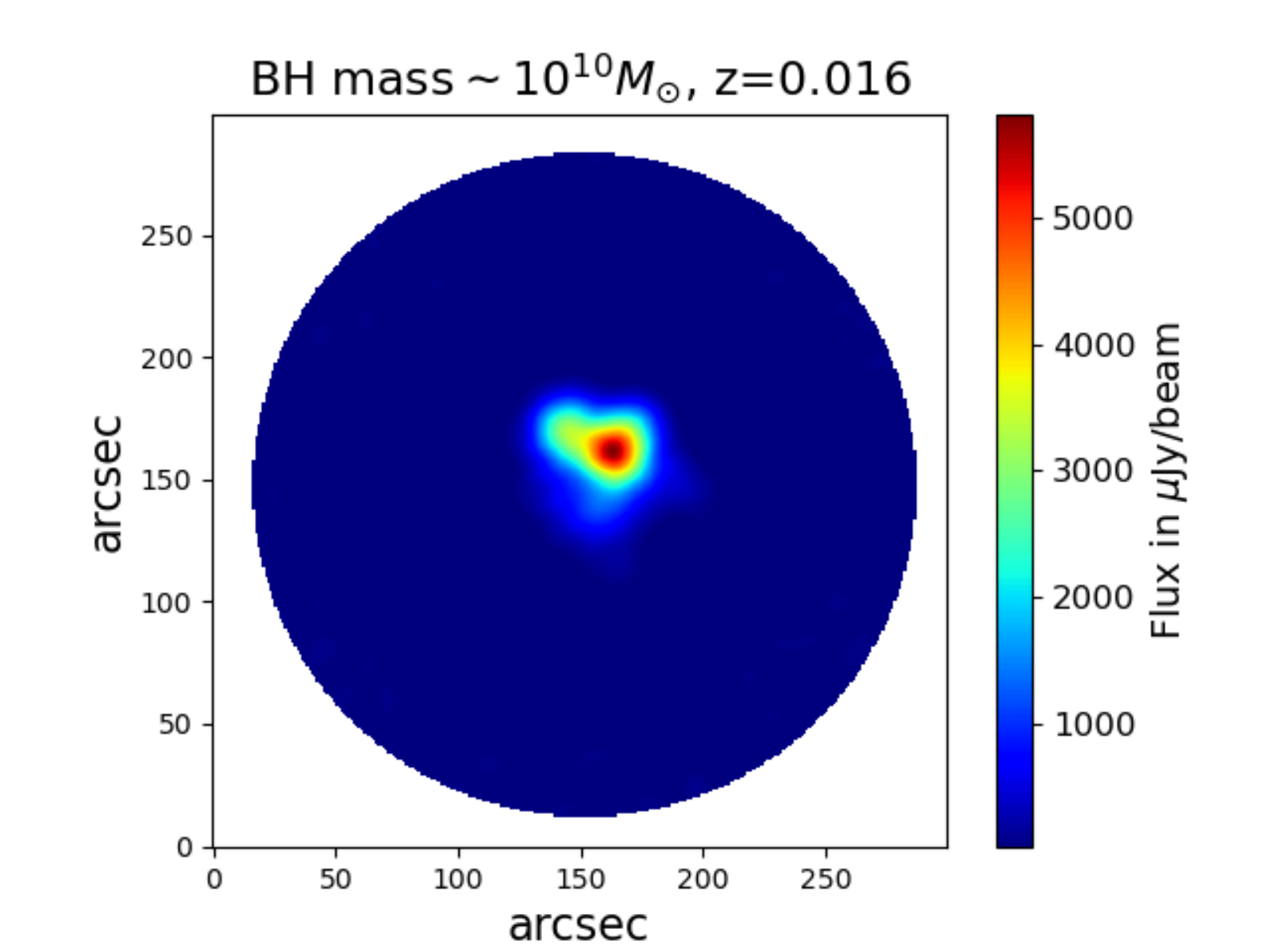}&\includegraphics[width=4.5cm]{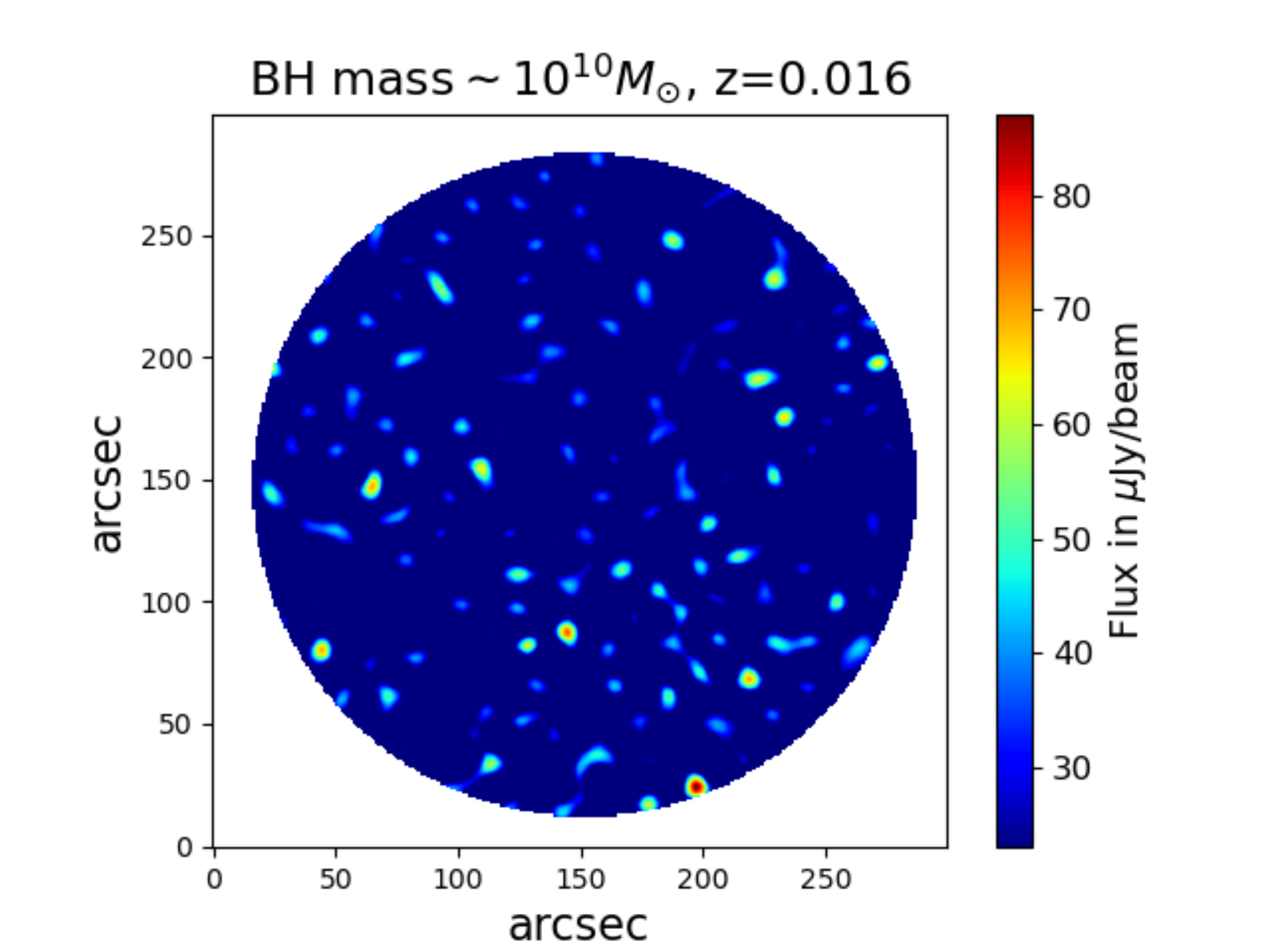}&\includegraphics[width=4.5cm]{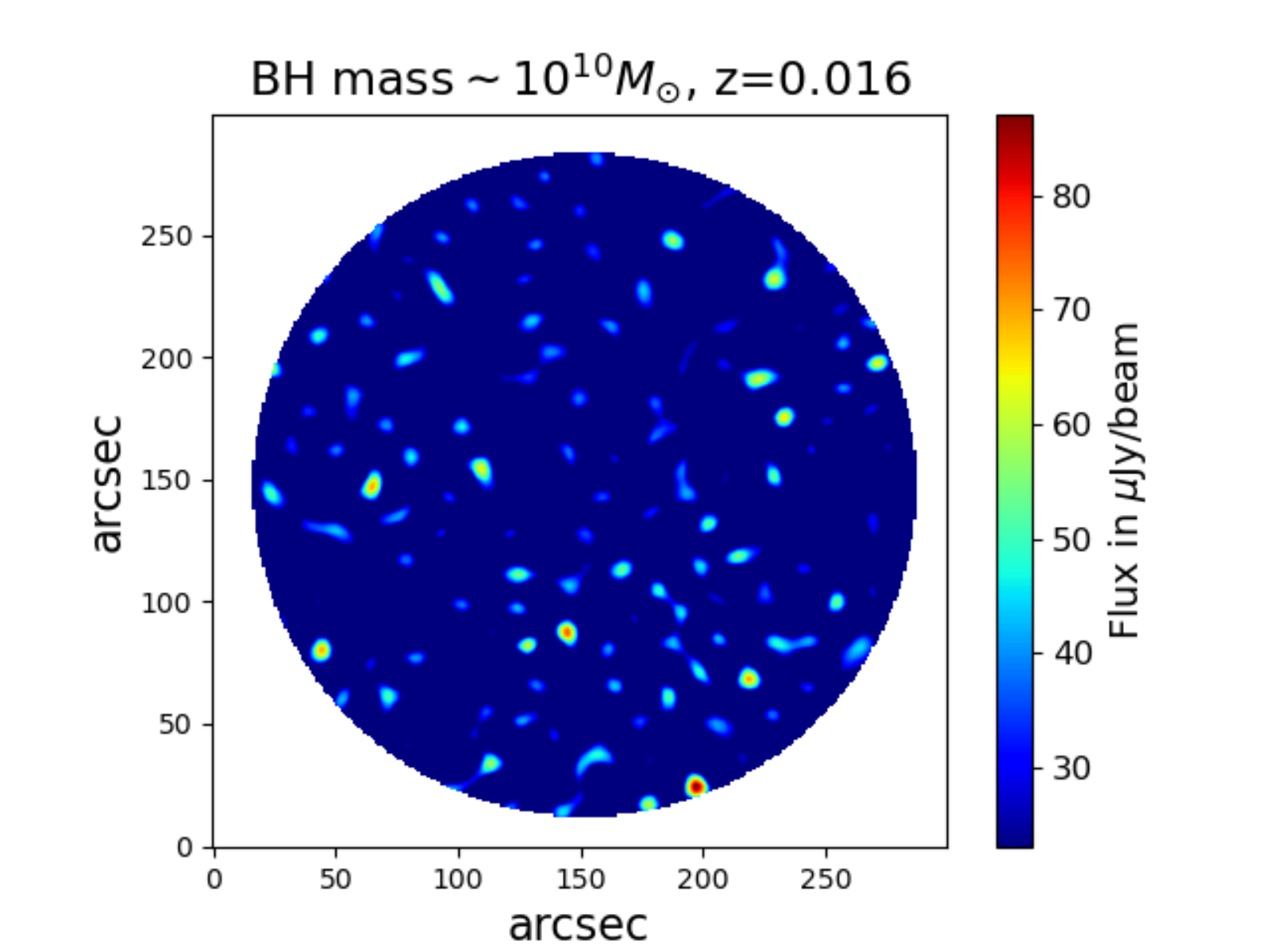}\\
      \includegraphics[width=4.5cm]{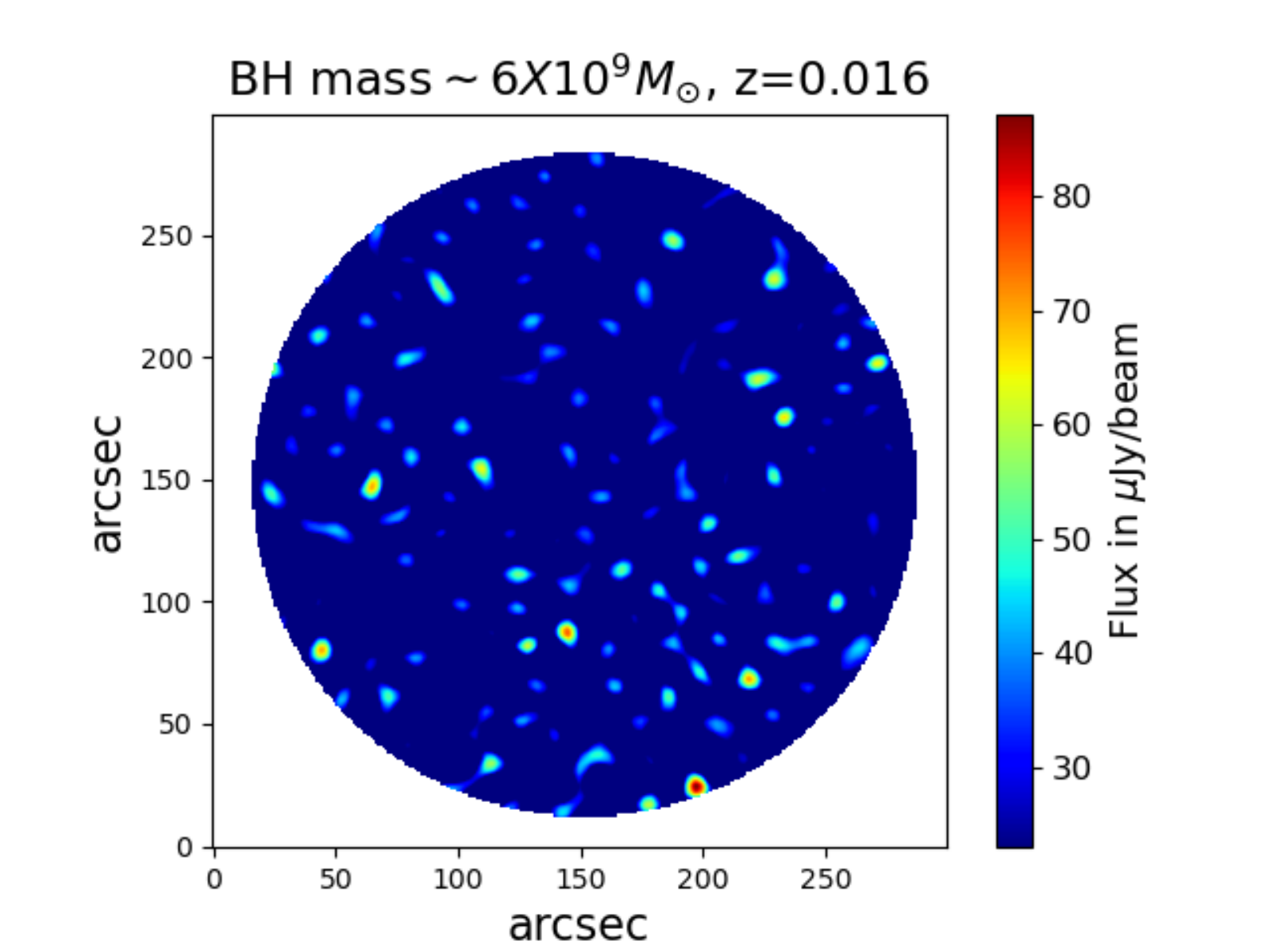}&\includegraphics[width=4.5cm]{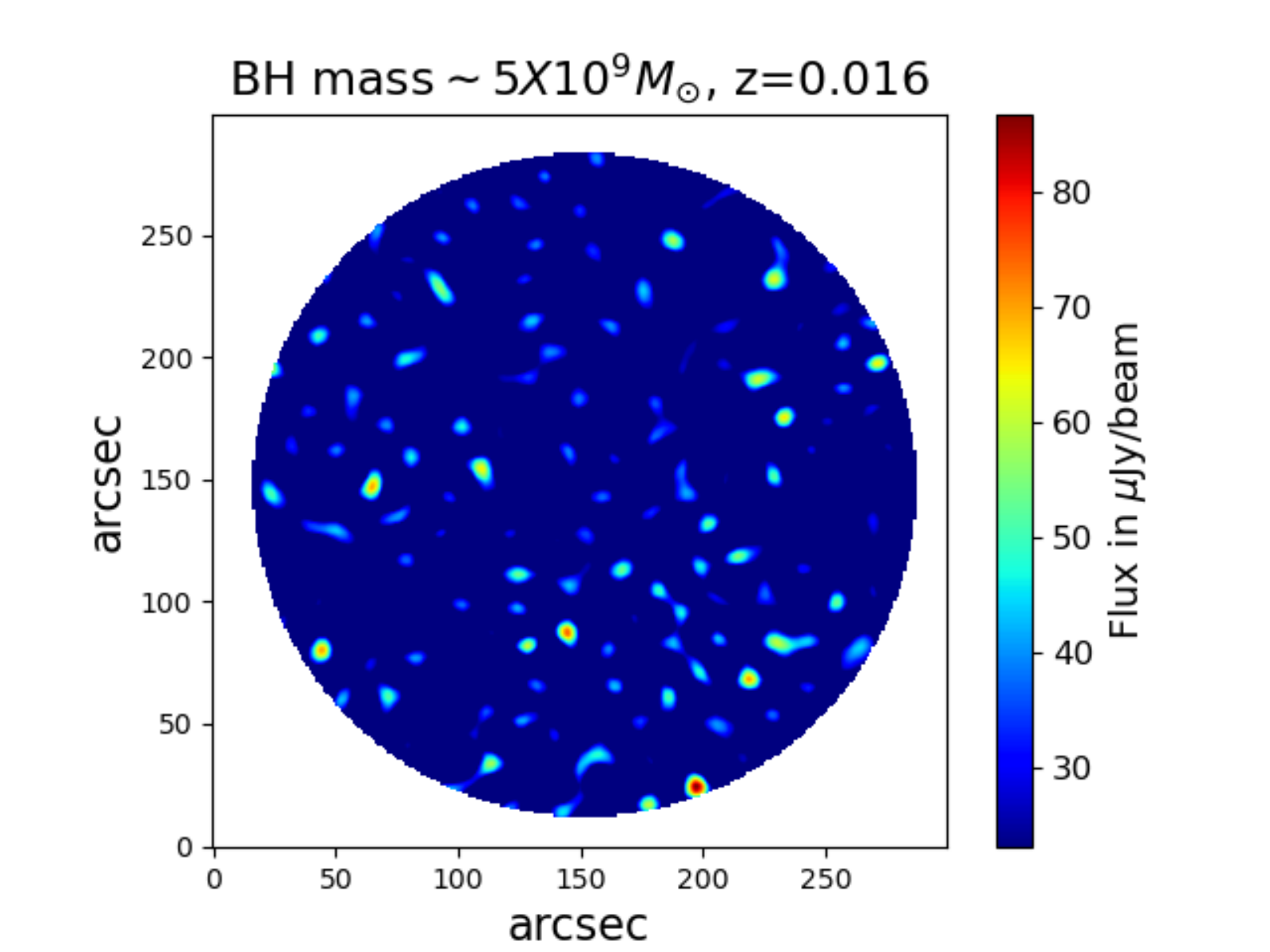}&\includegraphics[width=4.5cm]{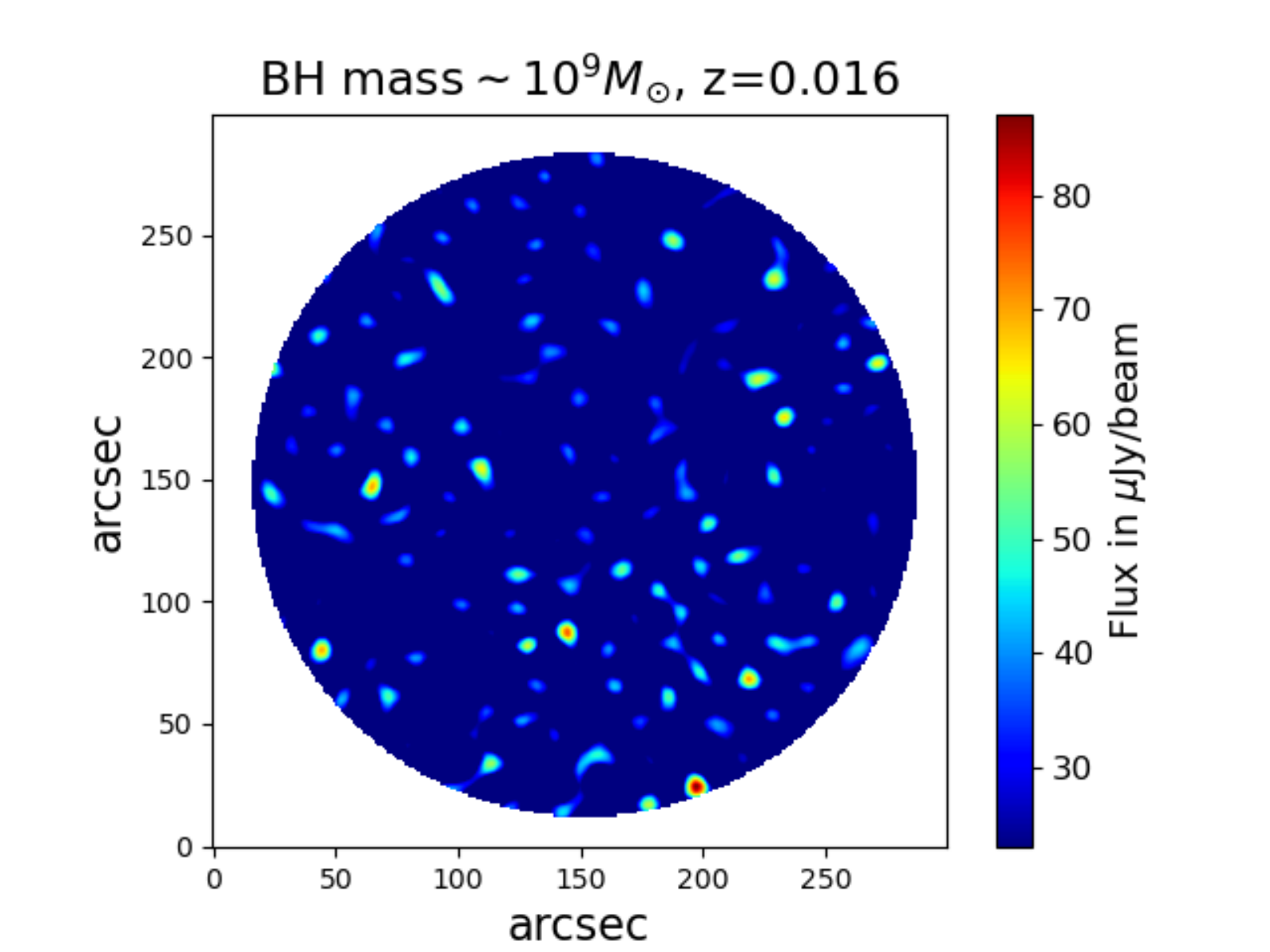}&\includegraphics[width=4.5cm]{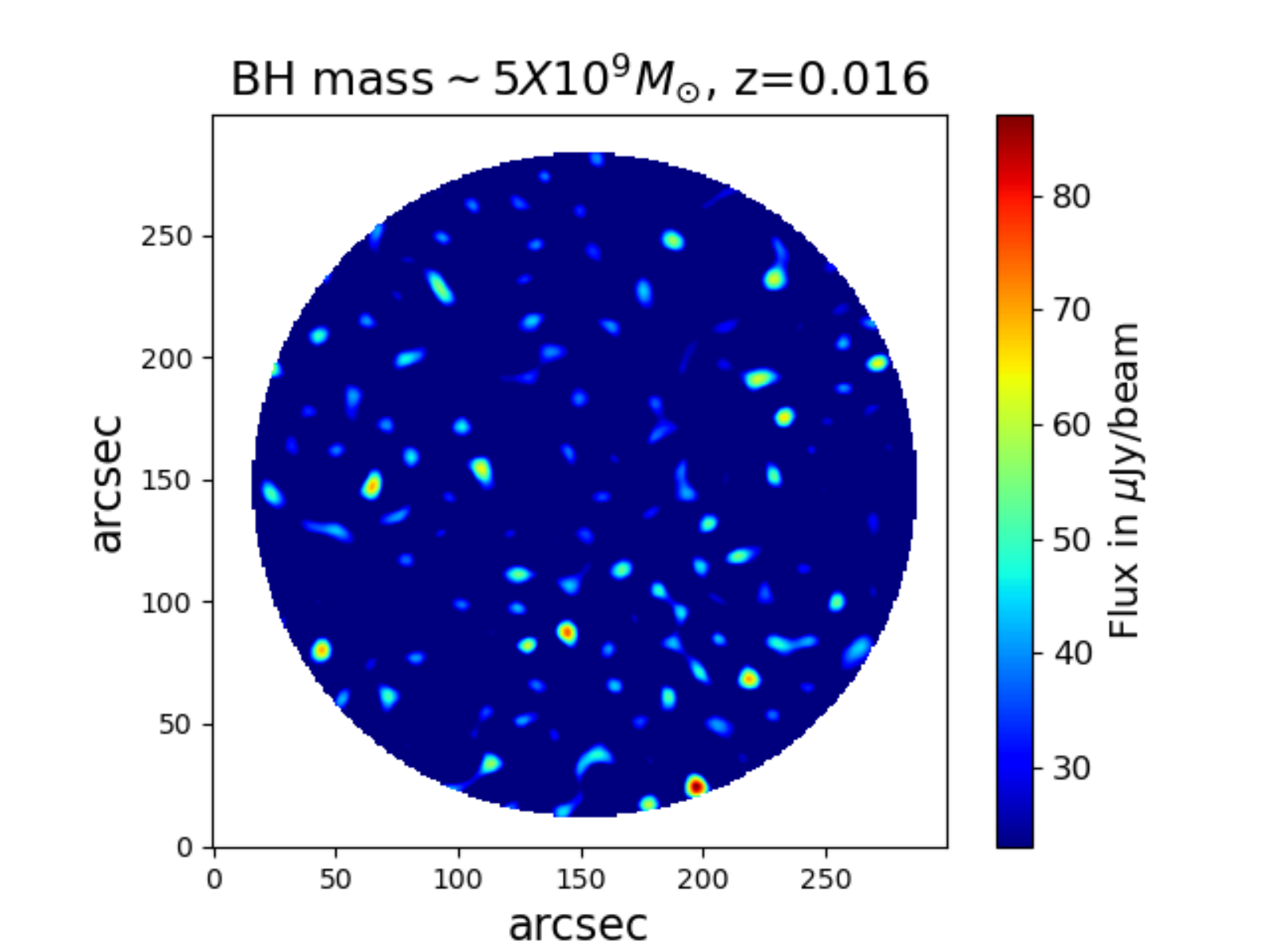}\\
      \end{tabular}
        \caption{The mock ALMA tSZ maps at 320 GHz (Band 7) for different feedback modes around most massive and most active BHs at two different redshifts from \cite{dave19} using the observational parameters in Table 3. {\bf Top Panel} The mock ALMA tSZ maps for no feedback ({\bf left most column}), no-jet feedback, no X-ray feedback, and all feedback modes ({\bf right most column}) respectively around the most massive BH at z$\sim$1. {\bf Second Panel} The mock ALMA tSZ maps for no feedback, no-jet feedback, no X-ray feedback, and all feedback modes respectively around the most active BH at z$\sim$1. {\bf Third Panel} Same as the top panel but now at z=0.016. {\bf Fourth Panel} Same as the second panel but now at z=0.016. We note that ALMA has the capability to detect the SZ signal for the no feedback and no-jet feedback modes. The signal gets enhanced when, only the radiative mode of feedback is added. For no X-ray feedback and all feedback modes the signal gets suppressed below the noise threshold of ALMA. The different feedback models and the black hole properties are summarized in Tables 1 and 2 respectively. The ALMA results are summarized in Table 4.}
        \label{fig:4}
    \end{center}
\end{figure*}

To obtain the SZ map around the black holes, a projected direction is chosen and then the SZ signal arising from all the elements along the line of sight is integrated for all the pixels in the map. The change in the intensity of the CMB due to the tSZ effect is given by \cite{sazanovetal98} :
\begin{equation}
    \Delta I(x) = \frac{2k_{B}T_{CMB}}{\lambda^{2}}\frac{x^{2}e^{x}}{(e^{x}-1)^{2}} \frac{k_{B}\sigma_{T}}{m_{e}c^{2}}f_{1}(x)\int \mathrm{d}l\, n_{e}(l)T_{e}(l) 
\end{equation}
where $x=h\nu/(k_{B}T_{CMB})$, and the integral is along the line of sight direction specified by the direction of projection and $f_{1}(x)=x\,\mathrm{coth}(x/2)-4$ stands for the frequency dependence of the tSZ effect. $n_{e}$ and $T_{e}$ are the electron number density and temperature along the line-of-sight. The flux density is given as $$S_{\nu} = I_{\nu} \int \mathrm{d} \Omega,$$ where $I_{\nu}$ is the intensity and $\Omega$ is the solid angle subtended by the region. As flux density is in Jansky (Jy) we obtain the intensity in Jy/Steradian. In a region within the smoothing length, the energy due to feedback from the black hole is assumed to be distributed isotropically among the gas particles surrounding them and a common B spline kernel is used to compute the smoothed density and temperature \citep{RKC22}.

\begin{figure*}
    \begin{center}
        \begin{tabular}{cccc}
        \hline
         {No}&{No-Jet}&{No X-ray}&{All}\\[0.1pt]
         \hline
      \includegraphics[width=4.5cm]{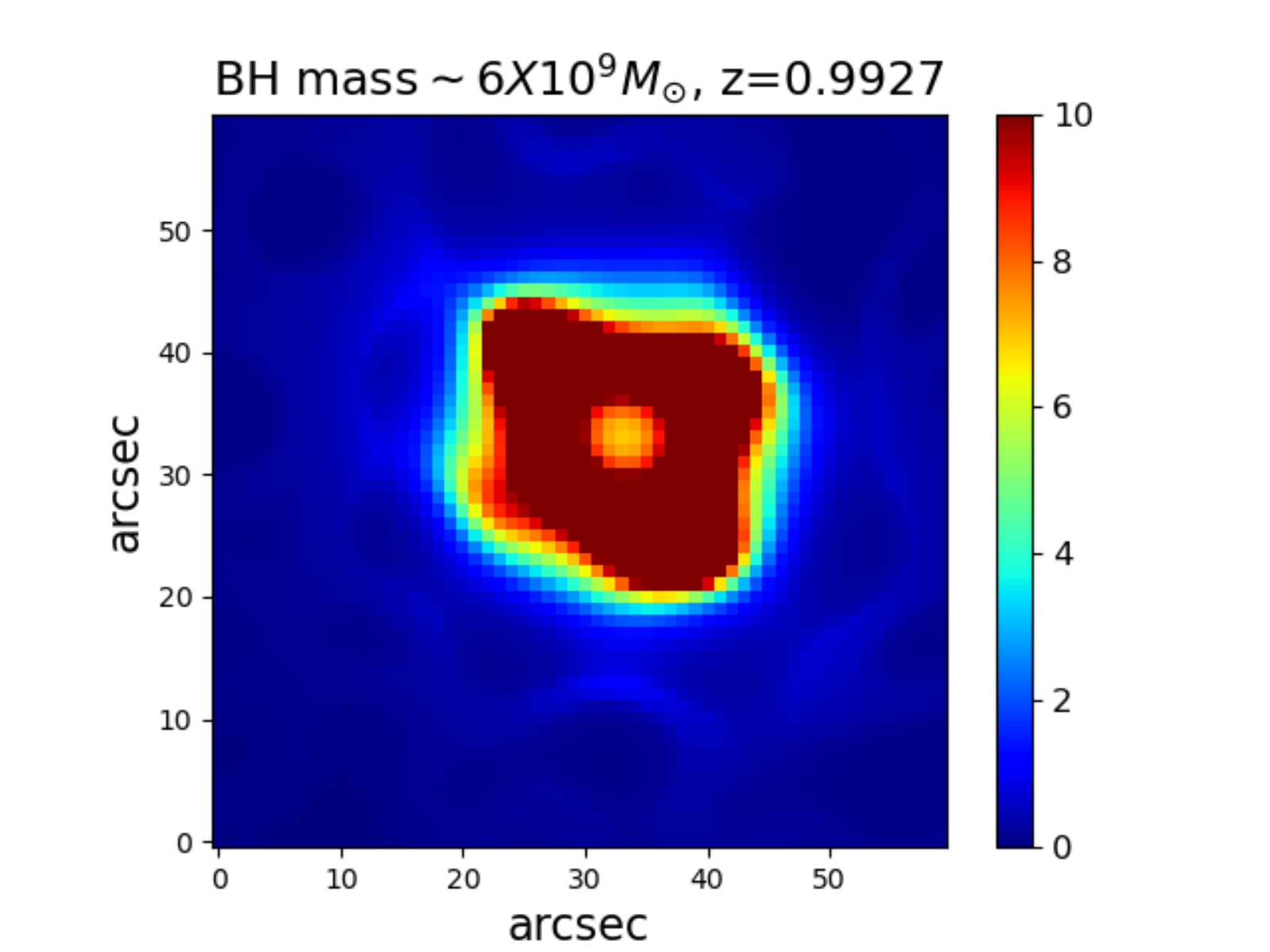}&\includegraphics[width=4.5cm]{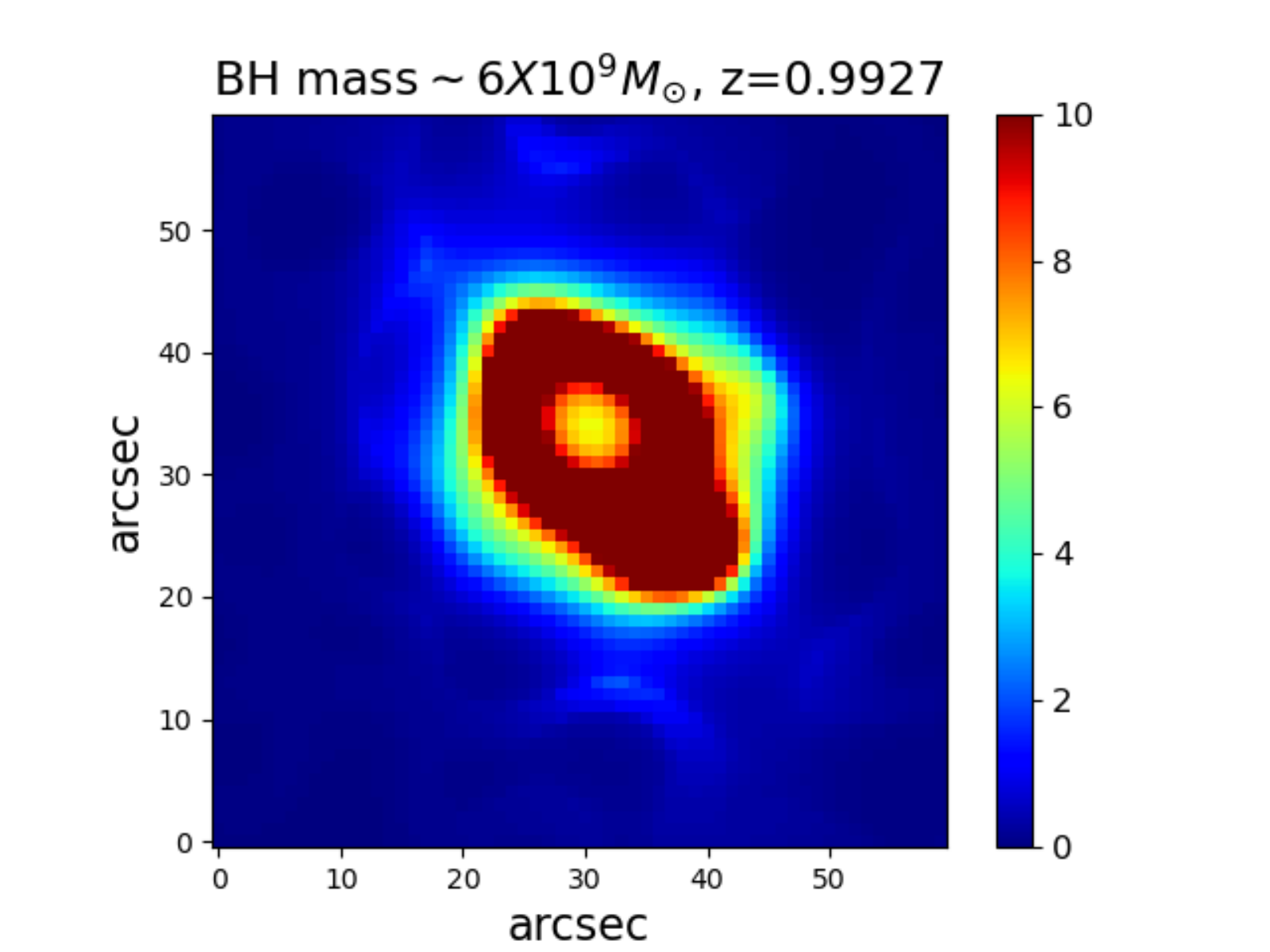}&\includegraphics[width=4.5cm]{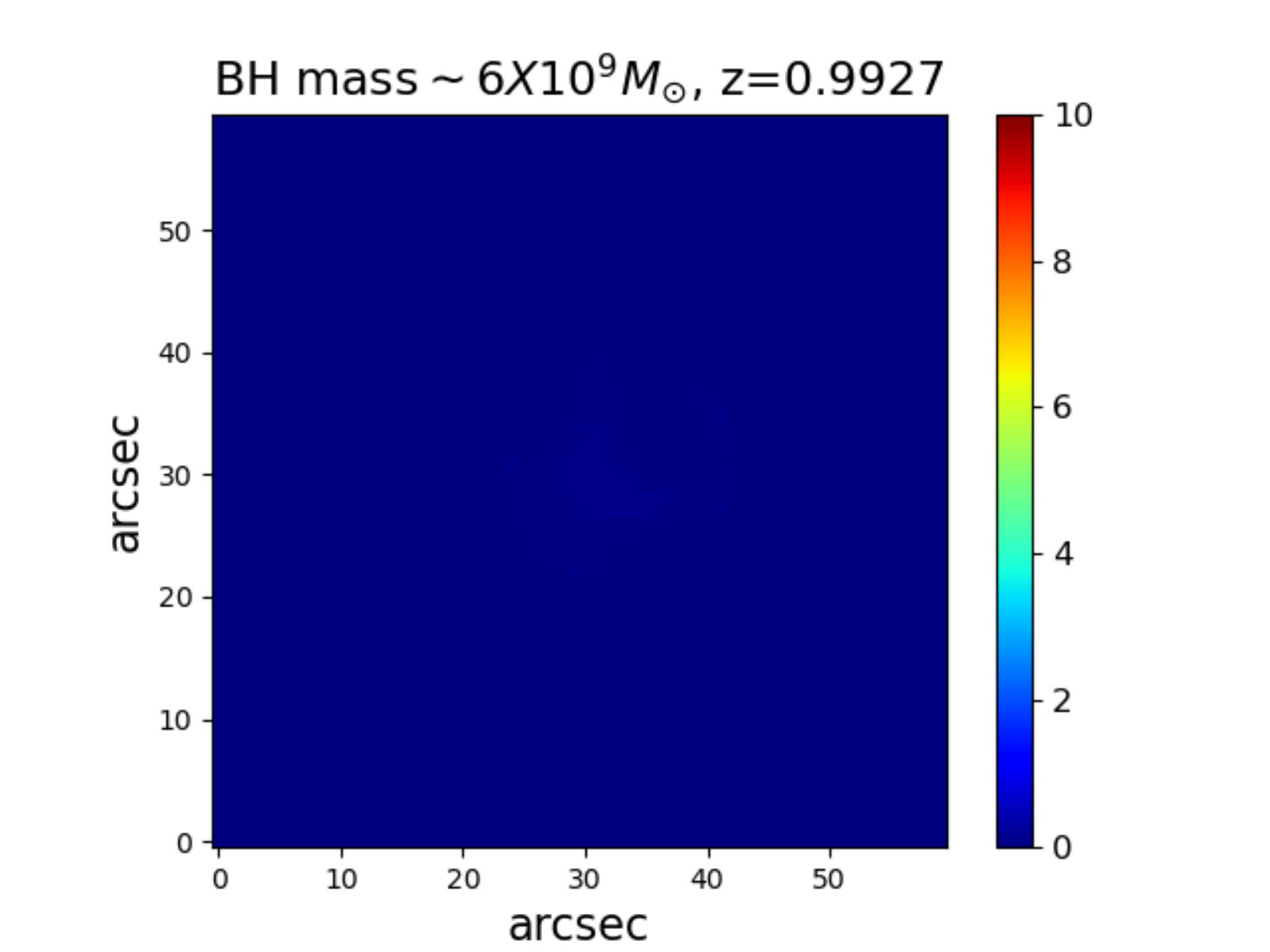}&\includegraphics[width=4.5cm]{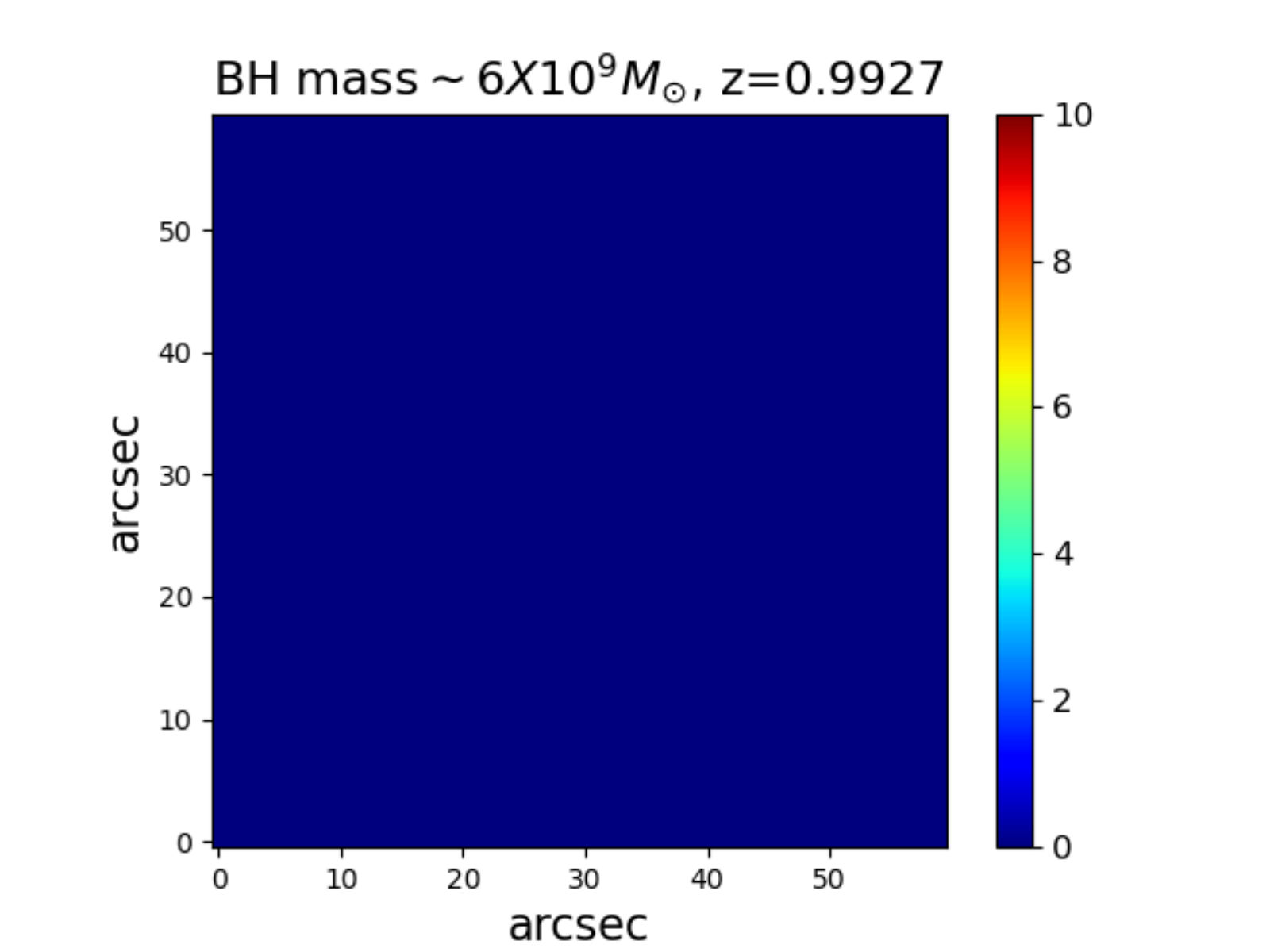}\\
         \includegraphics[width=4.5cm]{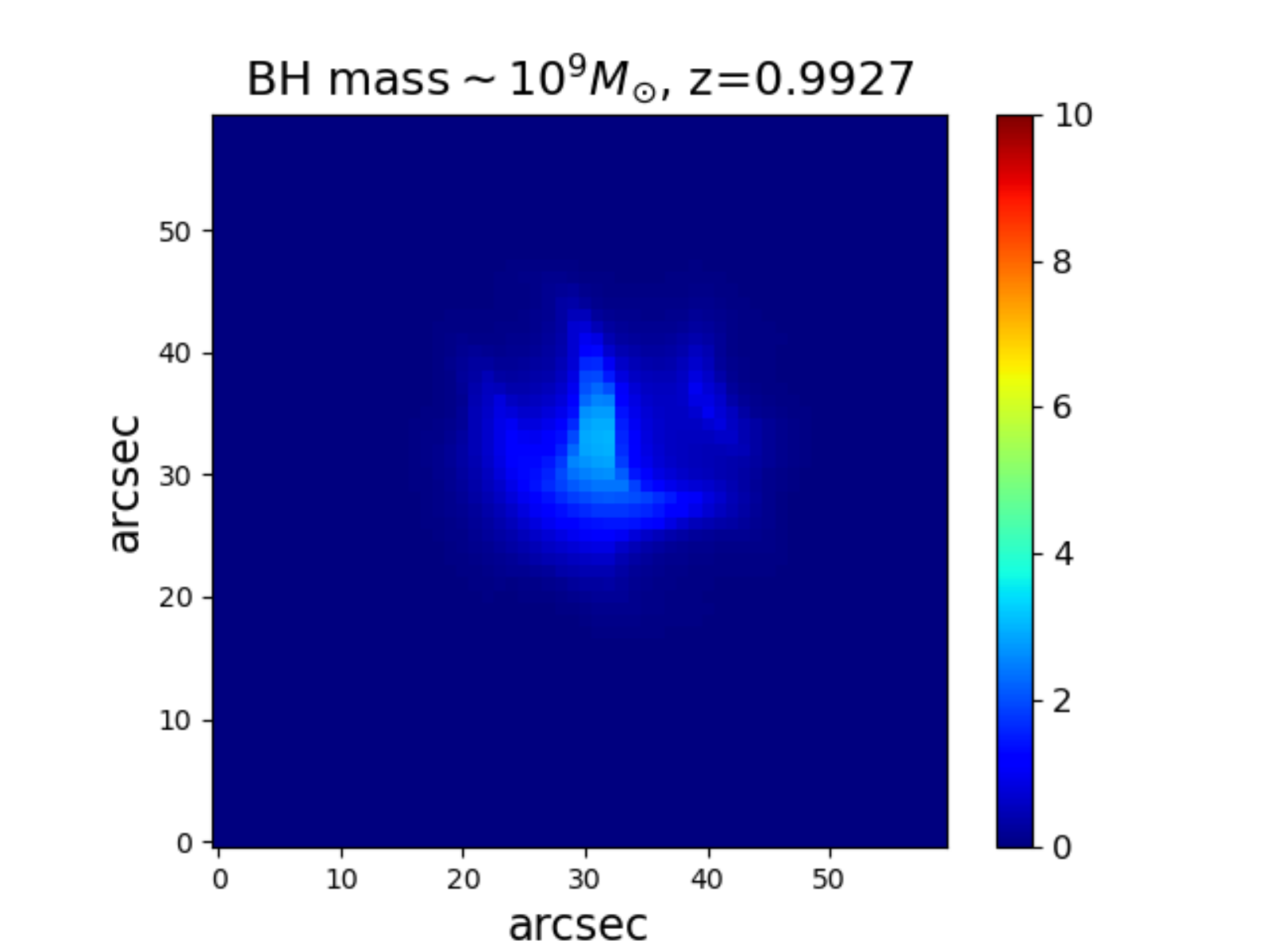}&\includegraphics[width=4.5cm]{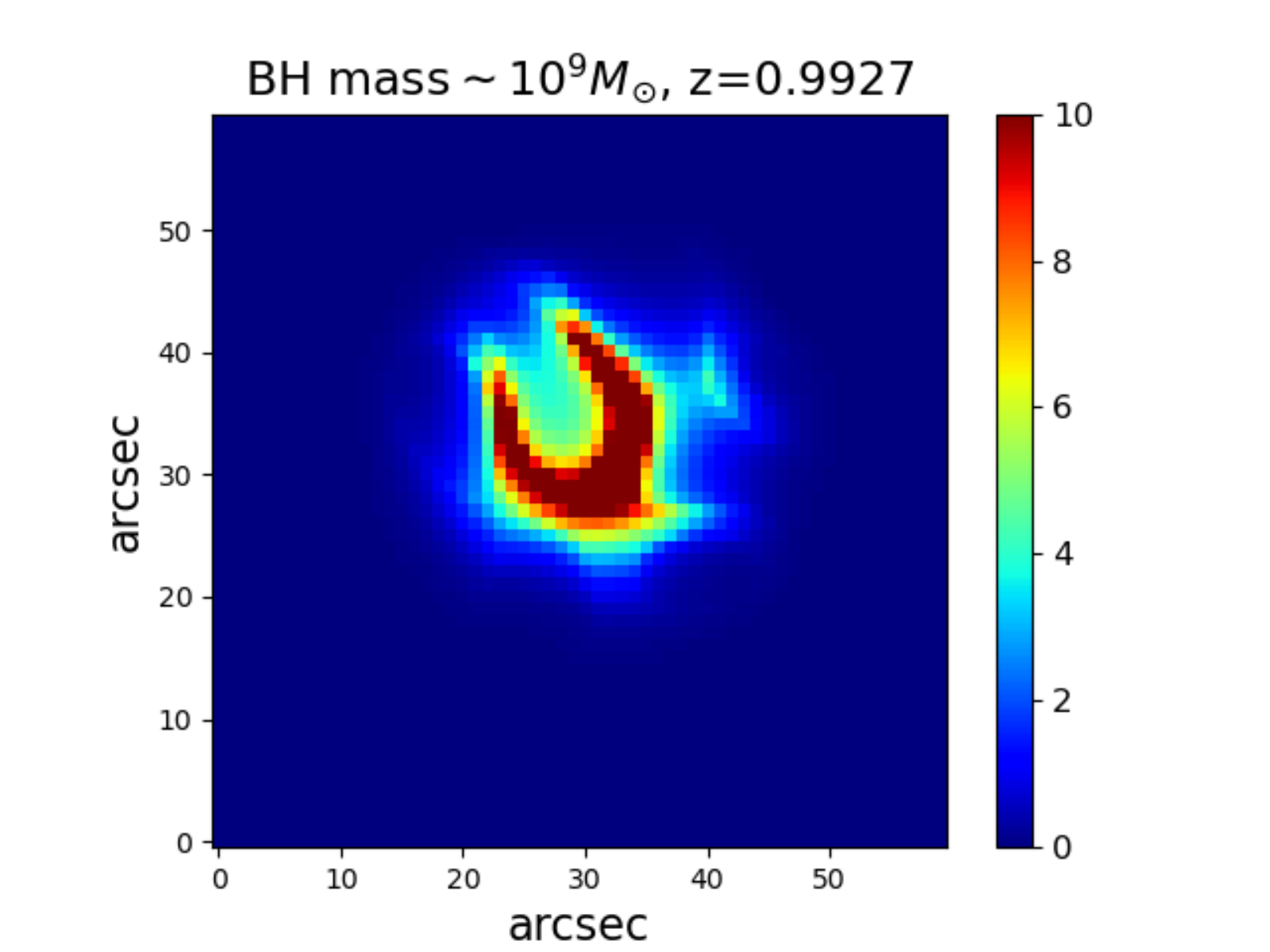}&\includegraphics[width=4.5cm]{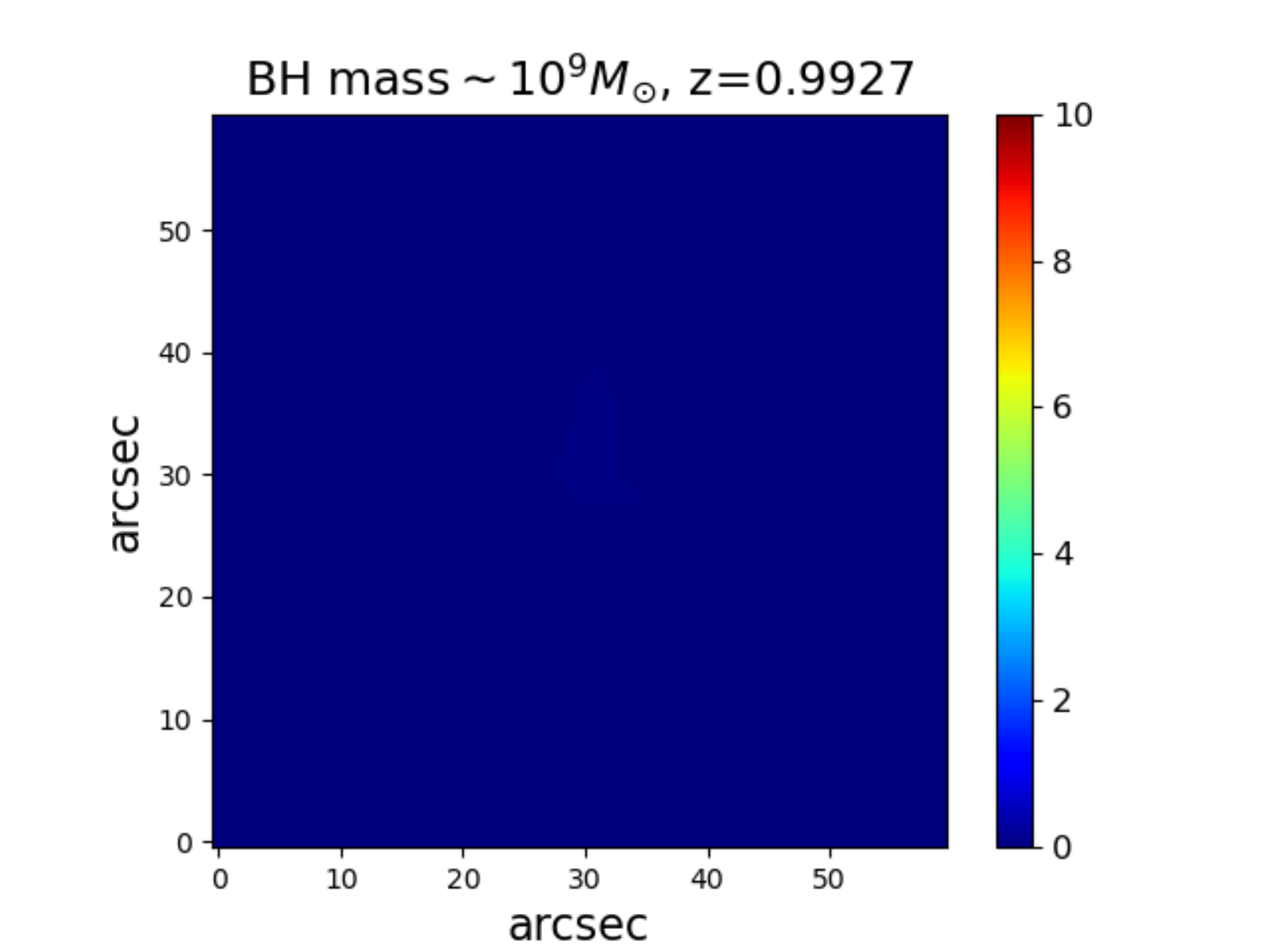}&\includegraphics[width=4.5cm]{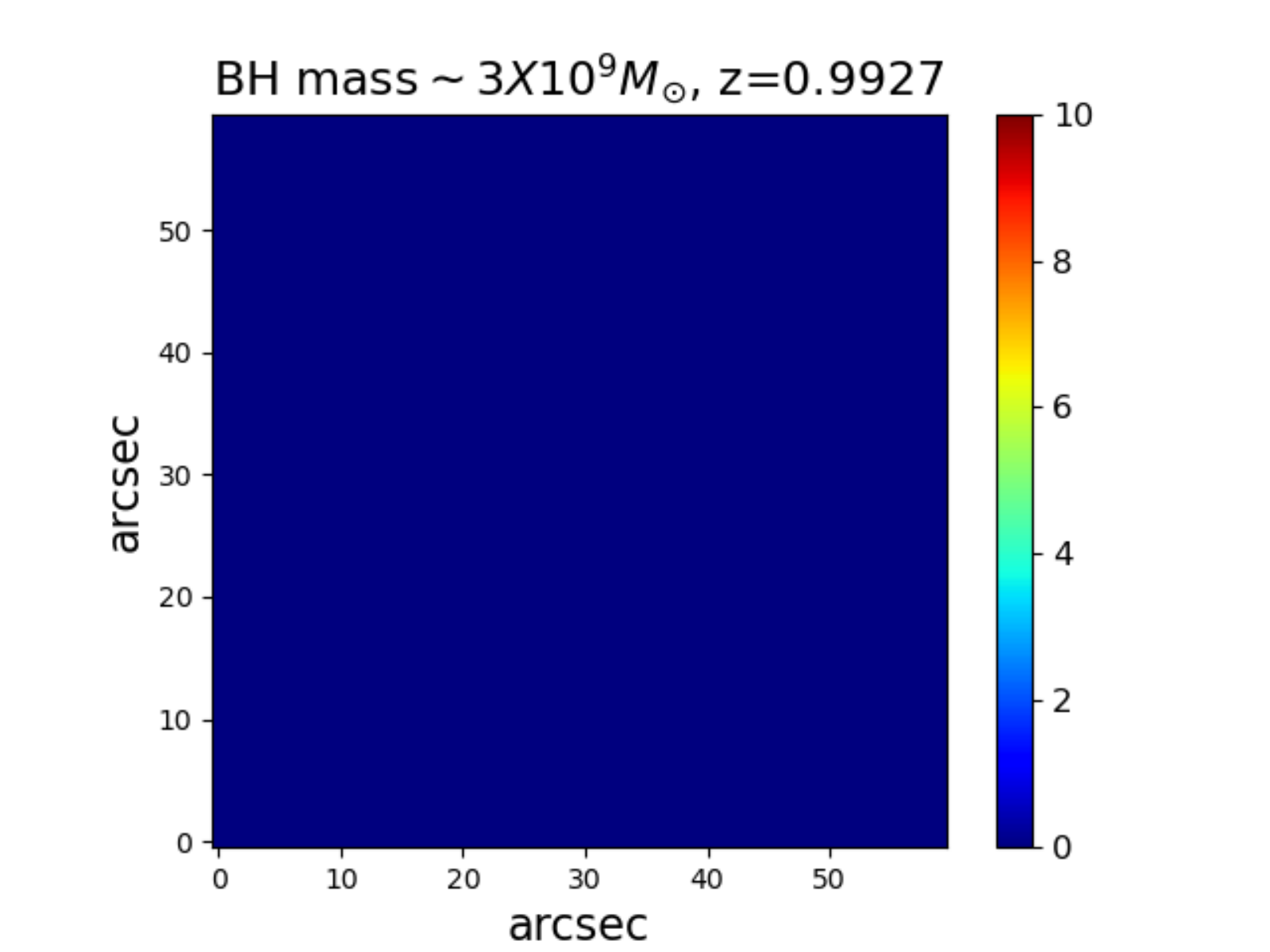}\\
      \includegraphics[width=4.5cm]{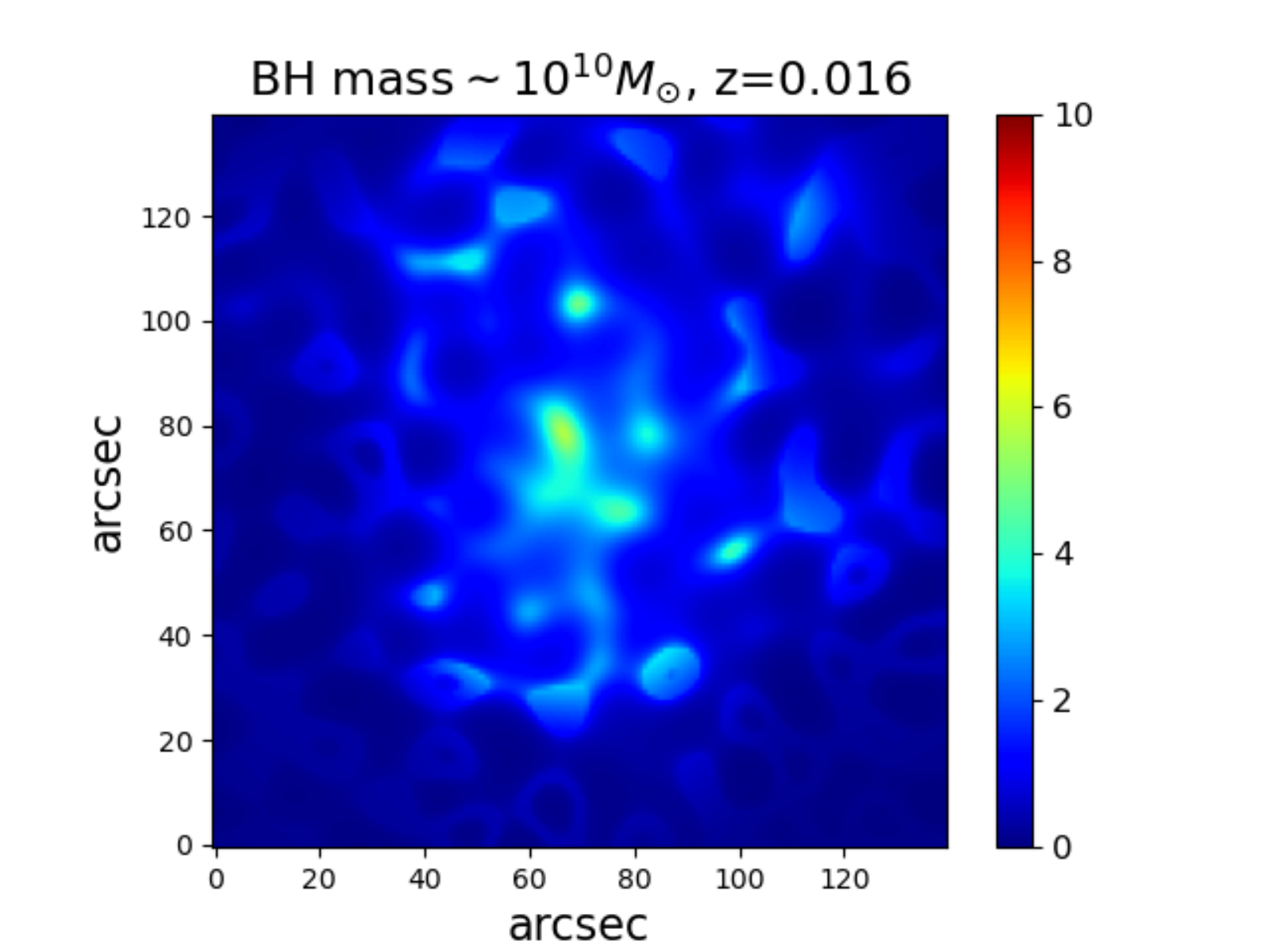}&\includegraphics[width=4.5cm]{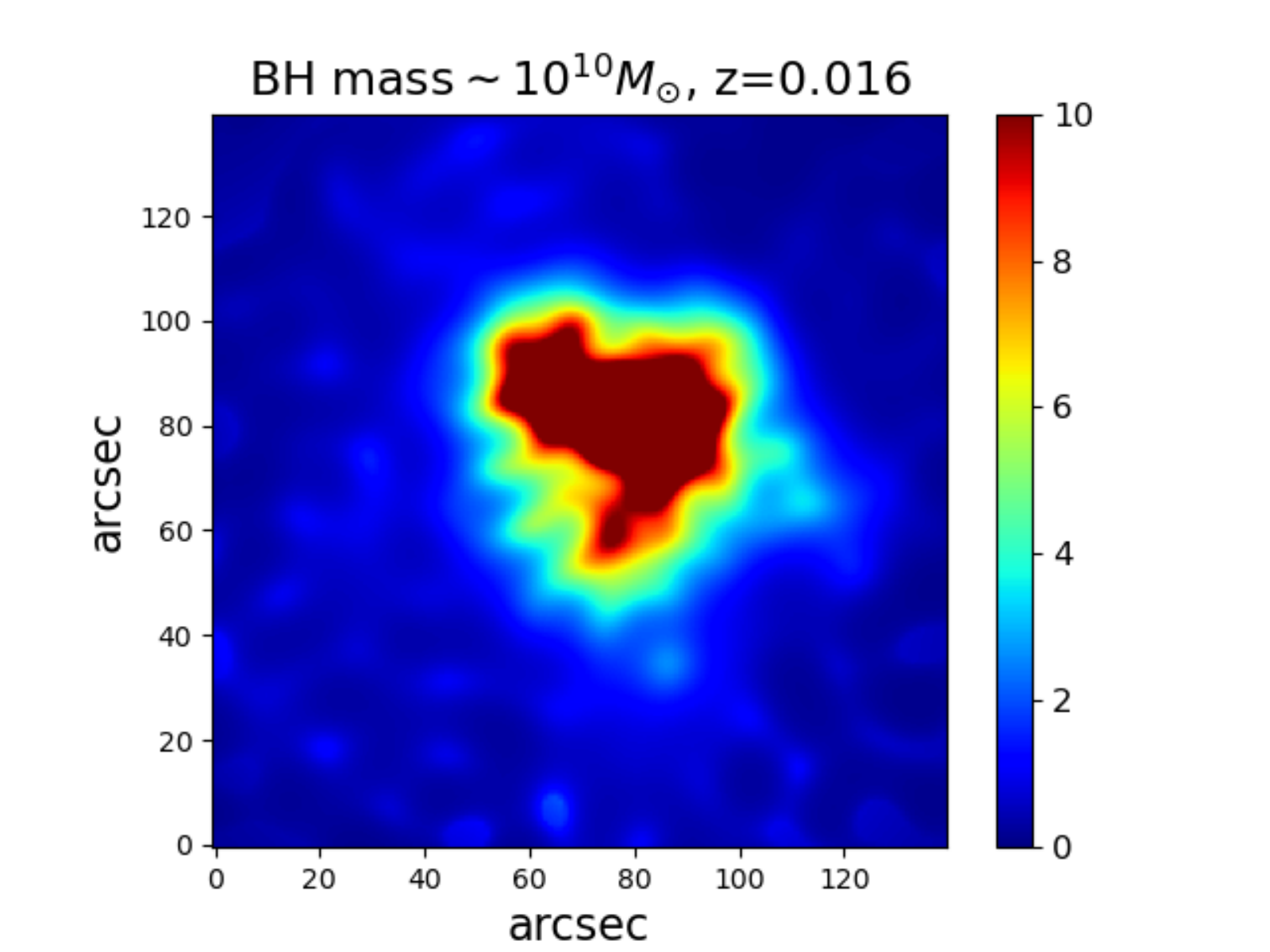}&\includegraphics[width=4.5cm]{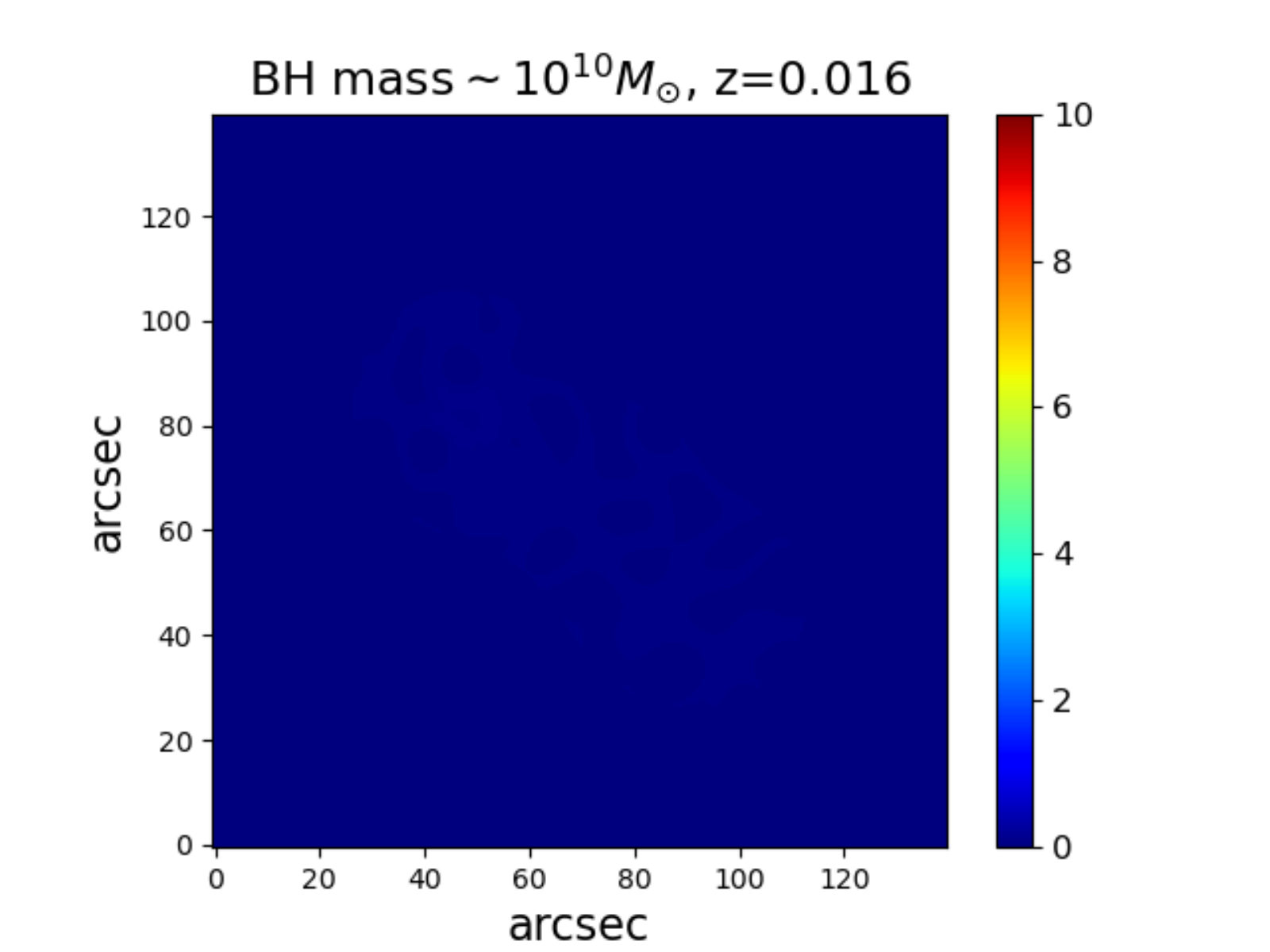}&\includegraphics[width=4.5cm]{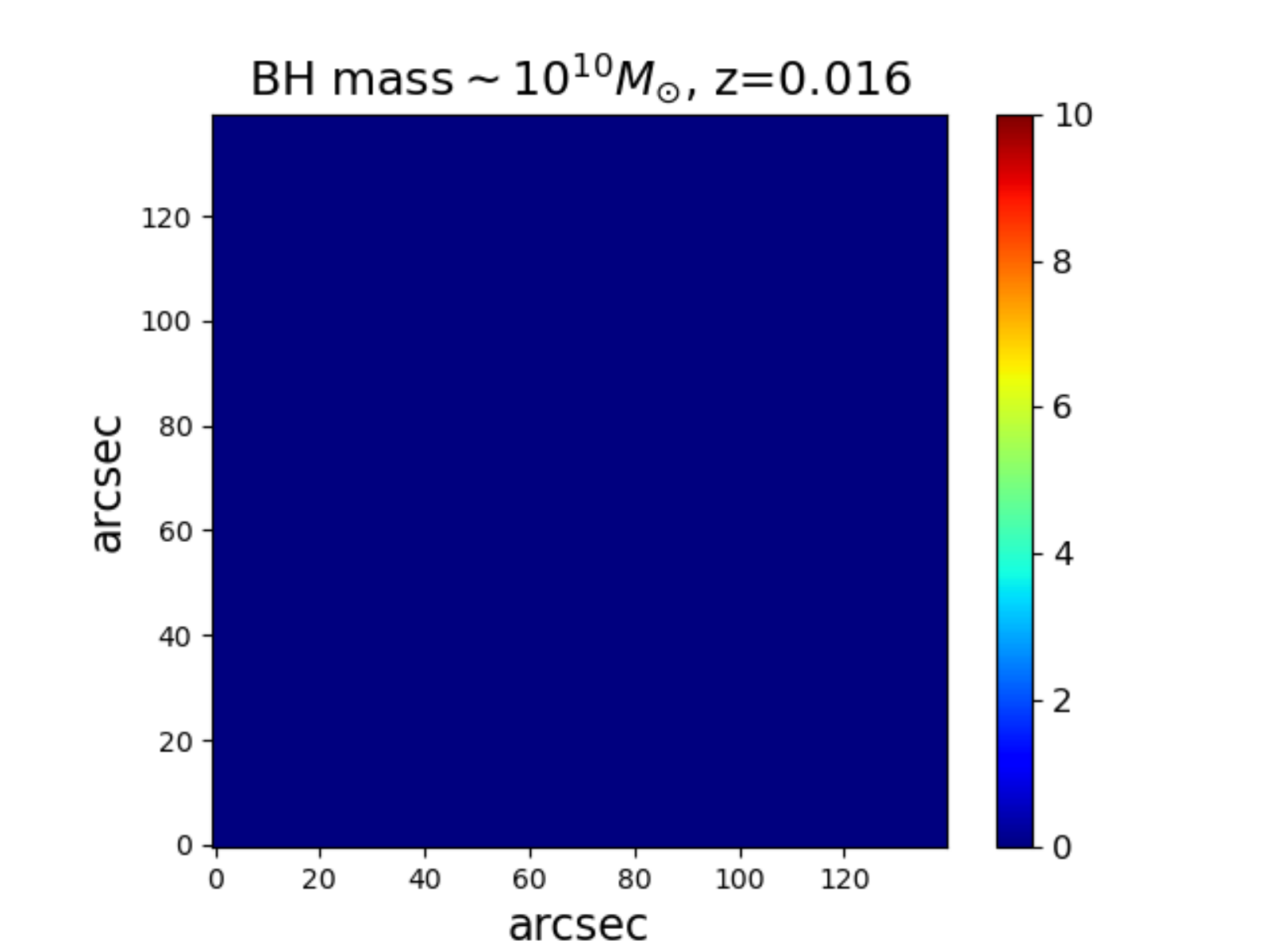}\\
      \includegraphics[width=4.5cm]{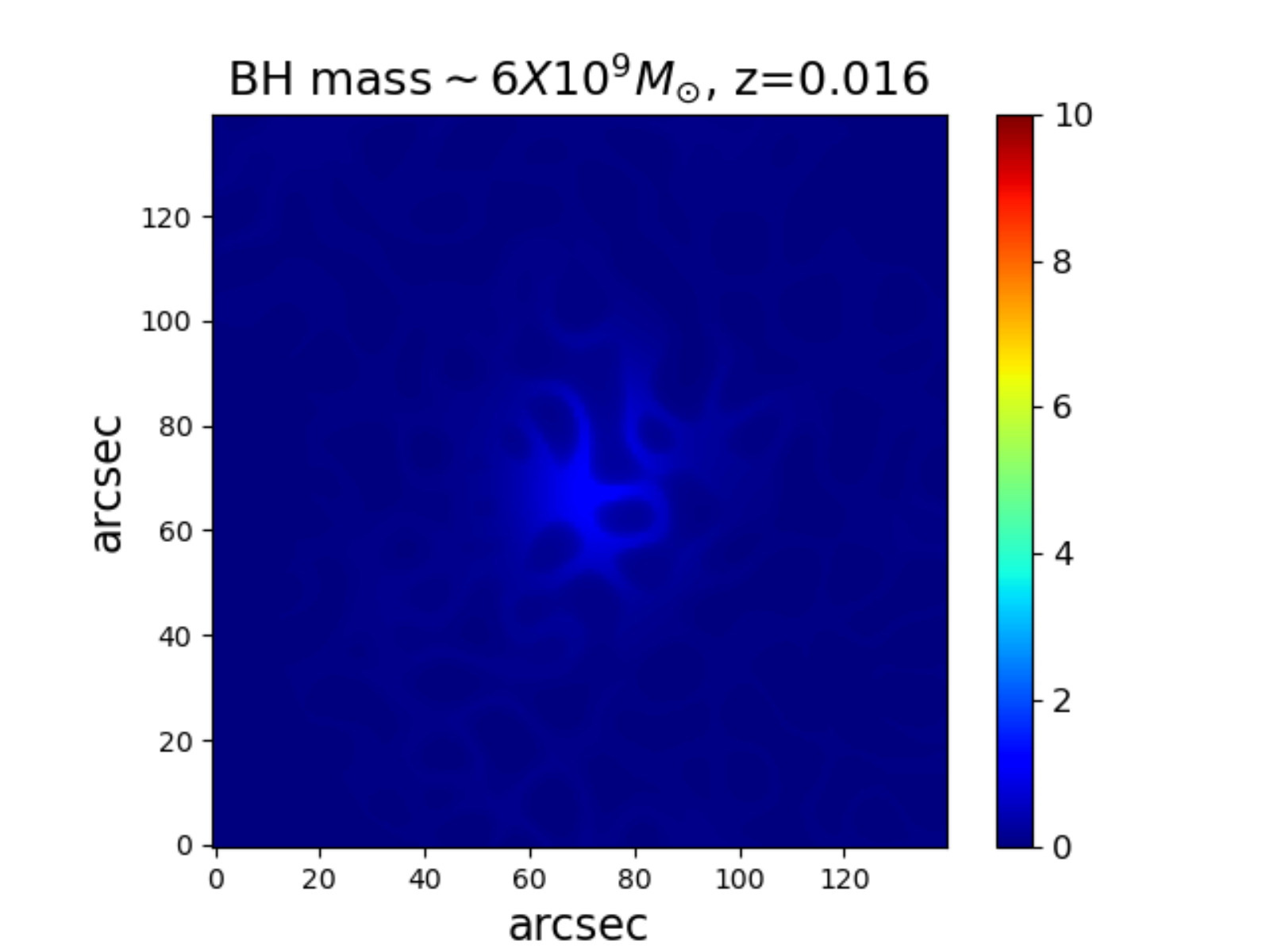}&\includegraphics[width=4.5cm]{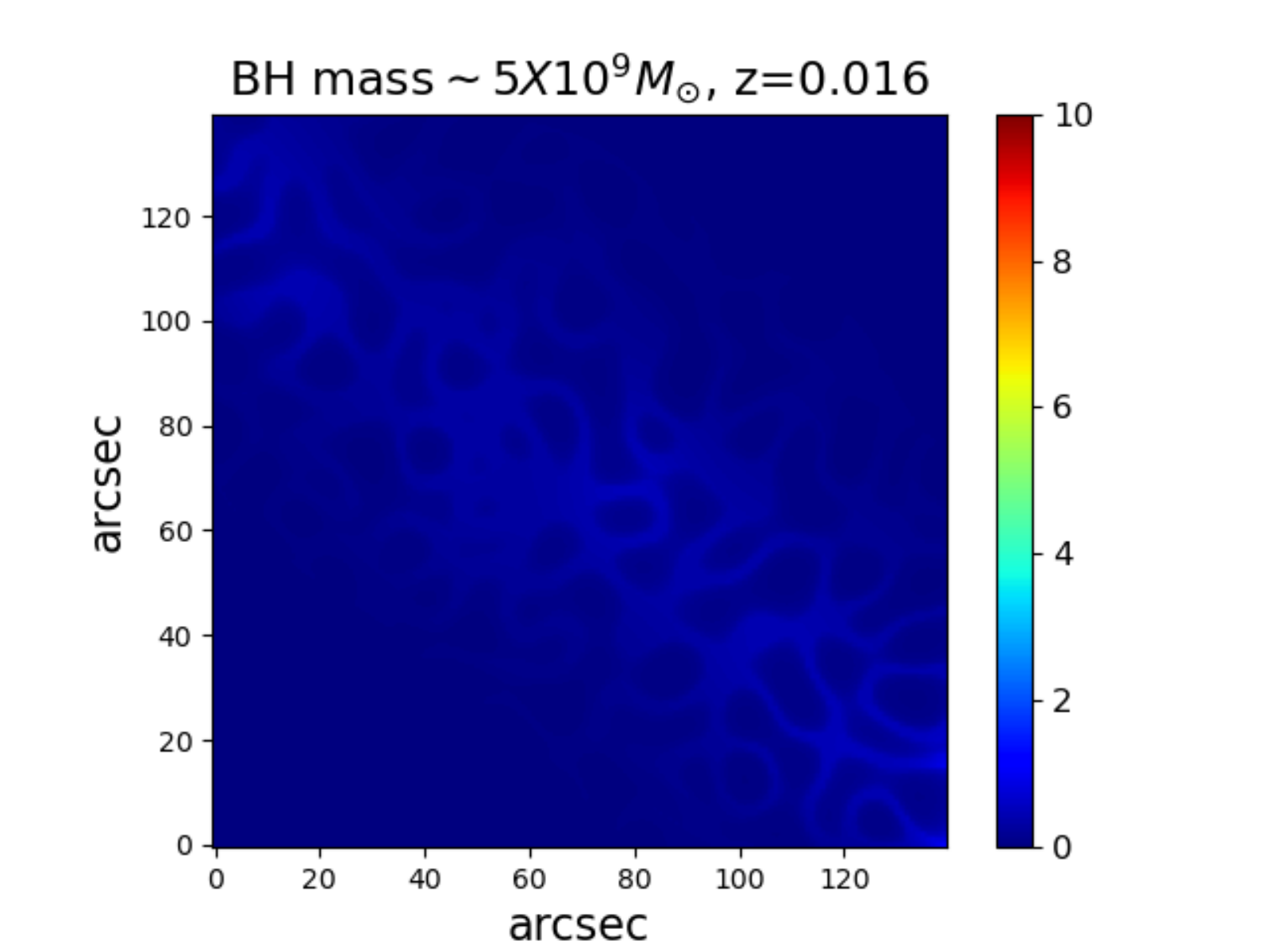}&\includegraphics[width=4.5cm]{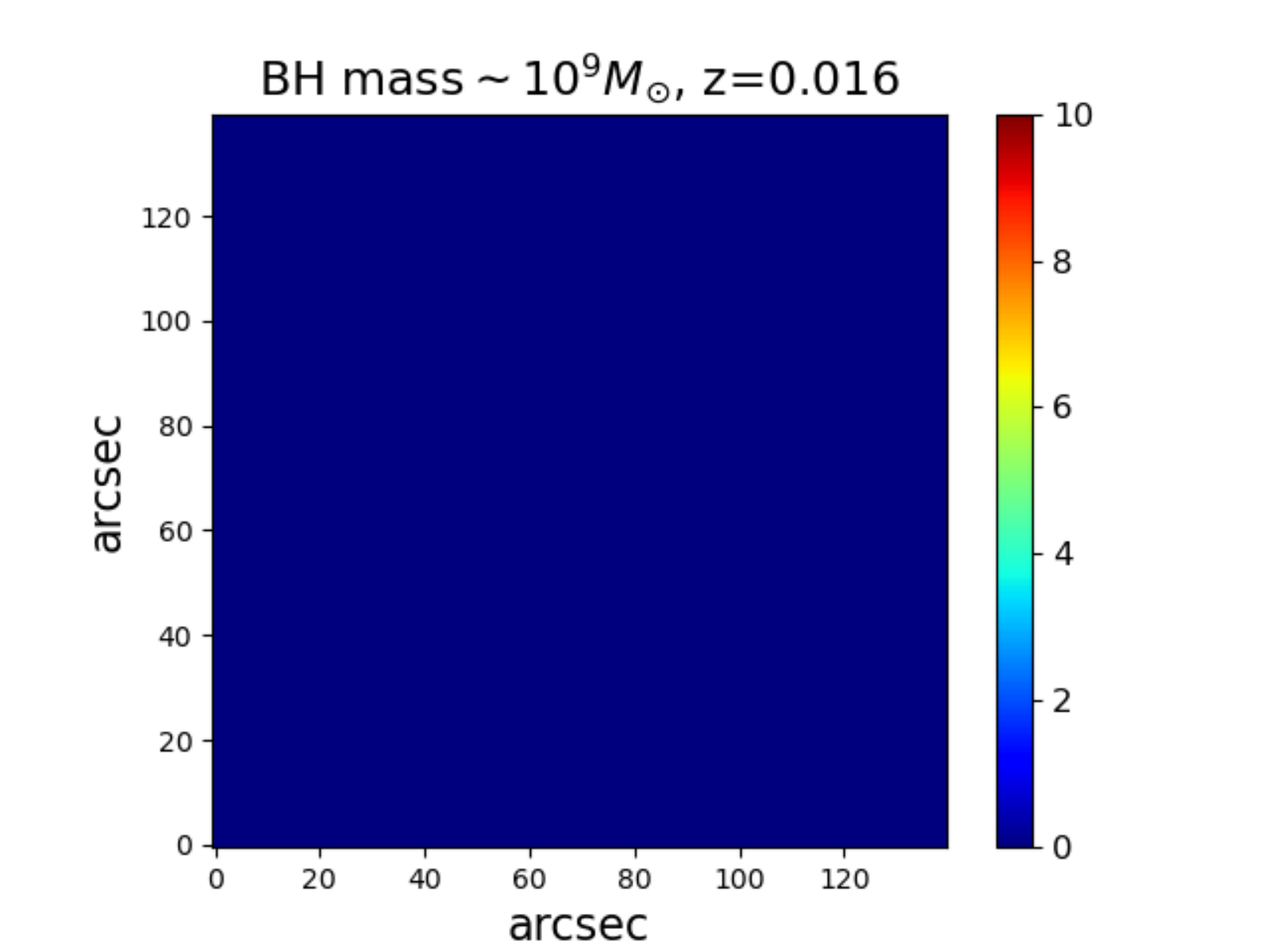}&\includegraphics[width=4.5cm]{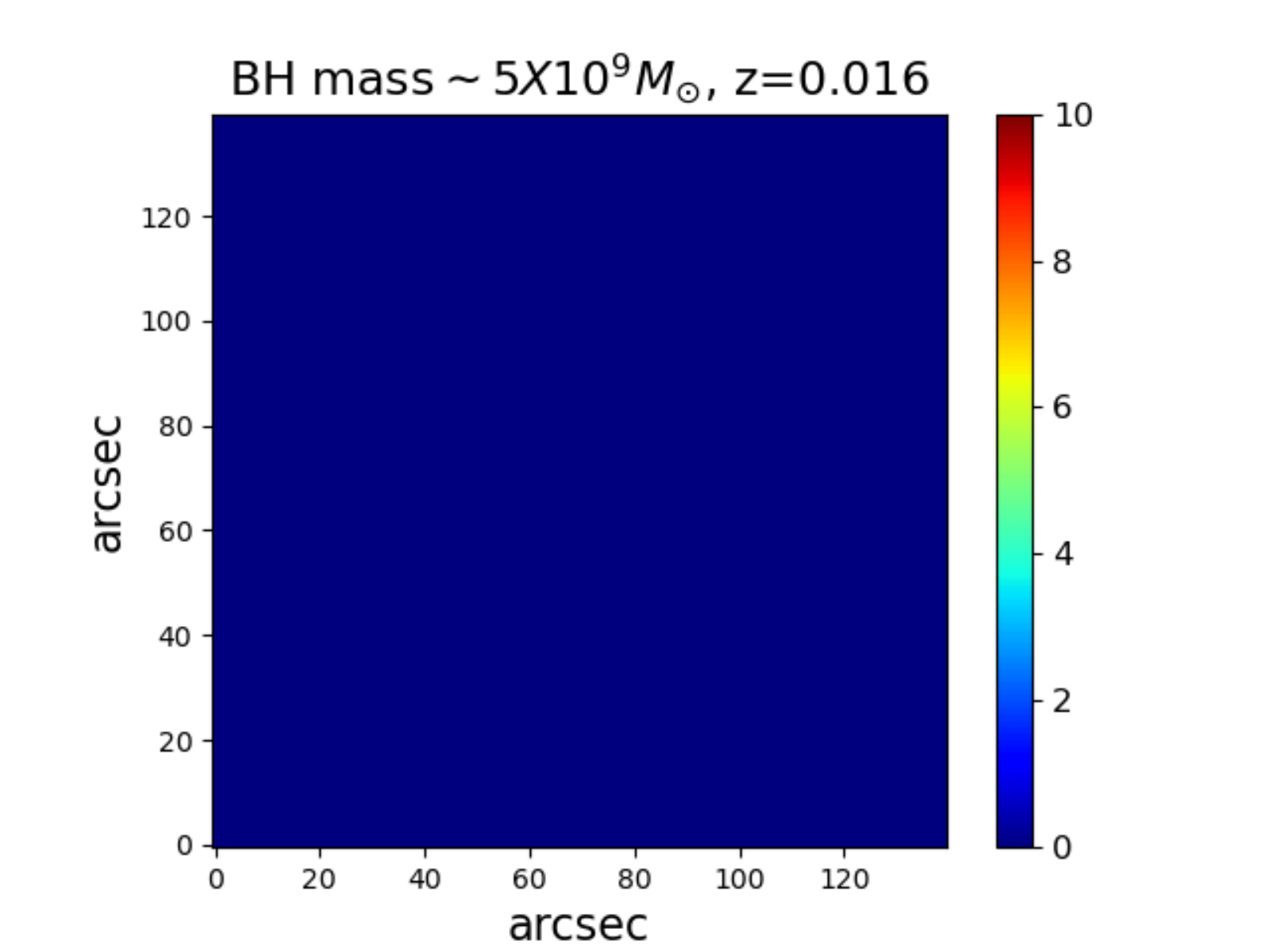}\\
      \end{tabular}
        \caption{The fidelity (signal to noise) maps of the mock ALMA tSZ simulations at 320 GHz for different feedback modes around the most massive and the most active BHs at two different redshifts from \cite{dave19} corresponding to Fig.\ 4. {\bf Top Panel} The fidelity maps for the no feedback, no jet feedback, no X-ray feedback, and all feedback modes respectively around most massive BH at z$\sim$1. {\bf Second Panel} The mock ALMA tSZ maps for no feedback, no jet feedback, no X-ray feedback, and all feedback modes respectively around the most active BH at z$\sim$1. {\bf Third Panel} Same as the top panel but now at z=0.016. {\bf Fourth Panel} Same as the second panel but now at z=0.016.}
        \label{fig:5}
    \end{center}
\end{figure*}

To construct the synthetic observations, we use the Common Astronomy Software Application (CASA; \citealt{jager08}) which is a suite of tools for calibration, imaging and analysis in radio astronomy for both interferometric and single dish configurations. The package takes a model image for a patch of the sky as input and turn it into an observation from multiple viewing angles. For our purpose we employed the most compact configuration for ALMA cycle 8.1. Our observational specifications are summarized in Table 3. 

The theoretical tSZ maps and the telescope configuration are convolved through `simalma' task in CASA. `simalma' is a combination of both `simobserve' and `simanalyze'. `simobserve' is used to simulate observations with ALMA and the Atacama Compact Array (ACA), and generate simulated visibility maps, whereas `simanalyze' is used to generate images from the simulated visibility results. We used central observing frequency at 320 GHz (ALMA band 7) with 7.5 GHz bandwidth. We also set a pointing position of the observation (Epoch J2000, RA 13:29:53.94, DEC -047:11:41.0) to prevent, the simulator from obtaining a  mosaic to cover the value of the mapsize parameter. Precipitable water vapor (pwv) is set to 0.5 mm to represent observations in nominal weather. Based on these settings, the simulation added noise to the data. For both $z\sim$1 and $z=0.016$, synthetic maps are constructed using a total observation time of 3 hours with an integration time of 30 seconds. The primary beam size for ALMA band 7 is $\sim$ 15 arcsec. The cell size was chosen to be 0.1 arcsec, about 20\% of the synthesized beamsize (1-arcsec). Each image is constructed with $300 \times 300$ pixels. 

\section{Results}\label{section_3}

We now present the tSZ signals predicted from our simulations for different modes of feedback. We note that in our results  `all feedback' mode includes radiative, jet and X-ray feedback, no-jet feedback mode turns off the jet and X-ray feedback and includes only radiative feedback. For the no X-ray feedback case, only X-ray feedback mode is turned off and it includes jet and radiative feedback. The nomenclature is summarized in Table 1. 

\begin{figure*}
    \begin{center}
     \begin{tabular}{c |ccc}
     \hline
         &{135 GHz}&{100 GHz}&{42 GHz}\\[0.1pt]
         \hline
      {No}&\includegraphics[width=4.5cm]{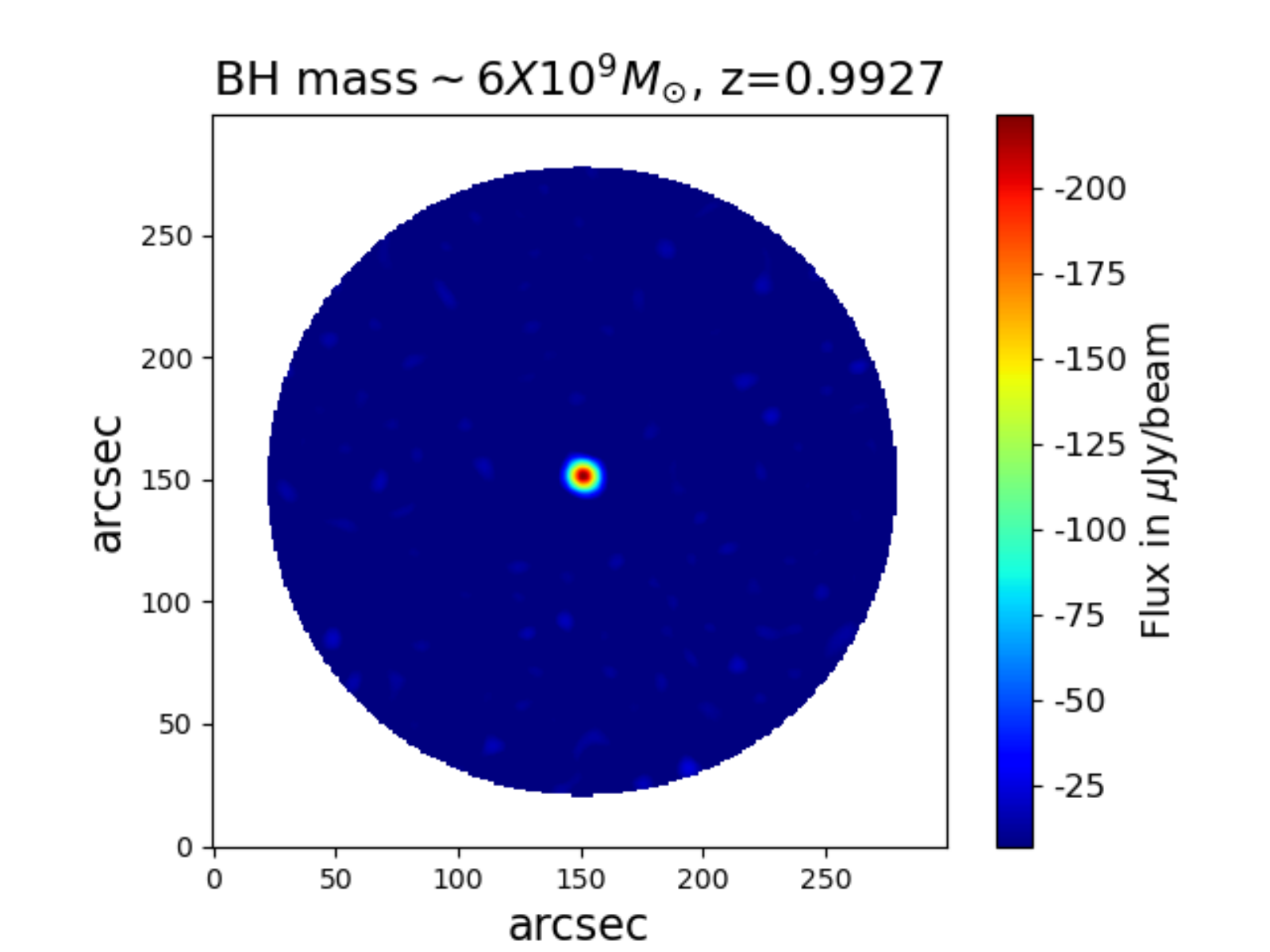}&\includegraphics[width=4.5cm]{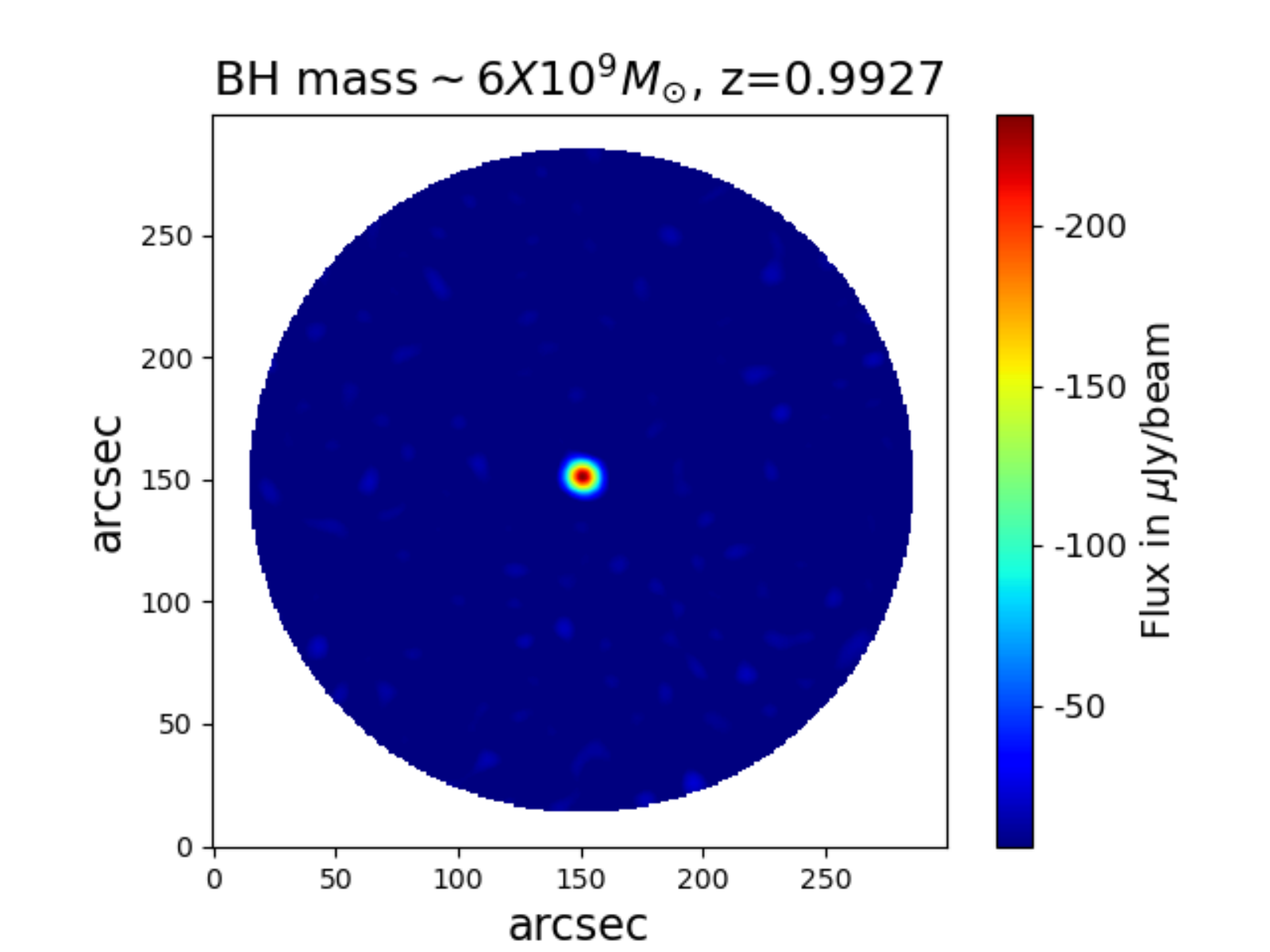}&\includegraphics[width=4.5cm]{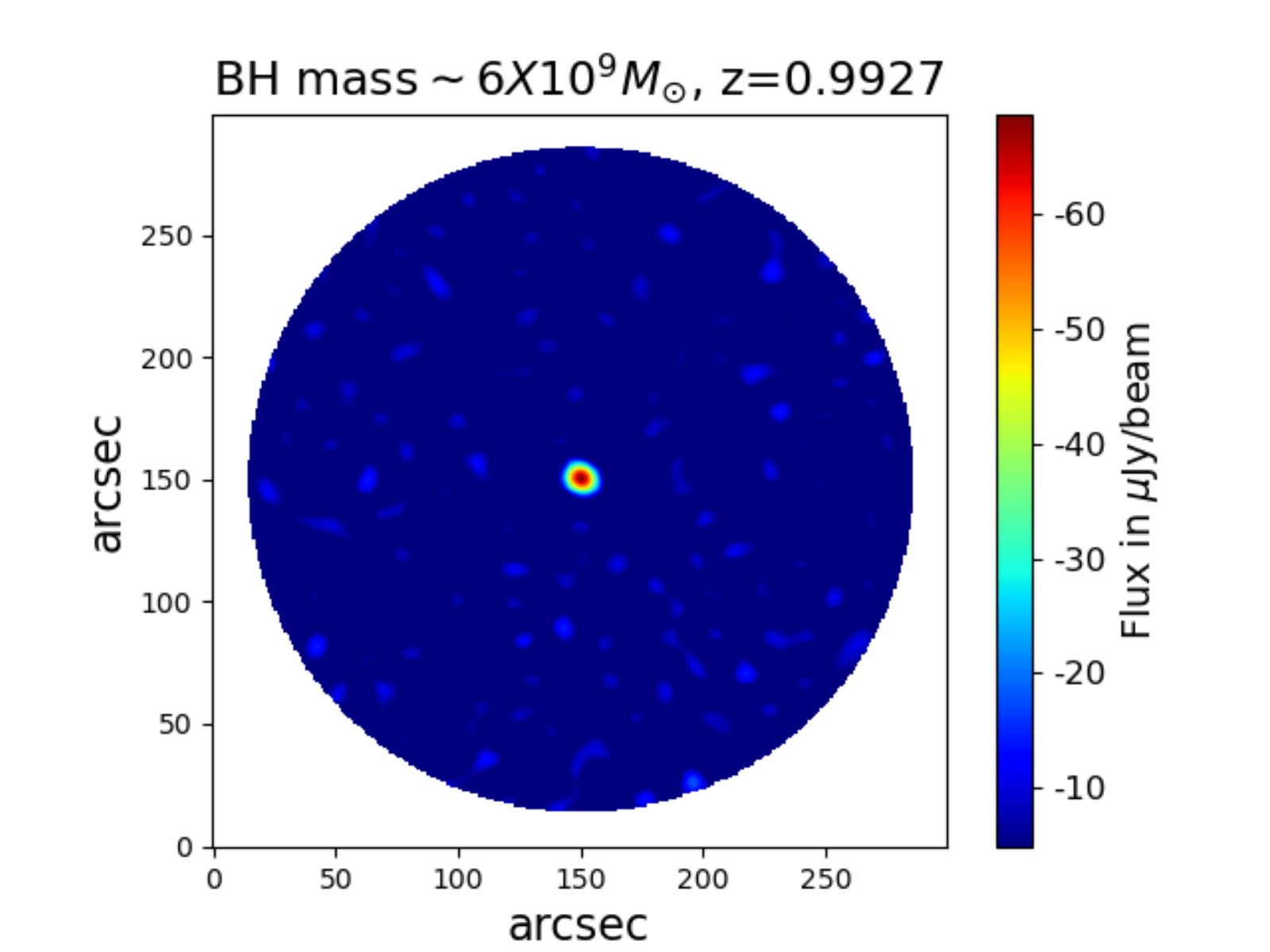}\\
       &\includegraphics[width=4.5cm]{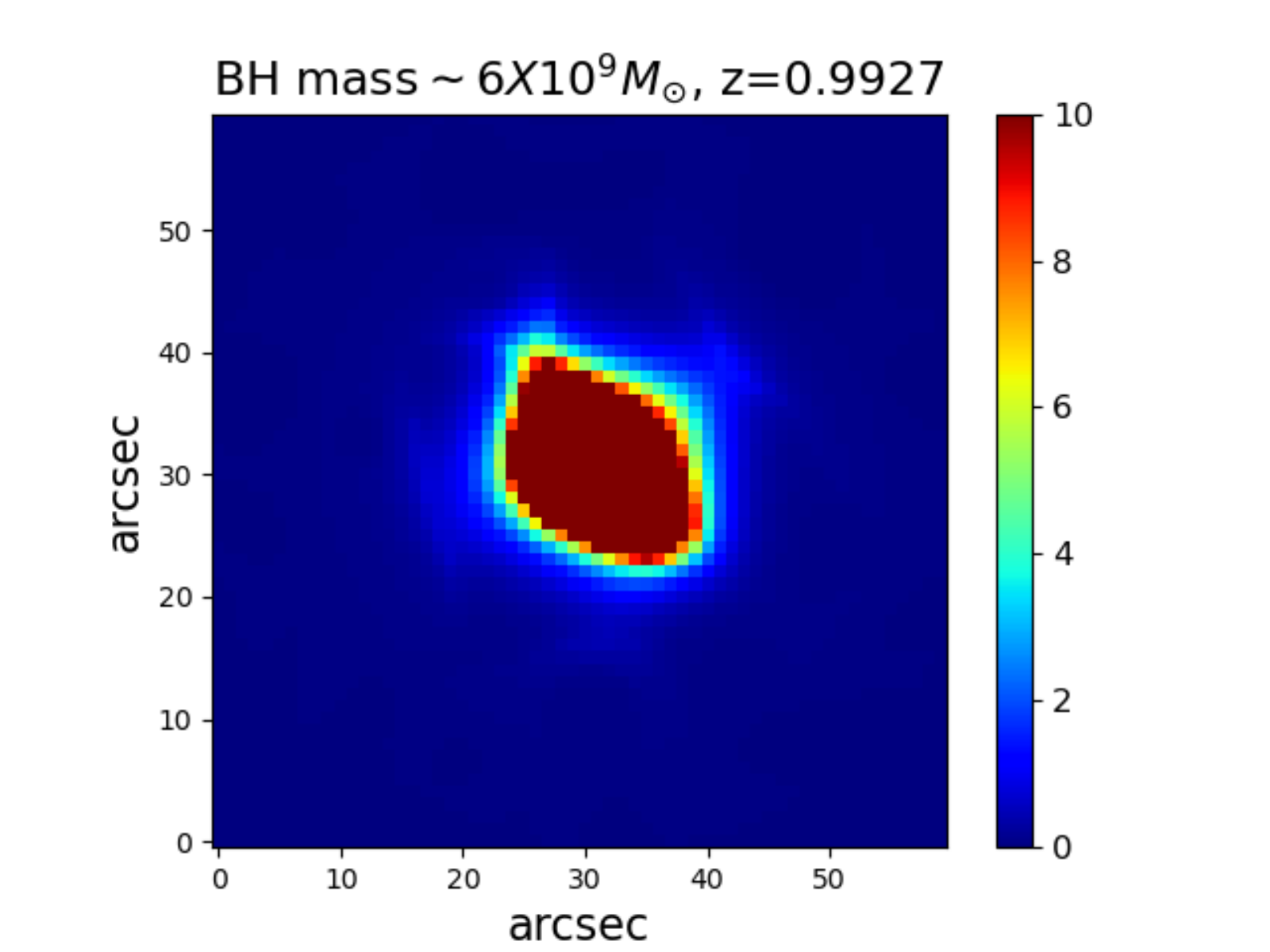}&\includegraphics[width=4.5cm]{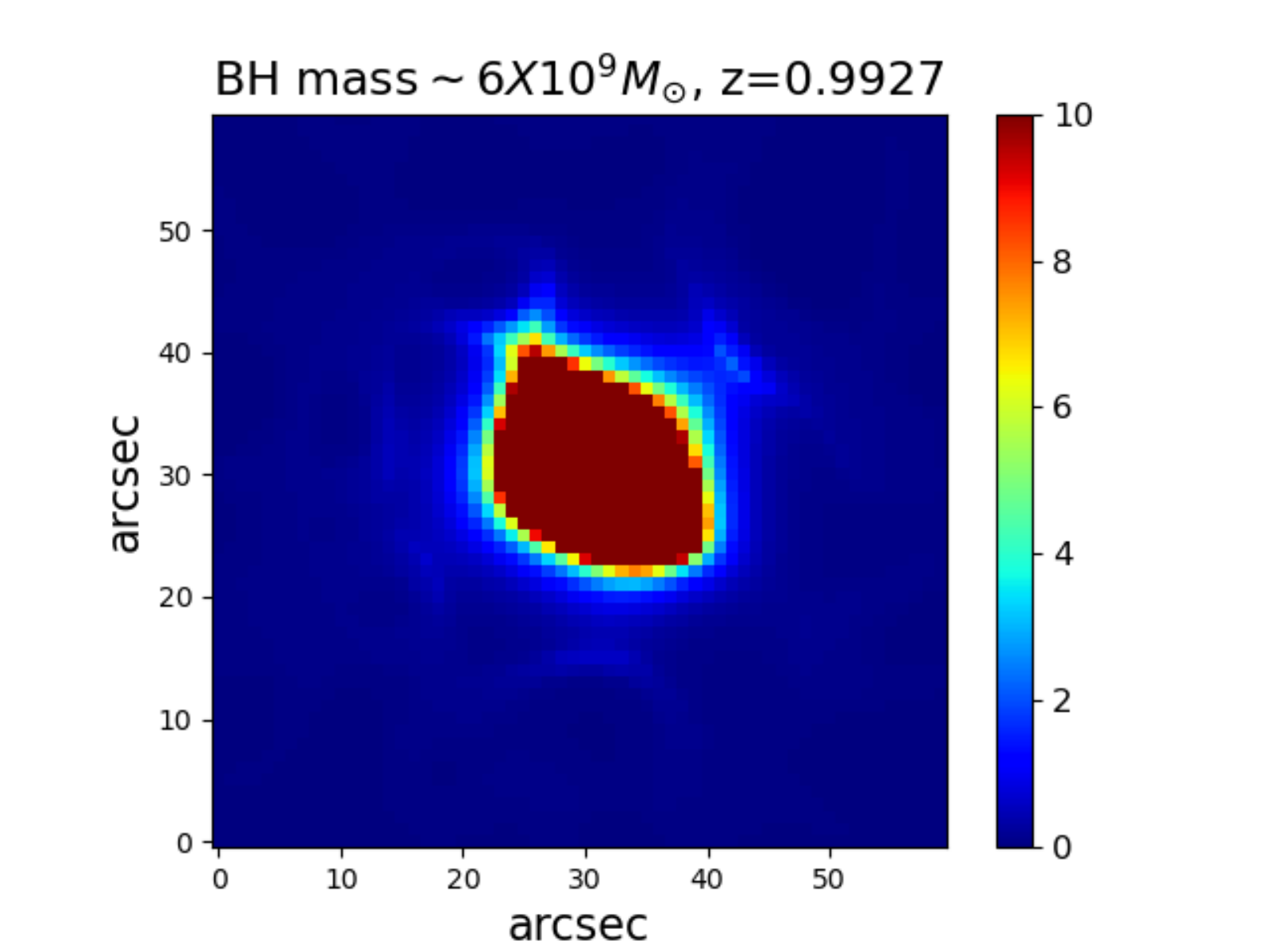}&\includegraphics[width=4.5cm]{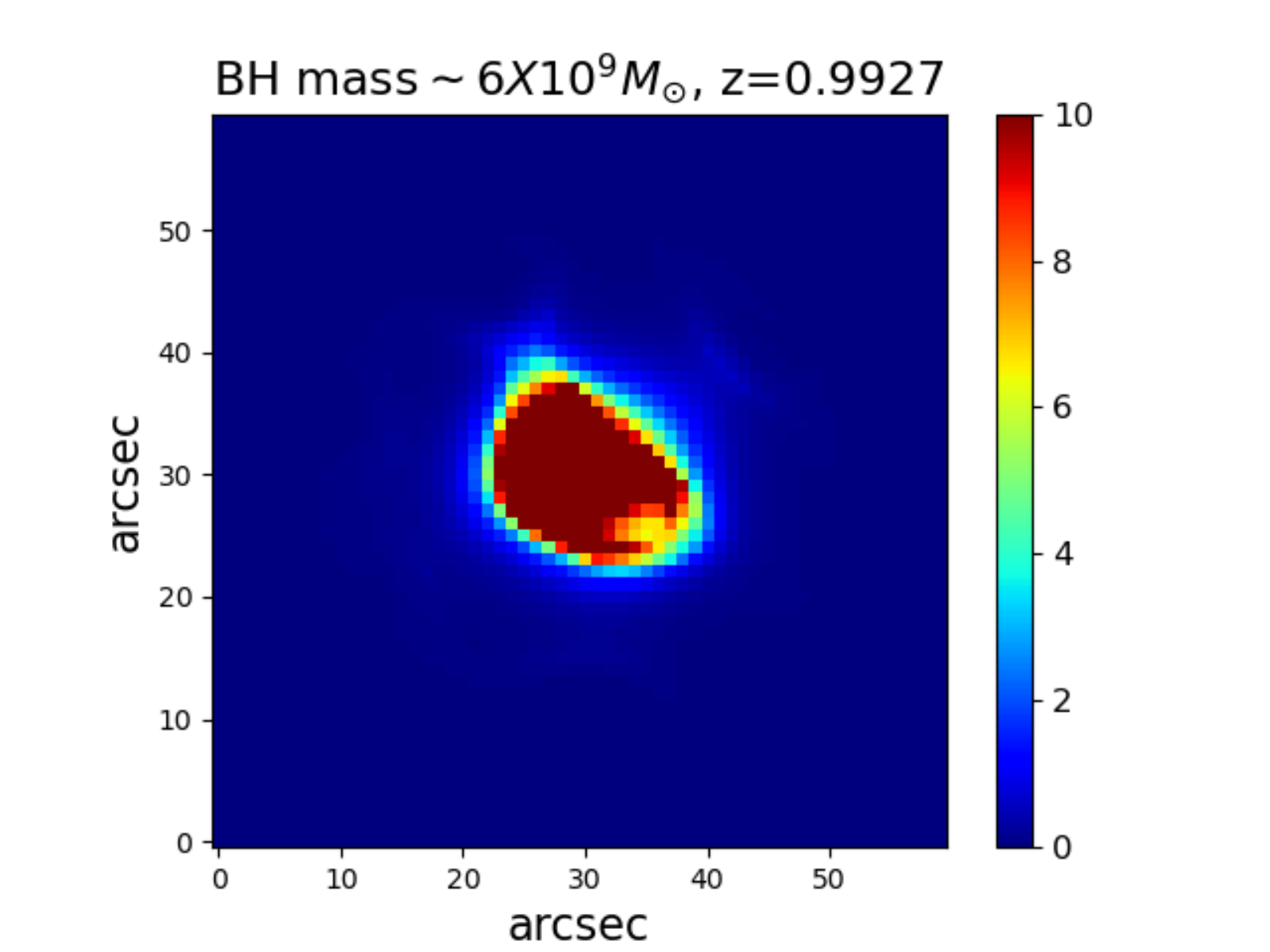}\\
       \hline
       {No-Jet}&\includegraphics[width=4.5cm]{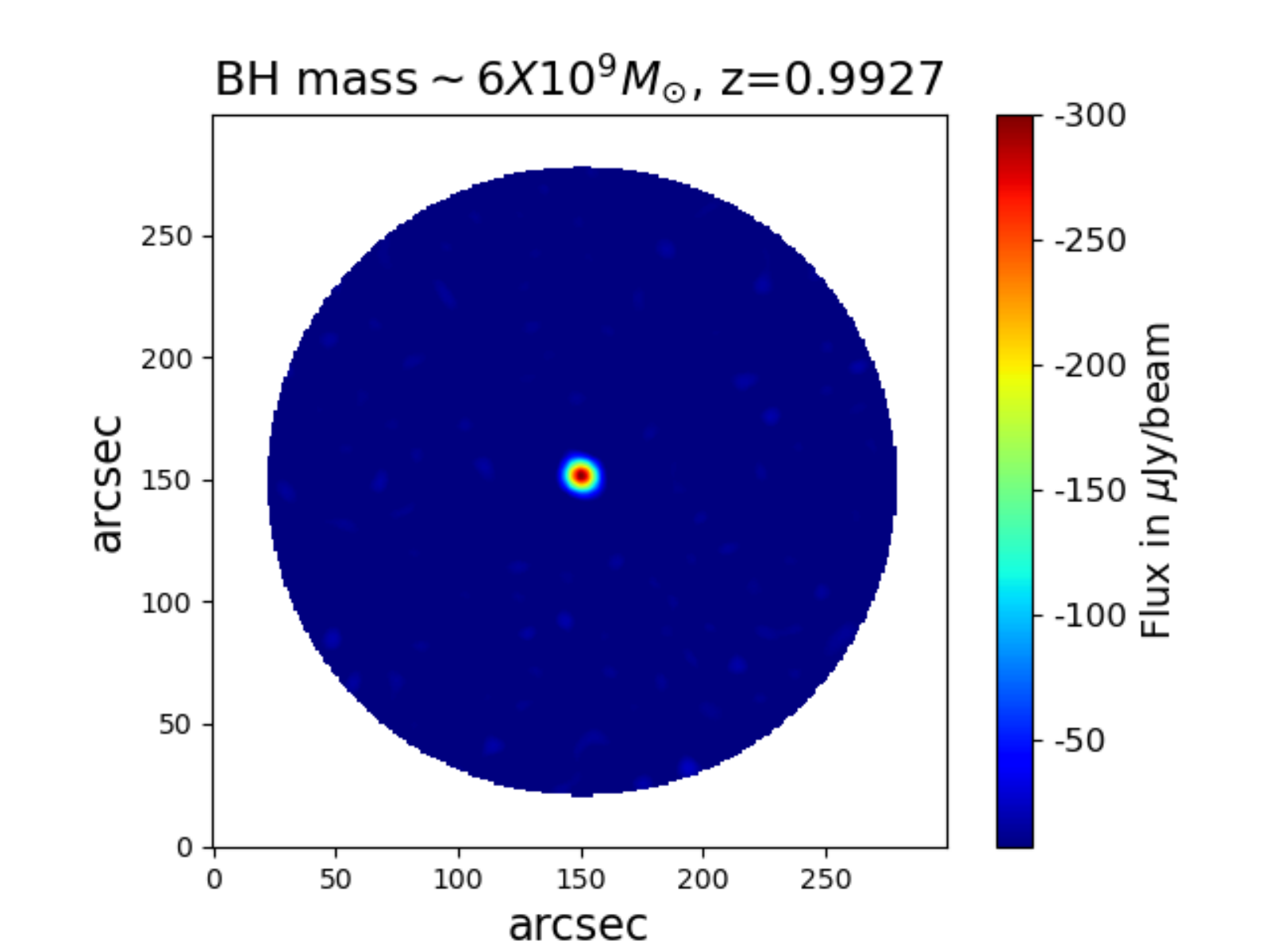}&\includegraphics[width=4.5cm]{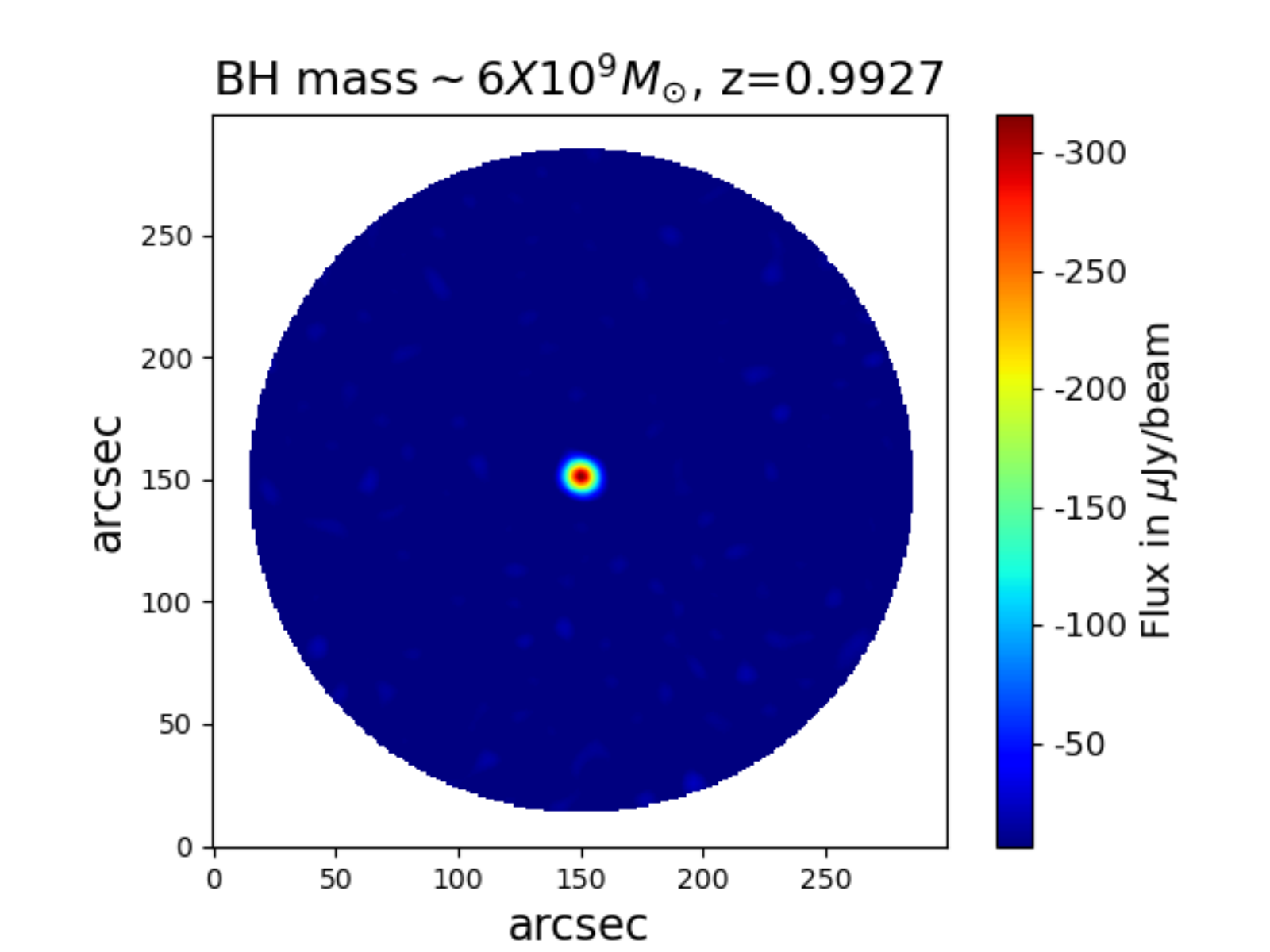}&\includegraphics[width=4.5cm]{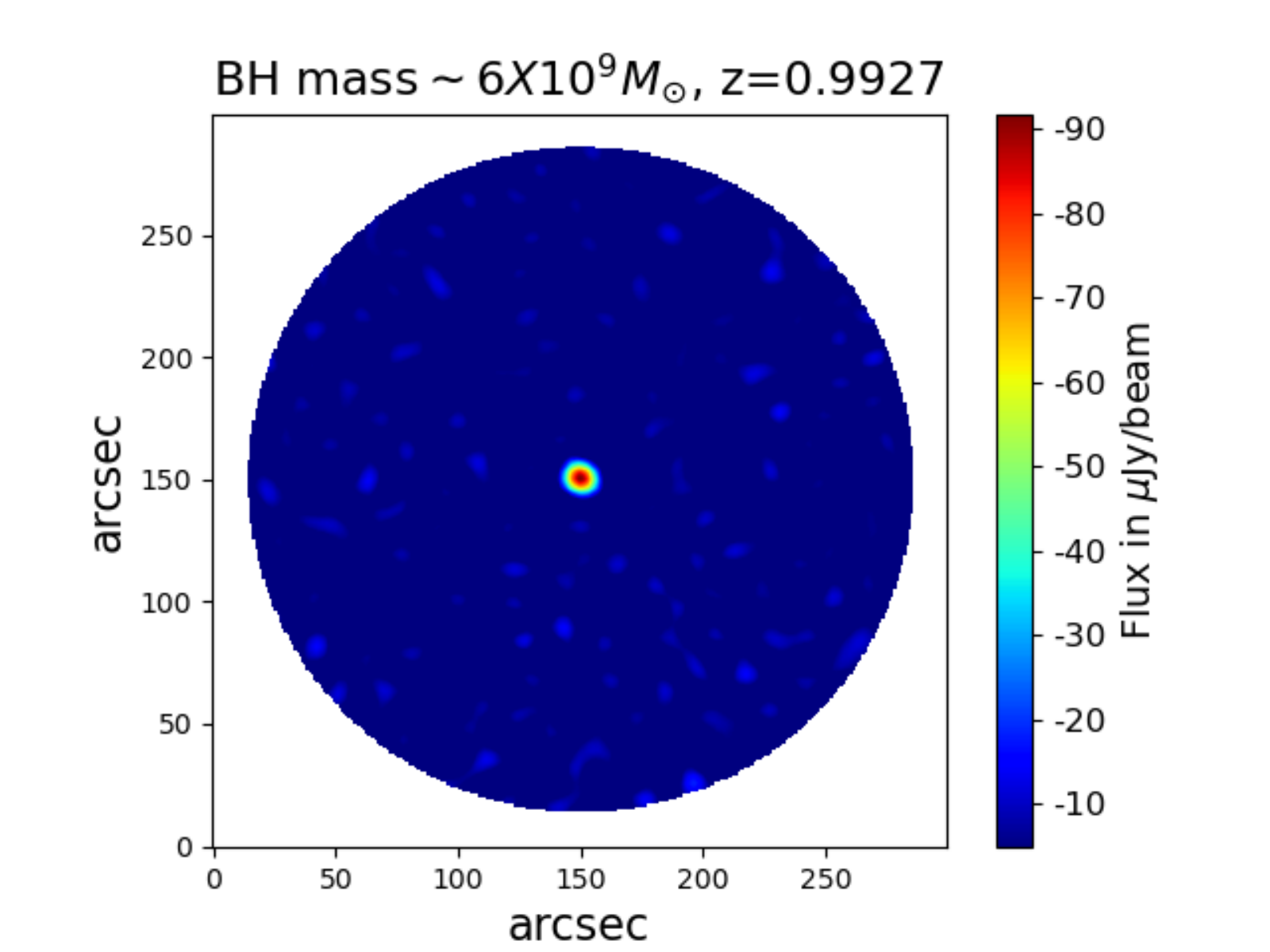}\\
        &\includegraphics[width=4.5cm]{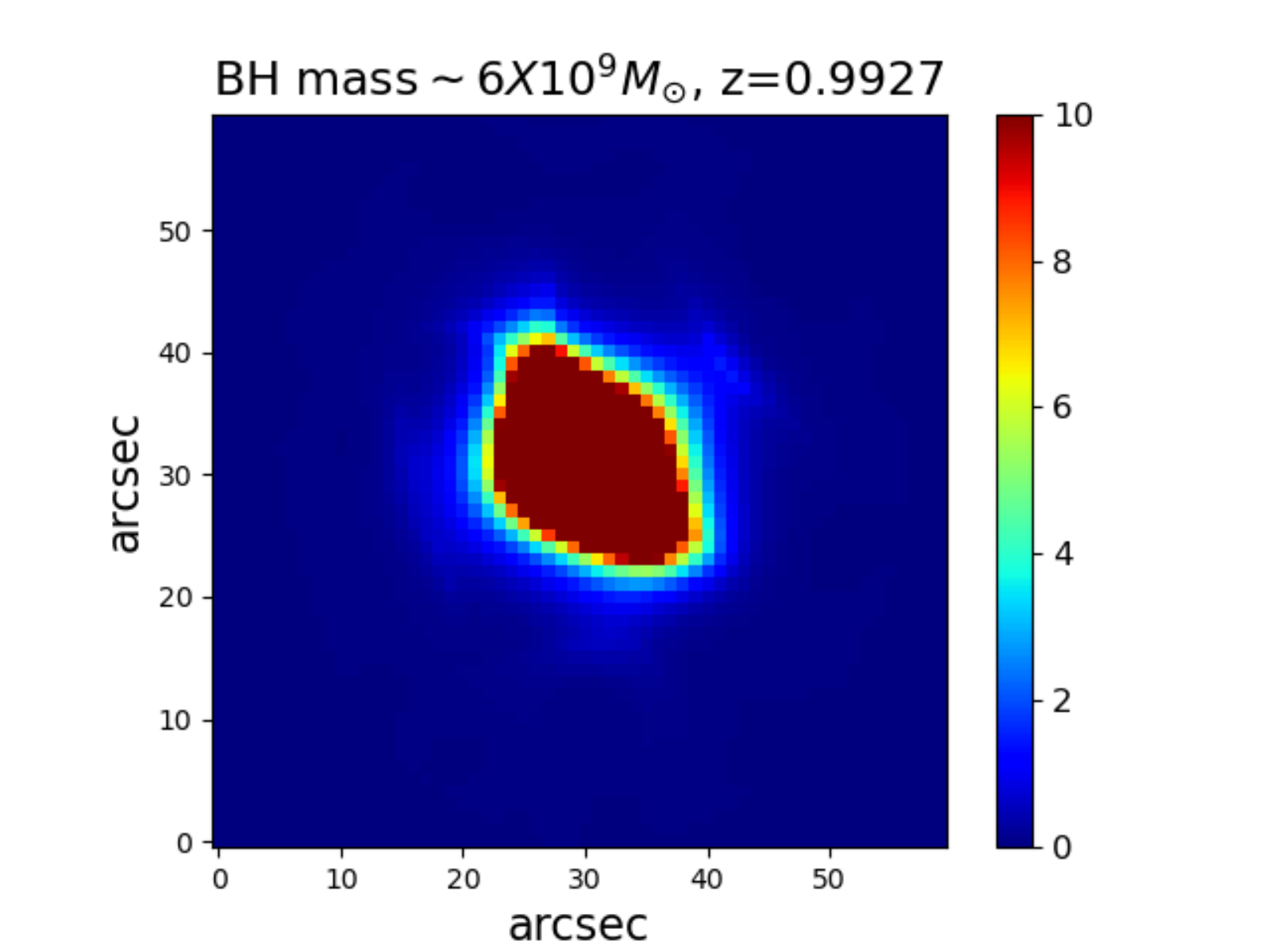}&\includegraphics[width=4.5cm]{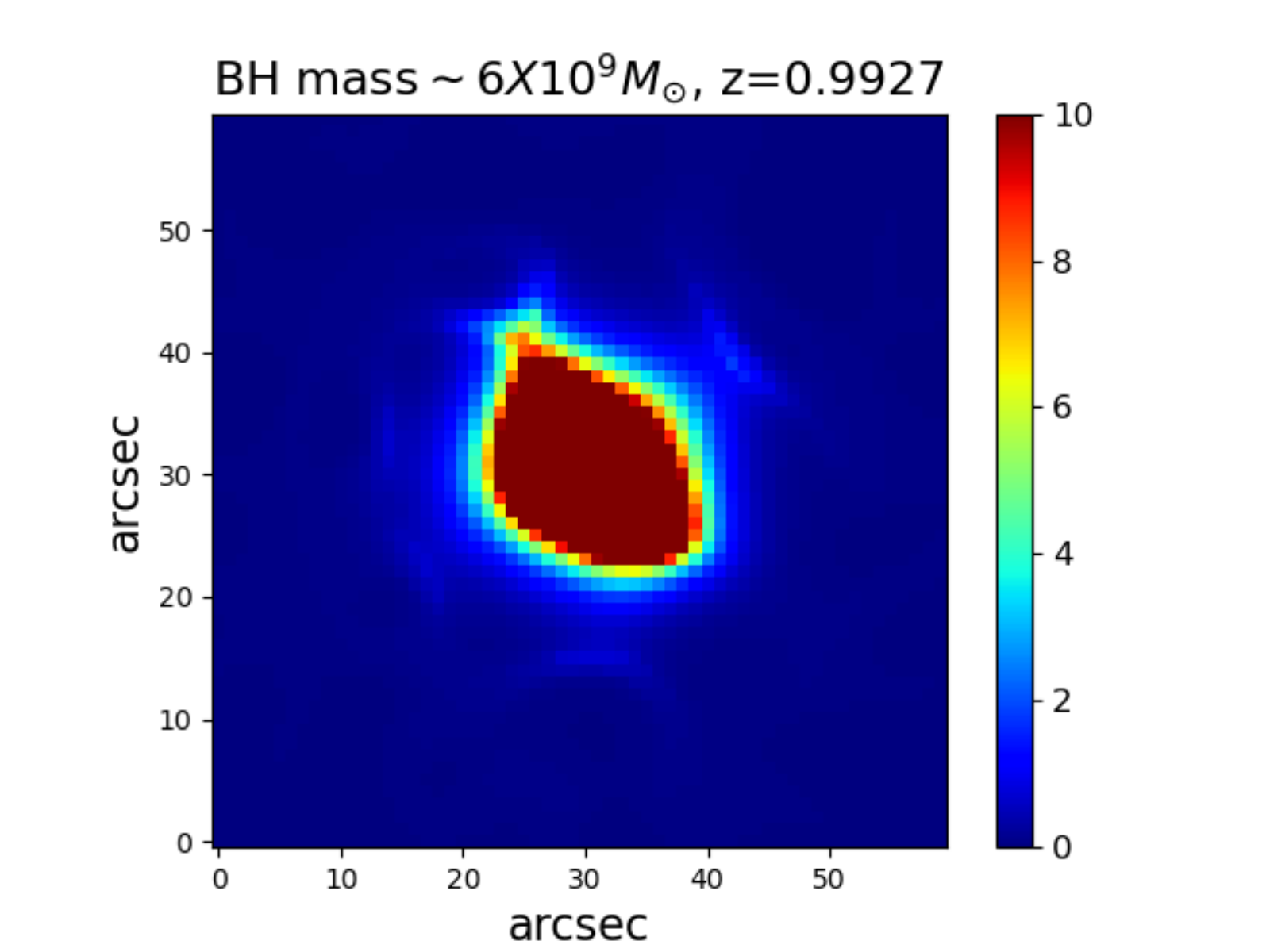}&\includegraphics[width=4.5cm]{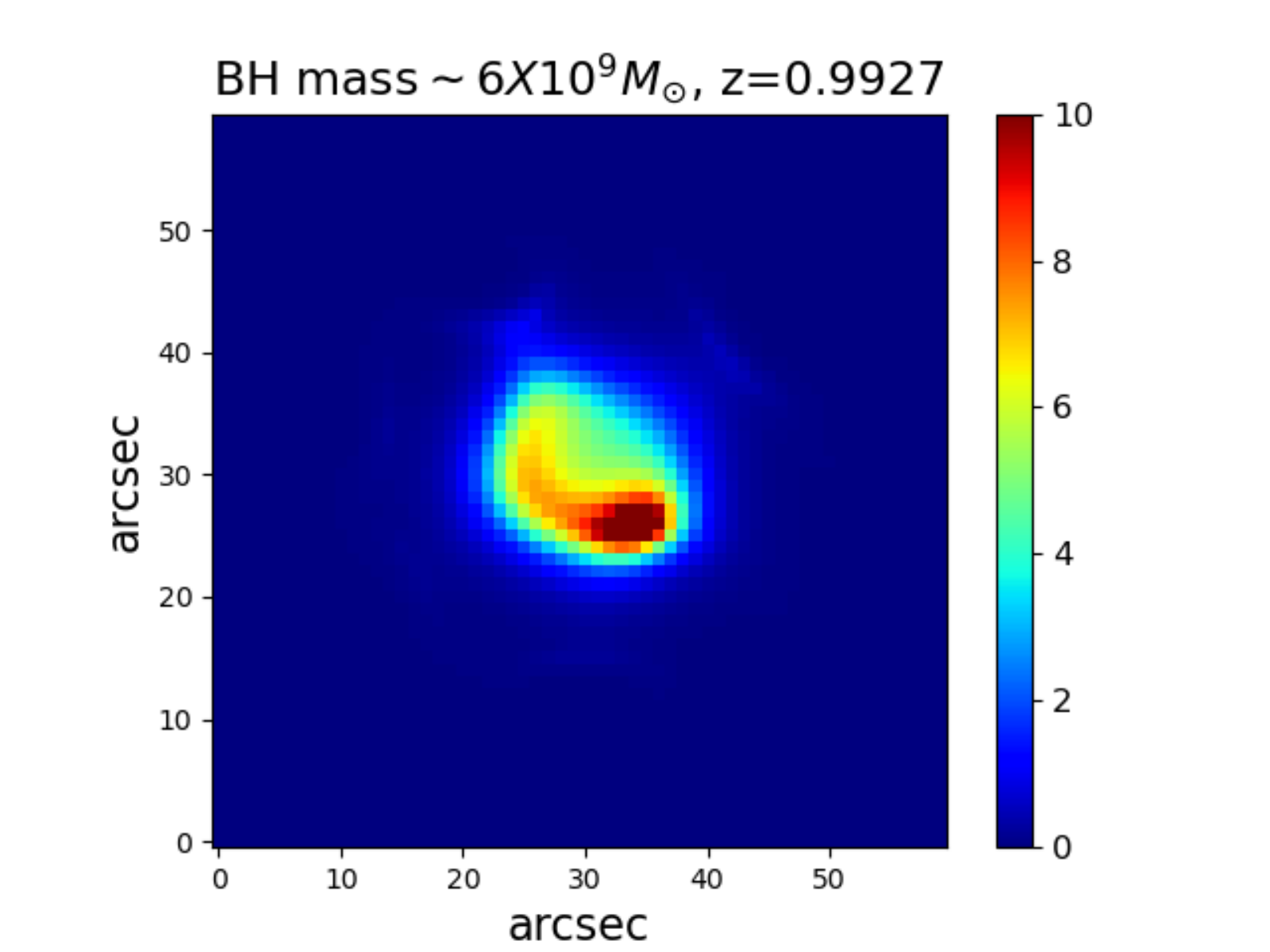}\\
       \end{tabular}
       \caption{Simulated ALMA maps constructed using the same observing parameters as Fig.\ 4 but at 135 GHz (Band 4, {\bf left most column}), 100 GHz (Band 3, {\bf middle column}), and 42 GHz (Band 1, {\bf right most column}) for the {\it most massive} high redshift ($z=1$) BH. {\bf Top Panel} The mock ALMA tSZ maps for no feedback {\bf Second Panel} The corresponding signal-to-noise maps, {\bf Third Panel} Same maps, but now for the no-jet feedback mode. {\bf Fourth Panel} The signal-to-noise maps corresponding to the third panel. The signal is enhanced when we have radiative feedback from the black holes.}
        \label{fig:6}
    \end{center}
\end{figure*}

\begin{figure*}
    \begin{center}
     \begin{tabular}{c |ccc}
     \hline
         &{135 GHz}&{100 GHz}&{42 GHz}\\[0.1pt]
         \hline
      {No}&\includegraphics[width=4.5cm]{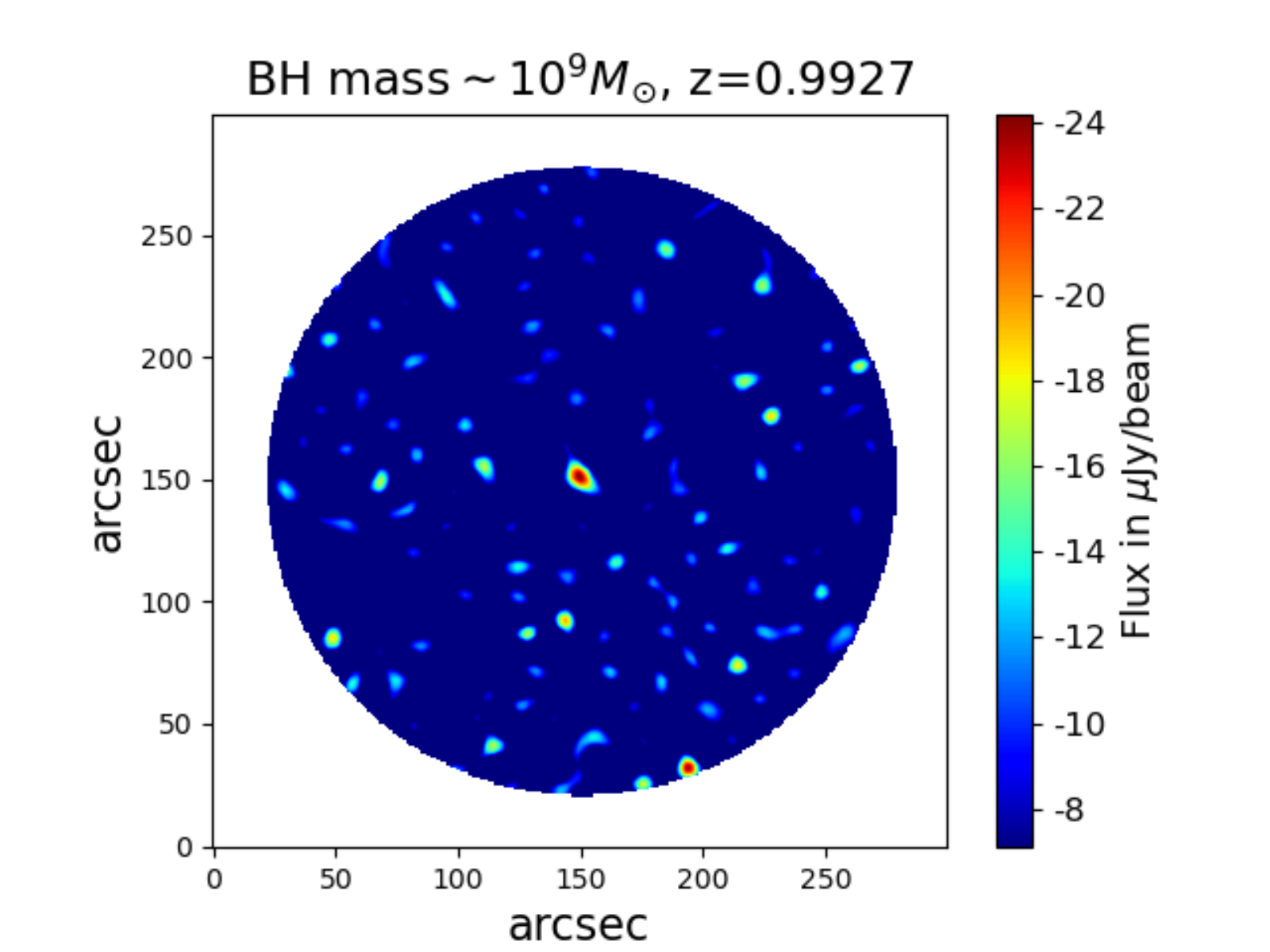}&\includegraphics[width=4.5cm]{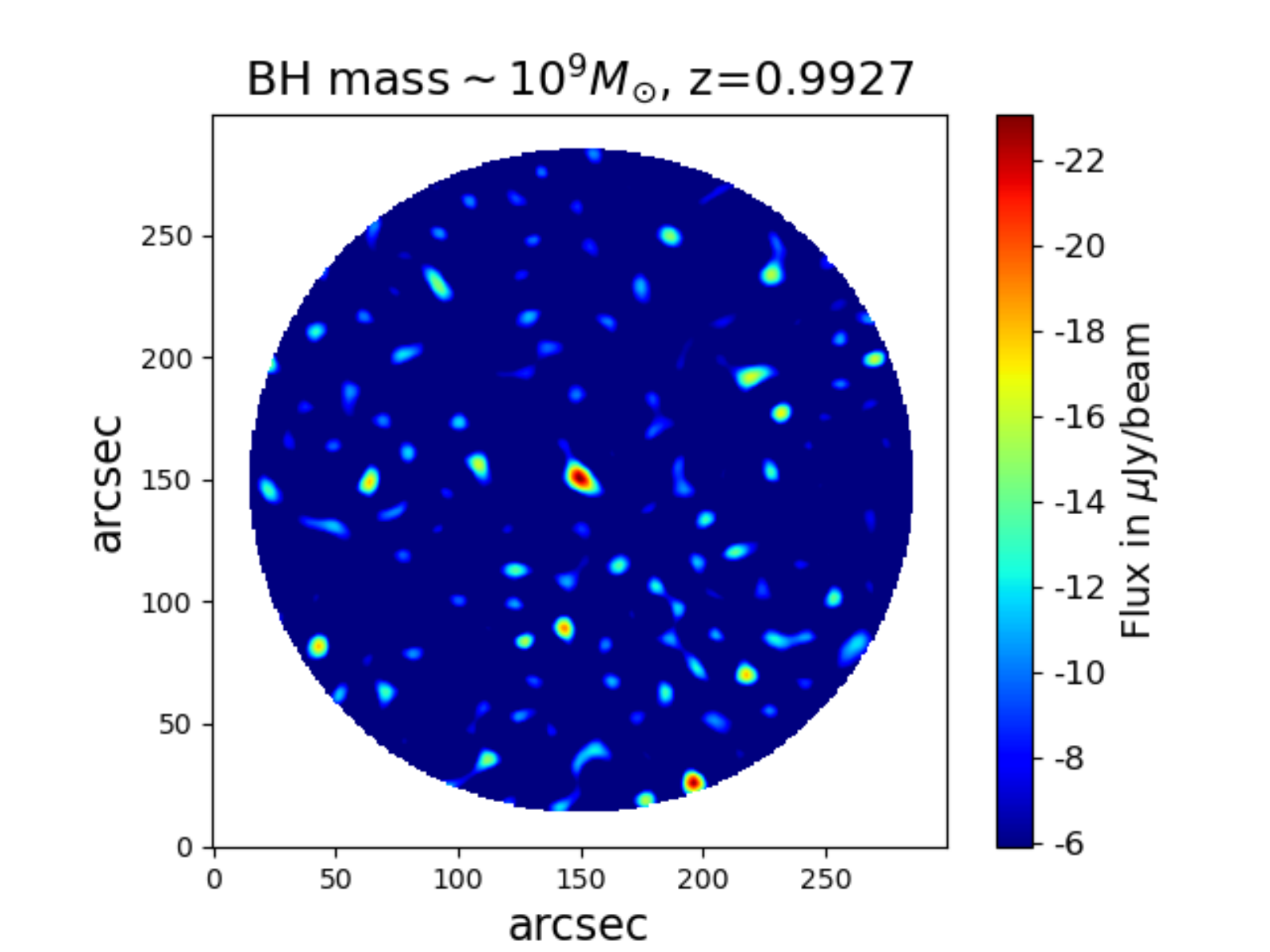}&\includegraphics[width=4.5cm]{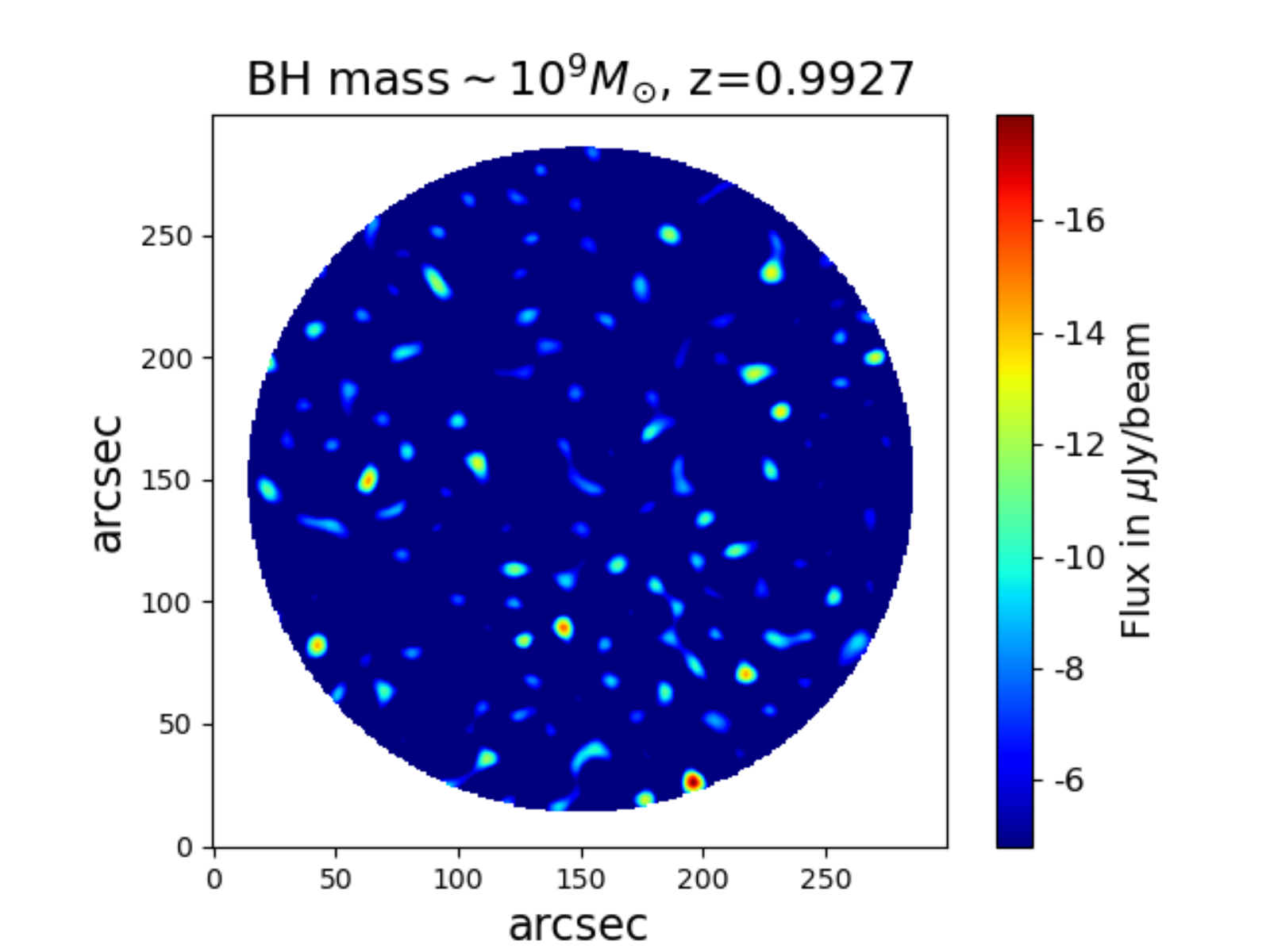}\\
    &\includegraphics[width=4.5cm]{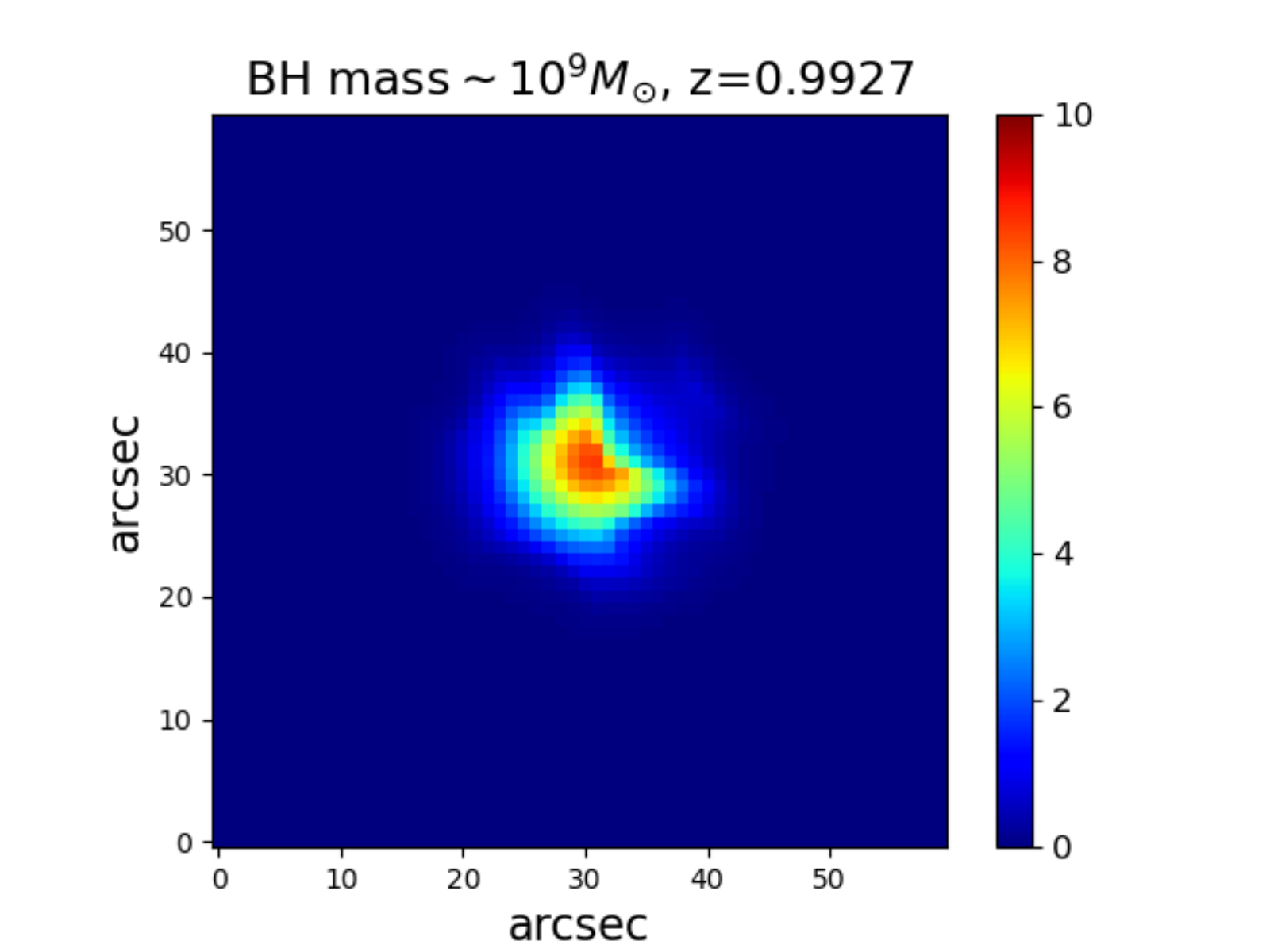}&\includegraphics[width=4.5cm]{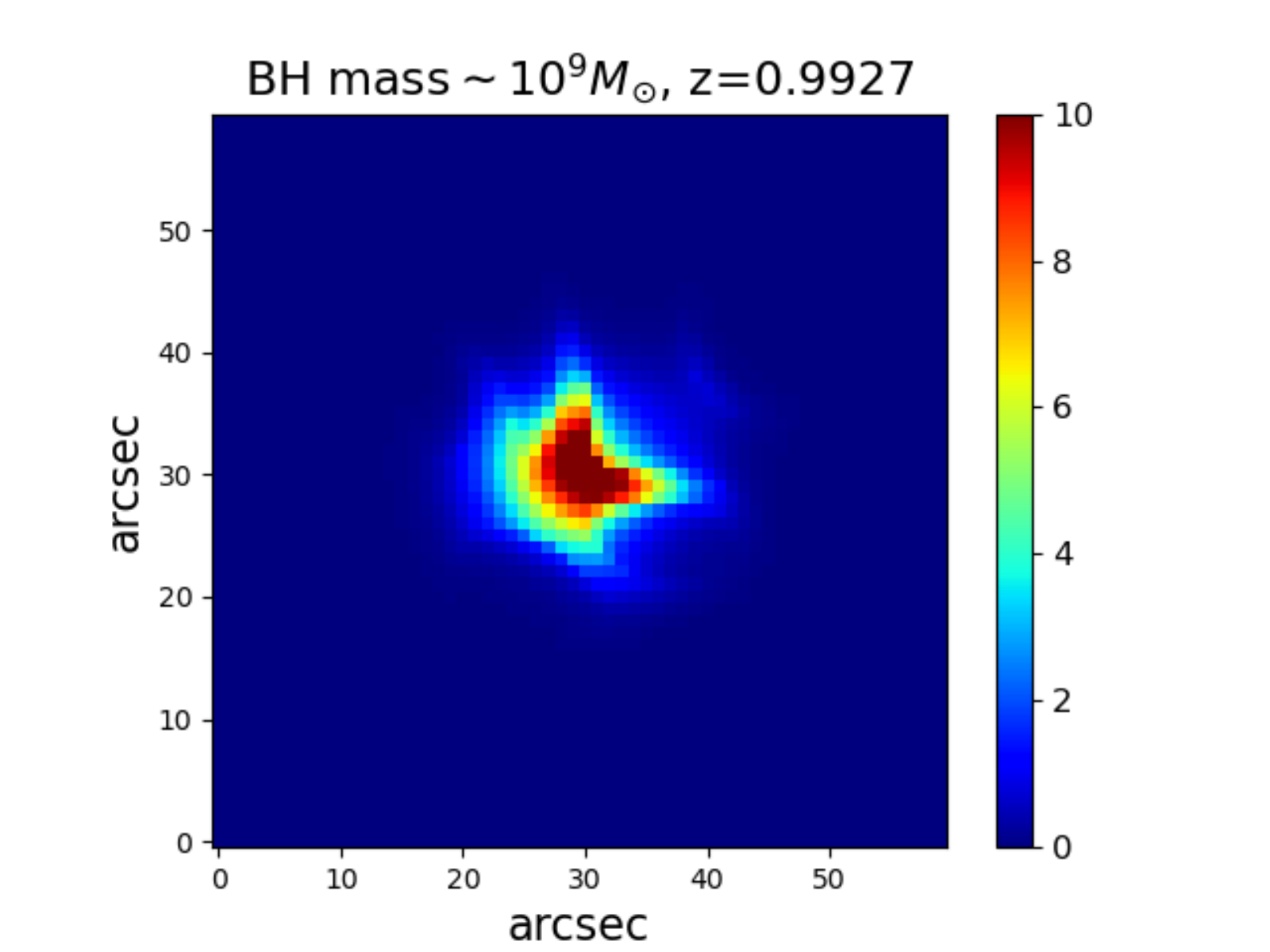}&\includegraphics[width=4.5cm]{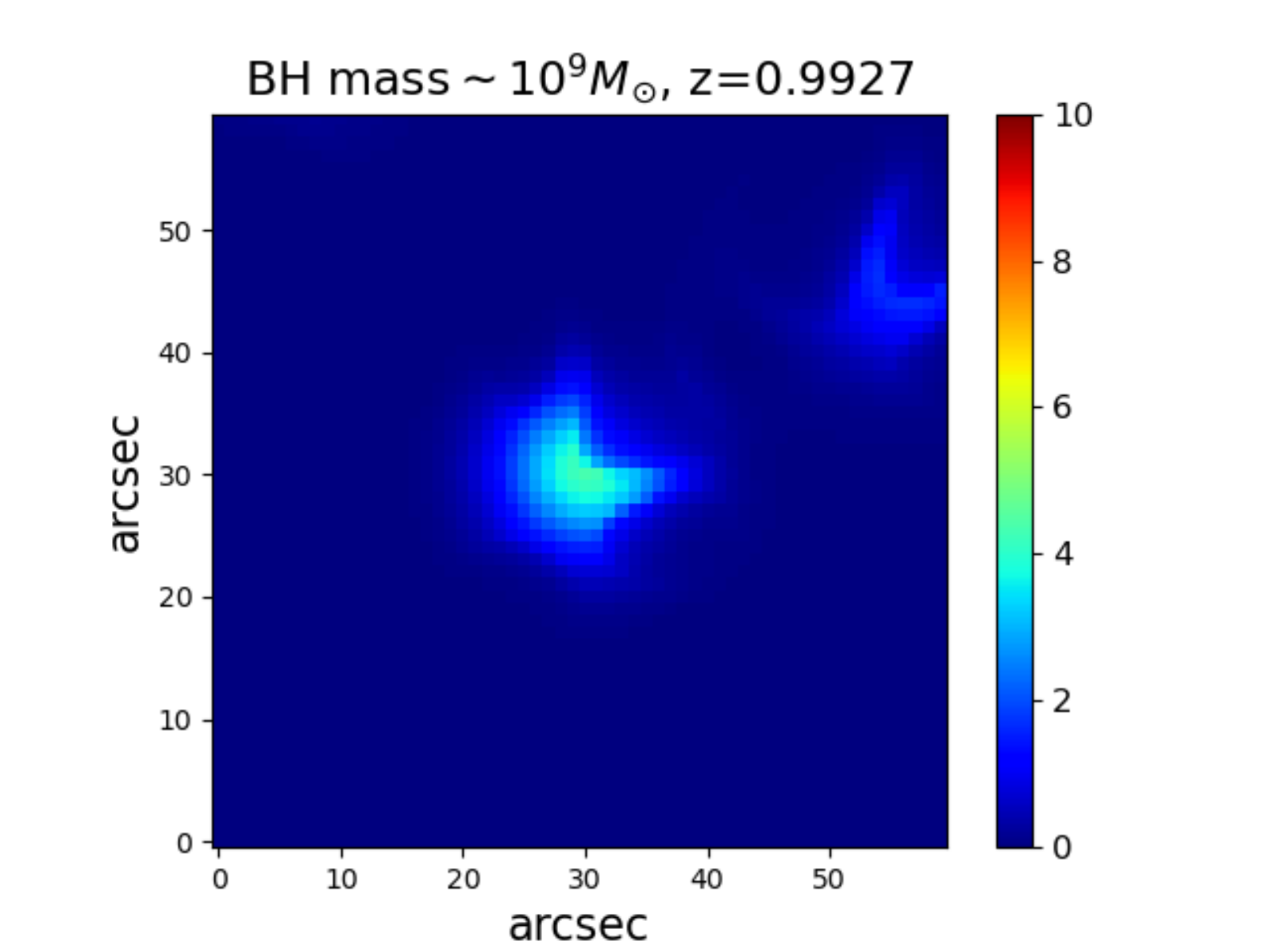}\\
    \hline
       {No-Jet}&\includegraphics[width=4.5cm]{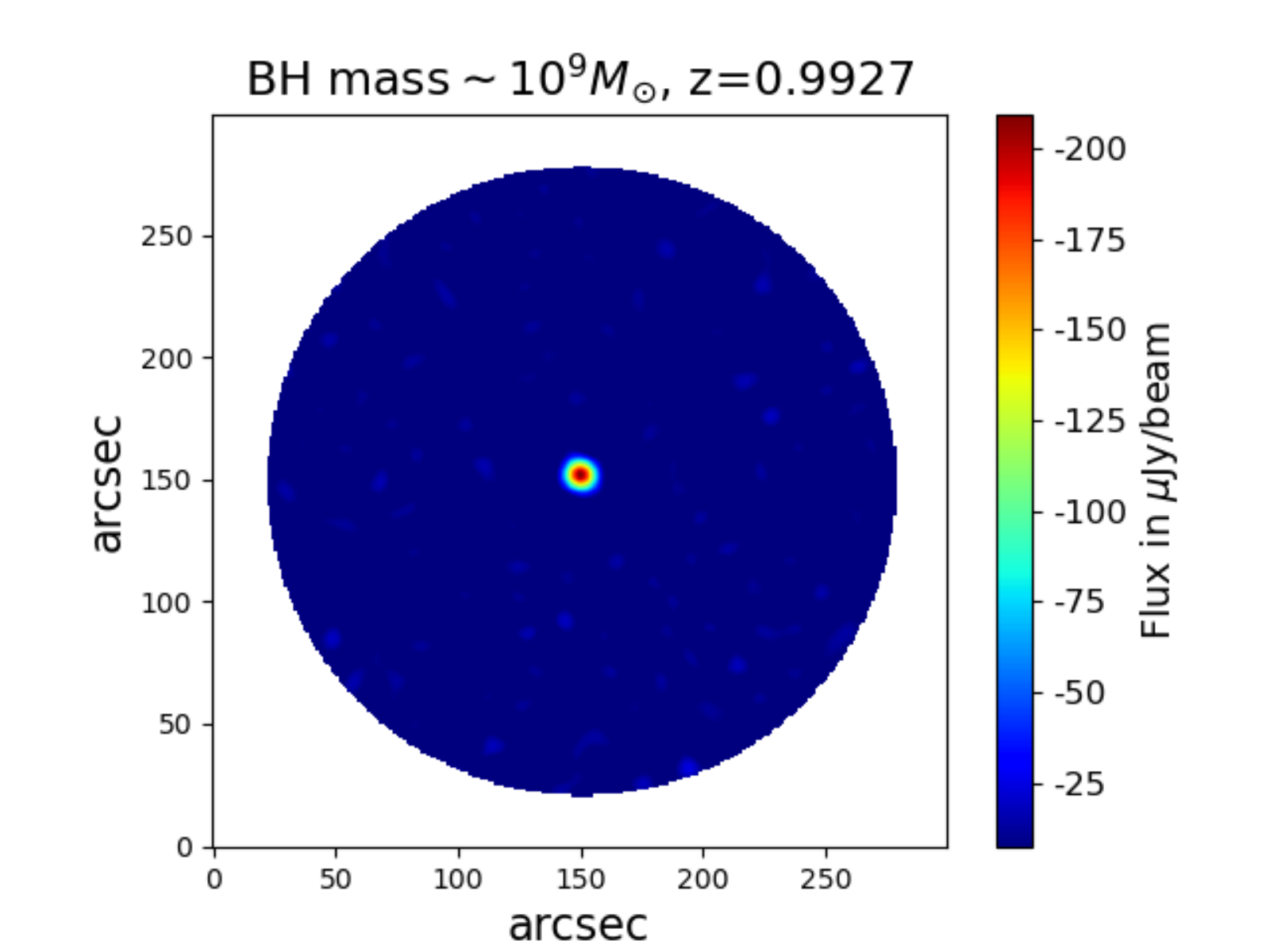}&\includegraphics[width=4.5cm]{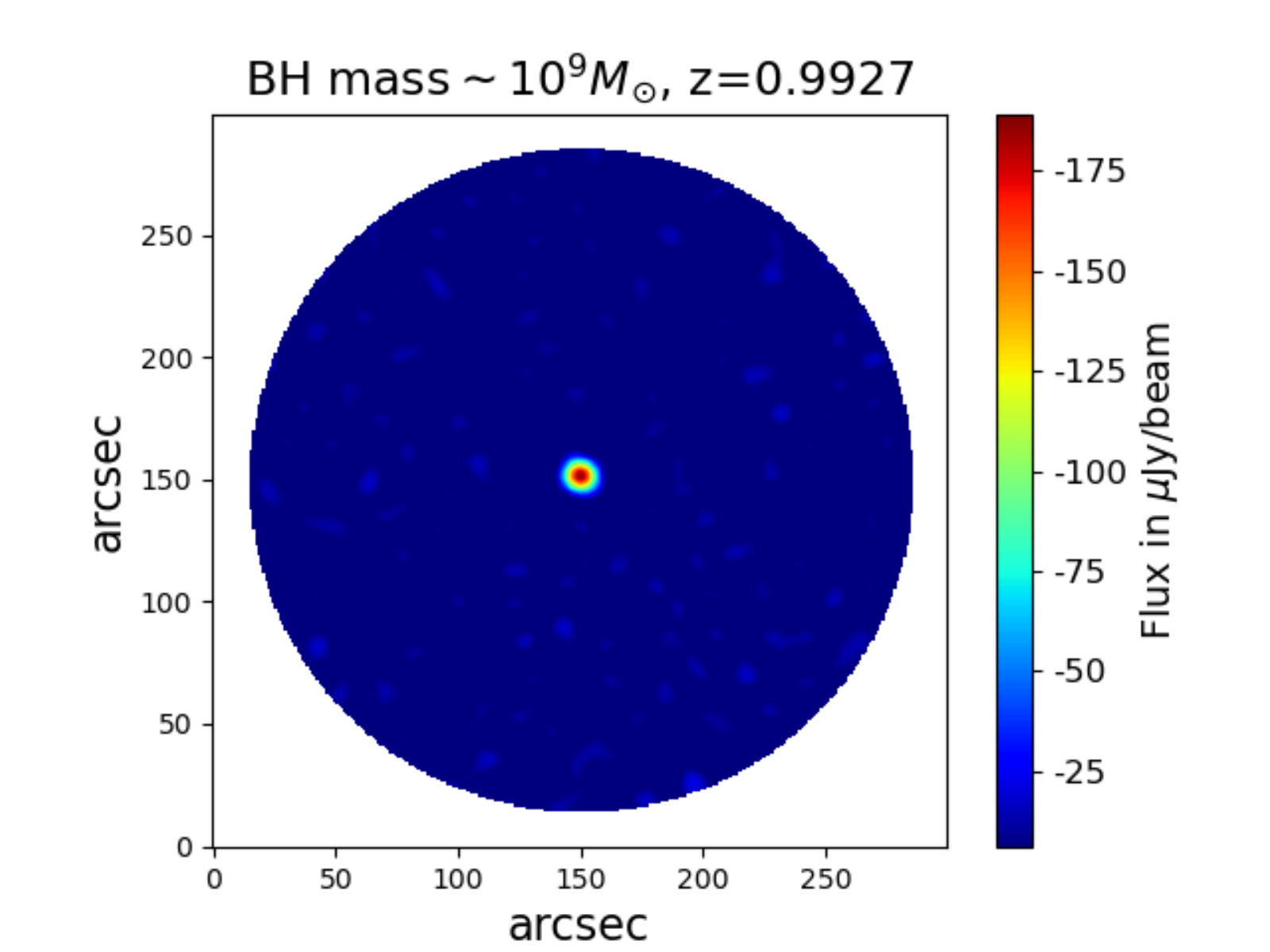}&\includegraphics[width=4.5cm]{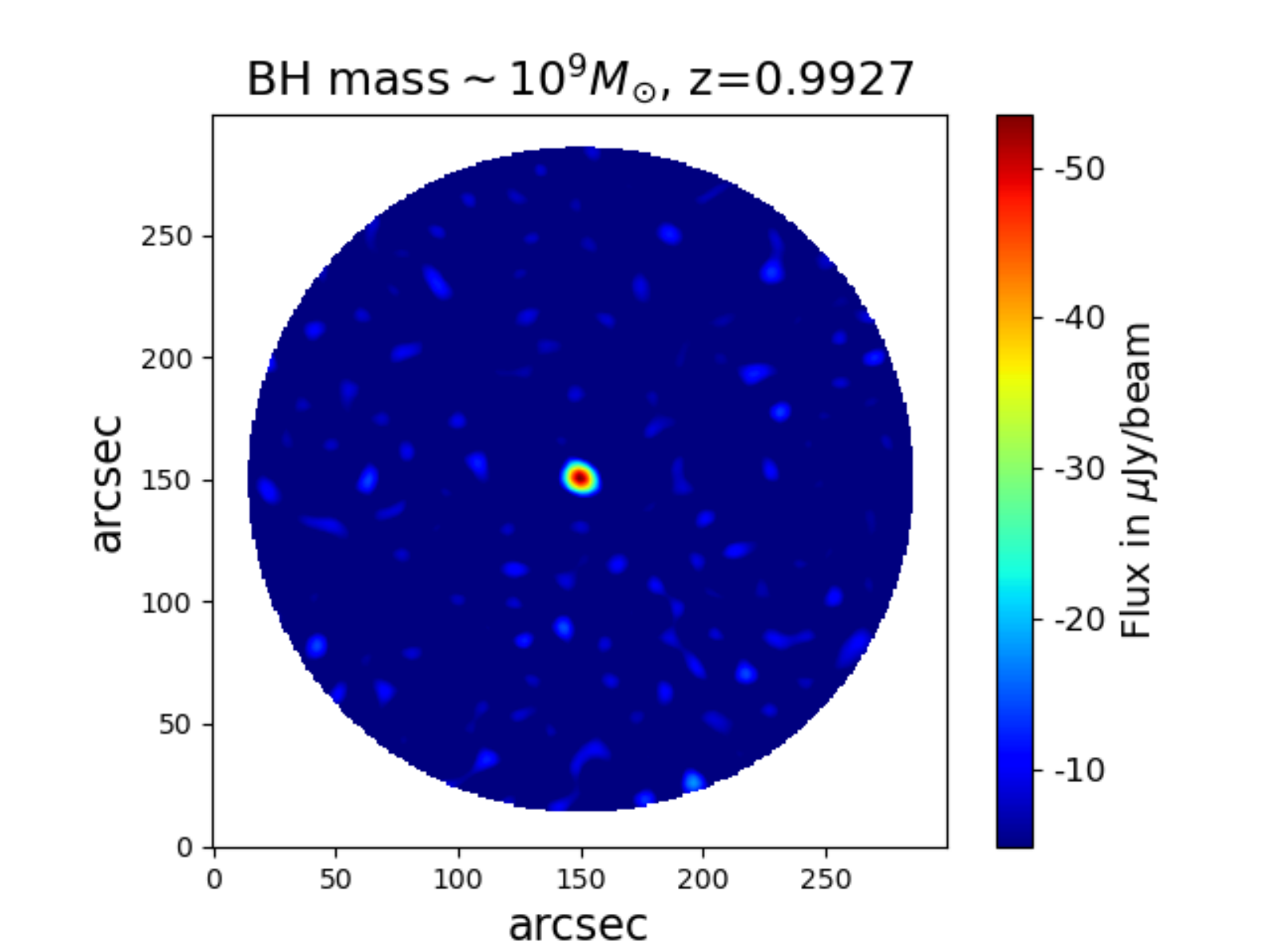}\\
        &\includegraphics[width=4.5cm]{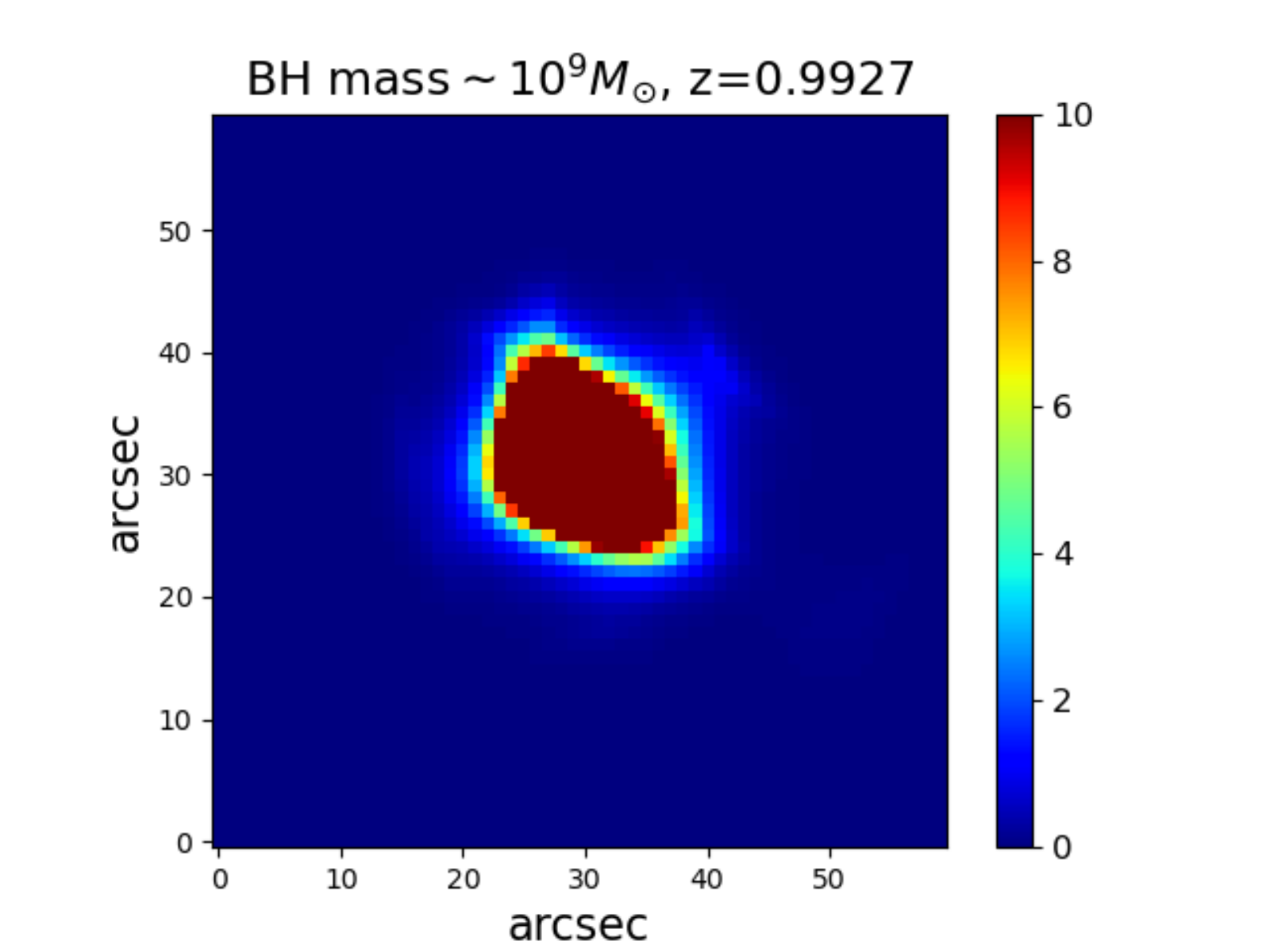}&\includegraphics[width=4.5cm]{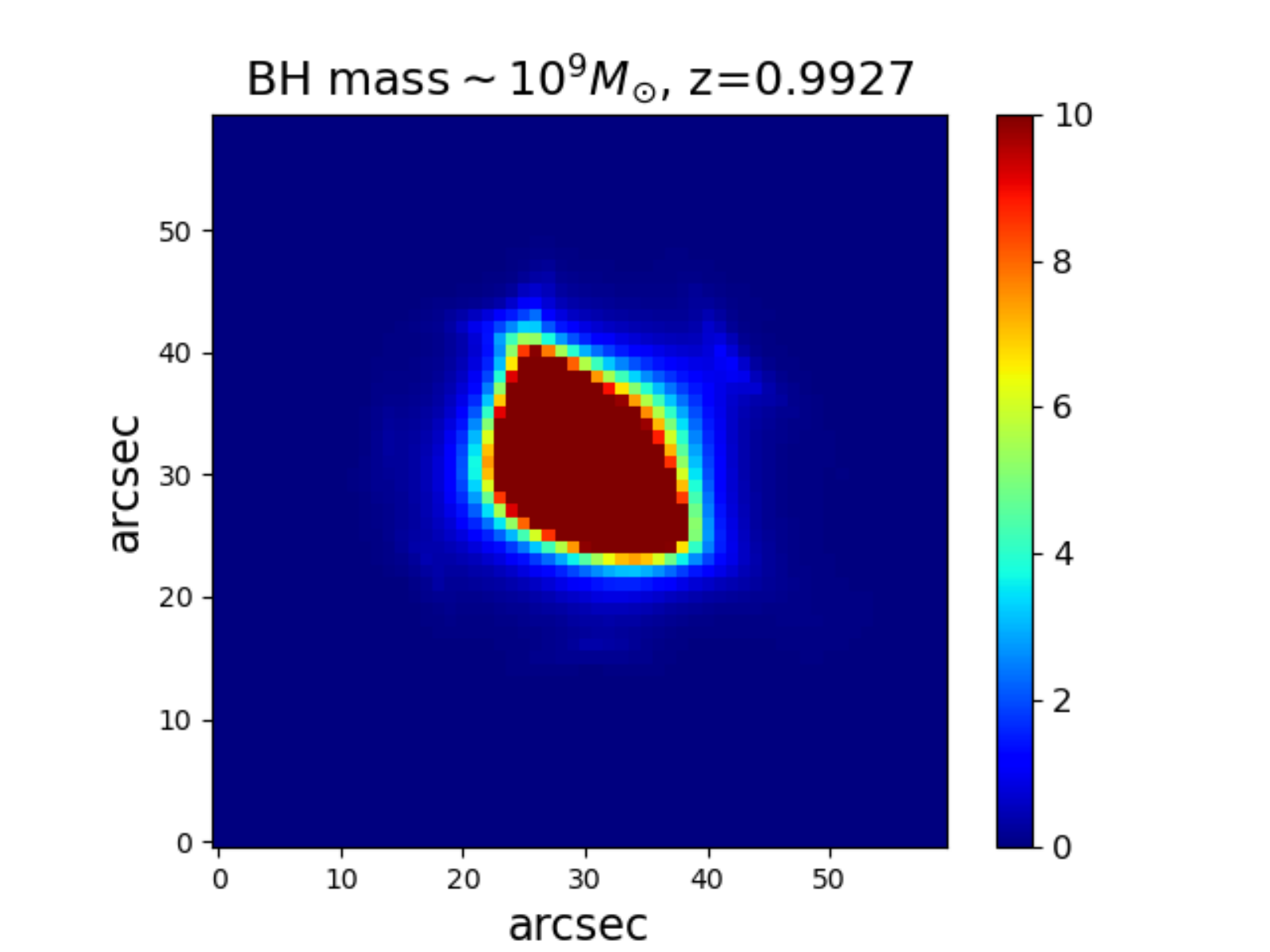}&\includegraphics[width=4.5cm]{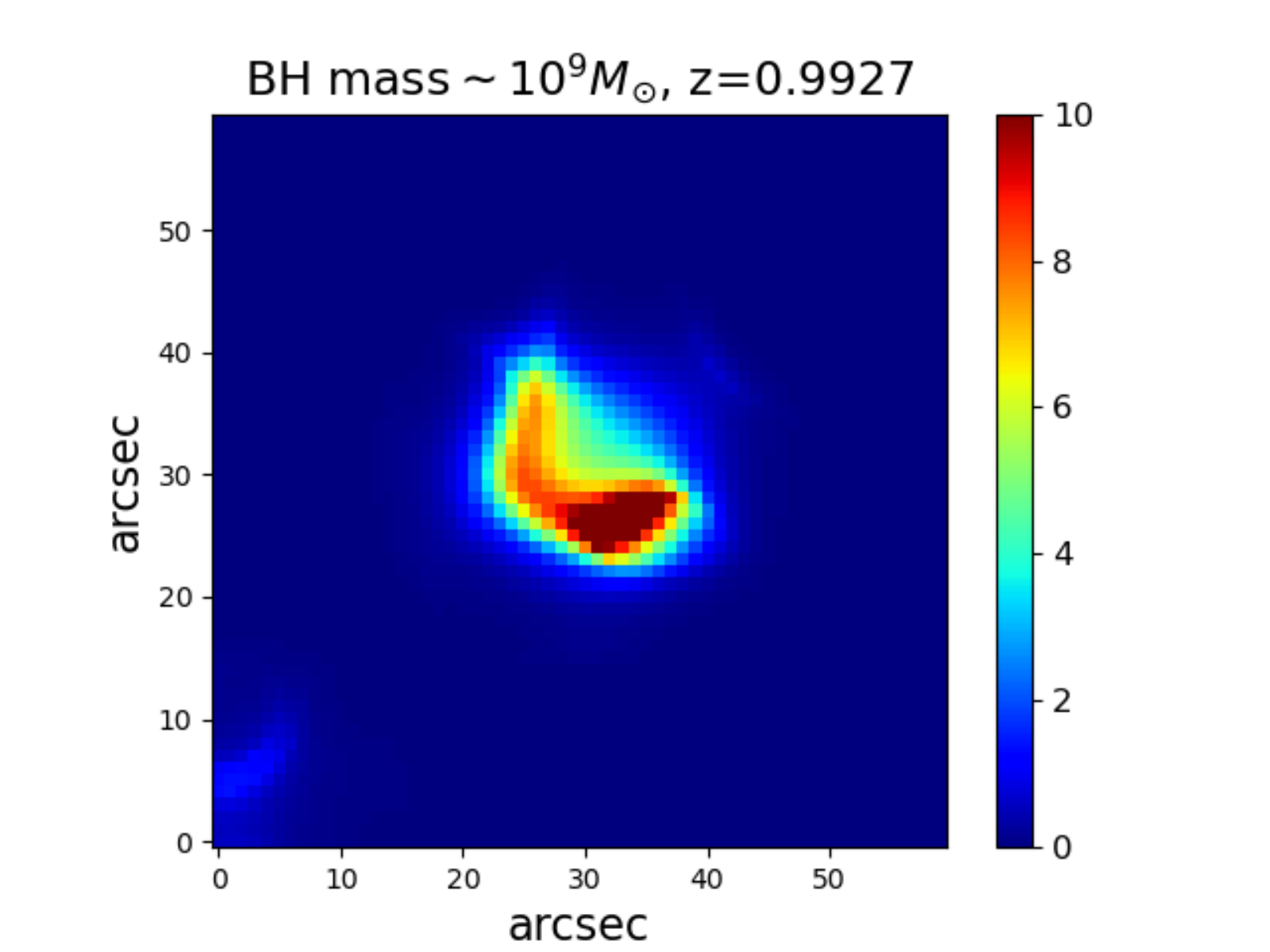}\\
       \end{tabular}
       \caption{Simulated ALMA maps constructed using the same observing parameters as Fig.\ 4 but at 135 GHz (Band 4), 100 GHz (Band 3), and 42 GHz (Band 1) for the {\it most active} high redshift ($z=1$) BHs. {\bf Top Panel} The mock ALMA tSZ maps for no feedback. {\bf Second Panel} The corresponding signal-to-noise maps, {\bf Third Panel} Same maps, but now for the no-jet feedback mode. {\bf Fourth Panel} The signal-to-noise maps corresponding to the third panel. See Table 1 and 2 for feedback nomenclature and black hole properties. }
        \label{fig:7}
    \end{center}
\end{figure*}

\begin{figure*}
    \begin{center}
     \begin{tabular}{c |ccc}
     \hline
         &{135 GHz}&{100 GHz}&{42 GHz}\\[0.1pt]
         \hline
      {No}&\includegraphics[width=4.5cm]{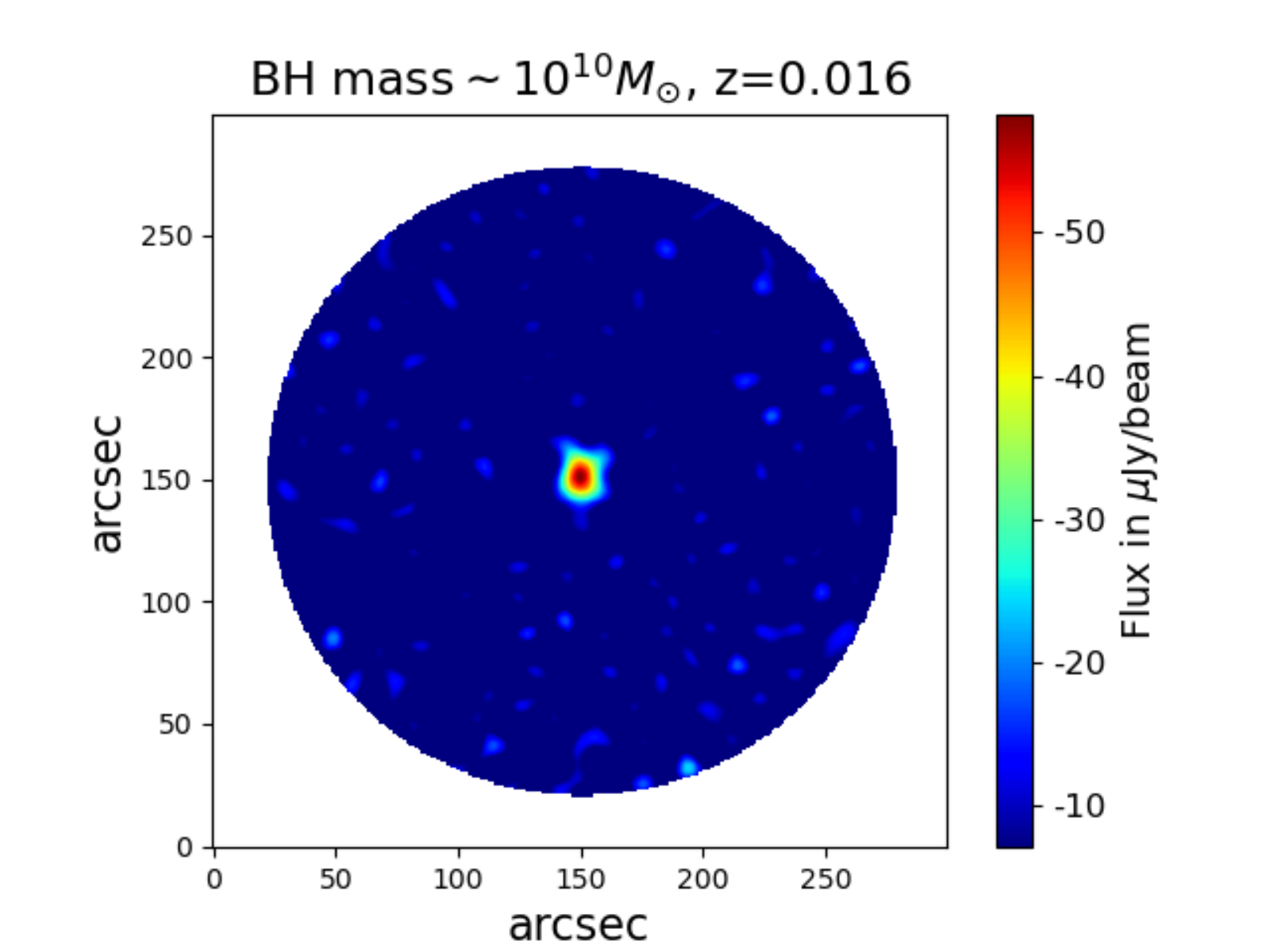}&\includegraphics[width=4.5cm]{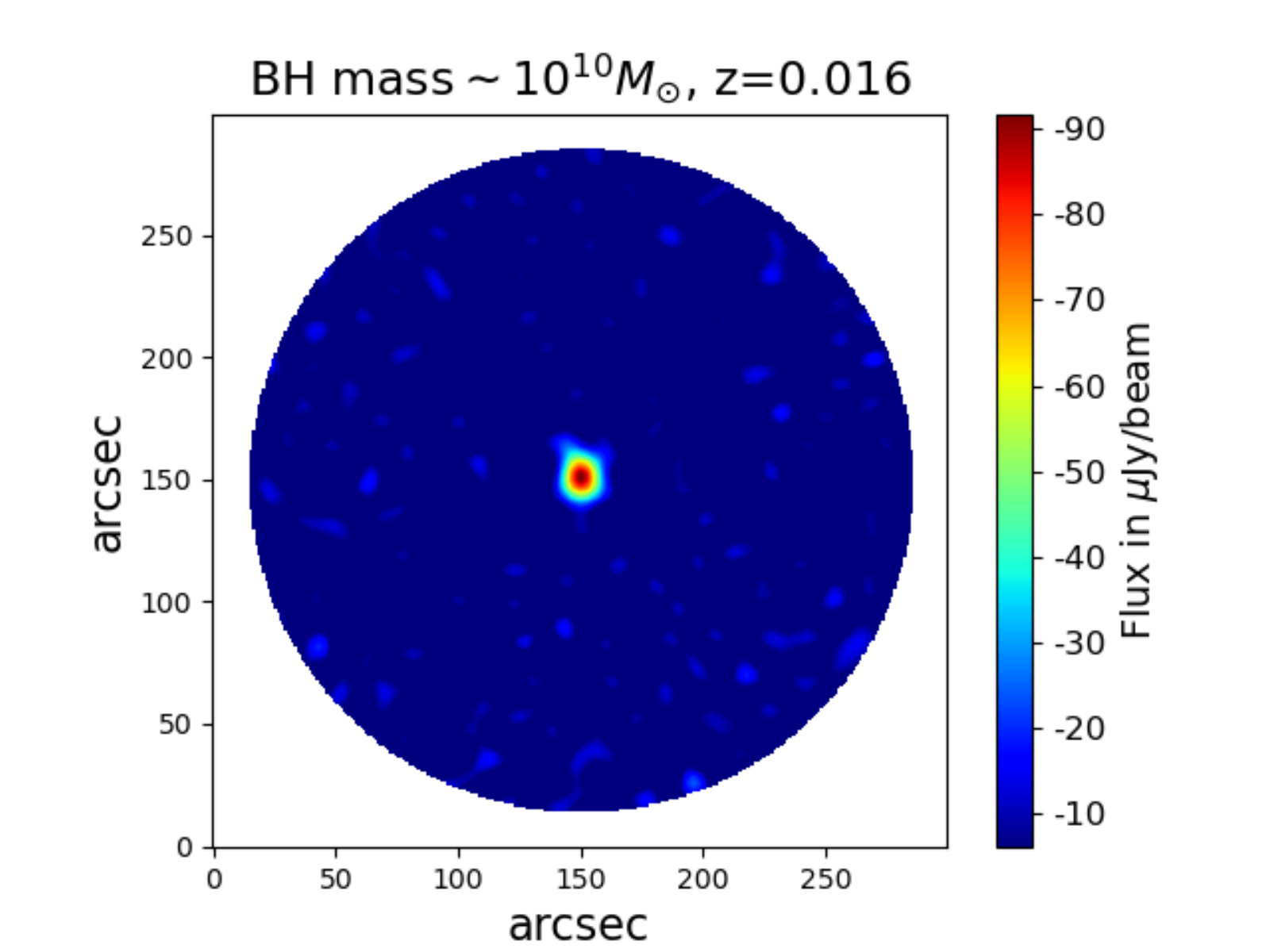}&\includegraphics[width=4.5cm]{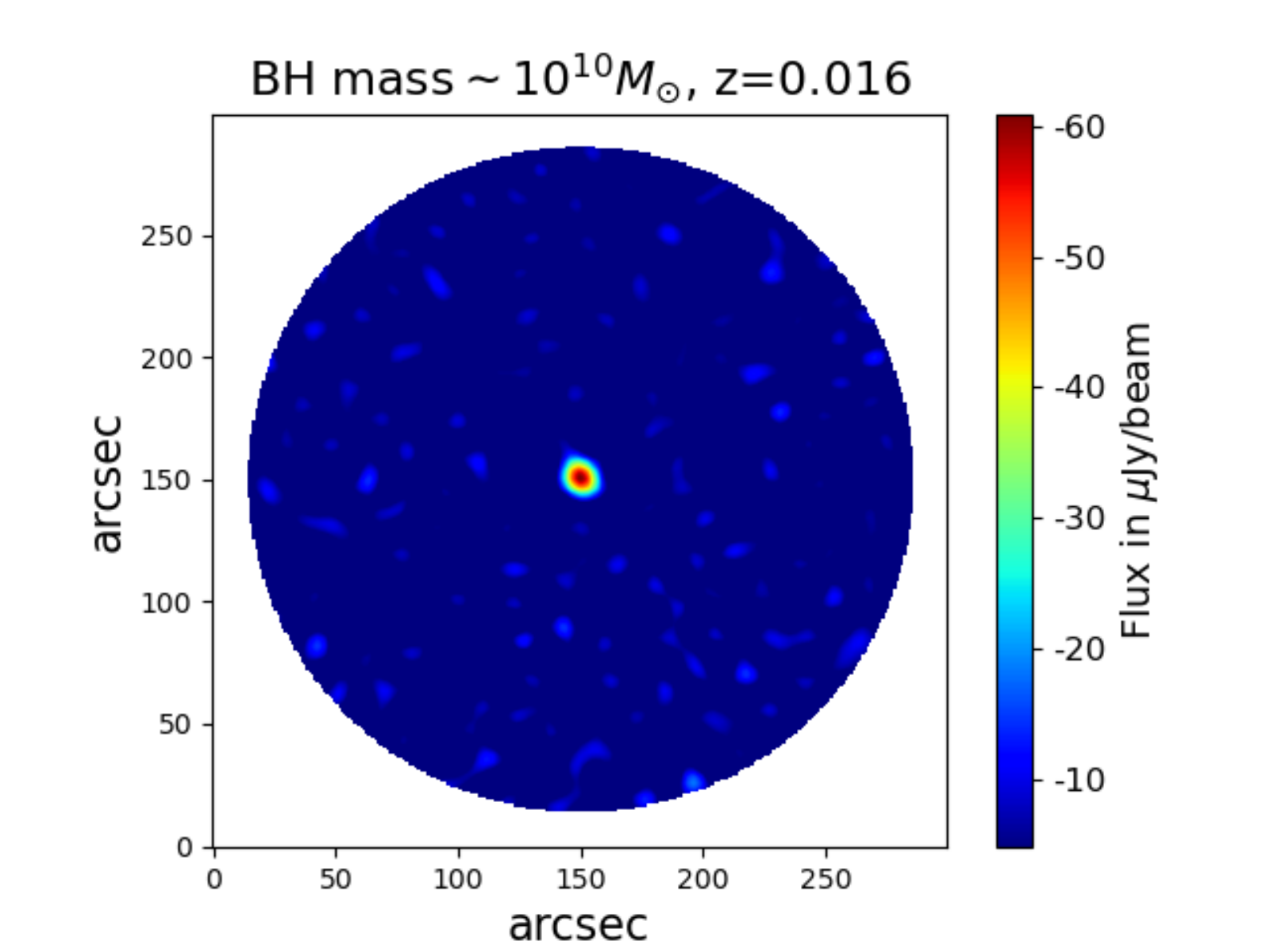}\\
       &\includegraphics[width=4.5cm]{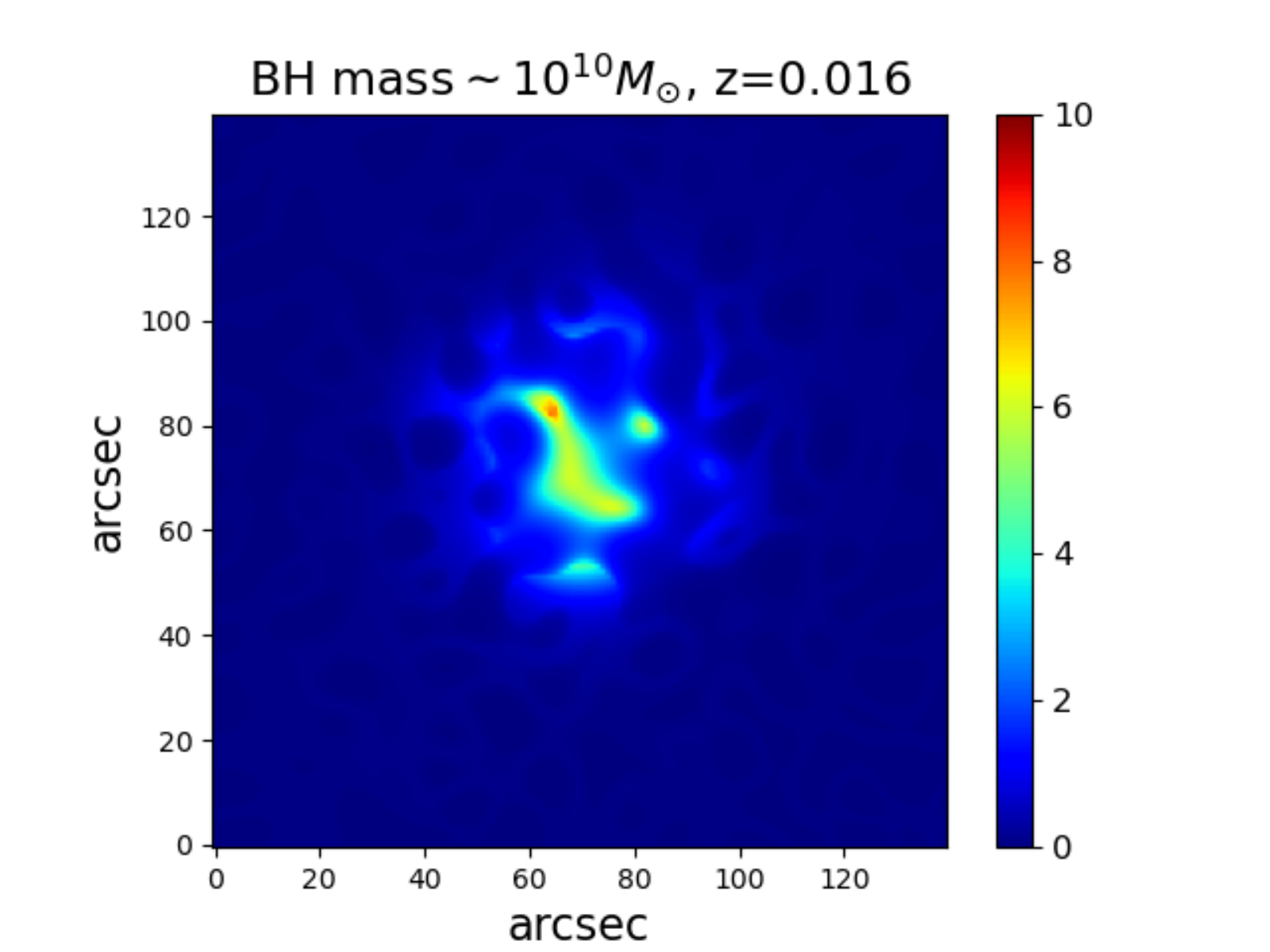}&\includegraphics[width=4.5cm]{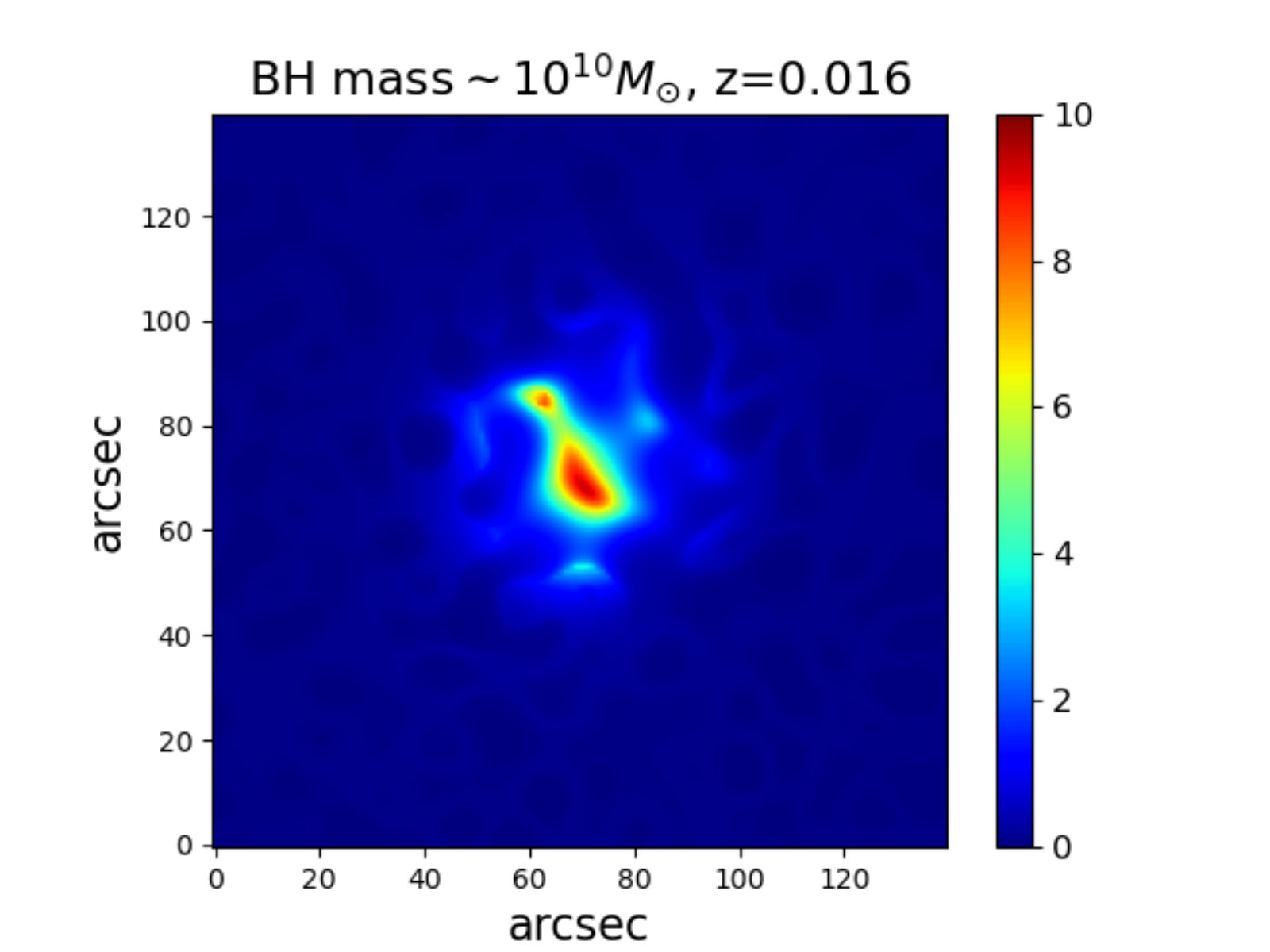}&\includegraphics[width=4.5cm]{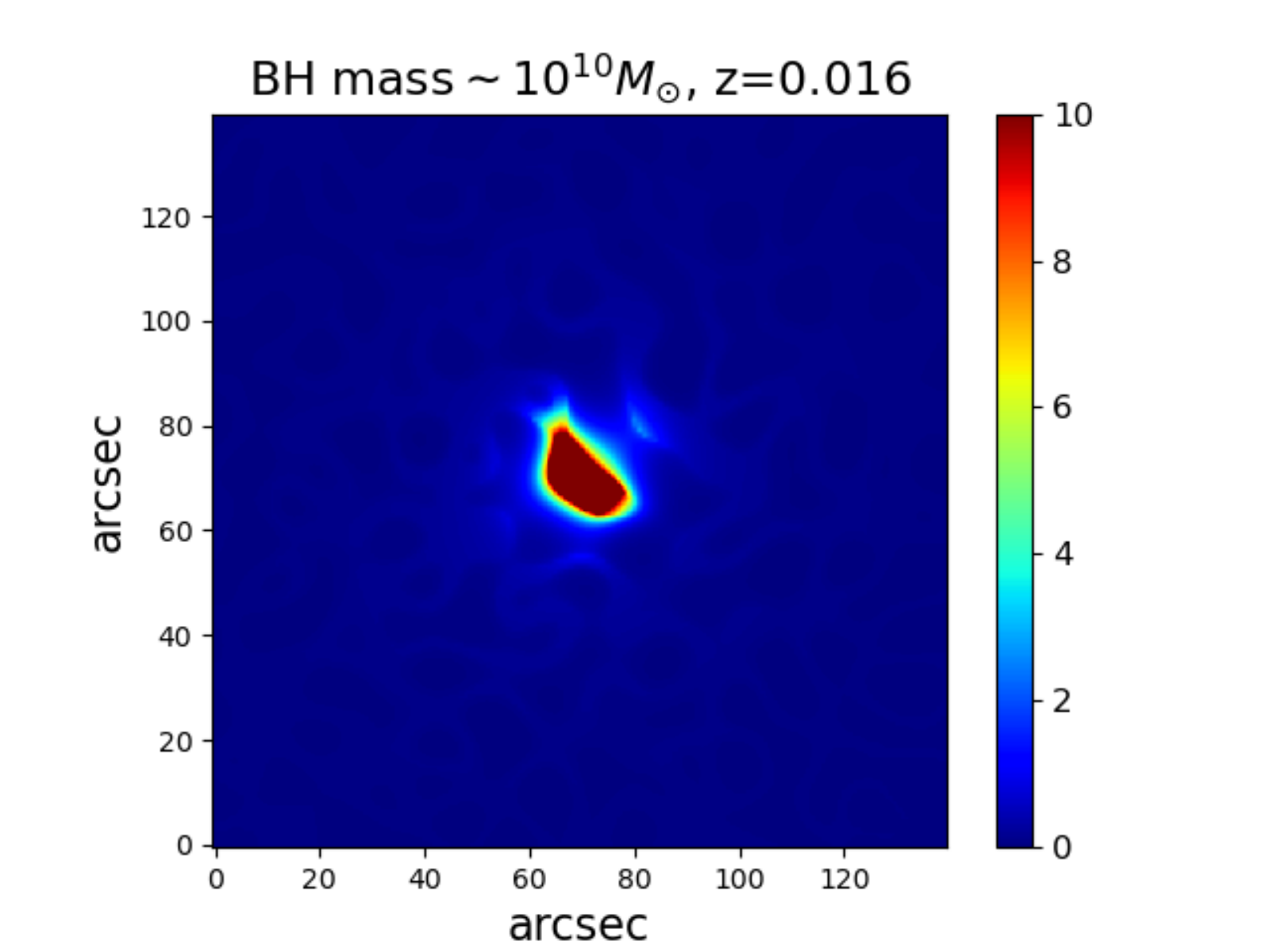}\\
       \hline
       {No-Jet}&\includegraphics[width=4.5cm]{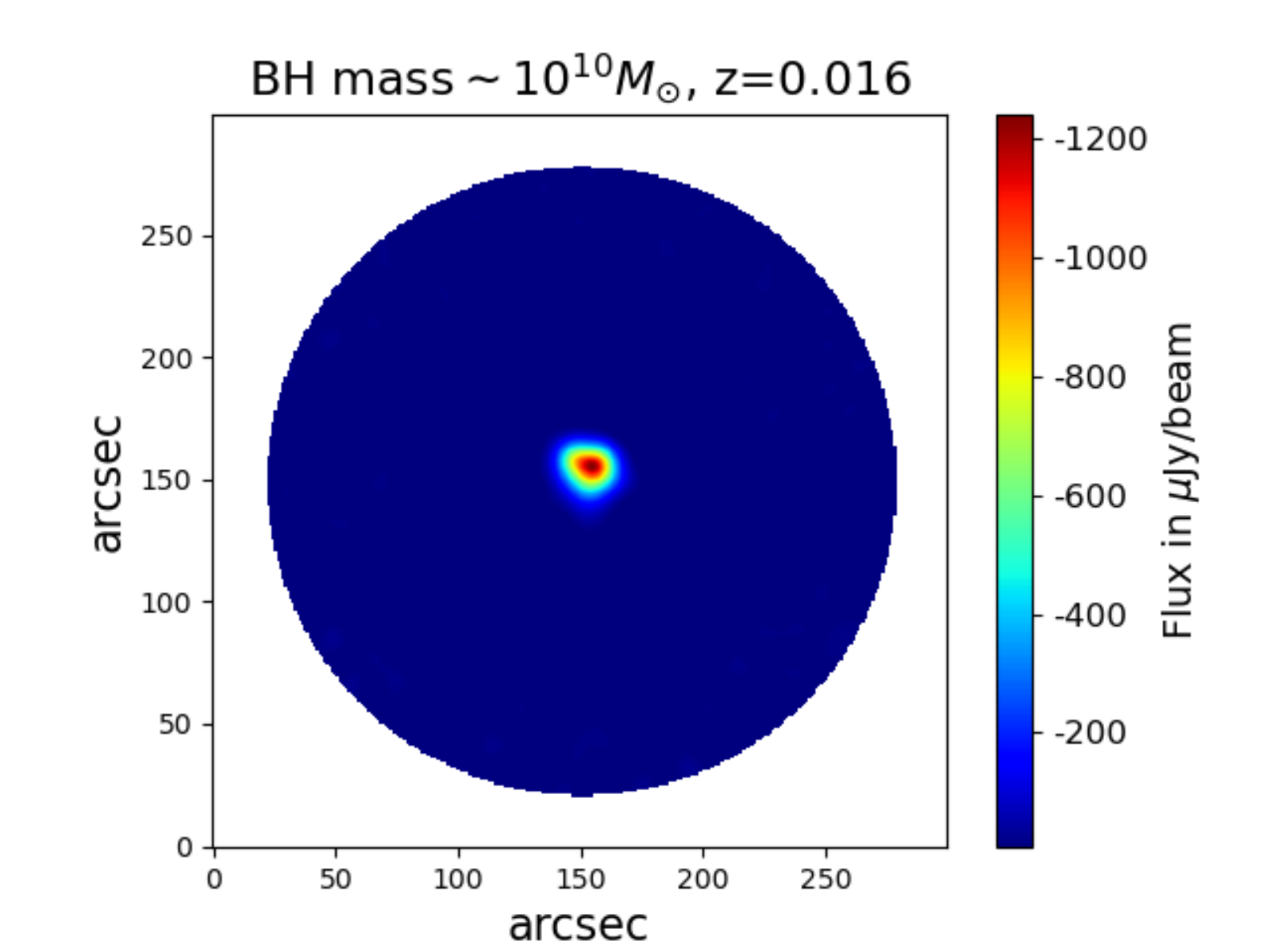}&\includegraphics[width=4.5cm]{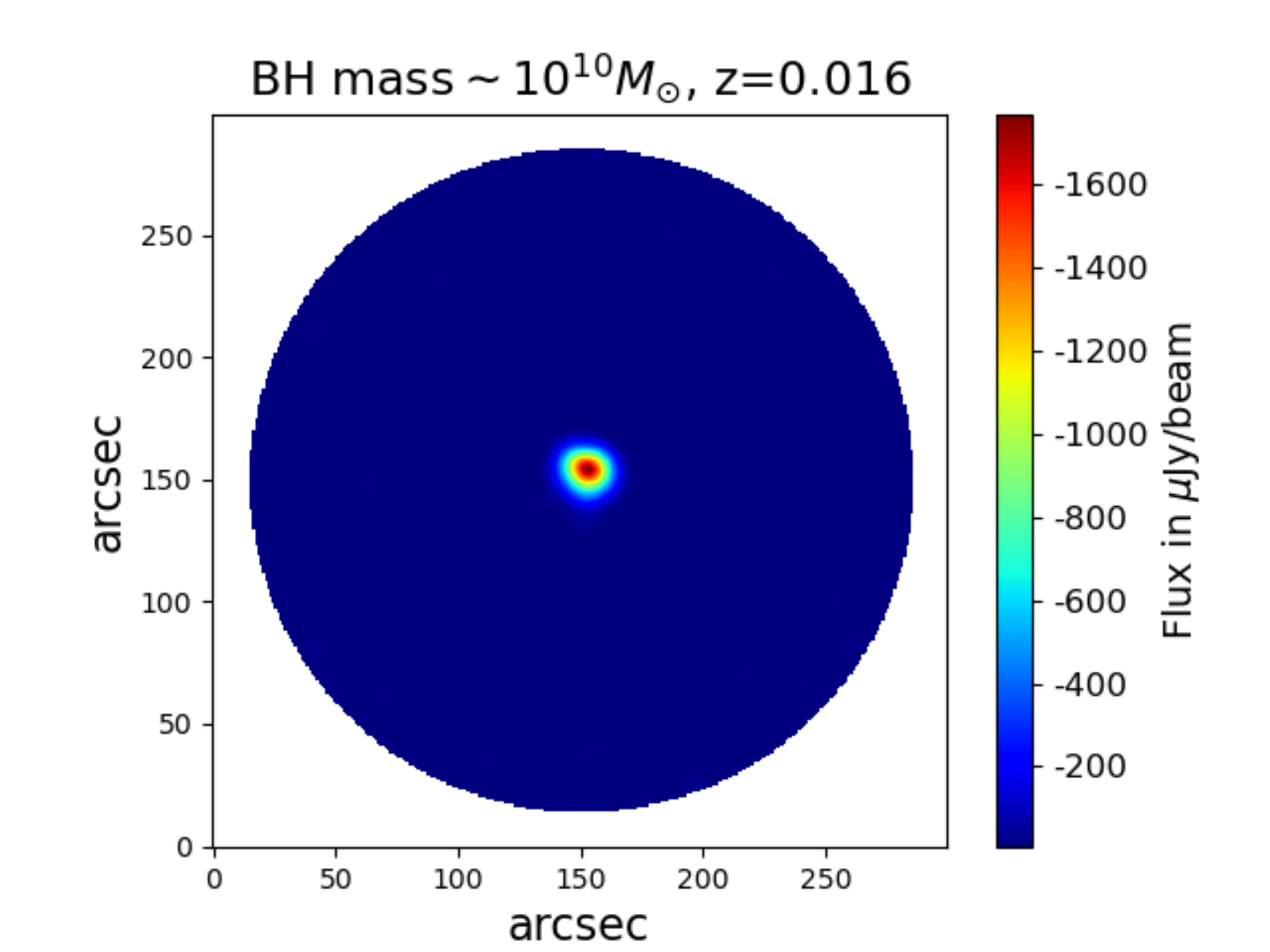}&\includegraphics[width=4.5cm]{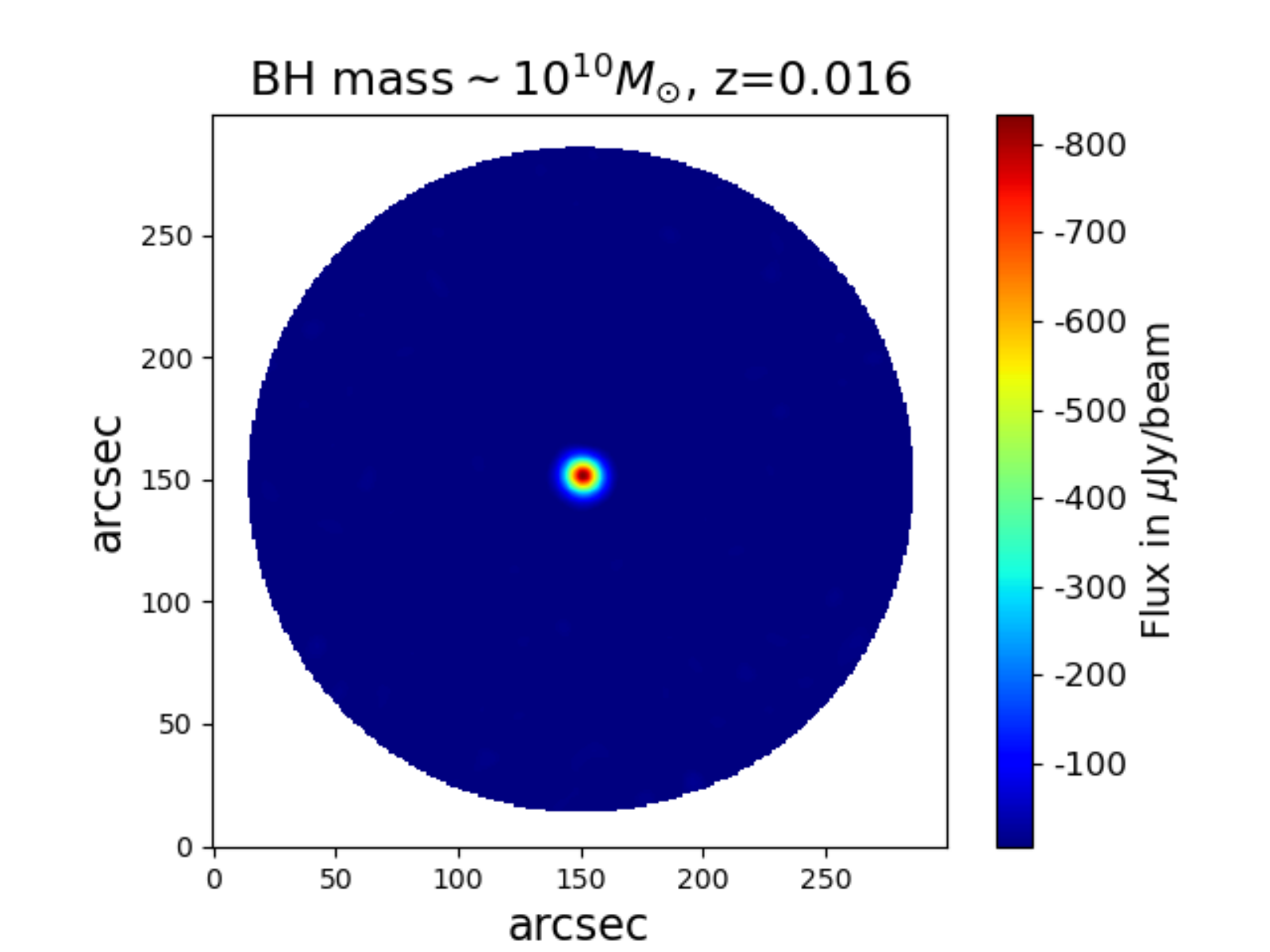}\\
        &\includegraphics[width=4.5cm]{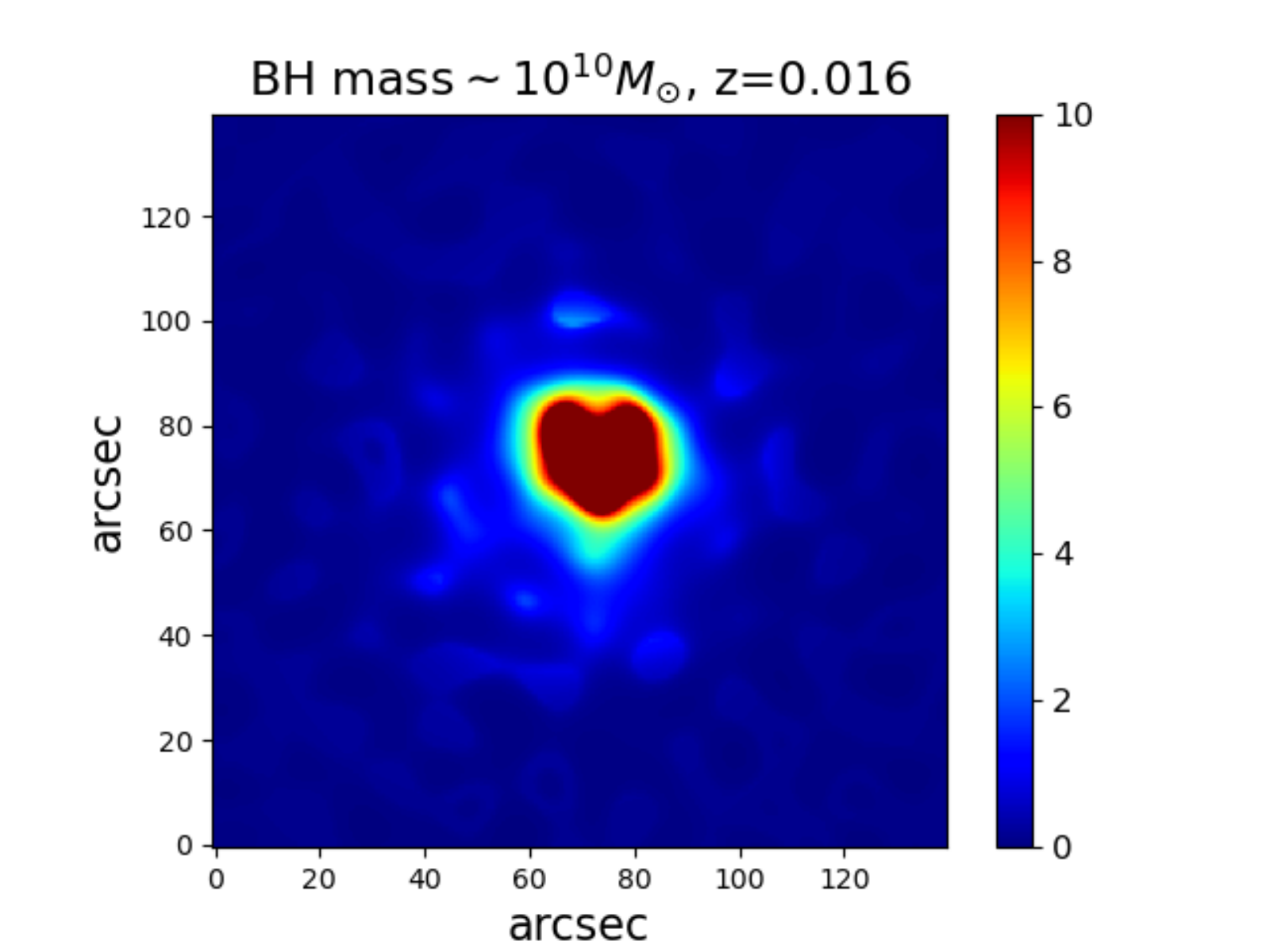}&\includegraphics[width=4.5cm]{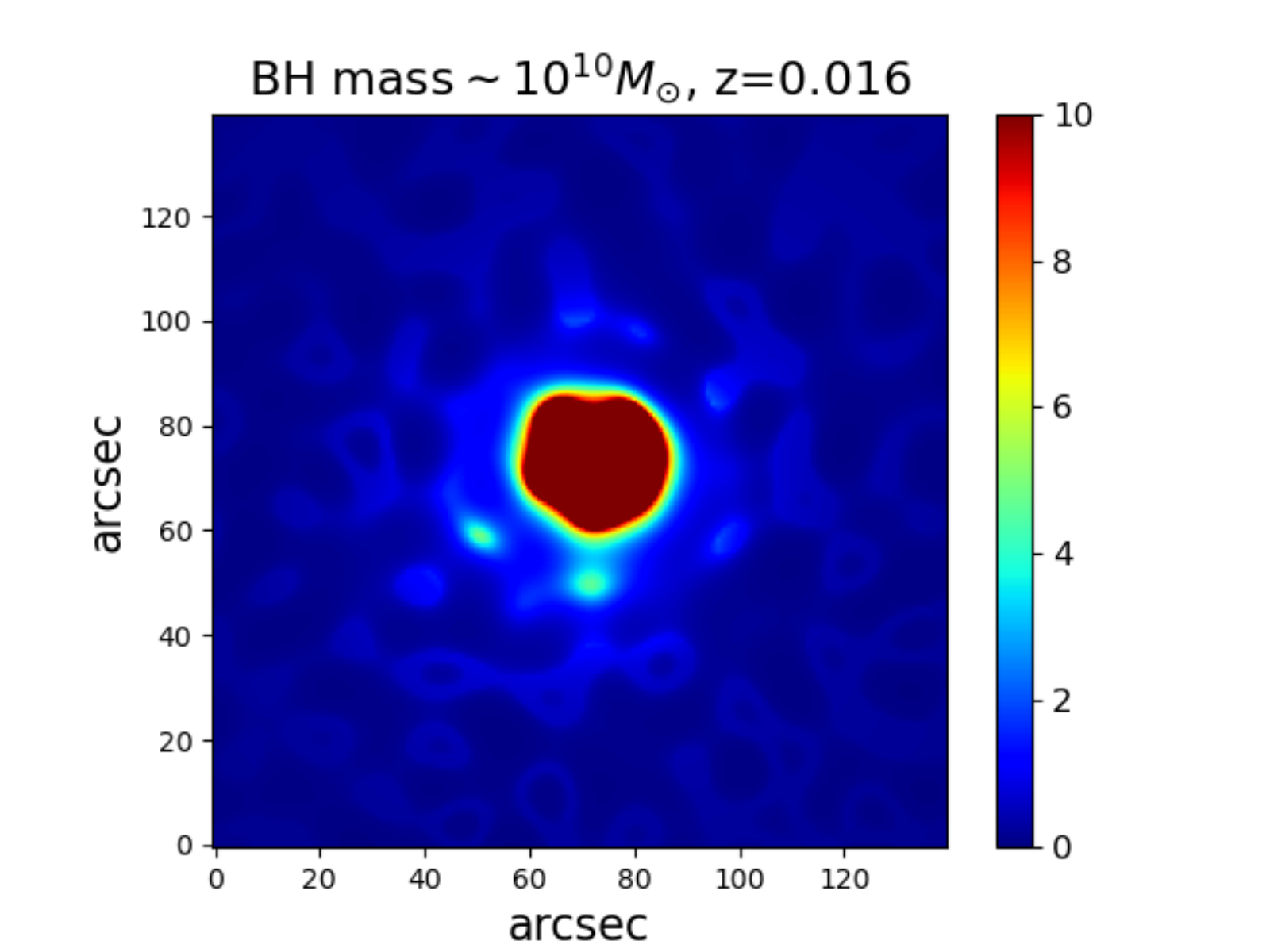}&\includegraphics[width=4.5cm]{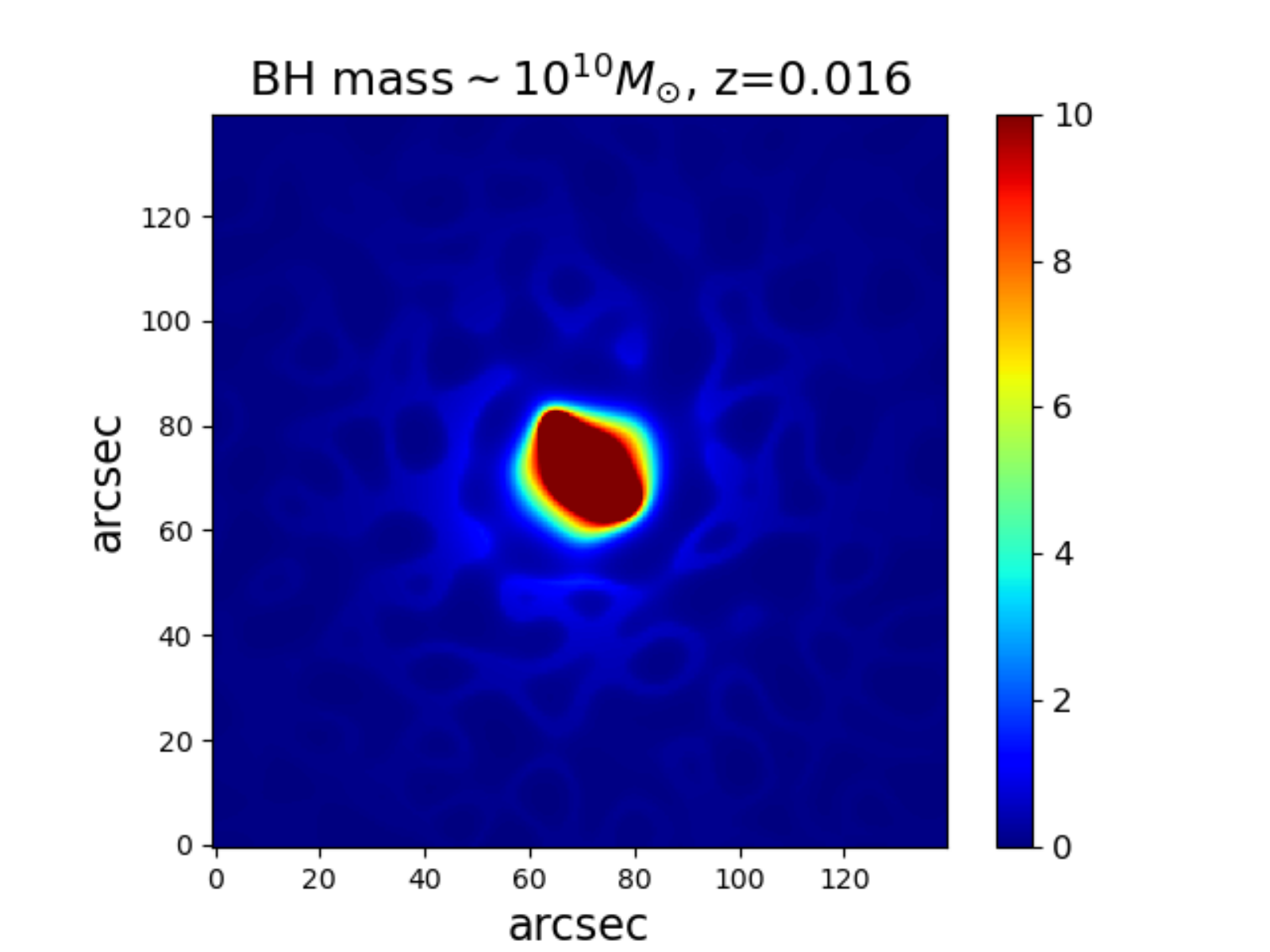}\\
       \end{tabular}
        \caption{Simulated ALMA maps constructed using the same observing parameters as Fig.\ 4 but at 135 GHz (Band 4), 100 GHz (Band 3), and 42 GHz (Band 1) for the {\it most massive} lower redshift ($z=.016$) BH. {\bf Top Panel} The mock ALMA tSZ maps for no feedback case. {\bf Second Panel} The corresponding signal-to-noise maps, {\bf Third Panel} Same maps, but now for the no-jet feedback mode. {\bf Fourth Panel} The signal-to-noise maps corresponding to the third panel. We note that the lower frequency maps reveal more structure at lower redshift. See Table 1 and 2 for feedback nomenclature and black hole properties.}
        \label{fig:8}
    \end{center}
\end{figure*}

\begin{figure*}
    \begin{center}
     \begin{tabular}{c |ccc}
     \hline
         &{135 GHz}&{100 GHz}&{42 GHz}\\[0.1pt]
         \hline
       {No}&\includegraphics[width=4.5cm]{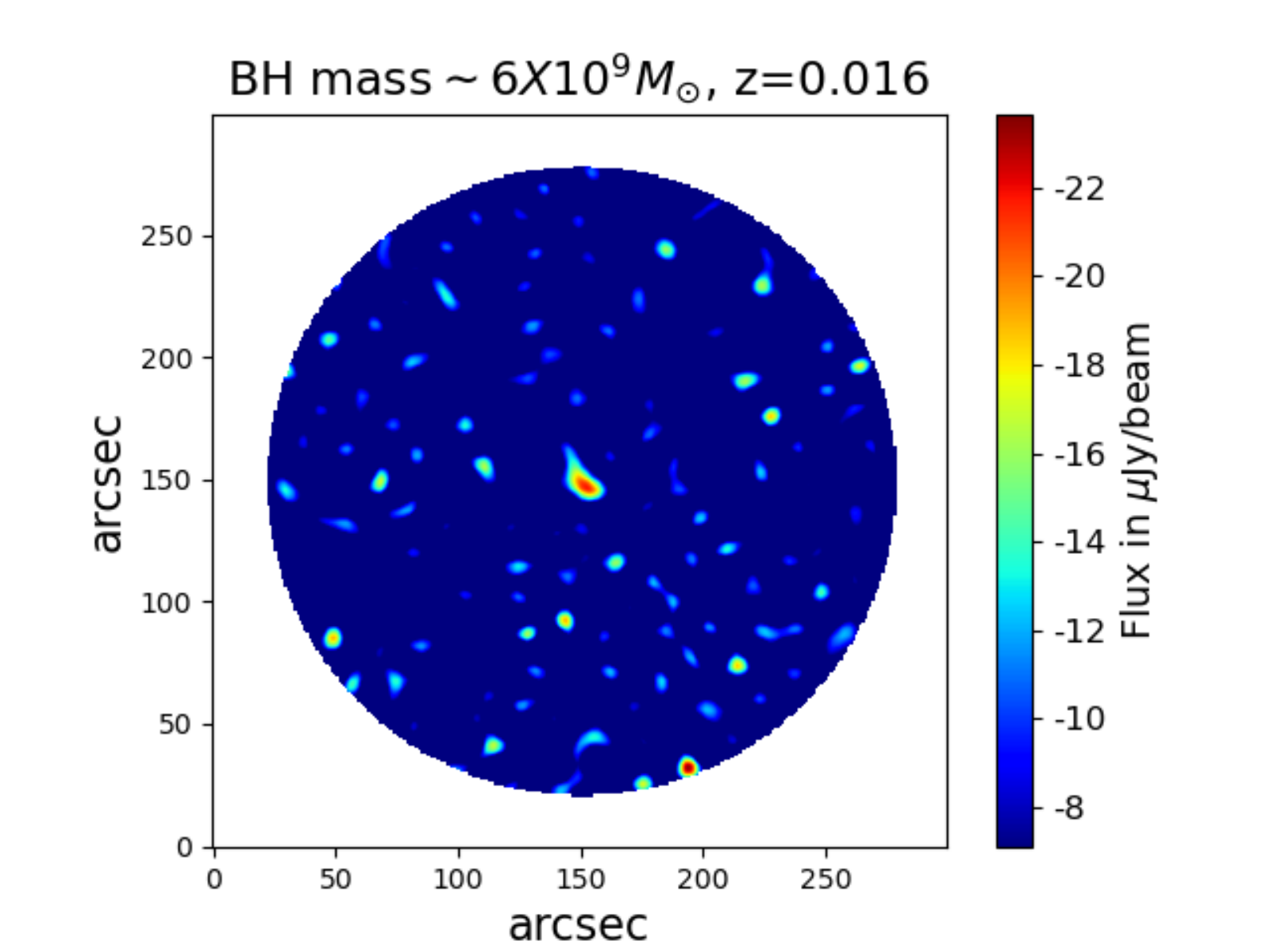}&\includegraphics[width=4.5cm]{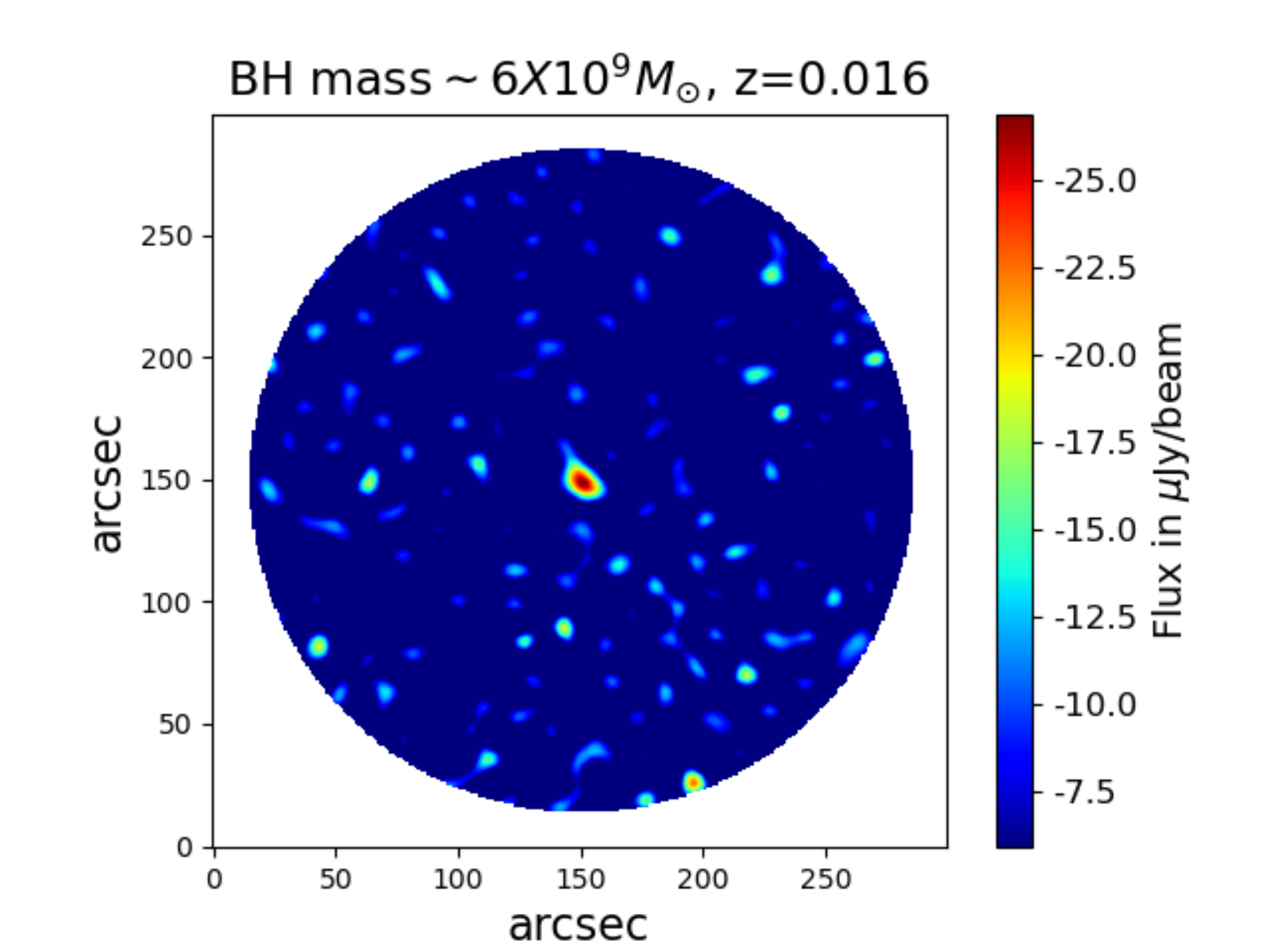}&\includegraphics[width=4.5cm]{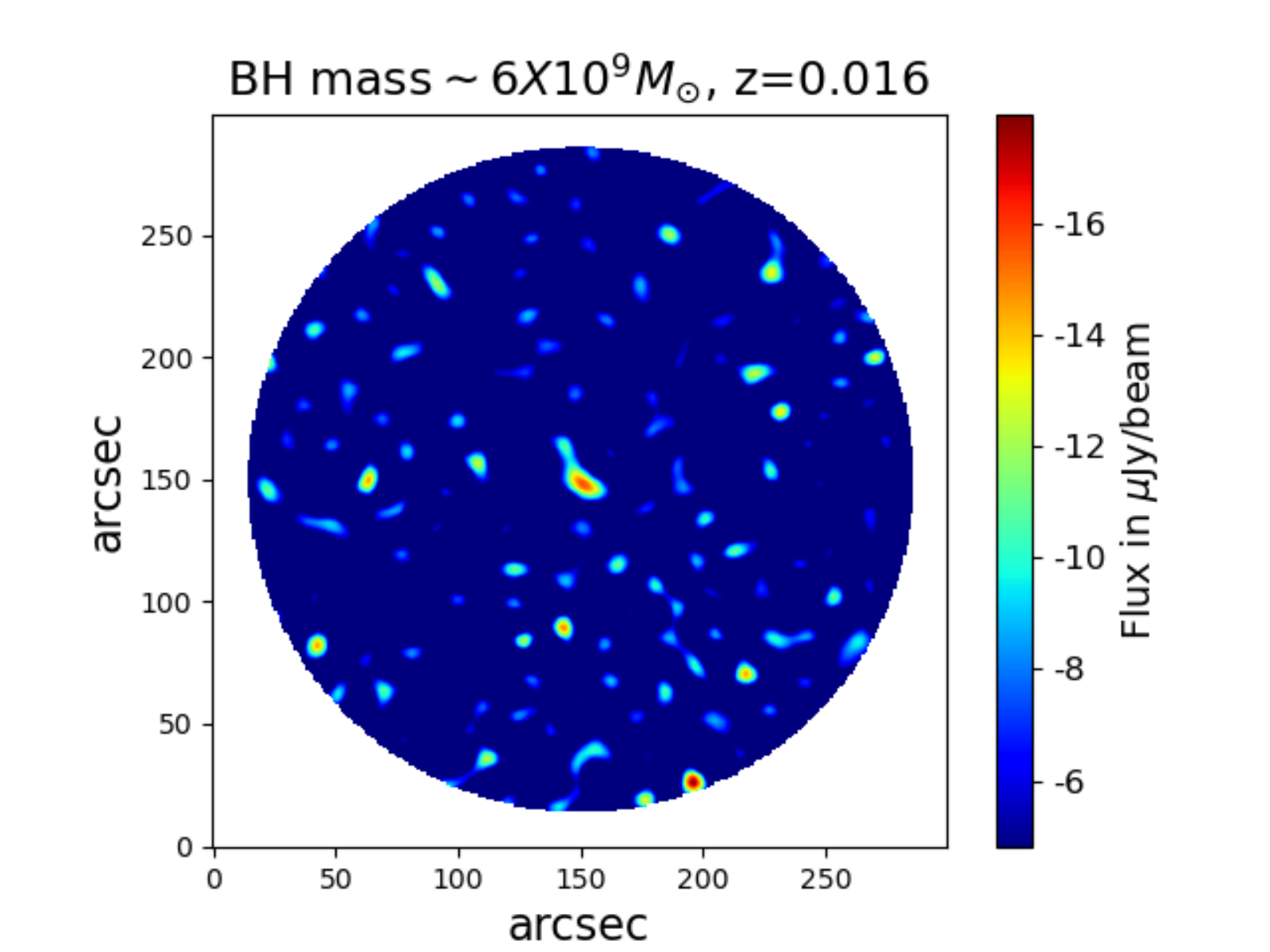}\\
       &\includegraphics[width=4.5cm]{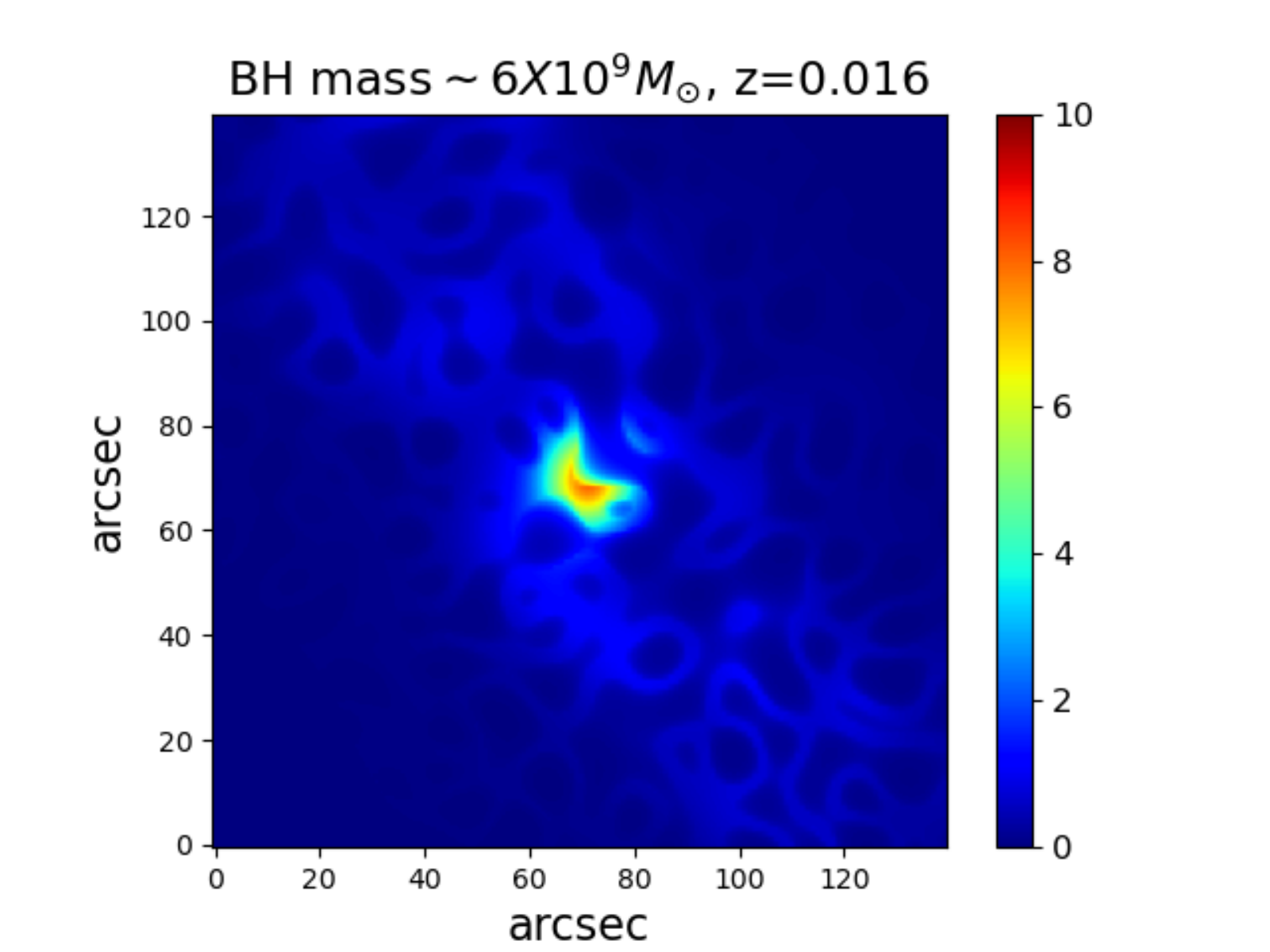}&\includegraphics[width=4.5cm]{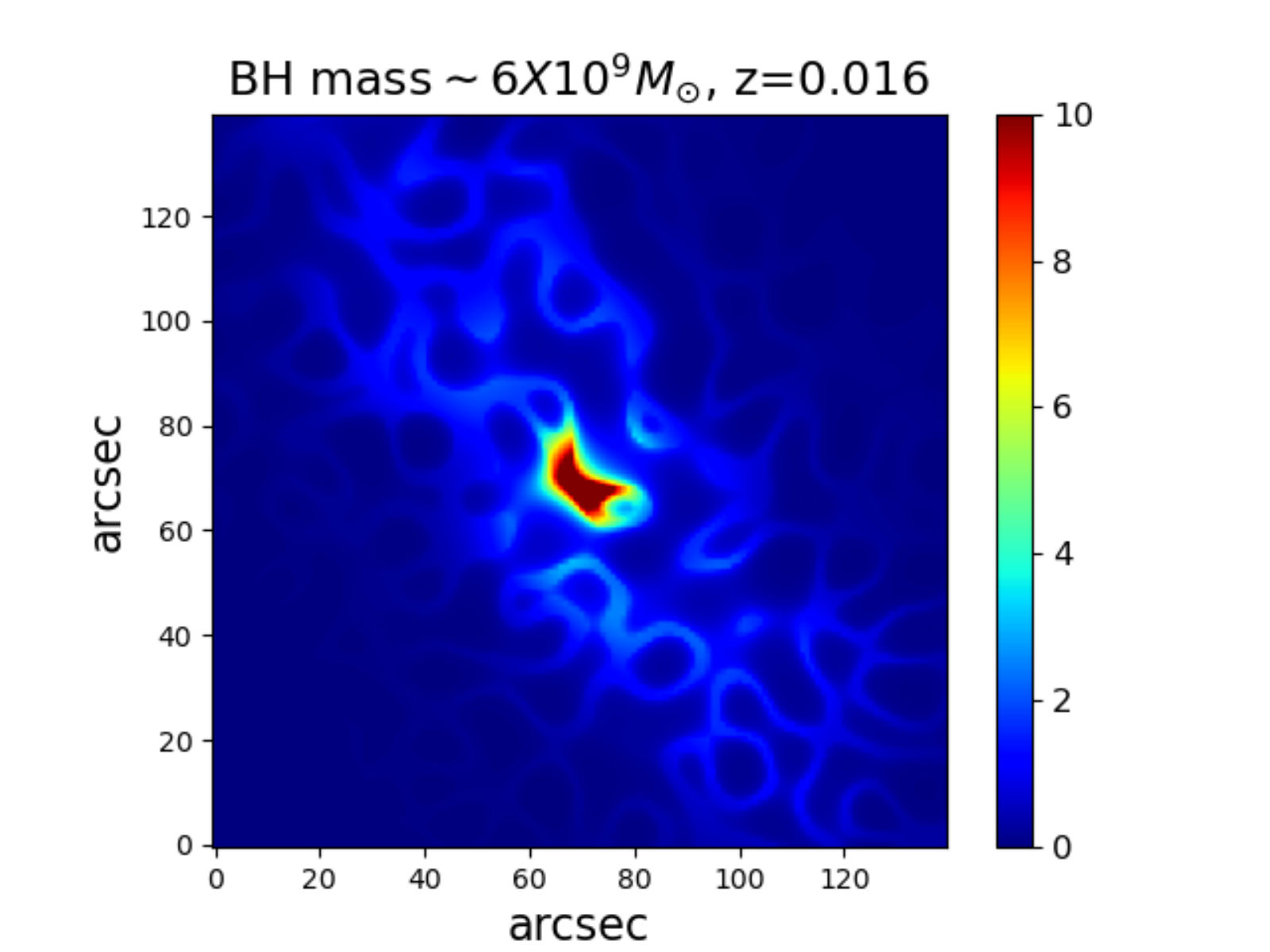}&\includegraphics[width=4.5cm]{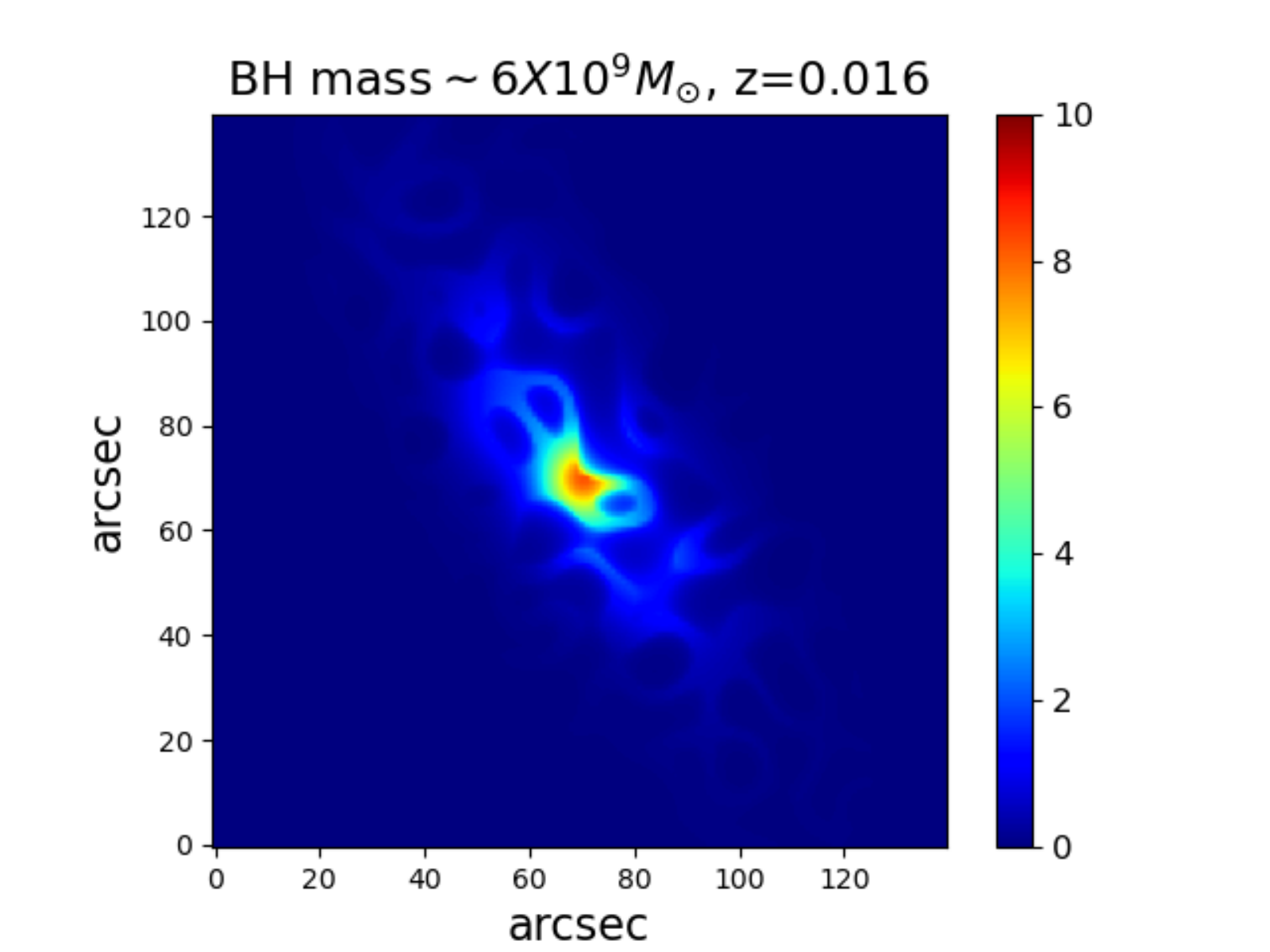}\\
       \hline
       {No-Jet}&\includegraphics[width=4.5cm]{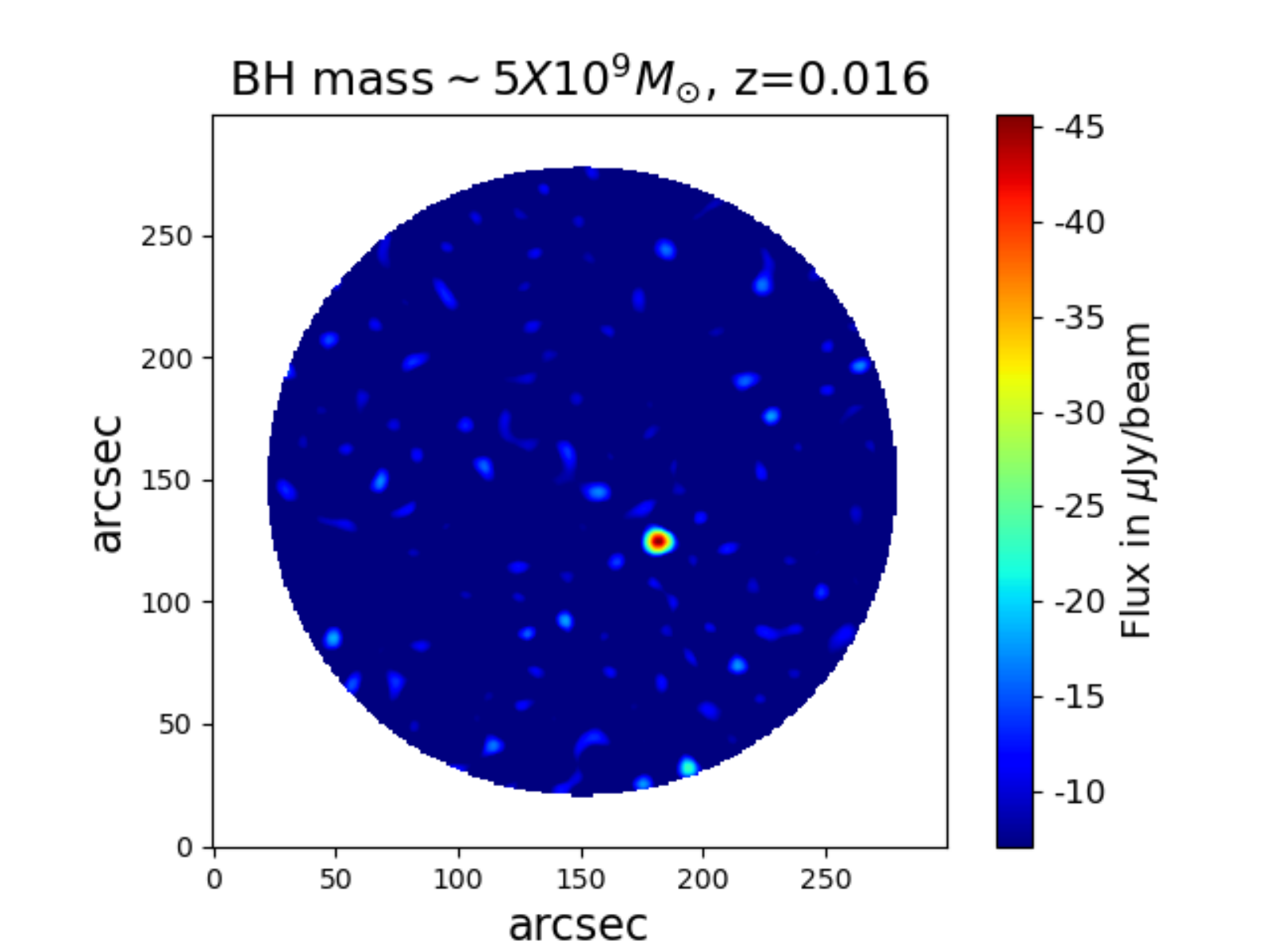}&\includegraphics[width=4.5cm]{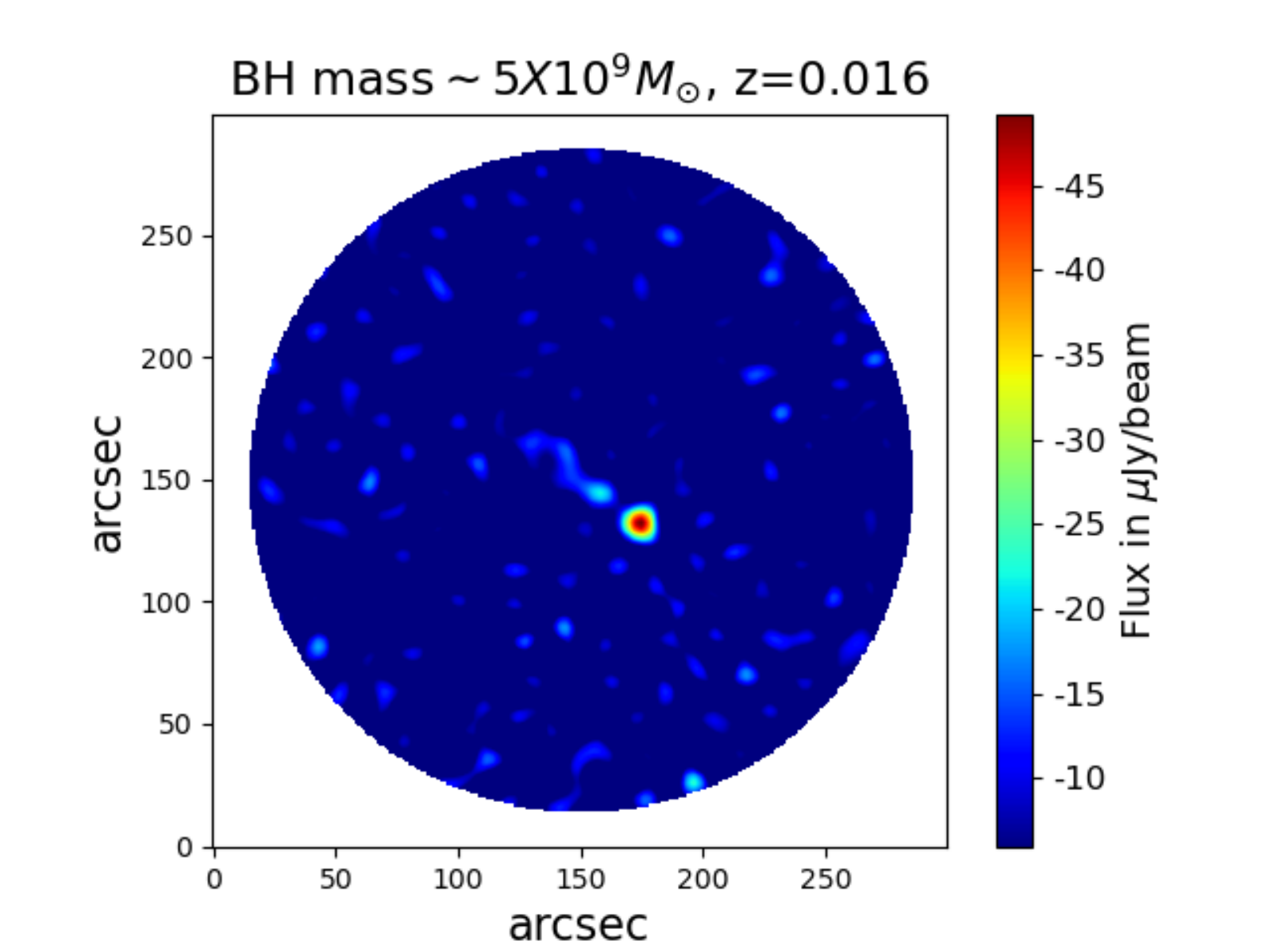}&\includegraphics[width=4.5cm]{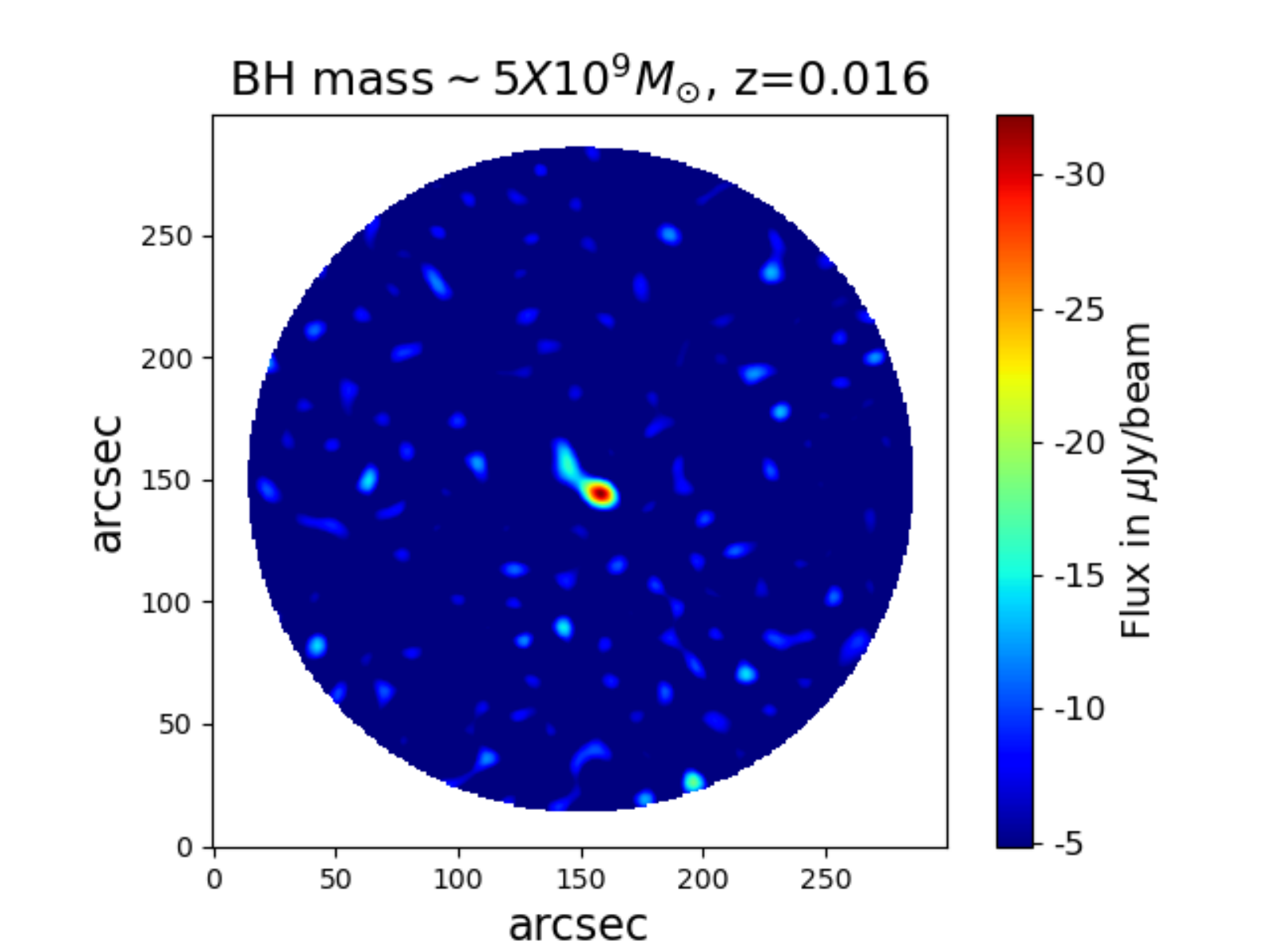}\\
        &\includegraphics[width=4.5cm]{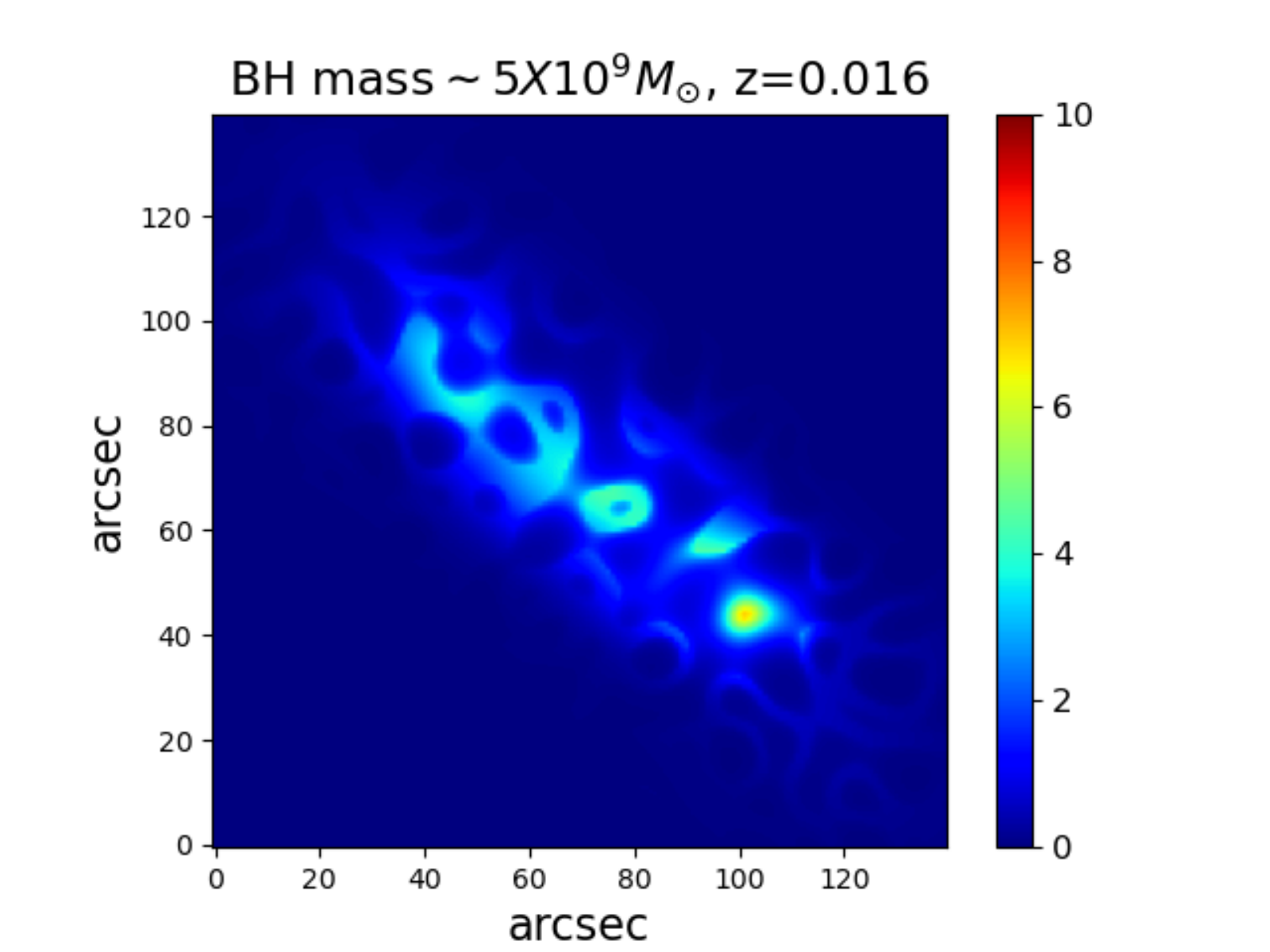}&\includegraphics[width=4.5cm]{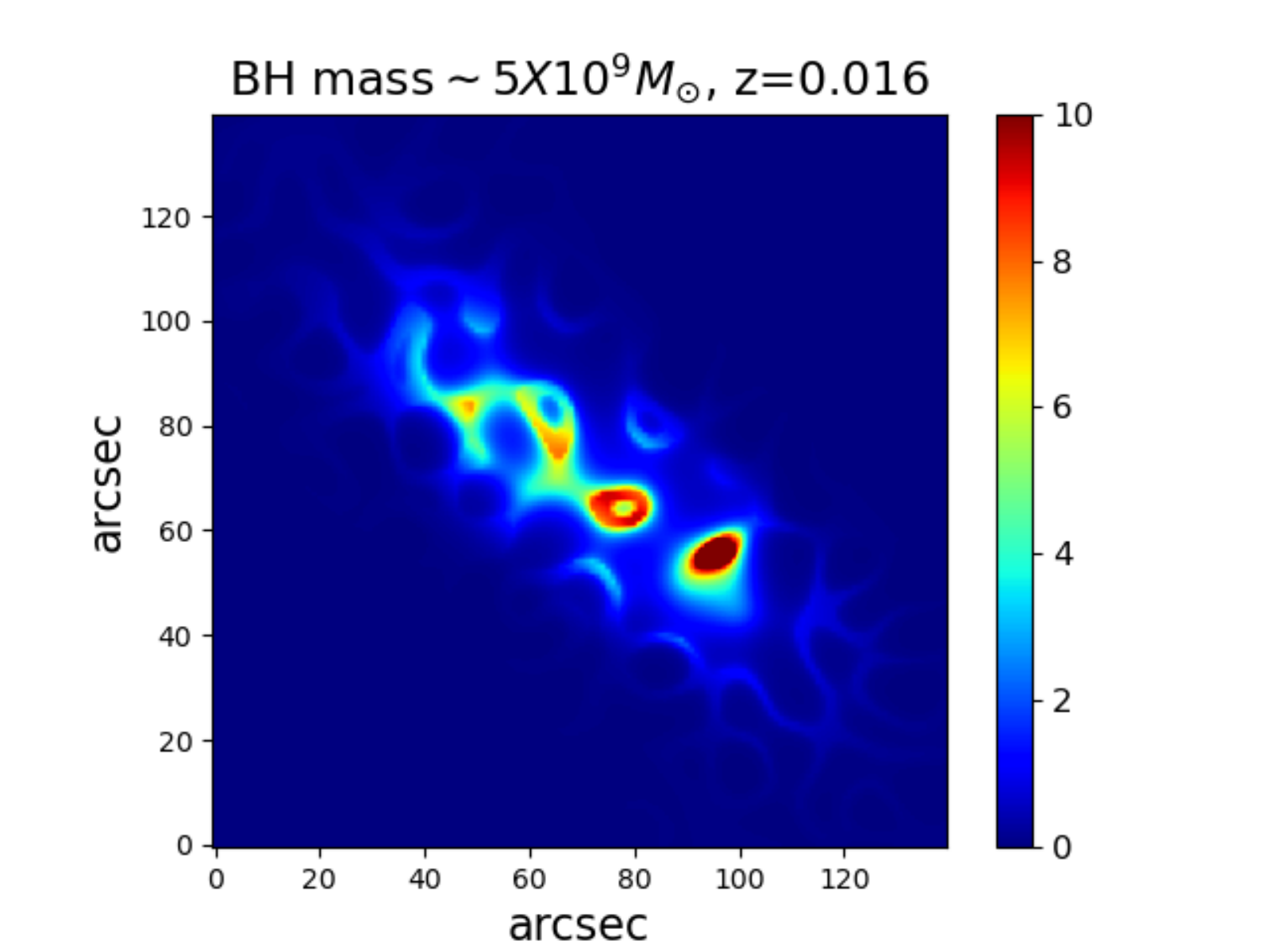}&\includegraphics[width=4.5cm]{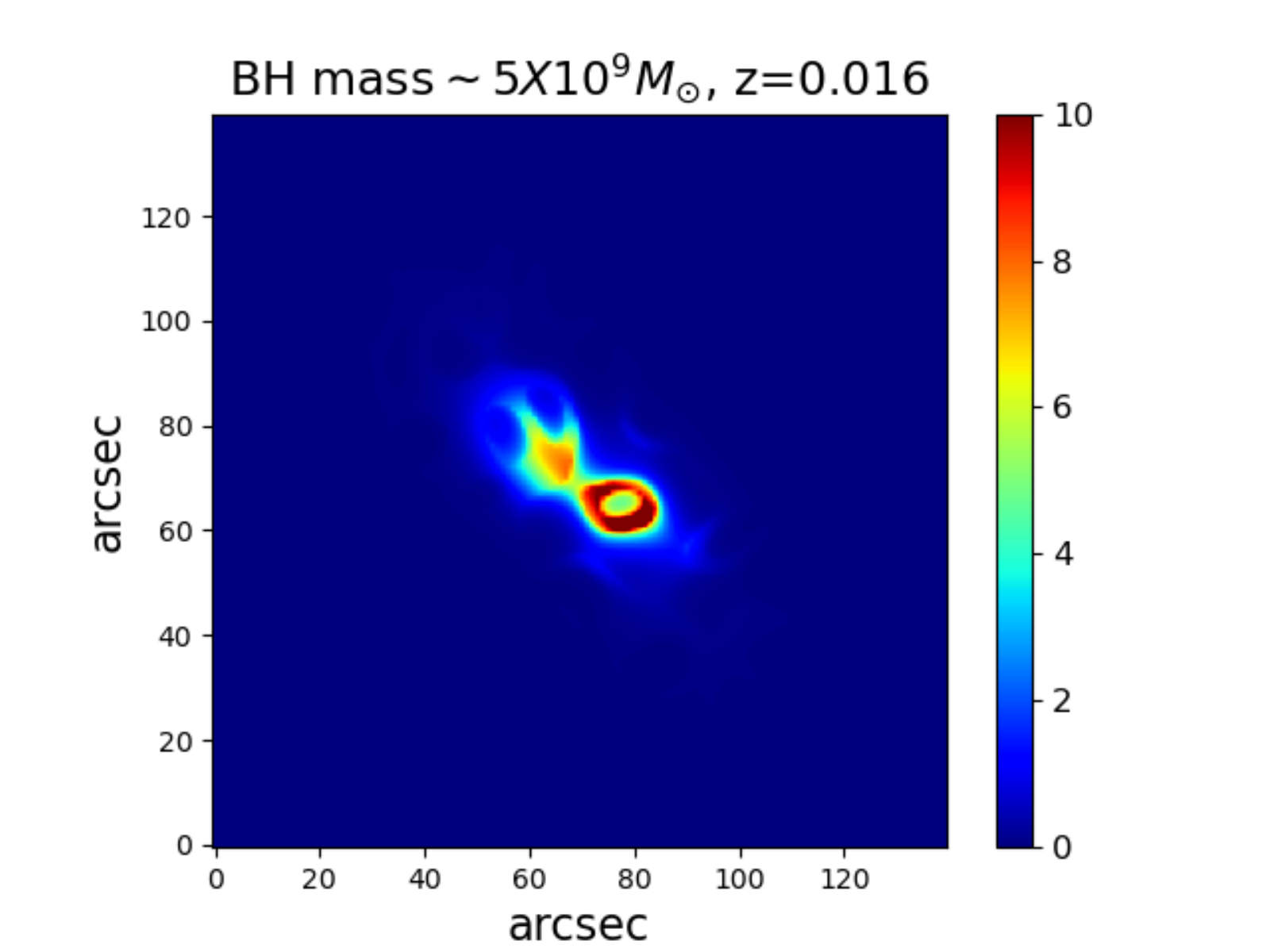}\\
       \end{tabular}
        \caption{Simulated ALMA maps constructed using the same observing parameters as Fig.\ 4 but at 135 GHz (Band 4), 100 GHz (Band 3), and 42 GHz (Band 1) for the {\it most active} lower redshift ($z=.016$) BH. {\bf Top Panel} The mock ALMA tSZ maps for no feedback. {\bf Second Panel} The corresponding signal-to-noise maps, {\bf Third Panel} Same maps, but now for the no-jet feedback mode. {\bf Fourth Panel} The signal-to-noise maps corresponding to the third panel. See Table 1 and 2 for feedback nomenclature and black hole properties. The ALMA results are summarized in Table 4.}
        \label{fig:9}
    \end{center}
\end{figure*}

\subsection{Theoretical Maps}\label{section_3.1}
Figure \ref{fig:1} shows the simulated tSZ maps ($500 \times 500$ pixels) at 320 GHz for different feedback modes around the BHs (identified in the simulations) in a 100 square $h^{-1}$kpc region at z$\sim$1 for both most massive and most active cases. Here we compare tSZ signal for different feedback modes, namely no feedback, no-jet feedback, no X-ray feedback, and all feedback at z$\sim$1 for two different cases. Properties of the most massive and most active BHs are listed in Table 2. The maps reveal that, for both cases tSZ signal is the lowest for all feedback mode, and the suppression of the signal is driven by the jet mode of feedback. The SZ signal gets enhanced for the no-jet feedback (radiative feedback only) case compared to the no-feedback model. 

Figure \ref{fig:2} shows the theoretical simulated tSZ maps ($500 \times 500$ pixels) at 320 GHz for different feedback modes around similar BHs in a 100 square kpc $h^{-1}$ region at $z=0.016$. Here we compare tSZ signal for the same feedback modes as Figure \ref{fig:1} but at $z=0.016$. Properties of the BHs are listed in Table 2. We find that the lower redshift results exhibit similar trends to that of higher redshifts. From our results we observe that for both redshifts, `jet' is the main driver of feedback and radiative and X-ray feedback have comparatively less effect on altering the SZ signal from the no-feedback case. The implications of these results are discussed in \S 4. We also constructed theoretical tSZ maps for 135 GHz, 100 GHz, and 42 GHz respectively, where we observe decrement in the tSZ signal instead of increment since those frequencies are below the null frequency of the tSZ effect (Eq.\ 1).

Figure \ref{fig:3} shows the profiles of the theoretical tSZ signal computed from Figs.\ 1 and 2. From the figure we can observe that for all four cases, no feedback mode and no-jet feedback mode are almost indistinguishable but there is a significant suppression of the signal for all feedback and no X-ray feedback modes at both redshifts. The implications of these differences in the suppression and enhancement of signals will be evident in the observational feasibility.

\subsection{Mock Observational Maps}\label{section_3.2}
Figure \ref{fig:4} shows the simulated ALMA maps constructed using the observing parameters shown in Table 3 where the minimum flux of the maps is set to the rms value of the noise. The maps are for the same black holes represented in Figure \ref{fig:1} and Figure \ref{fig:2} respectively. The black hole parameters are listed in Table 2. Each map is constructed for a $30^{''} \times 30^{''}$ region in the sky and the flux is reported in $\mu$Jy. The maps are made at 320 GHz (Band 7). The leftmost column refers to the case when there is no feedback while the rightmost column represents the all feedback case for all the black holes at the two representative redshifts ($z\sim$1 and $z=0.016$). The second and the third columns represent the cases where only the radiative feedback and the radiative and the jet modes of feedback are present (see Table 1 for the nomenclature). The corresponding signal-to-noise maps are presented in Figure \ref{fig:5}. Figures \ref{fig:6}, \ref{fig:7}, \ref{fig:8} and \ref{fig:9} represent the simulated ALMA maps constructed using the same observing parameters as Fig.\ 4 but at 135 GHz (Band 4), 100 GHz (Band 3), and 42 GHz (Band 1). We note that for all cases , turning on the jet feedback suppresses the signal below the ALMA detection threshold and hence we have only included the no-feedback and the radiative feedback simulations for the lower frequency bands.

From Figs.\ 4 and 5, we note that once the jet mode of feedback is turned on, the SZ signal drops significantly and goes below the detection threshold of ALMA. It is also seen that for all systems, the signal is most prominent when the radiative mode of feedback is turned on. The signal is more pronounced at higher redshifts as seen from the results presented in Fig.\ 5. It has been noted in other studies that AGN feedback generally enhances the temperature of the IGM/ICM gas while suppressing the gas density \citep[e.g.,][]{RKC22, RKC21, chatterjeeetal08}. Since the SZ signal comes from the integrated line-of-sight gas pressure, there is always an optimization between temperature and density in the magnitude of the effect. The increase in SZ signal, as reported is predominantly, due to the presence of hot gas from AGN feedback \citep[e.g.,][] {scannapiecoetal08, chatterjeeetal08}. Our results show that such is the case for the radiative mode only, but the suppression of density is much higher in the jet mode resulting in a decrement of the SZ signal. 

ALMA observations may serve as important probes in understanding feedback effects at higher redshifts as well as constraining theoretical models of feedback. In Figs.\ 6 through 9 we present the ALMA mock SZ simulations for the four black holes (listed in Table 2) for the lower frequency bands. Our results show that at lower frequencies we generally tend to resolve more structures in the low redshift case. The ALMA simulation results are summarized in Table 4. We discuss the implications of these results in the next section.

\section{Discussion of Results}\label{section_4}
In recent years the SZ effect has become a powerful tool to probe the warm hot universe at multiple length scales \citep{mroczkowskietal19}. In the pioneering work of \citet{c&k07}, it was suggested that AGN feedback can be probed using the SZ effect through future high resolution submm experiments. The first detection of the SZ effect from quasar feedback with the Atacama Large Millimeter Array \citep{lacyetal19} served as a validation to the proposed studies \citep[e.g][]{n&s99, yamadaetal99, chatterjeeetal08, scannapiecoetal08}. Following the first detection, other groups followed \citep{brownsonetal19, halletal19}.

In a series of recent works, SZ and X-ray studies have been suggested to constrain models of baryonic physics, and specifically the role of AGN feedback in determining the thermodynamic properties of the ICM \citep[e.g.,][]{eckertetal21, chadayammurietal22, acharyaetal21, RKC22, yang22, junhanetal22}. In a previous study \citet{brownsonetal19} proposed using the FABLE cosmological simulations, that ALMA has the potential to constrain AGN feedback models through SZ observations. The study proposed that the best observational band for detecting the SZ signal would be band 3 ($\sim$ 100 GHz). Recently \citet{junhanetal22} used the Illustris TNG, EAGLE and FIRE simulations to infer that the properties of the circumgalactic medium (CGM) and the ICM are highly sensitive to feedback prescriptions, and concluded that SZ measurements of the gas can potentially put constraints on theoretical models. A detection of the SZ effect in the CGM has been recently reported by \citet{dasetal23}.

\begin{table}
\caption{Summary of Mock ALMA SZ Results} 
\centering 
\begin{tabular}{c |c |c | c c c c} 
\hline \hline
& Mode & z & Band  & Beam   & SNR & Signal ($\mu$Jy)\\ 
&      &   &            &(arcsec)& Peak           & Peak \\ 
[0.5ex] 
\hline 
Massive & None & 0.99 & 7 & 1.04 & 12 & 2604.8\\
&  &  & 4 & 2.6 & 3 & -221.5\\
&  &  & 3 & 3.4 & 6 & -234.8\\
&  &  & 1 & 8.2 & 37 & -68.7\\
&  & 0.016 & 7 & 1.04 & 3 & 211\\
&  &  & 4 & 2.6 & 4 & -58.3\\
&  &  & 3 & 3.4 & 4 & -91.8\\
&  &  & 1 & 8.2 & 2 & -60.9\\
& No-Jet & 0.99 &  7 & 1.04 & 10 & 3692.2\\ 
&  &  & 4 & 2.6 & 4 & -300.3\\
&  &  & 3 & 3.4 & 5 & -316.6\\
&  &  & 1 & 8.2 & 13 & -91.8\\
&  & 0.016 & 7 & 1.04 & 19 & 5817.7\\
&  &  & 4 & 2.6 & 12 & -1242.4\\
&  &  & 3 & 3.4 & 20 & -1767.5\\
&  &  & 1 & 8.2 & 9 & -834.4\\[0.5ex] 
\hline 
Active & None & 0.99 &  7 & 1.04 & 4 & 87.3\\
&  &  &  4 & 2.6 & 1 & -24.2\\
&  &  &  3 & 3.4 & 2 & -23.1\\
&  &  &  1 & 8.2 & 4 & -17.9\\
&  & 0.016 &  7 & 1.04 & 1 & 86.7\\
&  &  &  4 & 2.6 & 2 & -23.7\\
&  &  &  3 & 3.4 & 3 & -26.9\\
&  &  &  1 & 8.2 & 2 & -18\\
& No-Jet & 0.99 &  7 & 1.04 & 8 & 290.6\\ 
&  &  & 4 & 2.6 & 3 & -209.7\\
&  &  & 3 & 3.4 & 4 & -189.3\\
&  &  & 1 & 8.2 & 20 & -53.6\\
&  & 0.016 & 7 & 1.04 & 1 & 87.6\\
&  &  & 4 & 2.6 & 3 & -45.6\\
&  &  & 3 & 3.4 & 5 & -49.2\\
&  &  & 1 & 8.2 & 4 & -32.3\\[0.5ex] 
\hline\hline
\end{tabular}
\end{table}

In the current work we use the cosmological simulation SIMBA (D19), to constrain models of AGN feedback using ALMA compact array observations. As discussed before, our results show that certain models of AGN feedback go below the detection threshold of ALMA, and that provides an upper limit from SZ observables, on the amount of feedback that is allowed in cosmological simulations. Using the same set of simulations, in a companion paper, \citet{RKC22} show that feedback suppresses the X-ray flux from the vicinity of the black hole and the jet mode of feedback plays the most significant role in evacuating the gas from the centers of groups and clusters, resulting in suppression of X-ray signal. 

\citet{RKC22} did a thorough study of the thermodynamic properties of the ICM for both individual objects as well as statistical samples. It is observed that once radiative feedback is introduced in the simulation, the temperature of the gas gets slightly enhanced, however the gas density is suppressed in the vicinity of the black hole. It is noted that the scale of evacuation due to the radiative wind feedback is not high enough and hence a density enhancement is observed at distances further away from the black hole. The drop in gas density, closer to the black hole stays responsible for the drop in X-ray flux in the vicinity of the black hole, compared to the scenario where there is no feedback. Previously \citet{RKC21} noted similar effects with the MassiveBlack-II simulation \citep{khandaietal15} where, only the radiative mode of feedback was used in the modeling. However, in this work we compute the SZ flux which is an integrated line-of-sight pressure and thus the increase in gas temperature manifests more strongly in the SZ signal, and for the systems under consideration, we observe an enhancement of SZ effect compared to the no feedback case. Previously, enhancement of SZ signal due to the radiative mode of feedback was also reported by other groups using different simulations \citep[e.g.,][]{chatterjeeetal08,scannapiecoetal08}.

The situation is dramatically reversed, once the jet feedback is turned on. It has been noted that the temperature as well as the density of gas becomes significantly lower once the jet feedback is turned on \citep{RKC22,Robson20,Robson23}. The powerful jet evacuates the gas to larger length scales resulting in drastic drop of the temperature as well as the density of hot gas near to the black hole. This effect results in strong suppression of both the X-ray and the SZ fluxes. Although in simulations it is possible to turn on and turn-off different modes of feedback and assess their impact on the ICM, it was noted in \citet{RKC22} that detection of X-ray cavities due to feedback effects with current and future telescopes will only be useful in disentangling the combined effects of jet versus radiative modes. It is not possible to quantify their individual contributions with X-ray observations only.  

Our SZ results reveal that unlike X-rays, SZ signals can get both enhanced or suppressed due to feedback effects, depending on the mode of feedback. This provides a unique route to combine X-ray and SZ measurements to disentangle the mode of feedback that is dominant in a given system. In a future work we wish to perform X-ray-SZ cross studies to address the effect of different modes of feedback. Recently \citet{chadayammurietal22} concluded using eROSITA stacks of the circumgalactic medium (CGM), that numerical simulations include higher levels of feedback from AGN while quenching star formation in the CGM. We note that adding the SZ component will substantially improve the constraining power of such observations. 

In addition to the unique feature in SZ signals where we can get both enhancement and decrement from AGN feedback there is an added advantage of SZ effect being redshift independent. \citet{RKC22} reported that X-ray studies of feedback effects will only be restricted to low redshift systems. This is due to the fact that the X-ray dimming happens due to the distance effects and hence with similar physical conditions it is harder to detect high redshift systems with X-rays. SZ effect, on the other hand is a distortion in the CMB and hence is a redshift independent observable \citep{carlstrometal02}. Further, AGN are more active at higher redshifts, resulting in higher level of energy injection into the IGM/ICM. Thus the effect is generally seen to be more pronounced at high redshifts \citep{chatterjeeetal08}. Our results show that the SZ signal is indeed enhanced at higher redshifts confirming previous work, and hence it is possible to detect the high redshift signal with current ALMA configuration (evident from Figs.\ 4, 5, 6 and 7). 

Recently \citet{wadekaretal23} made use of extensive machine learning tools to study the effect of feedback parameters on the Y-M (integrated Y distortion and halo mass) relation of galaxy groups and clusters. They used the SIMBA and the Illustris-TNG simulations and demonstrated that, stronger the AGN feedback parameter (e.g., jet speed in the case of SIMBA) higher is the deviation of the Y-M scaling from self-similarity. They also demonstrate that with the SIMBA prescriptions the Y parameter for the entire halo gets substantially suppressed due to the inclusion of jet- feedback in the simulation. The statistical studies of \citet{wadekaretal23}, along with the results obtained from the AGN feedback parameter space exploration, clearly indicate the strong role of jet feedback in suppressing the SZ signal in halos at all mass scales. This is completely consistent with our current results, where we focused on the thermodynamic parameters of individual systems. As discussed before, our analysis provide yet another robust tool, namely the SZ effect to observationally constrain the models used in current cosmological simulations.  

In this study we use the most compact antenna configuration for ALMA and
four frequency bands (320 GHz, 135 GHz, 100 GHz, and 42 GHz,
corresponding to ALMA Bands 7, 4, 3 and 1) to investigate how the
detection of the SZ signal depends on the frequency and spatial sampling
of the array. We note that the lower frequency bands capture more of the
signal on large angular scales, particularly important for our low
redshift simulation (Fig.\ 8 and Fig.\ 9). Simulations can thus be a
useful guide to optimizing observing strategies. We note that our study
focuses on the thermal SZ distortions from quasar feedback and hence
detection of the spectral distortion from the CMB in a single band is
adequate. In future, we wish to include simulations of the kinetic SZ
effect, and to distinguish the kinetic and thermal SZ components we need
to consider multiband observations with similar uv-plane coverage.
Finally, although visibility-plane model-fitting has been used for
analysis of the SZ effect in ALMA data \citep[e.g.,][]{brownsonetal19}, we note that the good uv-plane sampling of ALMA results in high fidelity images that are able to accurately represent the details of the SZ signal
without resorting to model fitting of the visibility data.

The importance of AGN/quasar feedback in galaxy evolution has been very well studied in the literature. The prospect of detecting AGN/quasar feedback via the SZ effect serves as a potential probe to observe the signal to high redshifts. In this work we show that the compact configuration of the Atacama Large Millimeter Array not only provides an avenue to detect the signal at higher redshifts, it also potentially has the constraining power to observationally distinguish between current feedback models in numerical simulations. Our results also establish that the best route to constrain feedback models lies in a combined X-ray-SZ analysis of the same system.

\section{Acknowledgements}
The authors would like to thank the referee for making very important suggestions which greatly helped in improving the draft. A.C. and S.C. thank Inter-University Centre for Astronomy and Astrophysics
(IUCAA) for providing computational support through the Pegasus supercomputing facility. S.C. acknowledges support from the Department of Science and Technology, GOI, through the SERB- CRG-2020-002064 grant and from the Department of Atomic Energy, GOI, for the 57/14/10/2019-BRNS grant. RKC thanks National Natural Science Foundation of China (HKU12122309) for financial support. The National Radio Astronomy Observatory is a facility of the National Science Foundation operated under cooperative agreement by Associated Universities, Inc.

\bibliography{arXiv}{}

\begin{thebibliography}{}
\expandafter\ifx\csname natexlab\endcsname\relax\def\natexlab#1{#1}\fi
\providecommand{\url}[1]{\href{#1}{#1}}
\providecommand{\dodoi}[1]{doi:~\href{http://doi.org/#1}{\nolinkurl{#1}}}
\providecommand{\doeprint}[1]{\href{http://ascl.net/#1}{\nolinkurl{http://ascl.net/#1}}}
\providecommand{\doarXiv}[1]{\href{https://arxiv.org/abs/#1}{\nolinkurl{https://arxiv.org/abs/#1}}}

\bibitem[{{Acharya} {et~al.}(2021){Acharya}, {Majumdar}, \&
  {Nath}}]{acharyaetal21}
{Acharya}, S.~K., {Majumdar}, S., \& {Nath}, B.~B. 2021, \mnras, 503, 5473,
  \dodoi{10.1093/mnras/stab810}

\bibitem[{{Aghanim} {et~al.}(1999){Aghanim}, {Desert}, {Puget}, \&
  {Gispert}}]{aghanimetal99}
{Aghanim}, N., {Desert}, F.~X., {Puget}, J.~L., \& {Gispert}, R. 1999, \aap,
  341, 640

\bibitem[{{Aller} \& {Richstone}(2007)}]{alleretal07}
{Aller}, M.~C., \& {Richstone}, D.~O. 2007, \apj, 665, 120,
  \dodoi{10.1086/519298}

\bibitem[{{Andersson} {et~al.}(2009){Andersson}, {Peterson}, {Madejski}, \&
  {Goobar}}]{Andersson09}
{Andersson}, K., {Peterson}, J.~R., {Madejski}, G., \& {Goobar}, A. 2009, \apj,
  696, 1029, \dodoi{10.1088/0004-637X/696/1/1029}

\bibitem[{{Angl{\'e}s-Alc{\'a}zar}
  {et~al.}(2017{\natexlab{a}}){Angl{\'e}s-Alc{\'a}zar}, {Dav{\'e}},
  {Faucher-Gigu{\`e}re}, {{\"O}zel}, \& {Hopkins}}]{angle17}
{Angl{\'e}s-Alc{\'a}zar}, D., {Dav{\'e}}, R., {Faucher-Gigu{\`e}re}, C.-A.,
  {{\"O}zel}, F., \& {Hopkins}, P.~F. 2017{\natexlab{a}}, \mnras, 464, 2840,
  \dodoi{10.1093/mnras/stw2565}

\bibitem[{{Angl{\'e}s-Alc{\'a}zar}
  {et~al.}(2017{\natexlab{b}}){Angl{\'e}s-Alc{\'a}zar}, {Dav{\'e}},
  {Faucher-Gigu{\`e}re}, {{\"O}zel}, \& {Hopkins}}]{Angles-Alcazar17}
---. 2017{\natexlab{b}}, \mnras, 464, 2840, \dodoi{10.1093/mnras/stw2565}

\bibitem[{{Angl{\'e}s-Alc{\'a}zar} {et~al.}(2013){Angl{\'e}s-Alc{\'a}zar},
  {{\"O}zel}, \& {Dav{\'e}}}]{Angles-Alcazar13}
{Angl{\'e}s-Alc{\'a}zar}, D., {{\"O}zel}, F., \& {Dav{\'e}}, R. 2013, \apj,
  770, 5, \dodoi{10.1088/0004-637X/770/1/5}

\bibitem[{{Angl{\'e}s-Alc{\'a}zar} {et~al.}(2015){Angl{\'e}s-Alc{\'a}zar},
  {{\"O}zel}, {Dav{\'e}}, {Katz}, {Kollmeier}, \&
  {Oppenheimer}}]{Angles-Alcazar15}
{Angl{\'e}s-Alc{\'a}zar}, D., {{\"O}zel}, F., {Dav{\'e}}, R., {et~al.} 2015,
  \apj, 800, 127, \dodoi{10.1088/0004-637X/800/2/127}

\bibitem[{{Bondi}(1952)}]{Bondi52}
{Bondi}, H. 1952, \mnras, 112, 195, \dodoi{10.1093/mnras/112.2.195}

\bibitem[{{Bondi} \& {Hoyle}(1944)}]{BondiHoyle}
{Bondi}, H., \& {Hoyle}, F. 1944, \mnras, 104, 273,
  \dodoi{10.1093/mnras/104.5.273}

\bibitem[{{Brownson} {et~al.}(2019){Brownson}, {Maiolino}, {Tazzari},
  {Carniani}, \& {Henden}}]{brownsonetal19}
{Brownson}, S., {Maiolino}, R., {Tazzari}, M., {Carniani}, S., \& {Henden}, N.
  2019, \mnras, 490, 5134, \dodoi{10.1093/mnras/stz2945}

\bibitem[{{Carlstrom} {et~al.}(2002){Carlstrom}, {Holder}, \&
  {Reese}}]{carlstrometal02}
{Carlstrom}, J.~E., {Holder}, G.~P., \& {Reese}, E.~D. 2002, \araa, 40, 643,
  \dodoi{10.1146/annurev.astro.40.060401.093803}

\bibitem[{{Cattaneo} {et~al.}(2009){Cattaneo}, {Faber}, {Binney}, {Dekel},
  {Kormendy}, {Mushotzky}, {Babul}, {Best}, {Br{\"u}ggen}, {Fabian}, {Frenk},
  {Khalatyan}, {Netzer}, {Mahdavi}, {Silk}, {Steinmetz}, \&
  {Wisotzki}}]{Cattaneo09}
{Cattaneo}, A., {Faber}, S.~M., {Binney}, J., {et~al.} 2009, \nat, 460, 213,
  \dodoi{10.1038/nature08135}

\bibitem[{{Chadayammuri} {et~al.}(2022){Chadayammuri}, {Bogd{\'a}n},
  {Oppenheimer}, {Kraft}, {Forman}, \& {Jones}}]{chadayammurietal22}
{Chadayammuri}, U., {Bogd{\'a}n}, {\'A}., {Oppenheimer}, B.~D., {et~al.} 2022,
  \apjl, 936, L15, \dodoi{10.3847/2041-8213/ac8936}

\bibitem[{{Chatterjee} {et~al.}(2008){Chatterjee}, {Di Matteo}, {Kosowsky}, \&
  {Pelupessy}}]{chatterjeeetal08}
{Chatterjee}, S., {Di Matteo}, T., {Kosowsky}, A., \& {Pelupessy}, I. 2008,
  \mnras, 390, 535, \dodoi{10.1111/j.1365-2966.2008.13784.x}

\bibitem[{{Chatterjee} {et~al.}(2010){Chatterjee}, {Ho}, {Newman}, \&
  {Kosowsky}}]{chatterjeeetal10}
{Chatterjee}, S., {Ho}, S., {Newman}, J.~A., \& {Kosowsky}, A. 2010, \apj, 720,
  299, \dodoi{10.1088/0004-637X/720/1/299}

\bibitem[{{Chatterjee} \& {Kosowsky}(2007)}]{c&k07}
{Chatterjee}, S., \& {Kosowsky}, A. 2007, \apjl, 661, L113,
  \dodoi{10.1086/518860}

\bibitem[{{Choi} {et~al.}(2012){Choi}, {Ostriker}, {Naab}, \&
  {Johansson}}]{choi12}
{Choi}, E., {Ostriker}, J.~P., {Naab}, T., \& {Johansson}, P.~H. 2012, \apj,
  754, 125, \dodoi{10.1088/0004-637X/754/2/125}

\bibitem[{{Costa} {et~al.}(2015){Costa}, {Sijacki}, \& {Haehnelt}}]{Costa15}
{Costa}, T., {Sijacki}, D., \& {Haehnelt}, M.~G. 2015, \mnras, 448, L30,
  \dodoi{10.1093/mnrasl/slu193}

\bibitem[{{Crichton} {et~al.}(2016){Crichton}, {Gralla}, {Hall}, {Marriage},
  {Zakamska}, {Battaglia}, {Bond}, {Devlin}, {Hill}, {Hilton}, {Hincks},
  {Huffenberger}, {Hughes}, {Kosowsky}, {Moodley}, {Niemack}, {Page},
  {Partridge}, {Sievers}, {Sif{\'o}n}, {Staggs}, {Viero}, \&
  {Wollack}}]{crichtonetal16}
{Crichton}, D., {Gralla}, M.~B., {Hall}, K., {et~al.} 2016, \mnras, 458, 1478,
  \dodoi{10.1093/mnras/stw344}

\bibitem[{{Das} {et~al.}(2023){Das}, {Chiang}, \& {Mathur}}]{dasetal23}
{Das}, S., {Chiang}, Y.-K., \& {Mathur}, S. 2023, arXiv e-prints,
  arXiv:2305.12353, \dodoi{10.48550/arXiv.2305.12353}

\bibitem[{{Dav{\'e}} {et~al.}(2019){Dav{\'e}}, {Angl{\'e}s-Alc{\'a}zar},
  {Narayanan}, {Li}, {Rafieferantsoa}, \& {Appleby}}]{dave19}
{Dav{\'e}}, R., {Angl{\'e}s-Alc{\'a}zar}, D., {Narayanan}, D., {et~al.} 2019,
  \mnras, 486, 2827, \dodoi{10.1093/mnras/stz937}

\bibitem[{{David} {et~al.}(2001){David}, {Nulsen}, {McNamara}, {Forman},
  {Jones}, {Ponman}, {Robertson}, \& {Wise}}]{David01}
{David}, L.~P., {Nulsen}, P.~E.~J., {McNamara}, B.~R., {et~al.} 2001, \apj,
  557, 546, \dodoi{10.1086/322250}

\bibitem[{{de Nicola} {et~al.}(2019){de Nicola}, {Marconi}, \&
  {Longo}}]{deNicola19}
{de Nicola}, S., {Marconi}, A., \& {Longo}, G. 2019, \mnras, 490, 600,
  \dodoi{10.1093/mnras/stz2472}

\bibitem[{{Di Matteo} {et~al.}(2008){Di Matteo}, {Colberg}, {Springel},
  {Hernquist}, \& {Sijacki}}]{dimatteoetal08}
{Di Matteo}, T., {Colberg}, J., {Springel}, V., {Hernquist}, L., \& {Sijacki},
  D. 2008, \apj, 676, 33, \dodoi{10.1086/524921}

\bibitem[{{Di Matteo} {et~al.}(2005){Di Matteo}, {Springel}, \&
  {Hernquist}}]{dimatteoetal05}
{Di Matteo}, T., {Springel}, V., \& {Hernquist}, L. 2005, \nat, 433, 604,
  \dodoi{10.1038/nature03335}

\bibitem[{{Dressler} \& {Richstone}(1988)}]{D&R88}
{Dressler}, A., \& {Richstone}, D.~O. 1988, \apj, 324, 701,
  \dodoi{10.1086/165930}

\bibitem[{{Dutta Chowdhury} \& {Chatterjee}(2017)}]{d&c17}
{Dutta Chowdhury}, D., \& {Chatterjee}, S. 2017, \apj, 839, 34,
  \dodoi{10.3847/1538-4357/aa64d6}

\bibitem[{{Eckert} {et~al.}(2021){Eckert}, {Gaspari}, {Gastaldello}, {Le Brun},
  \& {O'Sullivan}}]{eckertetal21}
{Eckert}, D., {Gaspari}, M., {Gastaldello}, F., {Le Brun}, A. M.~C., \&
  {O'Sullivan}, E. 2021, Universe, 7, 142, \dodoi{10.3390/universe7050142}

\bibitem[{{Fabian}(2012)}]{Fabian12}
{Fabian}, A.~C. 2012, Annual Review of Astronomy and Astrophysics, 50, 455,
  \dodoi{10.1146/annurev-astro-081811-125521}

\bibitem[{{Ferrarese} \& {Ford}(2005)}]{ferrareseetal05}
{Ferrarese}, L., \& {Ford}, H. 2005, \ssr, 116, 523,
  \dodoi{10.1007/s11214-005-3947-6}

\bibitem[{{Ferrarese} \& {Merritt}(2000)}]{F&M00}
{Ferrarese}, L., \& {Merritt}, D. 2000, \apjl, 539, L9, \dodoi{10.1086/312838}

\bibitem[{{Fiore} {et~al.}(2017){Fiore}, {Feruglio}, {Shankar}, {Bischetti},
  {Bongiorno}, {Brusa}, {Carniani}, {Cicone}, {Duras}, {Lamastra}, {Mainieri},
  {Marconi}, {Menci}, {Maiolino}, {Piconcelli}, {Vietri}, \&
  {Zappacosta}}]{Fiore17}
{Fiore}, F., {Feruglio}, C., {Shankar}, F., {et~al.} 2017, \aap, 601, A143,
  \dodoi{10.1051/0004-6361/201629478}

\bibitem[{{Gebhardt} {et~al.}(2000{\natexlab{a}}){Gebhardt}, {Bender}, {Bower},
  {Dressler}, {Faber}, {Filippenko}, {Green}, {Grillmair}, {Ho}, {Kormendy},
  {Lauer}, {Magorrian}, {Pinkney}, {Richstone}, \& {Tremaine}}]{Gebhardt00}
{Gebhardt}, K., {Bender}, R., {Bower}, G., {et~al.} 2000{\natexlab{a}}, \apjl,
  539, L13, \dodoi{10.1086/312840}

\bibitem[{{Gebhardt} {et~al.}(2000{\natexlab{b}}){Gebhardt}, {Bender}, {Bower},
  {Dressler}, {Faber}, {Filippenko}, {Green}, {Grillmair}, {Ho}, {Kormendy},
  {Lauer}, {Magorrian}, {Pinkney}, {Richstone}, \& {Tremaine}}]{gebhardtetal00}
---. 2000{\natexlab{b}}, \apjl, 539, L13, \dodoi{10.1086/312840}

\bibitem[{{Gitti} {et~al.}(2012){Gitti}, {Brighenti}, \&
  {McNamara}}]{gittietal12}
{Gitti}, M., {Brighenti}, F., \& {McNamara}, B.~R. 2012, Advances in Astronomy,
  2012, \dodoi{10.1155/2012/950641}

\bibitem[{{Graham} {et~al.}(2001){Graham}, {Erwin}, {Caon}, \&
  {Trujillo}}]{Graham01}
{Graham}, A.~W., {Erwin}, P., {Caon}, N., \& {Trujillo}, I. 2001, \apjl, 563,
  L11, \dodoi{10.1086/338500}

\bibitem[{{Hall} {et~al.}(2019){Hall}, {Zakamska}, {Addison}, {Battaglia},
  {Crichton}, {Devlin}, {Dunkley}, {Gralla}, {Hill}, {Hilton}, {Hubmayr},
  {Hughes}, {Huffenberger}, {Kosowsky}, {Marriage}, {Maurin}, {Moodley},
  {Niemack}, {Page}, {Partridge}, {D{\"u}nner Planella}, {Schillaci},
  {Sif{\'o}n}, {Staggs}, {Wollack}, \& {Xu}}]{halletal19}
{Hall}, K.~R., {Zakamska}, N.~L., {Addison}, G.~E., {et~al.} 2019, \mnras, 490,
  2315, \dodoi{10.1093/mnras/stz2751}

\bibitem[{{H{\"a}ring} \& {Rix}(2004{\natexlab{a}})}]{h&r04}
{H{\"a}ring}, N., \& {Rix}, H.-W. 2004{\natexlab{a}}, \apjl, 604, L89,
  \dodoi{10.1086/383567}

\bibitem[{{H{\"a}ring} \& {Rix}(2004{\natexlab{b}})}]{haringtal04}
---. 2004{\natexlab{b}}, \apjl, 604, L89, \dodoi{10.1086/383567}

\bibitem[{{Harrison}(2017)}]{Harrison17}
{Harrison}, C.~M. 2017, Nature Astronomy, 1, 0165,
  \dodoi{10.1038/s41550-017-0165}

\bibitem[{{Heckman} \& {Best}(2014)}]{heckman14}
{Heckman}, T.~M., \& {Best}, P.~N. 2014, \araa, 52, 589,
  \dodoi{10.1146/annurev-astro-081913-035722}

\bibitem[{{Hopkins} {et~al.}(2014){Hopkins}, {Kere{\v{s}}}, {O{\~n}orbe},
  {Faucher-Gigu{\`e}re}, {Quataert}, {Murray}, \& {Bullock}}]{hop14}
{Hopkins}, P.~F., {Kere{\v{s}}}, D., {O{\~n}orbe}, J., {et~al.} 2014, \mnras,
  445, 581, \dodoi{10.1093/mnras/stu1738}

\bibitem[{{Hopkins} \& {Quataert}(2011)}]{H&Q11}
{Hopkins}, P.~F., \& {Quataert}, E. 2011, \mnras, 415, 1027,
  \dodoi{10.1111/j.1365-2966.2011.18542.x}

\bibitem[{{Hopkins} {et~al.}(2006){Hopkins}, {Robertson}, {Krause},
  {Hernquist}, \& {Cox}}]{hopkinsetal06}
{Hopkins}, P.~F., {Robertson}, B., {Krause}, E., {Hernquist}, L., \& {Cox},
  T.~J. 2006, \apj, 652, 107, \dodoi{10.1086/508055}

\bibitem[{{Hopkins} {et~al.}(2018){Hopkins}, {Wetzel}, {Kere{\v{s}}},
  {Faucher-Gigu{\`e}re}, {Quataert}, {Boylan-Kolchin}, {Murray}, {Hayward},
  {Garrison-Kimmel}, {Hummels}, {Feldmann}, {Torrey}, {Ma},
  {Angl{\'e}s-Alc{\'a}zar}, {Su}, {Orr}, {Schmitz}, {Escala}, {Sanderson},
  {Grudi{\'c}}, {Hafen}, {Kim}, {Fitts}, {Bullock}, {Wheeler}, {Chan},
  {Elbert}, \& {Narayanan}}]{hop18}
{Hopkins}, P.~F., {Wetzel}, A., {Kere{\v{s}}}, D., {et~al.} 2018, \mnras, 480,
  800, \dodoi{10.1093/mnras/sty1690}

\bibitem[{{Hoyle} \& {Lyttleton}(1939)}]{hoyleetal39}
{Hoyle}, F., \& {Lyttleton}, R.~A. 1939, Proceedings of the Cambridge
  Philosophical Society, 35, 405, \dodoi{10.1017/S0305004100021150}

\bibitem[{{Jaeger}(2008)}]{jager08}
{Jaeger}, S. 2008, in Astronomical Society of the Pacific Conference Series,
  Vol. 394, Astronomical Data Analysis Software and Systems XVII, ed. R.~W.
  {Argyle}, P.~S. {Bunclark}, \& J.~R. {Lewis}, 623

\bibitem[{{Jiang} {et~al.}(2014){Jiang}, {Stone}, \& {Davis}}]{jiang14}
{Jiang}, Y.-F., {Stone}, J.~M., \& {Davis}, S.~W. 2014, \apj, 796, 106,
  \dodoi{10.1088/0004-637X/796/2/106}

\bibitem[{Kar~Chowdhury {et~al.}(2020)Kar~Chowdhury, Chatterjee, Lonappan,
  Khandai, \& Di~Matteo}]{RKC19}
Kar~Chowdhury, R., Chatterjee, S., Lonappan, A.~I., Khandai, N., \& Di~Matteo,
  T. 2020, The Astrophysical Journal, 889, 60, \dodoi{10.3847/1538-4357/ab5b96}

\bibitem[{{Kar Chowdhury} {et~al.}(2022){Kar Chowdhury}, {Chatterjee}, {Paul},
  {Sarazin}, \& {Dai}}]{RKC22}
{Kar Chowdhury}, R., {Chatterjee}, S., {Paul}, A., {Sarazin}, C.~L., \& {Dai},
  J.~L. 2022, \apj, 940, 47, \dodoi{10.3847/1538-4357/ac951c}

\bibitem[{{Kar Chowdhury} {et~al.}(2021){Kar Chowdhury}, {Roy}, {Chatterjee},
  {Khandai}, {Sarazin}, \& {Di Matteo}}]{RKC21}
{Kar Chowdhury}, R., {Roy}, S., {Chatterjee}, S., {et~al.} 2021, Astronomische
  Nachrichten, 342, 164, \dodoi{10.1002/asna.202113898}

\bibitem[{{Kauffmann} \& {Haehnelt}(2000)}]{k&h00}
{Kauffmann}, G., \& {Haehnelt}, M. 2000, \mnras, 311, 576,
  \dodoi{10.1046/j.1365-8711.2000.03077.x}

\bibitem[{Khandai {et~al.}(2015)Khandai, Di~Matteo, Croft, Wilkins, Feng,
  Tucker, DeGraf, \& Liu}]{khandaietal15}
Khandai, N., Di~Matteo, T., Croft, R., {et~al.} 2015, Monthly Notices of the
  Royal Astronomical Society, 450, 1349, \dodoi{10.1093/mnras/stv627}

\bibitem[{{Kim} {et~al.}(2022){Kim}, {Golwala}, {Bartlett}, {Amodeo},
  {Battaglia}, {Benson}, {Hill}, {Hopkins}, {Hummels}, {Moser}, \&
  {Orr}}]{junhanetal22}
{Kim}, J., {Golwala}, S., {Bartlett}, J.~G., {et~al.} 2022, \apj, 926, 179,
  \dodoi{10.3847/1538-4357/ac4750}

\bibitem[{{Kormendy}(1993)}]{Kormendy93}
{Kormendy}, J. 1993, in The Nearest Active Galaxies, ed. J.~{Beckman},
  L.~{Colina}, \& H.~{Netzer}, 197--218

\bibitem[{{Kormendy} \& {Ho}(2013)}]{K&H13}
{Kormendy}, J., \& {Ho}, L.~C. 2013, \araa, 51, 511,
  \dodoi{10.1146/annurev-astro-082708-101811}

\bibitem[{{Lacy} {et~al.}(2019){Lacy}, {Mason}, {Sarazin}, {Chatterjee},
  {Nyland}, {Kimball}, {Rocha}, {Rowe}, \& {Surace}}]{lacyetal19}
{Lacy}, M., {Mason}, B., {Sarazin}, C., {et~al.} 2019, \mnras, 483, L22,
  \dodoi{10.1093/mnrasl/sly215}

\bibitem[{{Lapi} {et~al.}(2003){Lapi}, {Cavaliere}, \& {De Zotti}}]{lapietal03}
{Lapi}, A., {Cavaliere}, A., \& {De Zotti}, G. 2003, \apjl, 597, L93,
  \dodoi{10.1086/380106}

\bibitem[{Liu {et~al.}(2016)Liu, Di Matteo, \& Feng}]{liuetal16}
Liu, M., Di Matteo, T., \& Feng, Y. 2016, Monthly Notices of the Royal
  Astronomical Society, 458, 1402, \dodoi{10.1093/mnras/stw342}

\bibitem[{{Magorrian} {et~al.}(1998){Magorrian}, {Tremaine}, {Richstone},
  {Bender}, {Bower}, {Dressler}, {Faber}, {Gebhardt}, {Green}, {Grillmair},
  {Kormendy}, \& {Lauer}}]{magorrianetal98}
{Magorrian}, J., {Tremaine}, S., {Richstone}, D., {et~al.} 1998, \aj, 115,
  2285, \dodoi{10.1086/300353}

\bibitem[{{Marsden} {et~al.}(2020){Marsden}, {Shankar}, {Ginolfi}, \&
  {Zubovas}}]{Marsden20}
{Marsden}, C., {Shankar}, F., {Ginolfi}, M., \& {Zubovas}, K. 2020, Frontiers
  in Physics, 8, 61, \dodoi{10.3389/fphy.2020.00061}

\bibitem[{{Mart{\'\i}nez-Aldama} {et~al.}(2018){Mart{\'\i}nez-Aldama}, {del
  Olmo}, {Marziani}, {Sulentic}, {Negrete}, {Dultzin}, {D'Onofrio}, \&
  {Perea}}]{mart18}
{Mart{\'\i}nez-Aldama}, M.~L., {del Olmo}, A., {Marziani}, P., {et~al.} 2018,
  \aap, 618, A179, \dodoi{10.1051/0004-6361/201833541}

\bibitem[{{Maughan} {et~al.}(2012){Maughan}, {Giles}, {Randall}, {Jones}, \&
  {Forman}}]{Maughan12}
{Maughan}, B.~J., {Giles}, P.~A., {Randall}, S.~W., {Jones}, C., \& {Forman},
  W.~R. 2012, \mnras, 421, 1583, \dodoi{10.1111/j.1365-2966.2012.20419.x}

\bibitem[{{Merritt} \& {Ferrarese}(2001)}]{m&f01}
{Merritt}, D., \& {Ferrarese}, L. 2001, \apj, 547, 140, \dodoi{10.1086/318372}

\bibitem[{{Molham} {et~al.}(2020){Molham}, {Clerc}, {Takey}, {Sadibekova},
  {Morcos}, {Yousef}, {Hayman}, {Lieu}, {Raychaudhury}, \&
  {Gaynullina}}]{Molham20}
{Molham}, M., {Clerc}, N., {Takey}, A., {et~al.} 2020, \mnras, 494, 161,
  \dodoi{10.1093/mnras/staa677}

\bibitem[{{Mroczkowski} {et~al.}(2019){Mroczkowski}, {Nagai}, {Basu}, {Chluba},
  {Sayers}, {Adam}, {Churazov}, {Crites}, {Di Mascolo}, {Eckert},
  {Macias-Perez}, {Mayet}, {Perotto}, {Pointecouteau}, {Romero}, {Ruppin},
  {Scannapieco}, \& {ZuHone}}]{mroczkowskietal19}
{Mroczkowski}, T., {Nagai}, D., {Basu}, K., {et~al.} 2019, \ssr, 215, 17,
  \dodoi{10.1007/s11214-019-0581-2}

\bibitem[{{Muratov} {et~al.}(2015){Muratov}, {Kere{\v{s}}},
  {Faucher-Gigu{\`e}re}, {Hopkins}, {Quataert}, \& {Murray}}]{mur15}
{Muratov}, A.~L., {Kere{\v{s}}}, D., {Faucher-Gigu{\`e}re}, C.-A., {et~al.}
  2015, \mnras, 454, 2691, \dodoi{10.1093/mnras/stv2126}

\bibitem[{{Mutlu-Pakdil} {et~al.}(2018){Mutlu-Pakdil}, {Seigar}, {Hewitt},
  {Treuthardt}, {Berrier}, \& {Koval}}]{MP18}
{Mutlu-Pakdil}, B., {Seigar}, M.~S., {Hewitt}, I.~B., {et~al.} 2018, \mnras,
  474, 2594, \dodoi{10.1093/mnras/stx2935}

\bibitem[{{Natarajan} \& {Sigurdsson}(1999)}]{n&s99}
{Natarajan}, P., \& {Sigurdsson}, S. 1999, \mnras, 302, 288,
  \dodoi{10.1046/j.1365-8711.1999.02116.x}

\bibitem[{{Perna} {et~al.}(2017){Perna}, {Lanzuisi}, {Brusa}, {Mignoli}, \&
  {Cresci}}]{Perna17}
{Perna}, M., {Lanzuisi}, G., {Brusa}, M., {Mignoli}, M., \& {Cresci}, G. 2017,
  \aap, 603, A99, \dodoi{10.1051/0004-6361/201630369}

\bibitem[{{Peterson} {et~al.}(2003){Peterson}, {Kahn}, {Paerels}, {Kaastra},
  {Tamura}, {Bleeker}, {Ferrigno}, \& {Jernigan}}]{Peterson03}
{Peterson}, J.~R., {Kahn}, S.~M., {Paerels}, F.~B.~S., {et~al.} 2003, \apj,
  590, 207, \dodoi{10.1086/374830}

\bibitem[{{Planck Collaboration} {et~al.}(2016){Planck Collaboration}, {Ade},
  {Aghanim}, {Arnaud}, {Ashdown}, {Aumont}, {Baccigalupi}, {Banday},
  {Barreiro}, {Bartlett}, {Bartolo}, {Battaner}, {Battye}, {Benabed},
  {Beno{\^\i}t}, {Benoit-L{\'e}vy}, {Bernard}, {Bersanelli}, {Bielewicz},
  {Bock}, {Bonaldi}, {Bonavera}, {Bond}, {Borrill}, {Bouchet}, {Boulanger},
  {Bucher}, {Burigana}, {Butler}, {Calabrese}, {Cardoso}, {Catalano},
  {Challinor}, {Chamballu}, {Chary}, {Chiang}, {Chluba}, {Christensen},
  {Church}, {Clements}, {Colombi}, {Colombo}, {Combet}, {Coulais}, {Crill},
  {Curto}, {Cuttaia}, {Danese}, {Davies}, {Davis}, {de Bernardis}, {de Rosa},
  {de Zotti}, {Delabrouille}, {D{\'e}sert}, {Di Valentino}, {Dickinson},
  {Diego}, {Dolag}, {Dole}, {Donzelli}, {Dor{\'e}}, {Douspis}, {Ducout},
  {Dunkley}, {Dupac}, {Efstathiou}, {Elsner}, {En{\ss}lin}, {Eriksen},
  {Farhang}, {Fergusson}, {Finelli}, {Forni}, {Frailis}, {Fraisse},
  {Franceschi}, {Frejsel}, {Galeotta}, {Galli}, {Ganga}, {Gauthier}, {Gerbino},
  {Ghosh}, {Giard}, {Giraud-H{\'e}raud}, {Giusarma}, {Gjerl{\o}w},
  {Gonz{\'a}lez-Nuevo}, {G{\'o}rski}, {Gratton}, {Gregorio}, {Gruppuso},
  {Gudmundsson}, {Hamann}, {Hansen}, {Hanson}, {Harrison}, {Helou},
  {Henrot-Versill{\'e}}, {Hern{\'a}ndez-Monteagudo}, {Herranz}, {Hildebrandt},
  {Hivon}, {Hobson}, {Holmes}, {Hornstrup}, {Hovest}, {Huang}, {Huffenberger},
  {Hurier}, {Jaffe}, {Jaffe}, {Jones}, {Juvela}, {Keih{\"a}nen}, {Keskitalo},
  {Kisner}, {Kneissl}, {Knoche}, {Knox}, {Kunz}, {Kurki-Suonio}, {Lagache},
  {L{\"a}hteenm{\"a}ki}, {Lamarre}, {Lasenby}, {Lattanzi}, {Lawrence}, {Leahy},
  {Leonardi}, {Lesgourgues}, {Levrier}, {Lewis}, {Liguori}, {Lilje},
  {Linden-V{\o}rnle}, {L{\'o}pez-Caniego}, {Lubin}, {Mac{\'\i}as-P{\'e}rez},
  {Maggio}, {Maino}, {Mandolesi}, {Mangilli}, {Marchini}, {Maris}, {Martin},
  {Martinelli}, {Mart{\'\i}nez-Gonz{\'a}lez}, {Masi}, {Matarrese}, {McGehee},
  {Meinhold}, {Melchiorri}, {Melin}, {Mendes}, {Mennella}, {Migliaccio},
  {Millea}, {Mitra}, {Miville-Desch{\^e}nes}, {Moneti}, {Montier}, {Morgante},
  {Mortlock}, {Moss}, {Munshi}, {Murphy}, {Naselsky}, {Nati}, {Natoli},
  {Netterfield}, {N{\o}rgaard-Nielsen}, {Noviello}, {Novikov}, {Novikov},
  {Oxborrow}, {Paci}, {Pagano}, {Pajot}, {Paladini}, {Paoletti}, {Partridge},
  {Pasian}, {Patanchon}, {Pearson}, {Perdereau}, {Perotto}, {Perrotta},
  {Pettorino}, {Piacentini}, {Piat}, {Pierpaoli}, {Pietrobon}, {Plaszczynski},
  {Pointecouteau}, {Polenta}, {Popa}, {Pratt}, {Pr{\'e}zeau}, {Prunet},
  {Puget}, {Rachen}, {Reach}, {Rebolo}, {Reinecke}, {Remazeilles}, {Renault},
  {Renzi}, {Ristorcelli}, {Rocha}, {Rosset}, {Rossetti}, {Roudier},
  {Rouill{\'e} d'Orfeuil}, {Rowan-Robinson}, {Rubi{\~n}o-Mart{\'\i}n},
  {Rusholme}, {Said}, {Salvatelli}, {Salvati}, {Sandri}, {Santos},
  {Savelainen}, {Savini}, {Scott}, {Seiffert}, {Serra}, {Shellard}, {Spencer},
  {Spinelli}, {Stolyarov}, {Stompor}, {Sudiwala}, {Sunyaev}, {Sutton},
  {Suur-Uski}, {Sygnet}, {Tauber}, {Terenzi}, {Toffolatti}, {Tomasi},
  {Tristram}, {Trombetti}, {Tucci}, {Tuovinen}, {T{\"u}rler}, {Umana},
  {Valenziano}, {Valiviita}, {Van Tent}, {Vielva}, {Villa}, {Wade}, {Wandelt},
  {Wehus}, {White}, {White}, {Wilkinson}, {Yvon}, {Zacchei}, \&
  {Zonca}}]{planck16}
{Planck Collaboration}, {Ade}, P.~A.~R., {Aghanim}, N., {et~al.} 2016, \aap,
  594, A13, \dodoi{10.1051/0004-6361/201525830}

\bibitem[{{Platania} {et~al.}(2002){Platania}, {Burigana}, {De Zotti},
  {Lazzaro}, \& {Bersanelli}}]{plataniaetal02}
{Platania}, P., {Burigana}, C., {De Zotti}, G., {Lazzaro}, E., \& {Bersanelli},
  M. 2002, \mnras, 337, 242, \dodoi{10.1046/j.1365-8711.2002.05907.x}

\bibitem[{{Puchwein} {et~al.}(2010){Puchwein}, {Springel}, {Sijacki}, \&
  {Dolag}}]{puchweinetal10}
{Puchwein}, E., {Springel}, V., {Sijacki}, D., \& {Dolag}, K. 2010, \mnras,
  406, 936, \dodoi{10.1111/j.1365-2966.2010.16786.x}

\bibitem[{{Richstone} {et~al.}(1998){Richstone}, {Ajhar}, {Bender}, {Bower},
  {Dressler}, {Faber}, {Filippenko}, {Gebhardt}, {Green}, {Ho}, {Kormendy},
  {Lauer}, {Magorrian}, \& {Tremaine}}]{richstoneetal98}
{Richstone}, D., {Ajhar}, E.~A., {Bender}, R., {et~al.} 1998, \nat, 395, A14

\bibitem[{{Robson} \& {Dav{\'e}}(2023)}]{Robson23}
{Robson}, D., \& {Dav{\'e}}, R. 2023, \mnras, 518, 5826,
  \dodoi{10.1093/mnras/stac2982}

\bibitem[{Robson \& Davé(2020)}]{Robson20}
Robson, D., \& Davé, R. 2020, Monthly Notices of the Royal Astronomical
  Society, 498, 3061–3076, \dodoi{10.1093/mnras/staa2394}

\bibitem[{{Roy} {et~al.}(2021{\natexlab{a}}){Roy}, {Bundy}, {Nevin},
  {Belfiore}, {Yan}, {Campbell}, {Riffel}, {Riffel}, {Bershady}, {Westfall},
  {Drory}, \& {Zhang}}]{Roy21a}
{Roy}, N., {Bundy}, K., {Nevin}, R., {et~al.} 2021{\natexlab{a}}, \apj, 913,
  33, \dodoi{10.3847/1538-4357/abf1e6}

\bibitem[{{Roy} {et~al.}(2021{\natexlab{b}}){Roy}, {Moravec}, {Bundy},
  {Hardcastle}, {G{\"u}rkan}, {Diego Baldi}, {Leslie}, {Masters}, {Gelfand},
  {Riffel}, {Riffel}, {Mingo Fernandez}, \& {Drabent}}]{Roy21c}
{Roy}, N., {Moravec}, E., {Bundy}, K., {et~al.} 2021{\natexlab{b}}, \apj, 922,
  230, \dodoi{10.3847/1538-4357/ac24a0}

\bibitem[{{Roychowdhury}(2007)}]{SRC07}
{Roychowdhury}, S. 2007, in Groups of Galaxies in the Nearby Universe, ed.
  I.~{Saviane}, V.~D. {Ivanov}, \& J.~{Borissova}, 337,
  \dodoi{10.1007/978-3-540-71173-5_56}

\bibitem[{{Roychowdhury} {et~al.}(2005){Roychowdhury}, {Ruszkowski}, \&
  {Nath}}]{roychowdhuryetal05}
{Roychowdhury}, S., {Ruszkowski}, M., \& {Nath}, B.~B. 2005, \apj, 634, 90,
  \dodoi{10.1086/496910}

\bibitem[{{Ruan} {et~al.}(2015){Ruan}, {McQuinn}, \& {Anderson}}]{ruanetal15}
{Ruan}, J.~J., {McQuinn}, M., \& {Anderson}, S.~F. 2015, \apj, 802, 135,
  \dodoi{10.1088/0004-637X/802/2/135}

\bibitem[{Salviander {et~al.}(2015)Salviander, Shields, \&
  Bonning}]{Salviander15}
Salviander, S., Shields, G.~A., \& Bonning, E.~W. 2015, The Astrophysical
  Journal, 799, 173, \dodoi{10.1088/0004-637x/799/2/173}

\bibitem[{{Sazonov} \& {Sunyaev}(1998)}]{sazanovetal98}
{Sazonov}, S.~Y., \& {Sunyaev}, R.~A. 1998, Astronomy Letters, 24, 553

\bibitem[{{Scannapieco} {et~al.}(2008){Scannapieco}, {Thacker}, \&
  {Couchman}}]{scannapiecoetal08}
{Scannapieco}, E., {Thacker}, R.~J., \& {Couchman}, H.~M.~P. 2008, \apj, 678,
  674, \dodoi{10.1086/528948}

\bibitem[{Schellenberger {et~al.}(2017)Schellenberger, Vrtilek, David,
  O'Sullivan, Giacintucci, Johnston-Hollitt, Duchesne, \& Raychaudhury}]{SRC17}
Schellenberger, G., Vrtilek, J.~M., David, L., {et~al.} 2017, The Astrophysical
  Journal, 845, 84, \dodoi{10.3847/1538-4357/aa7f2e}

\bibitem[{{Schutte} {et~al.}(2019){Schutte}, {Reines}, \& {Greene}}]{Schutte19}
{Schutte}, Z., {Reines}, A.~E., \& {Greene}, J.~E. 2019, \apj, 887, 245,
  \dodoi{10.3847/1538-4357/ab35dd}

\bibitem[{{Sijacki} {et~al.}(2007){Sijacki}, {Springel}, {Di Matteo}, \&
  {Hernquist}}]{sijackietal07}
{Sijacki}, D., {Springel}, V., {Di Matteo}, T., \& {Hernquist}, L. 2007,
  \mnras, 380, 877, \dodoi{10.1111/j.1365-2966.2007.12153.x}

\bibitem[{Sijacki {et~al.}(2015)Sijacki, Vogelsberger, Genel, Springel, Torrey,
  Snyder, Nelson, \& Hernquist}]{sijackietal15}
Sijacki, D., Vogelsberger, M., Genel, S., {et~al.} 2015, Monthly Notices of the
  Royal Astronomical Society, 452, 575, \dodoi{10.1093/mnras/stv1340}

\bibitem[{{Silk} \& {Rees}(1998)}]{s&r98}
{Silk}, J., \& {Rees}, M.~J. 1998, \aap, 331, L1

\bibitem[{{Spacek} {et~al.}(2016){Spacek}, {Scannapieco}, {Cohen}, {Joshi}, \&
  {Mauskopf}}]{spaceketal16}
{Spacek}, A., {Scannapieco}, E., {Cohen}, S., {Joshi}, B., \& {Mauskopf}, P.
  2016, \apj, 819, 128, \dodoi{10.3847/0004-637X/819/2/128}

\bibitem[{{Sunyaev} \& {Zeldovich}(1972)}]{s&z72}
{Sunyaev}, R.~A., \& {Zeldovich}, Y.~B. 1972, Comments on Astrophysics and
  Space Physics, 4, 173

\bibitem[{{Tremaine} {et~al.}(2002){Tremaine}, {Gebhardt}, {Bender}, {Bower},
  {Dressler}, {Faber}, {Filippenko}, {Green}, {Grillmair}, {Ho}, {Kormendy},
  {Lauer}, {Magorrian}, {Pinkney}, \& {Richstone}}]{tremaineetal02}
{Tremaine}, S., {Gebhardt}, K., {Bender}, R., {et~al.} 2002, \apj, 574, 740,
  \dodoi{10.1086/341002}

\bibitem[{{Verdier} {et~al.}(2016){Verdier}, {Melin}, {Bartlett}, {Magneville},
  {Palanque-Delabrouille}, \& {Y{\`e}che}}]{verdieretal16}
{Verdier}, L., {Melin}, J.-B., {Bartlett}, J.~G., {et~al.} 2016, \aap, 588,
  A61, \dodoi{10.1051/0004-6361/201527431}

\bibitem[{{Vitale} {et~al.}(2013){Vitale}, {Mignoli}, {Cimatti}, {Lilly},
  {Carollo}, {Contini}, {Kneib}, {Le Fevre}, {Mainieri}, {Renzini},
  {Scodeggio}, {Zamorani}, {Bardelli}, {Barnes}, {Bolzonella}, {Bongiorno},
  {Bordoloi}, {Bschorr}, {Cappi}, {Caputi}, {Coppa}, {Cucciati}, {de la Torre},
  {de Ravel}, {Franzetti}, {Garilli}, {Iovino}, {Kampczyk}, {Knobel},
  {Koekemoer}, {Kova{\v{c}}}, {Lamareille}, {Le Borgne}, {Le Brun},
  {L{\'o}pez-Sanjuan}, {Maier}, {McCracken}, {Moresco}, {Nair}, {Oesch},
  {Pello}, {Peng}, {P{\'e}rez Montero}, {Pozzetti}, {Presotto}, {Silverman},
  {Tanaka}, {Tasca}, {Tresse}, {Vergani}, {Welikala}, \& {Zucca}}]{Vitale13}
{Vitale}, M., {Mignoli}, M., {Cimatti}, A., {et~al.} 2013, \aap, 556, A11,
  \dodoi{10.1051/0004-6361/201220258}

\bibitem[{{Vogelsberger} {et~al.}(2014){Vogelsberger}, {Genel}, {Springel},
  {Torrey}, {Sijacki}, {Xu}, {Snyder}, {Bird}, {Nelson}, \&
  {Hernquist}}]{vogelsbergeretal14}
{Vogelsberger}, M., {Genel}, S., {Springel}, V., {et~al.} 2014, \nat, 509, 177,
  \dodoi{10.1038/nature13316}

\bibitem[{{Wadekar} {et~al.}(2023){Wadekar}, {Thiele}, {Hill}, {Pandey},
  {Villaescusa-Navarro}, {Spergel}, {Cranmer}, {Nagai},
  {Angl{\'e}s-Alc{\'a}zar}, {Ho}, \& {Hernquist}}]{wadekaretal23}
{Wadekar}, D., {Thiele}, L., {Hill}, J.~C., {et~al.} 2023, \mnras, 522, 2628,
  \dodoi{10.1093/mnras/stad1128}

\bibitem[{{Yamada} {et~al.}(1999){Yamada}, {Sugiyama}, \&
  {Silk}}]{yamadaetal99}
{Yamada}, M., {Sugiyama}, N., \& {Silk}, J. 1999, \apj, 522, 66,
  \dodoi{10.1086/307604}

\bibitem[{{Yang} {et~al.}(2022){Yang}, {Cai}, {Cui}, {Dav{\'e}}, {Peacock}, \&
  {Sorini}}]{yang22}
{Yang}, T., {Cai}, Y.-C., {Cui}, W., {et~al.} 2022, arXiv e-prints,
  arXiv:2202.11430.
\newblock \doarXiv{2202.11430}

\bibitem[{{Zanni} {et~al.}(2005){Zanni}, {Murante}, {Bodo}, {Massaglia},
  {Rossi}, \& {Ferrari}}]{zannietal05}
{Zanni}, C., {Murante}, G., {Bodo}, G., {et~al.} 2005, \aap, 429, 399,
  \dodoi{10.1051/0004-6361:20041291}

\end{thebibliography}
\bibliographystyle{aasjournal}

\appendix
\section{The Process of Mock ALMA Observational Map using `simalma'}
\begin{figure*}
    \begin{center}
        \begin{tabular}{c}
        \includegraphics[width=7.5cm]{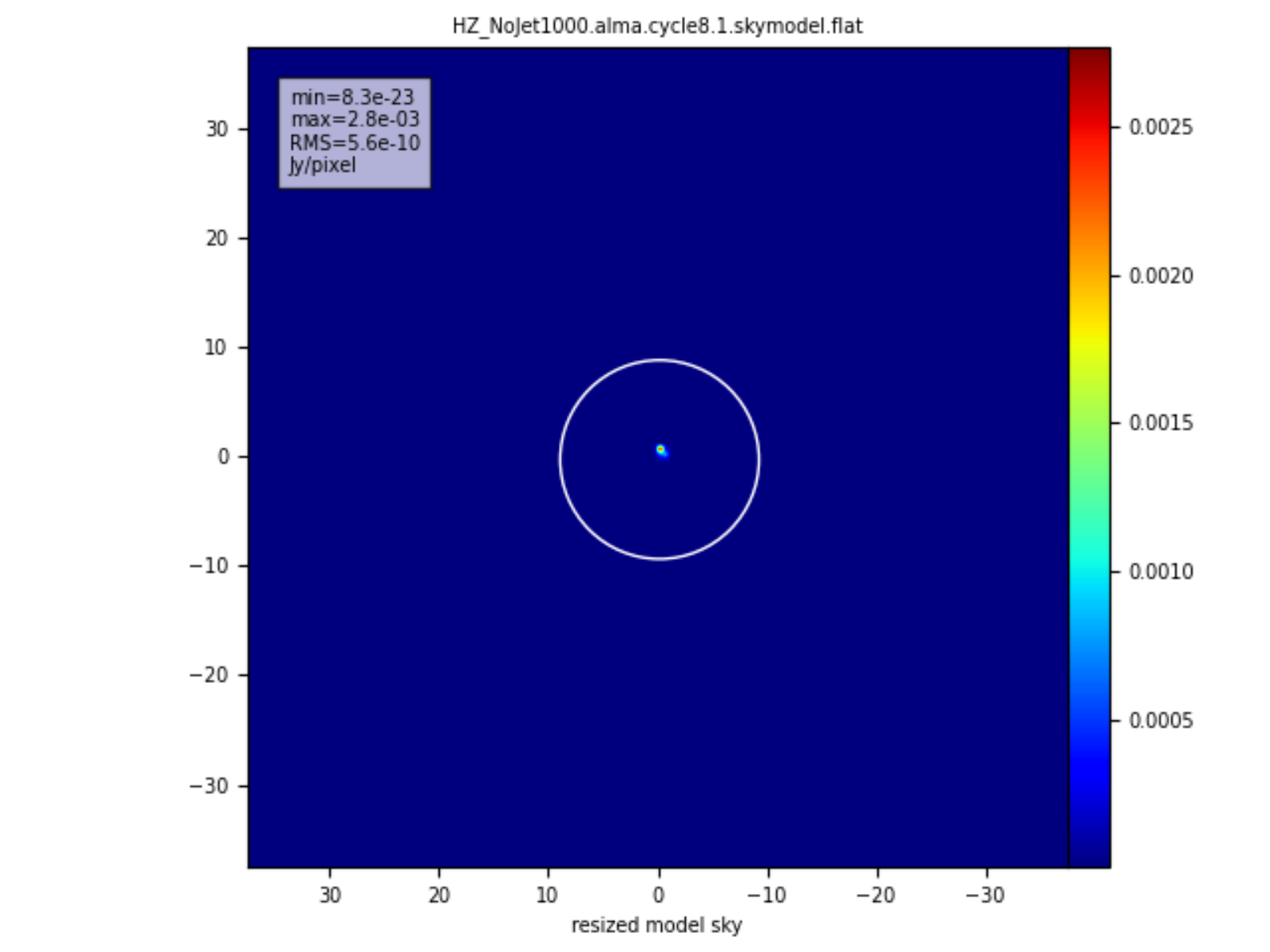}
        \includegraphics[width=7.5cm]{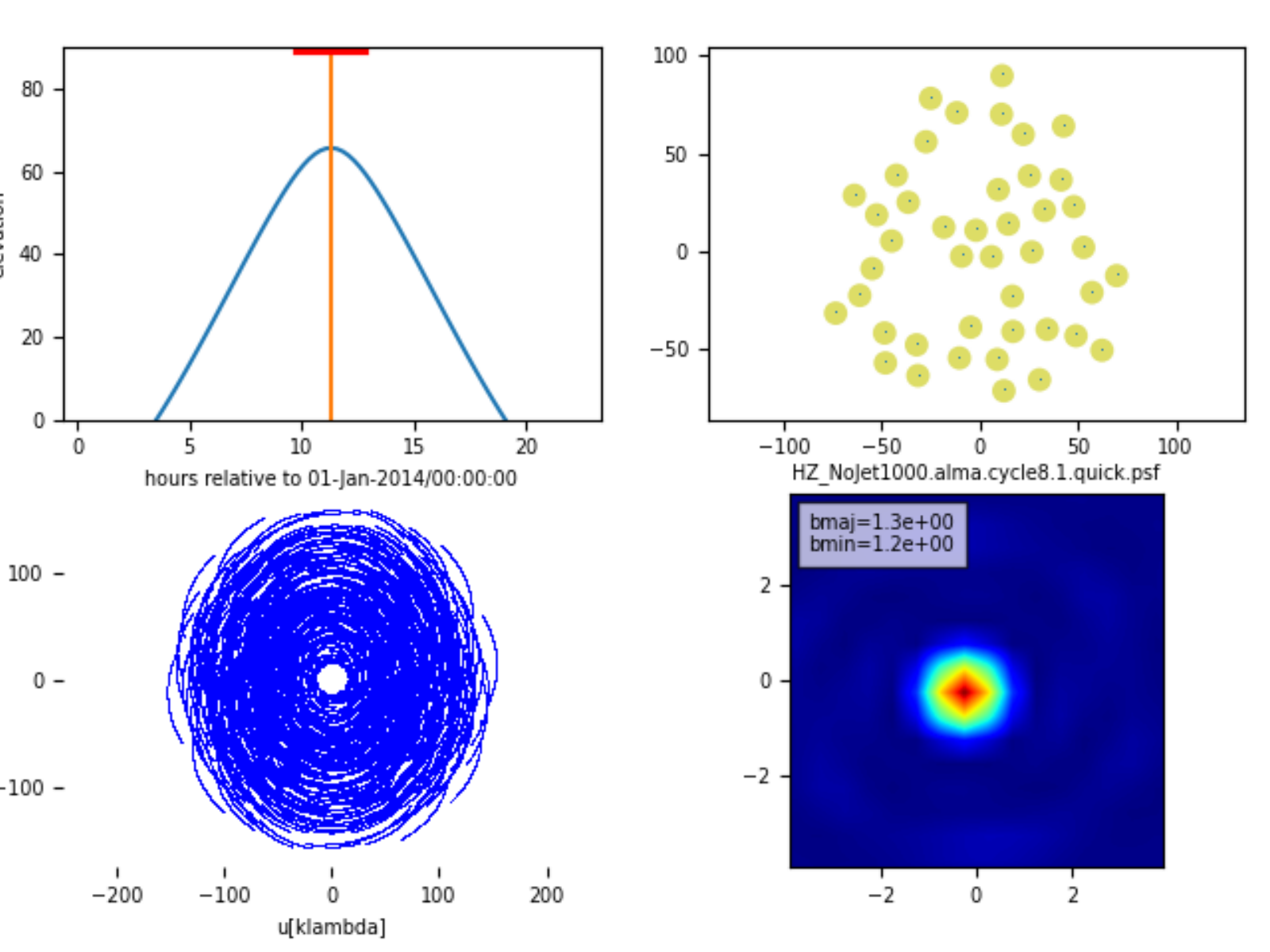}\\
      \includegraphics[width=8cm]{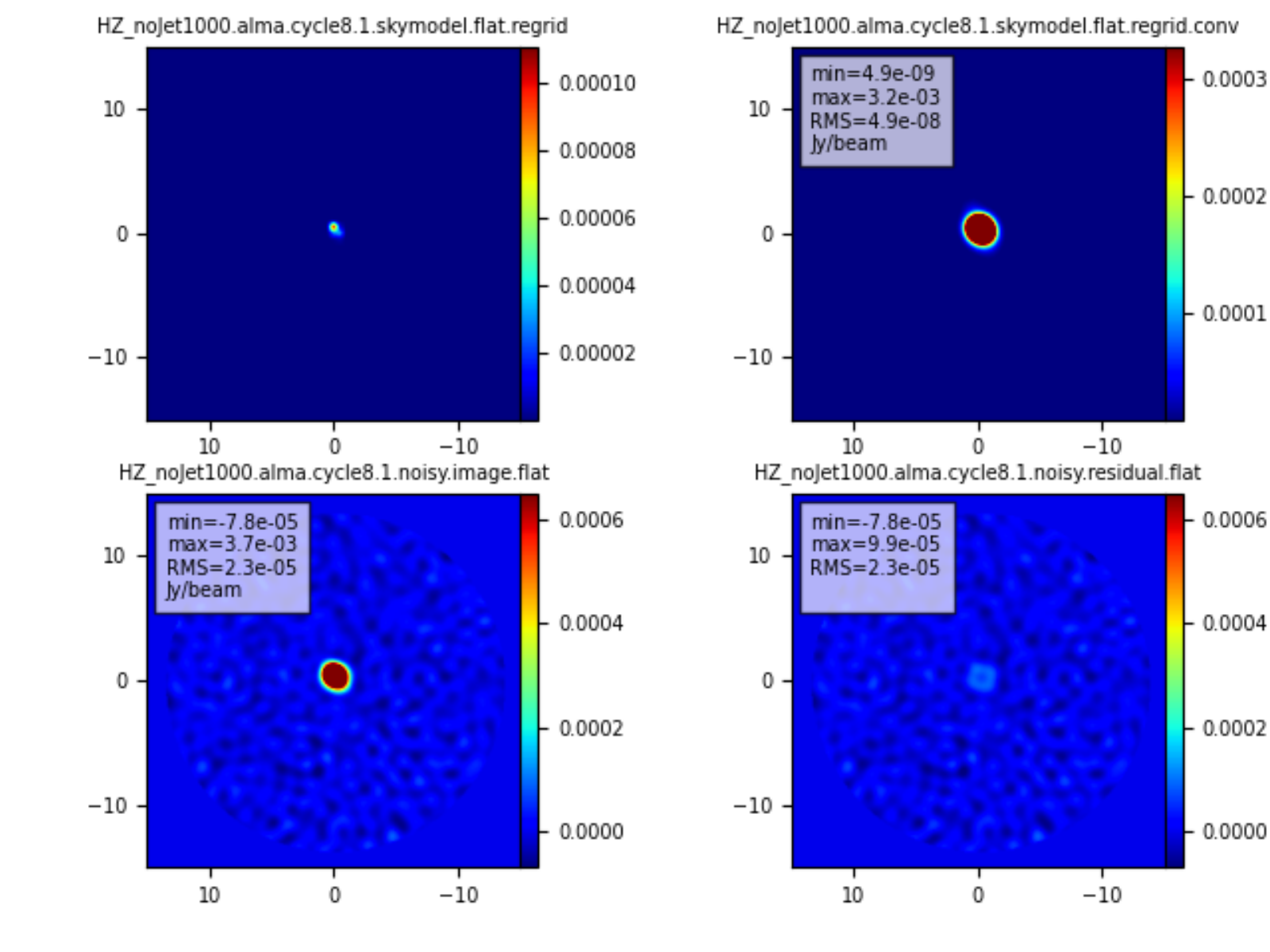}
      \includegraphics[width=8cm]{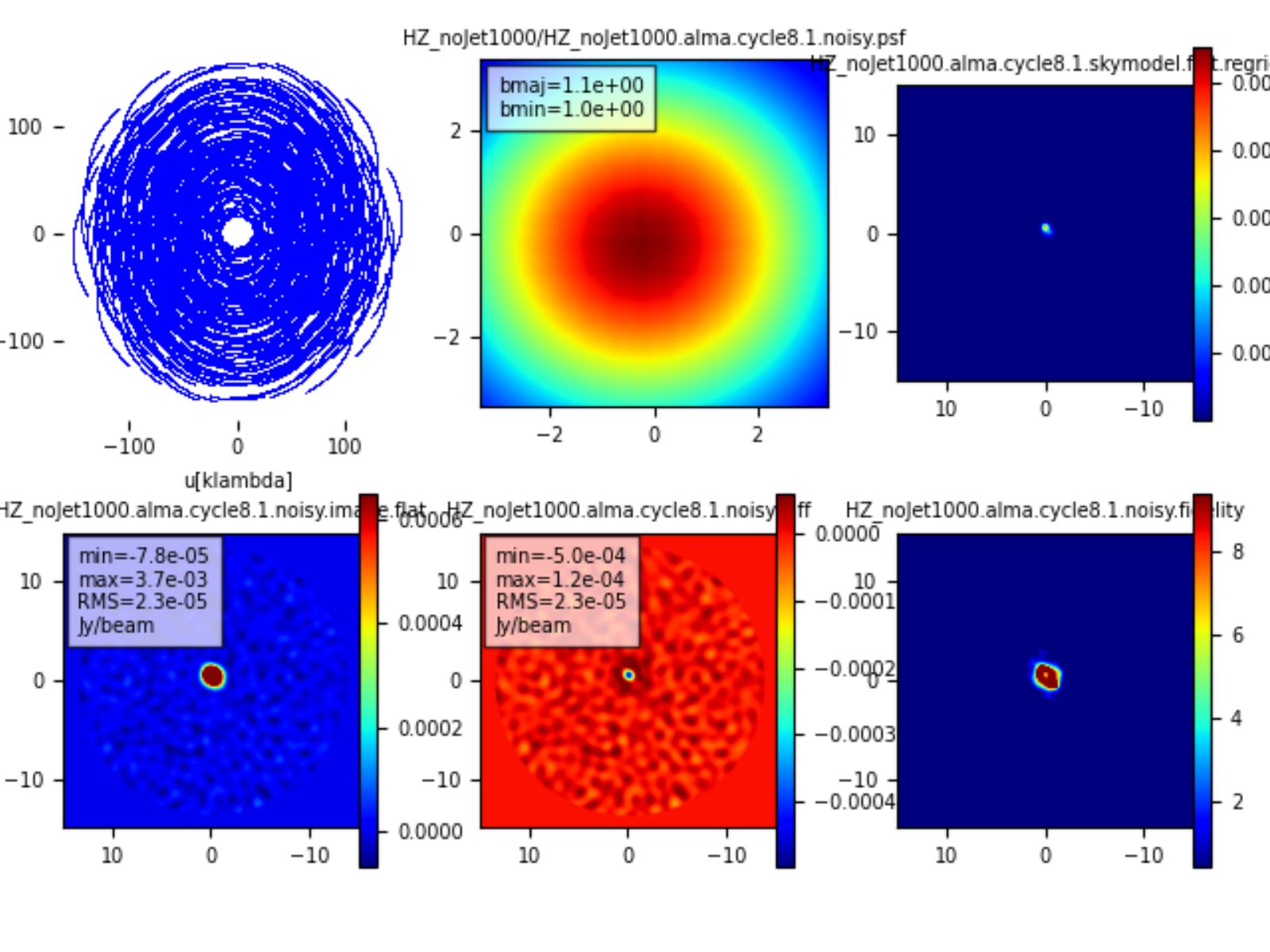}\\
      \end{tabular}
        \caption{An example of `simalma' outputs for one of the BHs. {\bf Upper Left Panel} Diagnostic figure of sky model with pointings. {\bf Upper Right Panel} Diagnostic figure of uv coverage and visibilities. {\bf Lower Left Panel} Diagnostic figure of clean image and residual. {\bf Lower Right Panel} Diagnostic figure of difference and fidelity.}
        \label{fig:10}
    \end{center}
\end{figure*}

We use the `simalma' task of CASA to make mock ALMA observational maps using the observing parameters shown in Table 3. Figure \ref{fig:10} shows an example of outputs generated from `simalma' for one of the black holes. The upper figures are the outputs from `simobserve' task where upper left figure shows the 4D input sky model image which is the input image at the original resolution and the upper right figure shows `Hour vs Elevation' plot, antenna positions, uv coverage and the primary beam respectively. Lower figures are the outputs from `simanalyze' task. The lower left figure shows four different images which are the sky model image, the image convolved with the output clean beam or primary beam where the input sky is re-gridded to match the output image, the synthesized image which is our mock ALMA image and the residual image after cleaning. Lastly, lower right figure is a more detailed output of `simanalyze' which consists of six images such as the uv coverage, the synthesized (unclean) beam calculated from weighted uv distribution, the sky model image, the synthesized image, the difference image between flattened convolved model and flattened output and the fidelity image which we considered as the signal to noise ratio plot.

\section{Description of Different Feedback Modes}

SIMBA models both the radiative and kinetic feedback from supermassive black holes (SMBH) in the simulation and also includes the X-ray mode of feedback. Although the models are sophisticated, compared to some of the previous cosmological simulations, they are still subgrid models and we require extrapolation of gas properties while calculating the large scale impact of feedback on host galaxy properties. We briefly describe the different modes of feedback here. We refer the reader to \citet{dave19} and \citet{RKC22} for more details on the simulation. 

\subsection{Modeling Accretion}
For modeling gas accretion onto the black hole, SIMBA uses a combination of the spherically symmetric Bondi model \citep{BondiHoyle, Bondi52}, as well as angular momentum transport through torque limited accretion \citep{H&Q11, Angles-Alcazar13, Angles-Alcazar15, Angles-Alcazar17}. The torque-limited accretion model is applied to the cold gas $\rm (T < 10^{5} K)$, while for the hotter gas the spherically symmetric Bondi model is used \citep{RKC22}. The accretion rate is given as  \begin{eqnarray}
\dot{M}_{Bondi} &=& \epsilon_{m}\frac{4 \pi G^2 M_{BH}^2 \rho}{(v^{2} + c_{s}^2)^{3/2}}, \\ \nonumber
\dot{M}_{Torque} &=& \epsilon_{T} f^{5/2}_d \left(\frac{M_{BH}}{10^8M_{\odot}}\right)^{1/6} \left(\frac{M_{enc}(R_{0})}{10^9M_{\odot}}\right)\left(\frac{R_0}{100 \rm {pc}}\right)^{-3/2}\left(1+ \frac{f_0}{f_{gas}}\right)^{-1} M_{\odot} yr^{-1}, \\ \nonumber
\dot{M}_{BH} &=& (1-\eta)\times(\dot{M}_{Torque} + \dot{M}_{Bondi})
\end{eqnarray}
The reader is referred to \citep{dave19} for the details of the models and the meanings of the terms. The key difference involves the inclusion of the torque which allows bipolar feedback. While the geometry of jets or winds are crucial in the determination of feedback from these systems, we note that they operate below the resolution threshold of the simulation and hence the large scale smoothed density, velocity and temperature fields appear more isotropic. 

\subsection{Modeling Feedback Modes}
As discussed before the three modes of feedback used in the simulation, incorporates {\it radiative wind mode}, {\it jet mode} and {\it X-ray feedback from the accretion disc/corona of the AGN}. For high Eddington ratio systems ($\geq 0.02$), the radiative mode of feedback is operative in the simulation. The outflow velocity is modeled in terms of black hole mass with the following form
\begin{equation}
\rm v_{wind} = 500 + 500 (\log M_{BH} - 6)/3 ~~km~s^{-1}
\end{equation}
$\rm v_{wind}$ is the wind velocity which drives the radiative feedback. The modeling is motivated by the observations of \citet{Perna17}.

For systems with lower Eddington ratios, the jet mode of feedback sets in. The jet speed is modeled as  
\begin{equation}
\rm v_{jet} = v_{wind} + 7000 \log(0.2/f_{Edd}) ~~km~s^{-1}
\end{equation}
Hence we note that for low Eddington systems, the jet feedback operates in addition to the wind mode. The scale of the jet velocities are motivated by observational studies \citep[e.g.,][]{Fabian12}. 

In addition to radiative winds and jets, feedback due to the photon pressure of the X-ray radiated off the accretion disk of the AGN \citep{choi12} is also incorporated in the simulation. Volume heating rate from the X-ray photons are calculated using the models of \cite{choi12}. The energy is distributed to the gas particles following a kernel and hence gas particles that are far away from the central black hole are hardly affected by the X-ray feedback from the accretion disc. Similar to the jet feedback the X-ray feedback is applied to the low Eddington rate systems. 


\end{document}